\documentclass[aps,prd,showpacs,eqsecnum,twocolumn,superscriptaddress]{revtex4-1}
\usepackage{amsmath,amssymb,graphicx,color,ulem,multirow, inputenc}
\usepackage{hyperref,ulem,url}
\hypersetup{colorlinks=true}

\begin{document}

\title{Implementation of advanced Riemann solvers in a neutrino-radiation magnetohydrodynamics code in numerical relativity and its application to a binary neutron star merger}

\author{Kenta Kiuchi}
\affiliation{Max Planck Institute for Gravitational Physics (Albert Einstein Institute), Am M\"{u}hlenberg, Potsdam-Golm, 14476, Germany}
\affiliation{Center for Gravitational Physics and Quantum Information, Yukawa Institute for Theoretical Physics, Kyoto University, Kyoto 606-8502, Japan}

\author{Loren E. Held}
\affiliation{Max Planck Institute for Gravitational Physics (Albert Einstein Institute), Am M\"{u}hlenberg, Potsdam-Golm, 14476, Germany}


\author{Yuichiro Sekiguchi}
\affiliation{Department of Physics, Toho University, Funabashi, Chiba 274-8510, Japan}
\affiliation{Center for Gravitational Physics and Quantum Information, Yukawa Institute for Theoretical Physics, Kyoto University, Kyoto 606-8502, Japan}

\author{Masaru Shibata}
\affiliation{Max Planck Institute for Gravitational Physics (Albert Einstein Institute), Am M\"{u}hlenberg, Potsdam-Golm, 14476, Germany}
\affiliation{Center for Gravitational Physics and Quantum Information, Yukawa Institute for Theoretical Physics, Kyoto University, Kyoto 606-8502, Japan}

\date{\today}

\begin{abstract}
We implement advanced Riemann solvers HLLC and HLLD~\cite{Mignone:2005ft,MUB:2009} together with an advanced constrained transport scheme~\cite{Gardiner:2007nc} in a numerical-relativity neutrino-radiation magnetohydrodynamics code. We validate our implementation by performing a series of one- and multi-dimensional test problems for relativistic hydrodynamics and magnetohydrodynamics in both Minkowski spacetime and a static black hole spacetime. 
We find that the numerical solutions with the advanced Riemann solvers are more accurate than those with the HLLE solver~\cite{DelZanna:2002rv}, which was originally implemented in our code. As an application to numerical relativity, we simulate an asymmetric binary neutron star merger leading to a short-lived massive neutron star both with and without magnetic fields. We find that the lifetime of the rotating massive neutron star formed after the merger and also the amount of the tidally-driven dynamical ejecta are overestimated when we employ the diffusive HLLE solver. 
We also find that the magnetorotational instability is less resolved when we employ the HLLE solver because of the solver's large numerical diffusivity. This causes a spurious enhancement both of magnetic winding resulting from large scale poloidal magnetic fields, and also of the energy of the outflow induced by magnetic pressure.
\end{abstract}

\maketitle

\section{Introduction}
The first direct detection of gravitational waves from a binary neutron star merger (GW170817) and its electromagnetic counterparts (AT 2017gfo/SGRB 170817A) heralded the beginning of multimessenger astronomy including gravitational waves~\cite{LIGOScientific:2017vwq,LIGOScientific:2017ync}.
In this event, the tidal deformability of the neutron star binary was measured for the first time and found to be in the interval $100 \lesssim \tilde{\Lambda}\lesssim 800$, with an accurate measurement of the total mass of the binary yielding $2.73^{+0.04}_{-0.01}M_\odot$~\cite{LIGOScientific:2017vwq,LIGOScientific:2018hze,LIGOScientific:2018cki,De:2018uhw} \footnote{The precise value of $\tilde{\Lambda}$ depends on the details of the analysis.}. Any viable neutron star matter equations of state must satisfy this observational constraint on tidal deformability. In this event it was also shown that the binary neutron star merger drives a short gamma-ray burst~\cite{Goldstein:2017mmi,LIGOScientific:2017ync,Savchenko:2017ffs,Mooley:2018qfh}, thus providing the first `smoking gun' for supporting the hypothesis that binary mergers can be the central engine of short gamma-ray bursts~\cite{Paczynski:1986,Goodman:1985,Eichler:1989,Narayan:1992iy}. Finally, this event indicated that neutron-rich matter is likely to be ejected during the merger and heavy elements are synthesized within these ejecta by means of the rapid neutron capture process on nuclei (the $r$-process) ~\cite{Metzger:2010,Lattimer:1974,Eichler:1989,Wanajo:2014wha}. It had been predicted that the $r$-process nucleosynthesis subsequently causes so-called kilonova emission via the radioactive decay of unstable $r$-process elements~\cite{Tanaka:2013ana,Barnes:2013wka}, and a kilonova was indeed observed after the merger in the near infrared, optical, and ultraviolet bands~\cite{Arcavi:2017xiz,Chornock:2017sdf,Coulter:2017wya,Cowperthwaite:2017dyu,Drout:2017ijr,Kasen:2017sxr,Kasliwal:2017ngb,Kilpatrick:2017mhz,McCully:2017lgx,Nicholl:2017ahq,Shappee:2017zly,Smartt:2017fuw,DES:2017kbs,Tanaka:2017qxj,Tanvir:2017pws}.

References~\cite{LIGOScientific:2020aai,LIGOScientific:2020ibl} reported the detection of a second binary neutron star merger event (GW190425) and measured a total binary mass of $3.4^{+0.1}_{-0.1}M_\odot$, which is much larger than the total mass measured in binary pulsars observed in our Galaxy~\cite{2019ApJ...876...18F}. The merger dynamics, mass ejection process, and resultant electromagnetic emission due to $r$-process nucleosynthesis could be different from those in GW170817 and AT 2017gfo~\cite{Kyutoku:2020xka,Dudi:2021abi}. 
Although an electromagnetic counterpart was \textit{not} observed in GW190425, either due to poor sky localization or due to intrinsically dimmer emission~\cite{Hosseinzadeh:2019ifm,Coughlin:2019xfb,Coughlin:2019zqi}, the existence of a massive binary neutron star suggests that the binary neutron star merger and associated mass ejection process could have a diversity of mechanisms. The new observation run O4 is planned to commence at the end of 2022~\cite{O4-link}, and could lead to the observation of binary neutron star mergers and associated electromagnetic counterparts that are qualitatively different from those observed in GW170817. This motivates  building binary neutron star merger models based on reliable numerical relativity simulations for predicting and interpreting gravitational wave events in preparation for the upcoming observational run~\cite{Shibata:2017xdx}.

Recent axisymmetric neutrino-radiation viscous-hydrodynamics simulations of binary neutron star merger remnants in numerical relativity suggest that the amount of post-merger ejecta launched from the merger remnant due to viscous effects (which were facilitated in these simulations by an effective `alpha' viscosity parameter) can be larger than the dynamical ejecta launched during the merger itself~\cite{Fujibayashi:2017puw,Fujibayashi:2020jfr,Fujibayashi:2020dvr,Fujibayashi:2020qda,Shibata:2017xdx}. The timescale of the post-merger mass ejection is $O(1)$ second, and depends on the value of the viscosity parameter. Plausible values of the viscosity parameter are inferred from three-dimensional magnetohydrodynamics simulations of the binary neutron star merger remnant in which angular momentum transport is facilitated in a self-consistent manner by the magnetorotational instability~\cite{Kiuchi:2017zzg} (see also Refs.~\cite{Christie:2019lim,Fernandez:2018kax} for magnetohydrodynamics simulations of a massive torus in a stationary black hole spacetime). The electron fraction of the post-merger ejecta and the resultant $r$-process nucleosynthesis also depends on this viscosity parameter~\cite{Fujibayashi:2020dvr,Fujibayashi:2020jfr,Fujibayashi:2020qda}, although the electron fraction of the post-merger ejecta is appreciably larger than that of the dynamical ejecta. 

Furthermore, very recently we performed neutrino-radiation magnetohydrodynamics simulations of \textit{black hole}-neutron star mergers in numerical relativity~\cite{Hayashi:2021oxy}. We found post-merger mass ejection due to magnetorotational instability-driven turbulence and the launch of a Poynting flux-dominated outflow. The post-merger mass ejection and the Poynting-flux dominated outflow sets in at several $100$\,ms after the merger and lasts for $1$--$2$ seconds after the merger. These timescales are determined by the strength of the effective viscosity associated with both magnetorotational-instability turbulence and neutrino cooling~\cite{Hayashi:2021oxy}. 

All these recent studies show that for modeling future gravitational wave events 
it is necessary to perform self-consistent (i.e. in which turbulence is sustained by the magnetorotational instability) three-dimensional neutrino-radiation magnetohydrodynamics simulations of binary neutron star mergers in general relativity for the durations of $O(1)$ second. In particular, it is crucial to reproduce a magneto-turbulent state driven by the magnetorotational instability inside the merger remnant because the resultant effective turbulent viscosity transports angular momentum outwards and heats up the matter via viscous heating~\cite{Balbus-Hawley:1998}. 

Finite volume methods are a popular combination of numerical schemes for simulations of astrophysical fluid dynamics due to their inherent conservation properties and ability to capture sharp discontinuities in the flow such as shocks~\cite{Toro-textbook}. Central to these schemes is the solution of the so-called \textit{Riemann problem} in which one considers two constant states separated by a discontinuity. The solution consists of three waves in hydrodynamics and seven waves in magnetohydrodynamics. As exact Riemann solvers are computationally expensive~\cite{Giacomazzo:2005jy}, approximate Riemann solvers are often used. One such family of approximate Riemann solvers is the HLL-based Riemann solvers, in which only a subset of the full seven waves in the Riemann fan are considered. The HLL(E) solver, for example, takes into account only shocks/rarefactions and omits the contact discontinuity~\cite{Harten:1983}.

At present, the Riemann solver and constrained transport scheme implemented in existing numerical relativity magnetohydrodynamics codes, e.g. ~\cite{Hayashi:2021oxy,Radice:2021jtw,Bernuzzi:2016pie,Most:2019kfe,Mosta:2013gwu,Etienne:2015cea,Vigano:2020ouc,Cipolletta:2020kgq,Foucart:2020xkt}, are based on the HLLE solver~\cite{Kurganov:2000,Harten:1983,DelZanna:2002rv}.  (An exception is the SpECTRE~\cite{Deppe:2021bhi} code, which is based on the discontinuous Galerkin method.) This Riemann solver is known to be very diffusive~\cite{Toro:1994,MUB:2009,Held:2018}. 
The numerical diffusion inherent in the Riemann solver adversely affects the accuracy of the numerical solution, in particular for long-term simulations of compact object mergers of $O(1)$ second. Although Refs.~\cite{Bernuzzi:2016pie,Most:2019kfe,Radice:2013hxh} reported the implementation of fourth-order accurate Riemann solvers in their numerical relativity codes, these solvers are based on the {\it finite difference} method. Therefore, it is unclear how accurate these finite difference-based Riemann solvers are for the problem of astrophysical turbulence. 

This paper reports a new implementation of advanced Riemann solvers in our neutrino-radiation magnetohydrodynamics numerical relativity code ~\cite{Kiuchi:2012qv,Sekiguchi:2012uc} based on the {\it finite volume} method. We implement the HLLC solver for relativistic hydrodynamics, which restores the contact discontinuity~\cite{Mignone:2005ft}, and the HLLD solver for relativistic magnetohydrodynamics, which takes into account five of the seven waves in the Riemann fan~\cite{MUB:2009}. Both these Riemann solves are known to be less diffusive than the HLLE solver~\cite{DelZanna:2002rv}. In addition, the constrained transport scheme in Ref.~\cite{Gardiner:2007nc}, which relies on the solution given by a Riemann solver, significantly suppresses numerical diffusion compared to the HLLE-constrained transport scheme proposed in Ref.~\cite{DelZanna:2002rv} (see Ref.~\cite{Mignone:2021} for a detailed comparison of different implementations of the constrained transport scheme). Thus, in addition to implementing the advanced Riemann solvers, HLLC and HLLD, we also implement the novel constrained transport scheme of Ref.~\cite{Gardiner:2007nc} in our code. 

This paper is organized as follows. Section~\ref{sec:basic_equation} summarizes the equations of motion for general relativistic neutrino-radiation magnetohydrodynamics. Section~\ref{sec:numerical_algorithm} is devoted to the numerical algorithm for general relativistic magnetohydrodynamics: the finite volume method, the constrained transport method (for enforcing divergence-free condition of the magnetic field), the tetrad transformation (which enables us to use Riemann solvers designed for special relativistic flows in full general relativity), the implementation of the HLLC solver ~\cite{Mignone:2005ft}, that of the HLLD solver~\cite{MUB:2009}, and the electric field evaluation (which is used by the constrained transport algorithm) ~\cite{Gardiner:2007nc}. In Sec.~\ref{sec:validation}, we validate our implementation of the new Riemann solvers by performing one- and multi-dimensional test problems both in Minkowski spacetime and in curved, but static, spacetime  in both relativistic hydrodynamics and magnetohydrodynamics. 
Finally, in Sec.~\ref{sec:dynamical_spacetime} we apply our new solvers in general relativity to a \textit{dynamical} spacetime. We first present the results of binary neutron star merger simulations in the absence of magnetic fields (which are run up to $\approx 40$--$50$ ms after the formation of the black hole), and subsequently the evolution of the merger remnant \textit{with} a magnetic field. Section~\ref{sec:conclusion} summarizes our results. Throughout this paper, we use geometrical units in which $c=G=1$. Greek and Latin indices without hats denote the spacetime and purely spatial components, respectively. Those with hats indicate tetrad components.

\section{Governing equations for general relativistic neutrino-radiation magnetohydrodynamics}\label{sec:basic_equation}

In this section, we briefly summarize the set of basic equations of general relativistic neutrino-radiation magnetohydrodynamics using the 3+1 formalism. The reader can find a more comprehensive derivation of these equations in, e.g., Ref.~\cite{Shibata-textbook}. 

We begin by introducing a unit vector normal to a spatial hypersurface of constant coordinate time, $t$,  
\begin{align}
n^\mu&=\left(\frac{1}{\alpha},-\frac{\beta^i}{\alpha}\right),~n_\mu=\left(-\alpha,0\right),
\end{align}
where $\alpha$ and $\beta^i$ are the lapse function and shift vector, respectively. With this vector, the four dimensional metric can be decomposed into 
\begin{align}
g_{\mu\nu} &= 
\left(
\begin{array}{cc}
-\alpha^2+\beta_i \beta^i     &  \beta_i \\
\beta_i     & \gamma_{ij}
\end{array}
\right),
\end{align}
where $\gamma_{ij}$ is the three-dimensional spatial metric. 

The stress-energy-momentum tensor for ideal magnetohydrodynamics and for a free-streaming neutrino-radiation field are, respectively, given by
\begin{align}
&T^{\mu\nu}_\text{(MHD)} = \rho h u^\mu u^\nu + P g^{\mu\nu} + \left( u^\mu u^\nu + \frac{1}{2}g^{\mu\nu} \right) b^2 - b^\mu b^\nu,\nonumber\\
&T^{\mu\nu}_\text{(Rad,s,$\nu_i$)} = E_{(\nu_i)} n^\mu n^\nu + F^\mu_{(\nu_i)} n^\nu + F^\nu_{(\nu_i)} n^\mu + P^{\mu\nu}_{(\nu_i)}, 
\end{align}
where $\rho$, $P$, $u^\mu$, $b^\mu$, $E_{(\nu_i)}$, $F^\mu_{(\nu_i)}$, and $P^{\mu\nu}_{(\nu_i)}$ are, respectively, the rest-mass density, pressure, four-velocity,  magnetic field (measured in the fluid rest frame), radiation energy density, radiation momentum, and radiation stress-energy-momentum tensor of the neutrino species $\nu_i$ in the Eulerian frame. $h=1 + \varepsilon + P/\rho$ denotes the relativistic specific enthalpy with $\varepsilon$ the specific internal energy. We consider the electron neutrino $\nu_e$, electron antineutrino $\bar{\nu}_e$, and the total of $\mu$ and $\tau$ neutrinos and antineutrinos collectively denoted by $\nu_x$~\cite{Sekiguchi:2012uc,Shibata-textbook}. Note that we assume that the stress-energy-momentum tensor of the neutrino-radiation field is split into a trapped component and a free-streaming component. The stress-energy-momentum tensor of the trapped neutrinos is then absorbed into that for the ideal magnetohydrodynamics fluid because trapped neutrinos are strongly coupled to the fluid~\cite{Sekiguchi:2012uc,Shibata-textbook}.  

The conserved mass density, total momentum density, and total energy density of an electrically conducting fluid are defined by
\begin{align}
D & \equiv \rho w\\
J_i & \equiv -{\gamma^\mu}_i n^\nu T_{\mu\nu}^{\text{(MHD)}}\nonumber\\
&=\rho w h u_i + \frac{B^2u_i-(B^ju_j)B_i}{w},\\
\rho_{\rm H} & \equiv n^\mu n^\nu T_{\mu\nu}^\text{(MHD)} \nonumber\\
&= \rho w \left ( h w - \frac{P}{\rho w} \right) + \left( w^2 - \frac{1}{2} \right) b^2 -(B^i u_i)^2, 
\end{align}
where $w \equiv - n_\mu u^\mu = \alpha u^t$ is the Lorentz factor measured by an Eulerian observer and $B^i$ is the magnetic field measured in the Eulerian frame and satisfies $B^\mu n_\mu=0$ (i.e., $B^t=0$). The relation between $b^\mu$ and $B^i$ is given by
\begin{align}
&b^t = \frac{B^iu_i}{\alpha},~b_i = \frac{B_i + \left(B^j u_j\right)u_i}{w},
\end{align}
and thus, 
\begin{align}
b^2 = \frac{B^2+(B^iu_i)^2}{w^2},
\end{align}
where $B^2=B_iB^i$.

The equations of motion of ideal magnetohydrodynamics and of the free-streaming neutrino-radiation field are derived from the conservation of the stress-energy-momentum tensor, the continuity equations for rest-mass density, electron fraction, electron neutrino fraction, electron antineutrino fraction, and heavy neutrino fraction, and the Maxwell equations. These conservation laws are written as
\begin{align}
&\nabla_\mu {\left(T_\text{(MHD)}\right)^\mu}_{\nu} = - \sum_{\nu_i=\nu_e,\bar\nu_e,\nu_x}G_\nu^\text{($\nu_i$,leak)}, \label{eq:MHD}\\
&\nabla_\mu {\left(T_\text{(Rad,s,$\nu_i$)}\right)^\mu}_{\nu} = G_\nu^\text{($\nu_i$,leak)}, \label{eq:Rad}\\
&\nabla_\mu (\rho u^\mu) = 0,\label{eq:MHD2}\\
&\nabla_\mu (\rho u^\mu Y_L) = \rho \gamma_L,\\
&\nabla_\mu {^*F}^{\mu\nu} = 0, \label{eq:MHD3}
\end{align}
where $L=e,\nu_e,\bar{\nu}_e,$ and $\nu_x$ denotes electrons, electron neutrinos, electron antineutrinos, and heavy neutrinos, respectively. $Y_L$ and $\gamma_L$ denote the fractions with respect to the baryon and the source term for the number of the species $L$, respectively. $G_\nu^\text{($\nu_i$,leak)}$ is an interaction term between the fluid and free-streaming neutrino-radiation field of the neutrino species $\nu_i$ in the framework of a general relativistic neutrino leakage scheme~\cite{Sekiguchi:2012uc,Sekiguchi:2010ep}. Here $^*F^{\mu\nu}$ is the Hodge dual of the Faraday tensor, which is given by $^*F^{\mu\nu}=b^\mu u^\nu - b^\nu u^\mu$ in ideal magnetohydrodynamics. 

Equations~(\ref{eq:MHD}), (\ref{eq:MHD2}), and (\ref{eq:MHD3}) can be written in conservative form as
\begin{align}
&\partial_t \left(\sqrt{\gamma}Q_A \right) + \partial_j \left(\sqrt{\gamma}F^j_A \right)= M_A,\label{eq:EOM}\\
&\partial_i\left(\sqrt{\gamma}B^i\right) = 0, \label{eq:divB}
\end{align}
where the flow quantities are given by the state vector $Q_A = (D, J_i, \rho_{\rm H}, B^k)^T$ with $A\in[0,7]$. The corresponding fluxes are given by
\begin{widetext}
\begin{equation}
F^j_A =\nonumber\\
\left(
\begin{array}{c}
D v^j\\
\displaystyle{
J_i v^j + \alpha \left( P + \frac{b^2}{2}\right){\delta^j}_i - \frac{\alpha}{w^2}B^j \left[B_i + (B^k u_k)u_i\right]}\\
\displaystyle{
\rho_{\rm H}v^j + \left(P+\frac{b^2}{2}\right)(v^j+\beta^j)-\frac{\alpha}{w}(B^ku_k)B^j}\\
B^k v^j - B^j v^k
\end{array}
\right),
\end{equation}
\end{widetext}
and the source terms are
\begin{widetext}
\begin{align}
M_A =\left(
\begin{array}{c}
0\\
-\sqrt{\gamma}\rho_{\rm H}\partial_i \alpha + \sqrt{\gamma}J_k \partial_i \beta^k+\frac{\alpha}{3}{S^k}_k\partial_i \sqrt{\gamma} - \frac{1}{2}\alpha\gamma^{1/6}S_{jk}\partial_i\tilde{\gamma}^{jk}-\alpha \sqrt{\gamma} G_\text{(leak)}^\mu {\gamma}_{\mu i}\\
\frac{\alpha}{3}\sqrt{\gamma}K{S_k}^k+\alpha\gamma^{1/6}\hat{S}_{ij}\tilde{A}^{ij} - \sqrt{\gamma}J_k D^k\alpha+\alpha \sqrt{\gamma}G_\text{(leak)}^\mu n_\mu\\
0
\end{array}
\right),
\end{align}
\end{widetext}
where $v^j\equiv u^j/u^t$, $G^\mu_\text{(leak)}=\sum_{\nu_i=\nu_e,\bar{\nu}_e,\nu_x}G^\mu_{(\nu_i,\text{leak})}$ , and the spatial components of the stress-energy-momentum tensor are given by
\begin{align}
&S_{ij} \equiv {\gamma^\mu}_i{\gamma^\nu}_j T^\text{(MHD)}_{\mu\nu}\nonumber\\
&=\left(\rho h + b^2 \right)u_i u_j + \left( P + \frac{b^2}{2} \right)\gamma_{ij} - b_i b_j,\\
&\hat{S}_{ij} = S_{ij} - \left(P+\frac{b^2}{2}\right)\gamma_{ij}.
\end{align}
We also introduce the conformal metric $\tilde{\gamma}_{ij} = \psi^{-4}\gamma_{ij}$ and the trace-free conformal extrinsic curvature $\tilde{A}_{ij}=\psi^{-4}\left(K_{ij}-\frac{1}{3}K\gamma_{ij}\right)$, where $\psi$ and $K_{ij}$ are the conformal factor and the extrinsic curvature, respectively.
The explicit forms for $\gamma_L$ and $G^\text{($\nu_i$,leak)}_\mu$ and for the equation of motion of the free-streaming neutrino-radiation field can be found in Refs.~\cite{Fujibayashi:2017xsz,Shibata-textbook}. 
The high resolution shock capturing scheme for the neutrino-radiation field Eq.~(\ref{eq:Rad}) is the same as that in Ref.~\cite{Shibata:2011kx}.

\section{Numerical algorithm}\label{sec:numerical_algorithm}

In this section, we describe the numerical algorithms which we implemented in our code. In Sec.~\ref{sec:finitevolumemethod} we present the finite volume algorithm and discretization scheme, and in Sec.~\ref{subsec:tetrad} we discuss the transformation to Minkowski spacetime used to implement the HLLC and HLLD solvers in general relativity. The implementation of the HLLC and HLLD solvers themselves is presented in Sec.~\ref{sec:HLLC} and \ref{sec:HLLD}, respectively. Finally, the evaluation of the electric field used by the constrained transport algorithm is discussed in Sec.~\ref{sec:electricfieldevaluation}.

\subsection{Finite volume method}
\label{sec:finitevolumemethod}
\subsubsection{Fluid and magnetic field at cell center}
Let $\Omega$ be a region of a given four-dimensional manifold $\cal M$, bounded by a closed three-dimensional surface $\partial \Omega$, where $\partial \Omega$ denotes the surface of a four-dimensional parallelepiped composed of two spacelike surfaces $\{\Sigma_{t},\Sigma_{t+\Delta t}\}$ and three sets of two timelike surfaces $\{\Sigma_{x^i},\Sigma_{x^i+\Delta x^i}\}$ that connect the two temporal slices~\cite{Font:2003}. The timelike surface, e.g., $\Sigma_{x}$, may also be regarded as a time series of constant-$(t,x)$ surfaces, $S_{x}(t)$.
We integrate Eq.~(\ref{eq:EOM}) over the domain of $\Omega$:
\begin{align}
&\int_\Omega \frac{1}{\sqrt{-g}}\partial_t\left(\sqrt{\gamma}Q_A\right)d\Omega
+\int_\Omega \frac{1}{\sqrt{-g}}\partial_i \left(\sqrt{\gamma}F^i_A\right)d\Omega \nonumber\\
&= \int_\Omega \frac{1}{\sqrt{-g}}M_A d\Omega,
\end{align}
where $d\Omega=\sqrt{-g}dtdxdydz$.

Using Gauss's theorem, this equation can be integrated to give
\begin{align}
&(\bar{Q}_A \Delta V)_{t+\Delta t} - (\bar{Q}_A \Delta V)_t =\nonumber\\
&-\left(\int_{\Sigma_{x+\Delta x}}\sqrt{\gamma}F^x_A dtdydz-\int_{\Sigma_{x}}\sqrt{\gamma}F^x_A dtdydz\right)\nonumber\\
&-\left(\int_{\Sigma_{y+\Delta y}}\sqrt{\gamma}F^y_A dtdxdz-\int_{\Sigma_{y}}\sqrt{\gamma}F^y_A dtdxdz\right)\nonumber\\
&-\left(\int_{\Sigma_{z+\Delta z}}\sqrt{\gamma}F^z_A dtdxdy-\int_{\Sigma_{z}}\sqrt{\gamma}F^z_A dtdxdy\right)\nonumber\\
&+\int_\Omega \frac{1}{\sqrt{-g}}M_A d\Omega,
\label{eq:EOM_FV}
\end{align}
where
\begin{align}
&\bar{Q}_A \equiv \frac{1}{\Delta V}\int \sqrt{\gamma} Q_A dxdydz,\\
&\Delta V \equiv \int \sqrt{\gamma}dxdydz,
\end{align}
are, respectively, the three-dimensional proper volume-averaged conserved quantities and the proper volume. 
Let us now define a cell consisting of $[x_j-\Delta x/2:x_j+\Delta x/2]\times [y_k-\Delta y/2:y_k+\Delta y/2]\times [z_l-\Delta z/2:z_l+\Delta z/2]$ (see Fig.~\ref{fig:Finite_volume_cell}). We next consider a numerical flux, which approximates a time-averaged flux at the cell interface and depends on the solution of the Riemann problem at the interface. For example, in the $x$-direction the flux across the right-hand interface is given by 
\begin{align}
({\tilde{F}^x}_A)_{j+\frac{1}{2},k,l} \approx \frac{1}{\Delta
t}\int^{t^{n+1}}_{t^n}F^x_A(Q_A(x_{j+1/2},y_k,z_l,t))dt, \label{eq:num_flux_formal}
\end{align}
where $t^{n+1}=t^n+\Delta t$. With this numerical flux, Eq.~(\ref{eq:EOM_FV}) can be discretized as
\begin{align}
&(\bar{Q}_A \Delta V)^{n+1}_{j,k,l} - (\bar{Q}_A \Delta V)^n_{j,k,l} =\nonumber\\
&-\Delta t \Big[ \left(\Delta A_x\right)_{j+\frac{1}{2},k,l} \left({\tilde{F}^x}_A\right)_{j+\frac{1}{2},k,l} \nonumber\\
&-\left(\Delta A_x\right)_{j-\frac{1}{2},k,l}
\left({\tilde{F}^x}_A\right)_{j-\frac{1}{2},k,l}\Big]\nonumber\\
&-\Delta t \Big[ \left(\Delta A_y\right)_{j,k+\frac{1}{2},l}\left({\tilde{F}^y}_A\right)_{j,k+\frac{1}{2},l}\nonumber\\
&-\left(\Delta A_y\right)_{j,k-\frac{1}{2},l}\left({\tilde{F}^y}_A\right)_{j,k-\frac{1}{2},l}\Big]\nonumber\\
&-\Delta t \Big [ \left(\Delta A_z \right)_{j,k,l+\frac{1}{2}}\left({\tilde{F}^z}_A\right)_{j,k,l+\frac{1}{2}}\nonumber\\
&-\left(\Delta A_z \right)_{j,k,l-\frac{1}{2}}\left({\tilde{F}^z}_A\right)_{j,k,l-\frac{1}{2}} \Big] \nonumber\\
&+\int M_A dtdxdydz,
\end{align}
where
\begin{align}
\left(\Delta A_x\right)_{j\pm\frac{1}{2},k,l} = \int \sqrt{\gamma(x_{j\pm\frac{1}{2}},y_k,z_l)} dydz,\\
\left(\Delta A_y\right)_{j,k\pm\frac{1}{2},l}= \int \sqrt{\gamma(x_j,y_{k\pm\frac{1}{2}},z_l)} dxdz,\\
\left(\Delta A_z\right)_{j,k,l\pm\frac{1}{2}} = \int \sqrt{\gamma(x_j,y_k,z_{l\pm\frac{1}{2}})} dxdy.
\end{align}

We also assume that the determinant of the spatial metric does not change significantly during the time step. 
If we introduce the volume- or surface area-averaged determinant of the spatial metric, denoted by $\bar{\gamma}$, this equation is reduced to
\begin{align}
&(\sqrt{\bar{\gamma}}\bar{Q}_A)^{n+1}_{j,k,l} - (\sqrt{\bar{\gamma}}\bar{Q}_A )^n_{j,k,l} =\nonumber\\
&-\frac{\Delta t}{\Delta x} \left[ \left(\sqrt{\bar{\gamma}}\right)_{j+\frac{1}{2},k,l}\left({\tilde{F}^x}_A\right)_{j+\frac{1}{2},k,l}-
\left(\sqrt{\bar{\gamma}}\right)_{j-\frac{1}{2},k,l}\left({\tilde{F}^x}_A\right)_{j-\frac{1}{2},k,l}\right]\nonumber\\
&-\frac{\Delta t}{\Delta y}\left[ \left(\sqrt{\bar{\gamma}}\right)_{j,k+\frac{1}{2},l}\left({\tilde{F}^y}_A\right)_{j,k+\frac{1}{2},l} -\left(\sqrt{\bar{\gamma}}\right)_{j,k-\frac{1}{2},l}\left({\tilde{F}^y}_A\right)_{j,k-\frac{1}{2},l}\right]\nonumber\\
&-\frac{\Delta t}{\Delta z}\left[ \left(\sqrt{\bar{\gamma}}\right)_{j,k,l+\frac{1}{2}}\left({\tilde{F}^z}_A\right)_{j,k,l+\frac{1}{2}} -\left(\sqrt{\bar{\gamma}}\right)_{j,k,l-\frac{1}{2}}\left({\tilde{F}^z}_A\right)_{j,k,l-\frac{1}{2}} \right]\nonumber\\
&+\left(\bar{M}_A\right)_{j,k,l}, \label{eq:FV}
\end{align}
where 
\begin{align}
&\left(\sqrt{\bar{\gamma}}\right)_{j,k,l} \equiv \frac{1}{\Delta x \Delta y \Delta z}\left(\Delta V\right)_{j,k,l},\\
&\left(\sqrt{\bar{\gamma}}\right)_{j\pm\frac{1}{2},k,l} \equiv \frac{1}{\Delta y \Delta z}\left(\Delta A_x\right)_{j\pm\frac{1}{2},k,l},\label{eq:Ax}\\
&\left(\sqrt{\bar{\gamma}}\right)_{j,k\pm\frac{1}{2},l} \equiv \frac{1}{\Delta x \Delta z}\left(\Delta A_y\right)_{j,k\pm\frac{1}{2},l},\label{eq:Ay}\\
&\left(\sqrt{\bar{\gamma}}\right)_{j,k,l\pm\frac{1}{2}} \equiv \frac{1}{\Delta x \Delta y}\left(\Delta A_z\right)_{j,k,l\pm\frac{1}{2}},\label{eq:Az}
\end{align}
and
\begin{align}
\left(\bar{M}_A\right)_{j,k,l} \equiv \frac{1}{\Delta x \Delta y \Delta z}\int M_A dtdxdydz.
\end{align}
\subsubsection{Magnetic fields at cell surface}
\begin{figure}[t]
 	 \includegraphics[width=0.99\linewidth]{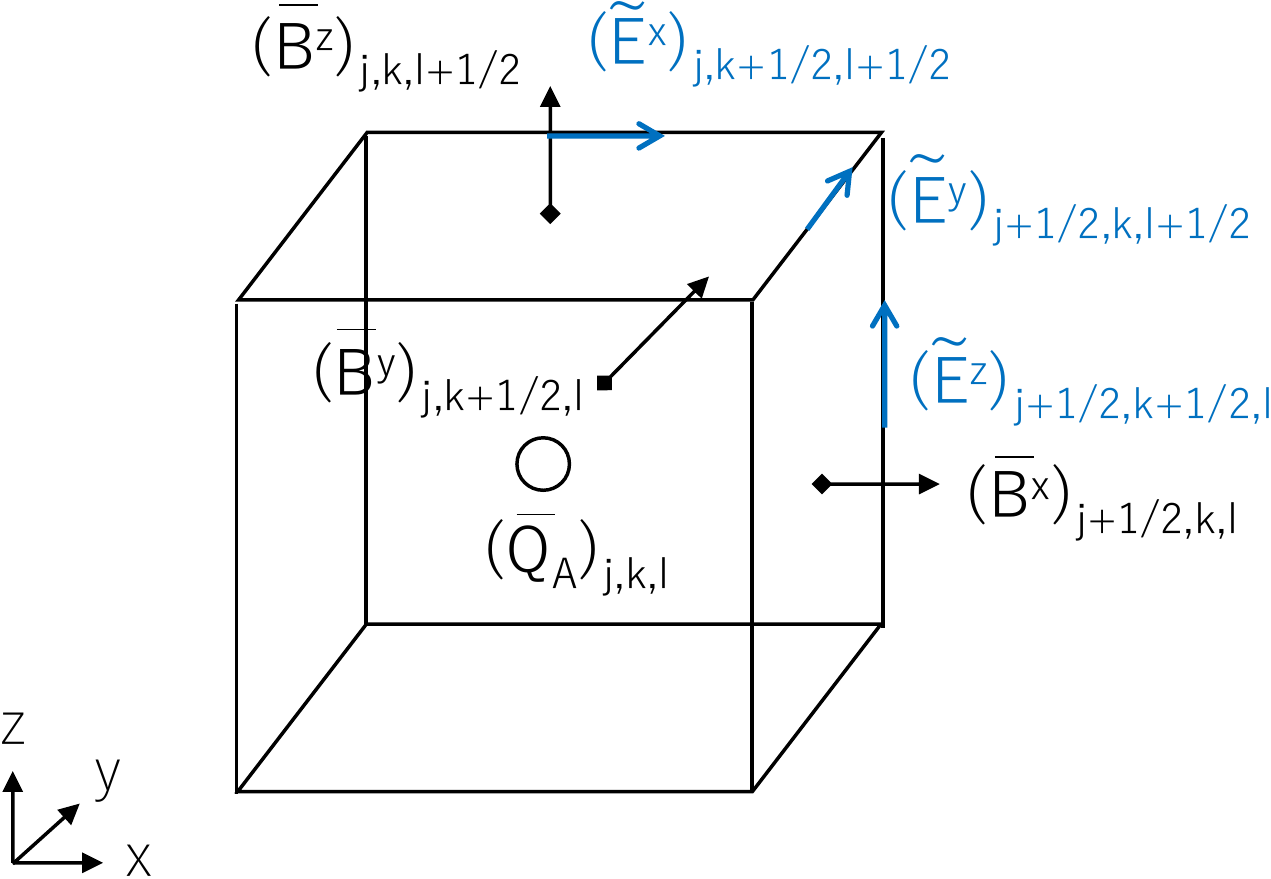}
 	 \caption{Schematic of a cell, cell interface, and cell edge for the finite volume method with the constrained transport method. Fluid quantities, $(\bar{Q}_A)_{j,k,l}$, are defined at the cell center. The magnetic field components, $(\bar{B}^x)_{j+\frac{1}{2},k,l},(\bar{B}^y)_{j,k+\frac{1}{2},l},(\bar{B}^z)_{j,k,l+\frac{1}{2}}$, are defined at the cell interfaces. The electric field components, $(\tilde{E}^x)_{j,k+\frac{1}{2},l+\frac{1}{2}},(\tilde{E}^y)_{j+\frac{1}{2},k,l+\frac{1}{2}},(\tilde{E}^z)_{j+\frac{1}{2},k+\frac{1}{2},l}$, are defined at the cell edges. 
         }\label{fig:Finite_volume_cell}
\end{figure}

To ensure that the divergence-free condition (\ref{eq:divB}) is maintained, we employ the constrained transport method introduced by Evans and Hawley~\cite{Evans:1988}. In this method, the magnetic-field components are defined at the cell surfaces, and the electric field components are defined at the cell edges (see Fig.~\ref{fig:Finite_volume_cell}). 

We then integrate Eq.~(\ref{eq:EOM}) for $A\in[5,7]$ on $\Sigma_{x^i+\Delta x^i}$. For example, through the surface $\Sigma_{z+\Delta z}$, we have
\begin{align}
&\int_{\Sigma_{z+\Delta z}}\frac{1}{\sqrt{-g}}
\partial_t\left(\sqrt{\gamma}B^z\right)
dS_{\Omega_z} \nonumber\\
&+\int_{\Sigma_{z+\Delta z}}\frac{1}{\sqrt{-g}}\partial_j\left(\sqrt{\gamma}\epsilon^{zjk}E_k\right)
dS_{\Omega_z}
=0,
\end{align}
where $E_k=-\epsilon_{kij}v^iB^j$, $\epsilon_{ijk}$ is the three-dimensional Levi-Civita tensor, and $dS_{\Omega_z}=\sqrt{-g}dt dx dy$. Using Stokes' theorem, this equation is integrated to give
\begin{align}
\label{eq:EOM_FV_EM}
&\left(\bar{B}^z\Delta A_z\right)_{t+\Delta t} 
- \left(\bar{B}^z\Delta A_z\right)_{t}\nonumber \\
&=-\int_{t}^{t+\Delta t} \oint_{\partial S_{z+\Delta z}}\sqrt{\gamma}E_i dx^i dt, 
\end{align}
where
\begin{align}
 \bar{B}^z \equiv \frac{1}{\Delta A_z} \int_{S_{z+\Delta z}} \sqrt{\gamma} B^z dxdy,\nonumber
\end{align}
is the surface-averaged magnetic field.
Similarly, through the surfaces $\Sigma_{x+\Delta x}$ and $\Sigma_{y+\Delta y}$, respectively, we have
\begin{align}
&\left(\bar{B}^x\Delta A_x\right)_{t+\Delta t} 
- \left(\bar{B}^x\Delta A_x\right)_{t}\nonumber \\
&=-\int_{t}^{t+\Delta t} \oint_{\partial S_{x+\Delta x}}\sqrt{\gamma}E_i dx^i dt,\label{eq:EOM_FV_EM2}\\
&\left(\bar{B}^y\Delta A_y\right)_{t+\Delta t} 
- \left(\bar{B}^y\Delta A_y\right)_{t}\nonumber \\
&=-\int_{t}^{t+\Delta t} \oint_{\partial S_{y+\Delta y}}\sqrt{\gamma}E_i dx^i dt,\label{eq:EOM_FV_EM3}
\end{align}
where 
\begin{align}
 &\bar{B}^x \equiv \frac{1}{\Delta A_x} \int_{S_{x+\Delta x}} \sqrt{\gamma} B^x dydz, \nonumber\\
 &\bar{B}^y \equiv \frac{1}{\Delta A_y} \int_{S_{y+\Delta y}} \sqrt{\gamma} B^y dxdz. \nonumber
\end{align}

We next consider a cell surface consisting of $[x_j-\Delta x/2:x_j+\Delta x/2]\times[y_k-\Delta y/2:y_k+\Delta y/2]$, $[y_k-\Delta y/2:y_k+\Delta y/2]\times[z_l-\Delta z/2:z_l+\Delta z/2]$, $[x_j-\Delta x/2:x_j+\Delta x/2]\times[z_l-\Delta z/2:z_l+\Delta z/2]$ and a numerical flux which approximates a time-averaged electric field
at the cell edge, given by
\begin{align}
&\left(\tilde{E}_x\right)_{j,k+\frac{1}{2},l+\frac{1}{2}} \approx \frac{1}{\Delta t}\int^{t^{n+1}}_{t^n} E_x(Q_A(x_j,y_{k+1/2},z_{l+1/2}))dt,\label{eq:EM_Num_flux1}\\
&\left(\tilde{E}_y\right)_{j+\frac{1}{2},k,l+\frac{1}{2}} \approx \frac{1}{\Delta t}\int^{t^{n+1}}_{t^n} E_y(Q_A(x_{j+1/2},y_k,z_{l+1/2}))dt,\label{eq:EM_Num_flux2}\\
&\left(\tilde{E}_z\right)_{j+\frac{1}{2},k+\frac{1}{2},l} \approx \frac{1}{\Delta t}\int^{t^{n+1}}_{t^n} E_z(Q_A(x_{j+1/2},y_{k+1/2},z_l))dt.\label{eq:EM_Num_flux3}
\end{align}
With these averaged electric fields,
Eqs.~(\ref{eq:EOM_FV_EM})--(\ref{eq:EOM_FV_EM3}) are discretized as
\begin{align}
&\left(\bar{B}^x\Delta A_x\right)^{n+1}_{j+\frac{1}{2},k,l}
- \left(\bar{B}^x\Delta A_x\right)^n_{j+\frac{1}{2},k,l}\nonumber\\
&=\Delta t\Big[
\left(\Delta l_y\right)_{j+\frac{1}{2},k,l+\frac{1}{2}}\left(\tilde{E}_y\right)_{j+\frac{1}{2},k,l+\frac{1}{2}}\nonumber\\
&~~~~~~~-\left(\Delta l_y\right)_{j+\frac{1}{2},k,l-\frac{1}{2}}\left(\tilde{E}_y\right)_{j+\frac{1}{2},k,l-\frac{1}{2}}\nonumber\\
&~~~~~~~-\left(\Delta l_z\right)_{j+\frac{1}{2},k+\frac{1}{2},l}\left(\tilde{E}_z\right)_{j+\frac{1}{2},k+\frac{1}{2},l}\nonumber\\
&~~~~~~~+\left(\Delta l_z\right)_{j+\frac{1}{2},k-\frac{1}{2},l}\left(\tilde{E}_z\right)_{j+\frac{1}{2},k-\frac{1}{2},l}
\Big],\label{eq:EOM_CT1}\\
&\left(\bar{B}^y\Delta A_y\right)^{n+1}_{j,k+\frac{1}{2},l} 
- \left(\bar{B}^y\Delta A_y\right)^n_{j,k+\frac{1}{2},l}\nonumber\\
&=\Delta t \Big[
\left(\Delta l_z\right)_{j+\frac{1}{2},k+\frac{1}{2},l}\left(\tilde{E}_z\right)_{j+\frac{1}{2},k+\frac{1}{2},l}\nonumber\\
&~~~~~~~-\left(\Delta l_z\right)_{j-\frac{1}{2},k+\frac{1}{2},l}\left(\tilde{E}_z\right)_{j-\frac{1}{2},k+\frac{1}{2},l}\nonumber\\
&~~~~~~~-\left(\Delta l_x\right)_{j,k+\frac{1}{2},l+\frac{1}{2}}
\left(\tilde{E}_x\right)_{j,k+\frac{1}{2},l+\frac{1}{2}}\nonumber\\
&~~~~~~~+\left(\Delta l_x\right)_{j,k+\frac{1}{2},l-\frac{1}{2}}\left(\tilde{E}_x\right)_{j,k+\frac{1}{2},l-\frac{1}{2}}
\Big],\\
&\left(\bar{B}^z\Delta A_z\right)^{n+1}_{j,k,l+\frac{1}{2}} - \left(\bar{B}^z\Delta A_z\right)^n_{j,k,l+\frac{1}{2}}\nonumber\\
&=\Delta t\Big[
\left(\Delta l_x\right)_{j,k+\frac{1}{2},l+\frac{1}{2}}\left(\tilde{E}_x\right)_{j,k+\frac{1}{2},l+\frac{1}{2}}\nonumber\\
&~~~~~-\left(\Delta l_x\right)_{j,k-\frac{1}{2},l+\frac{1}{2}}\left(\tilde{E}_x\right)_{j,k-\frac{1}{2},l+\frac{1}{2}}\nonumber\\
&~~~~~-\left(\Delta l_y\right)_{j+\frac{1}{2},k,l+\frac{1}{2}}\left(\tilde{E}_y\right)_{j+\frac{1}{2},k,l+\frac{1}{2}}\nonumber\\
&~~~~~+\left(\Delta l_y\right)_{j-\frac{1}{2},k,l+\frac{1}{2}}\left(\tilde{E}_y\right)_{j-\frac{1}{2},k,l+\frac{1}{2}}
\Big],\label{eq:EOM_CT3}
\end{align}
where
\begin{align}
&\left(\Delta l_x\right)_{j,k\pm\frac{1}{2},l\pm\frac{1}{2}}=\int \sqrt{\gamma(x_j,y_{k\pm 1/2},z_{l\pm 1/2})}dx,\\
&\left(\Delta l_y\right)_{j\pm\frac{1}{2},k,l\pm\frac{1}{2}}=\int \sqrt{\gamma(x_{j\pm 1/2},y_k,z_{l\pm 1/2})}dy,\\
&\left(\Delta l_z\right)_{j\pm\frac{1}{2},k\pm\frac{1}{2},l}=\int \sqrt{\gamma(x_{j\pm 1/2},y_{k\pm 1/2},z_l)}dz.
\end{align}
If we introduce a line-averaged determinant of the spatial metric by
\begin{align}
&\left(\sqrt{\bar{\gamma}}\right)_{j,k\pm\frac{1}{2},l\pm\frac{1}{2}}\equiv \frac{\left(\Delta l_x\right)_{j,k\pm\frac{1}{2},l\pm\frac{1}{2}} }{\Delta x},\nonumber\\
&\left(\sqrt{\bar{\gamma}}\right)_{j\pm\frac{1}{2},k,l\pm\frac{1}{2}} \equiv \frac{
\left(\Delta l_y\right)_{j\pm\frac{1}{2},k,l\pm\frac{1}{2}}}{\Delta y},\nonumber\\
&\left(\sqrt{\bar{\gamma}}\right)_{j\pm\frac{1}{2},k\pm\frac{1}{2},l}
\equiv \frac{\left(\Delta l_z\right)_{j\pm\frac{1}{2},k\pm\frac{1}{2},l}} {\Delta z},
\end{align}
then, together with the surface area-averaged spatial metric given by Eqs.~(\ref{eq:Ax})--(\ref{eq:Az}), Eqs.~(\ref{eq:EOM_CT1})--(\ref{eq:EOM_CT3}) are  reduced to
\begin{align}
&\left(\sqrt{\bar{\gamma}}\bar{B}^x\right)^{n+1}_{j+\frac{1}{2},k,l}
- \left(\sqrt{\bar{\gamma}}\bar{B}^x\right)^n_{j+\frac{1}{2},k,l}\nonumber\\
&=\frac{\Delta t}{\Delta z}\Big[
\left(\sqrt{\bar{\gamma}}\right)_{j+\frac{1}{2},k,l+\frac{1}{2}}\left(\tilde{E}_y\right)_{j+\frac{1}{2},k,l+\frac{1}{2}}\nonumber\\
&~~~~~~-\left(\sqrt{\bar{\gamma}}\right)_{j+\frac{1}{2},k,l-\frac{1}{2}}\left(\tilde{E}_y\right)_{j+\frac{1}{2},k,l-\frac{1}{2}}
\Big]\nonumber\\
&-\frac{\Delta t}{\Delta y}\Big[
\left(\sqrt{\bar{\gamma}}\right)_{j+\frac{1}{2},k+\frac{1}{2},l}\left(\tilde{E}_z\right)_{j+\frac{1}{2},k+\frac{1}{2},l}\nonumber\\
&~~~~~~-\left(\sqrt{\bar{\gamma}}\right)_{j+\frac{1}{2},k-\frac{1}{2},l}\left(\tilde{E}_z\right)_{j+\frac{1}{2},k-\frac{1}{2},l}
\Big], \label{eq:CT1}\\
&\left(\sqrt{\bar{\gamma}}\bar{B}^y\right)^{n+1}_{j,k+\frac{1}{2},l} 
- \left(\sqrt{\bar{\gamma}}\bar{B}^y\right)^n_{j,k+\frac{1}{2},l}\nonumber\\
&=\frac{\Delta t}{\Delta x}\Big[
\left(\sqrt{\bar{\gamma}}\right)_{j+\frac{1}{2},k+\frac{1}{2},l}\left(\tilde{E}_z\right)_{j+\frac{1}{2},k+\frac{1}{2},l}\nonumber\\
&~~~~~~-\left(\sqrt{\bar{\gamma}}\right)_{j-\frac{1}{2},k+\frac{1}{2},l}\left(\tilde{E}_z\right)_{j-\frac{1}{2},k+\frac{1}{2},l}
\Big],\nonumber\\
&-\frac{\Delta t}{\Delta z} \Big[
\left(\sqrt{\bar{\gamma}}\right)_{j,k+\frac{1}{2},l+\frac{1}{2}}\left(\tilde{E}_x\right)_{j,k+\frac{1}{2},l+\frac{1}{2}}
\nonumber\\
&~~~~~~~-\left(\sqrt{\bar{\gamma}}\right)_{j,k+\frac{1}{2},l-\frac{1}{2}}\left(\tilde{E}_x\right)_{j,k+\frac{1}{2},l-\frac{1}{2}}
\Big],\\
&\left(\sqrt{\bar{\gamma}}\bar{B}^z\right)^{n+1}_{j,k,l+\frac{1}{2}} 
- \left(\sqrt{\bar{\gamma}}\bar{B}^z\right)^n_{j,k,l+\frac{1}{2}}\nonumber\\
&=\frac{\Delta t}{\Delta y}\Big[
\left(\sqrt{\bar{\gamma}}\right)_{j,k+\frac{1}{2},l+\frac{1}{2}}\left(\tilde{E}_x\right)_{j,k+\frac{1}{2},l+\frac{1}{2}}
\nonumber\\
&~~~~~~-\left(\sqrt{\bar{\gamma}}\right)_{j,k-\frac{1}{2},l+\frac{1}{2}}\left(\tilde{E}_x\right)_{j,k-\frac{1}{2},l+\frac{1}{2}}
\Big]\nonumber\\
&-\frac{\Delta t}{\Delta x}\Big[\left(\sqrt{\bar{\gamma}}\right)_{j+\frac{1}{2},k,l+\frac{1}{2}}\left(\tilde{E}_y\right)_{j+\frac{1}{2},k,l+\frac{1}{2}}\nonumber\\
&~~~~~~-\left(\sqrt{\bar{\gamma}}\right)_{j-\frac{1}{2},k,l+\frac{1}{2}}\left(\tilde{E}_y\right)_{j-\frac{1}{2},k,l+\frac{1}{2}}
\Big]. \label{eq:CT3}
\end{align}
The magnetic-field distribution inside the cell is reconstructed from the magnetic fields at the cell surface. Practically, we reconstruct the magnetic field at the cell center in Eq.~(\ref{eq:FV}) by
\begin{align}
 &\left(\bar{B}^x\right)_{j,k,l} = \frac{1}{2}\left[\left(\bar{B}^x\right)_{j+\frac{1}{2},k,l}+\left(\bar{B}^x\right)_{j-\frac{1}{2},k,l}\right],\label{eq:Bcellcenter1}\\
 &\left(\bar{B}^y\right)_{j,k,l} = \frac{1}{2}\left[\left(\bar{B}^y\right)_{j,k+\frac{1}{2},l}+\left(\bar{B}^y\right)_{j,k-\frac{1}{2},l}\right],\\
 &\left(\bar{B}^z\right)_{j,k,l} = \frac{1}{2}\left[\left(\bar{B}^z\right)_{j,k,l+\frac{1}{2}}+\left(\bar{B}^z\right)_{j,k,l-\frac{1}{2}}\right].\label{eq:Bcellcenter3}
\end{align}
\subsection{Tetrad frame} \label{subsec:tetrad}
To evaluate the numerical fluxes through cell interfaces (e.g. Eq.~(\ref{eq:num_flux_formal})), we implement HLL-type Riemann solvers~\cite{Mignone:2005ft,MUB:2009}. Because these Riemann solvers are designed to solve a Riemann problem in Minkowski spacetime (except for the HLLE solver, which we have implemented directly in curved spacetime, see, e.g., Ref.~\cite{Shibata:2005gp}), it is necessary to transform all the equations into a tetrad frame in order to apply these methods to a general relativistic framework. 

Following Ref.~\cite{White:2015omx}, we define a tetrad basis in the $x$-direction, for example, by
\begin{align}
    &{e_{(\hat{t})}}^\mu = n^\mu \label{eq:tetrad_basis},\\
    &{e_{(\hat{x})}}^\mu = \hat{B}\left(0,\gamma^{xi}\right),\\
    &{e_{(\hat{y})}}^\mu = \hat{D}\left(0,0,\gamma_{zz},-\gamma_{yz}\right),\\
    &{e_{(\hat{z})}}^\mu = \hat{C}\left(0,0,0,1\right),
\end{align}
where
\begin{align}
&\hat{B} = \frac{1}{\sqrt{\gamma^{xx}}},\\
&\hat{C} = \frac{1}{\sqrt{\gamma_{zz}}},\\
&\hat{D}=\frac{1}{\sqrt{\gamma_{zz}\left(\gamma_{yy}\gamma_{zz}-\gamma_{yz}^2\right)}}.
\end{align}
With this basis, we can perform a transformation from the Eulerian frame to the tetrad frame by
\begin{align}
&V_{(\hat{\mu})} = {e_{(\hat{\mu})}}^\mu V_\mu,\\
&Q_{(\hat{\mu})(\hat{\nu})}={e_{(\hat{\mu})}}^\mu{e_{(\hat{\nu})}}^\nu Q_{\mu\nu},
\end{align}
where $V_\mu$ and $Q_{\mu\nu}$ denote a covariant vector and tensor, respectively, in the Eulerian frame. 
The covariant components of the tetrad basis are
\begin{align}
e_{(\hat{t})\mu} &= n_\mu,\\
e_{(\hat{x})\mu} &= \hat{B}\left(\beta^x,{\delta_i}^x\right),\\
e_{(\hat{y})\mu} &= \hat{D}\Big(\beta_y\gamma_{zz}-\beta_z\gamma_{yz},\gamma_{xy}\gamma_{zz}-\gamma_{xz}\gamma_{yz},\nonumber\\
&~~~~~~~~~~\gamma_{yy}\gamma_{zz}-\gamma_{yz}^2,0 \Big),\\
e_{(\hat{z})\mu} &= \hat{C}\left(\beta_z,\gamma_{iz}\right).
\end{align}
With this basis, we can then perform the transformation from the tetrad frame to the Eulerian frame by
\begin{align}
 &V_\mu = e_{(\hat{\mu})\mu}V^{(\hat{\mu})},\\
 &Q_{\mu\nu} = e_{(\hat{\mu})\mu}e_{(\hat{\nu})\nu}Q^{(\hat{\mu})(\hat{\nu})}.
\end{align}

With this tetrad basis the procedure to obtain the numerical flux $\left({\tilde{F}^x}_A\right)_{j+\frac{1}{2},k,l}$ is as follows: first, we calculate the tetrad component of $u_{(\hat{\imath})},v^{(\hat{\imath})}$, and $B^{(\hat{\imath})}$ by
\begin{align}
    & u_{(\hat{\imath})} = e_{(\hat{\imath})\mu}u^\mu = \frac{w}{\alpha}\left(e_{(\hat{\imath})t}+e_{(\hat{\imath})j}v^j\right),\\
    & v^{(\hat{\imath})}\equiv\frac{u^{(\hat{\imath})}}{u^{(\hat{t})}}=\frac{{e^{(\hat{\imath})}}_\mu u^\mu}{{e^{(\hat{t})}}_\nu u^\nu} =\frac{ e_{(\hat{\imath})t} + e_{(\hat{\imath})j} v^j}{\alpha},\\
    & B^{(\hat{\imath})}={e^{(\hat{\imath})}}_\mu B^\mu = {e^{(\hat{\imath})}}_j B^j.
\end{align}
Second, we solve a Riemann problem in the locally Minkowski spacetime to obtain the numerical flux  $\left({\tilde{f}^{(\hat{x})}}_A\right)_{j+\frac{1}{2},k,l}$ and the conserved quantities $\left(q_A\right)_{j+\frac{1}{2},k,l}$ at the cell interface (see the next section for more detail on the Riemann problem). Finally, we transform back to the Eulerian frame from the tetrad frame by
\begin{align}
&({\tilde{F}^x}_0)_{j+\frac{1}{2},k,l} = \left( D v^x \right)_{j+\frac{1}{2},k,l}
\nonumber\\
&=\left(\alpha
\left({e_{(\hat{t})}}^xD+{e_{(\hat{x})}}^x\tilde{f}^{(\hat{x})}_0
\right)
\right)_{j+\frac{1}{2},k,l}\label{eq:num_flux},\\
&({\tilde{F}^x}_1)_{j+\frac{1}{2},k,l}=\left(\alpha {T^x}_x\right)_{j+\frac{1}{2},k,l}
\nonumber\\
&=\Big(\alpha
\Big(
{e_{(\hat{t})}}^x e_{(\hat{\imath})x}J_{(\hat{\imath})}
+{e_{(\hat{x})}}^x e_{(\hat{\imath})x}\tilde{f}^{(\hat{x})}_i
\Big)
\Big)_{j+\frac{1}{2},k,l},\label{eq:tetrad2coordinate1}\\
&({\tilde{F}^x}_2)_{j+\frac{1}{2},k,l}
=\left(\alpha{T^x}_y\right)_{j+\frac{1}{2},k,l}
\nonumber\\
&=\Big(\alpha
\Big(
{e_{(\hat{t})}}^x e_{(\hat{\imath})y}J_{(\hat{\imath})}+{e_{(\hat{x})}}^x e_{(\hat{\imath})y}\tilde{f}^{(\hat{x})}_i
\Big)
\Big)_{j+\frac{1}{2},k,l},\label{eq:tetrad2coordinate2}\\
&({\tilde{F}^x}_3)_{j+\frac{1}{2},k,l}
=\left(\alpha{T^x}_z\right)_{j+\frac{1}{2},k,l}
\nonumber\\
&=\Big(\alpha
\Big(
{e_{(\hat{t})}}^x e_{(\hat{z})z}J_{(\hat{z})}+{e_{(\hat{x})}}^x e_{(\hat{z})z}\tilde{f}^{(\hat{x})}_3
\Big)
\Big)_{j+\frac{1}{2},k,l},\\
&({\tilde{F}^x}_4)_{j+\frac{1}{2},k,l}
=\left(-\alpha{T^x}_\mu n^\mu\right)_{j+\frac{1}{2},k,l}
\nonumber\\
&=\left(\alpha
\left(
{e_{(\hat{t})}}^x \rho_{\rm H}+{e_{(\hat{x})}}^x \tilde{f}^{(\hat{x})}_4
\right)
\right)_{j+\frac{1}{2},k,l},\\
&(\tilde{F}^x_5)_{j+\frac{1}{2},k,l}=0,\\
&(\tilde{F}^x_6)_{j+\frac{1}{2},k,l}=
\left(-\tilde{E}_z\right)_{j+\frac{1}{2},k,l}
=
\left(\alpha{^*F}^{yx}\right)_{j+\frac{1}{2},k,l}\label{eq:num_flux2}\nonumber\\
&=\Big(\alpha\Big(
{e_{(\hat{\imath})}}^y{e_{(\hat{t})}}^x\bar{B}^{(\hat{\imath})}-{e_{(\hat{t})}}^y{e_{(\hat{x})}}^x\bar{B}^{(\hat{x})}\nonumber\\
&~~~~~+{e_{(\hat{y})}}^y{e_{(\hat{x})}}^x\tilde{f}^{(\hat{x})}_6
\Big)\Big)_{j+\frac{1}{2},k,l},\\
&(\tilde{F}^x_7)_{j+\frac{1}{2},k,l}=
\left(\tilde{E}_y\right)_{j+\frac{1}{2},k,l}
=\left(\alpha{^*F}^{zx}\right)_{j+\frac{1}{2},k,l}\nonumber\\
&=\Big(\alpha\Big(
{e_{(\hat{\imath})}}^z{e_{(\hat{t})}}^x\bar{B}^{(\hat{\imath})}
-{e_{(\hat{t})}}^z{e_{(\hat{x})}}^x\bar{B}^{(\hat{x})}\nonumber\\
&~~~~~+{e_{(\hat{y})}}^z{e_{(\hat{x})}}^x\tilde{f}^{(\hat{x})}_6
+{e_{(\hat{z})}}^z{e_{(\hat{x})}}^x\tilde{f}^{(\hat{x})}_7
\Big)\Big)_{j+\frac{1}{2},k,l}, \label{eq:num_flux3}
\end{align}
where $\hat{\imath} = \hat{x}, \hat{y}, \hat{z}$ are contracted with $i=1,2,3$, respectively, in the second term of the right-hand side of Eqs.~(\ref{eq:tetrad2coordinate1}) and (\ref{eq:tetrad2coordinate2}). 
Note that, from now on, we do not distinguish the upper- and lower-spatial tetrad components, e.g., $B^{(\hat{\imath})}=B_{(\hat{\imath})}$. These numerical fluxes are used to update the conserved quantities in Eq.~(\ref{eq:FV}).
An interface velocity is calculated by~\cite{White:2015omx}
\begin{align}
v^{(\hat{x})}_\text{interface}=\frac{d\hat{x}}{d\hat{t}}=\frac{\beta^x}{\alpha\sqrt{\gamma^{xx}}}. 
\end{align}
This velocity is used to calculate a numerical flux at the interface (see Eqs.~(\ref{eq:num_flux_HLLC}) and (\ref{eq:num_flux_HLLD}) in the next section). The tetrad basis and numerical fluxes in the $y$- and $z$-directions are summarized in Appendix A.

\begin{figure*}[t]
 	 \includegraphics[width=0.95\linewidth]{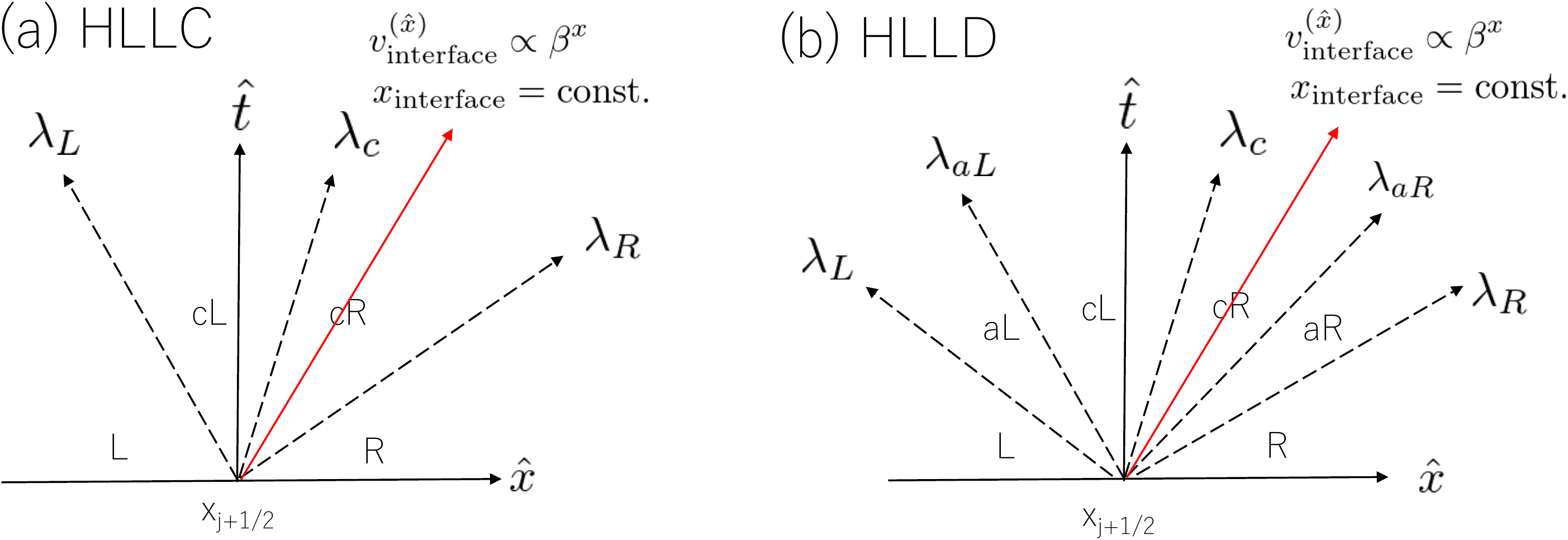}
 	 \caption{Riemann fan structure for the HLLC solver for relativistic hydrodynamics (left), and for the HLLD solver for relativistic magnetohydrodynamics (right) in the tetrad frame. 
 	 In the HLLC solver (left panel), the left-going nonlinear wave with $\lambda_L$, the contact discontinuity with $\lambda_c$, and the right-going nonlinear wave with $\lambda_R$, propagate from the discontinuity located at $\hat{x}_{j+\frac{1}{2}}$, where $\lambda_{L,c,R}$ denotes the characteristic speed of each wave. Consequently, the $L$, $cL$, $cR$, and $R$ states appear. 
 	 In the HLLD solver (right panel), the left/right-propagating fast wave with $\lambda_L$/$\lambda_R$, the left/right-propagating Alfv\'{e}n wave with characteristic speed $\lambda_{aL}$/$\lambda_{aR}$, and the contact discontinuity with $\lambda_c$, are taken into account. 
 	  Consequently, the $L$, $aL$, $cL$, $cR$, $aR$, and $R$ states appear. In the general relativistic case, the interface initially located at $\hat{x}_{j+\frac{1}{2}}$ may move with an interface velocity $v^{(\hat{x})}_\text{interface}$ which is proportional to the shift vector $\beta^x$.
         }\label{fig:Riemann_fan}
\end{figure*}

\subsection{HLLC solver for relativistic hydrodynamics}\label{sec:HLLC}
In the absence of electromagnetic fields, Eq.~(\ref{eq:FV}) with $A\in[0,4]$ are reduced to those of relativistic \textit{hydrodynamics}. In this case, one choice for the Riemann solver is the HLLC solver proposed in Ref.~\cite{Mignone:2005ft}. 
We calculate the HLLC flux  $\left(\tilde{f}^{(\hat{x})}_A\right)_{j+\frac{1}{2},k,l}$ in the tetrad frame by solving the source-free one-dimensional conservation law:
\begin{align}
&\partial_{(\hat{t})}q_A+\partial_{(\hat{x})}f^{(\hat{x})}_A = 0,\\
&q_A=\left(
\begin{array}{c}
   D              \\
   J_{(\hat{\imath})}  \\
   \rho_{\rm H}
\end{array}
\right),\\
&f^{(\hat{x})}_A=\left(
\begin{array}{c}
 D v^{(\hat{x})}\\
 J_{(\hat{\imath})} v^{(\hat{x})} + P {\delta^{(\hat{x})}}_{(\hat{\imath})}\\
\rho_{\rm H}v^{(\hat{x})} + P v^{(\hat{x})}
\end{array}
\right),
\end{align}
where $\partial_{(\hat{\mu})} \equiv {e_{(\hat{\mu})}}^{\mu} \partial_{\mu}$.
Given an initial condition described by
\begin{align}
 &q_A(x,0) = 
 \left\{
\begin{array}{l}
 \left(q_A\right)_L~\text{if $x<x_{j+\frac{1}{2}}$},\\
 \left(q_A\right)_R~\text{if $x>x_{j+\frac{1}{2}}$},
\end{array}
\right. \label{eq:initial}
\end{align}
for $x_j\le x \le x_{j+1}$, three characteristic speeds and therefore four states will appear in the Riemann fan (see Fig.~\ref{fig:Riemann_fan}).  In the HLLC solver, one needs to find the pressure in the intermediate states (the $cL$ and $cR$ states) which satisfies a jump condition. Then, the numerical flux is calculated by (see the left panel of Fig.~\ref{fig:Riemann_fan})
\begin{align}
&\left(\tilde{f}^{(\hat{x})}_A\right)_{j+\frac{1}{2}} \nonumber\\
&=\left\{
\begin{array}{cll}
& (f^{(\hat{x})}_A)_L   & \text{if } \lambda_L > v^{(\hat{x})}_\text{interface}\\
& (f^{(\hat{x})}_A)_{cL} & \text{if } \lambda_L < v^{(\hat{x})}_\text{interface} < \lambda_{c} \\
& (f^{(\hat{x})}_A)_{cR} & \text{if } \lambda_{c} < v^{(\hat{x})}_\text{interface} < \lambda_{R} \\
& (f^{(\hat{x})}_A)_R & \text{if } \lambda_R < v^{(\hat{x})}_\text{interface},\\
\end{array}
\right.\label{eq:num_flux_HLLC}
\end{align}
where
\begin{align}
&\left(f^{(\hat{x})}_A\right)_{L/R}=f^{(\hat{x})}_A\left(q_{L/R}\right),\\
&\left(f^{(\hat{x})}_A\right)_{cL/cR}=\left(f^{(\hat{x})}_A\right)_{L/R}\nonumber \\
&~~~~~~~~~~~~~~~+\lambda_{L/R}\left(\left(q_A\right)_{cL/cR}-\left(q_A\right)_{L/R}\right), \label{eq:flux_HLLC}
\end{align}
and $\lambda_{L/R}$ is the characteristic speed of the left/right-going nonlinear wave. Equation~(\ref{eq:flux_HLLC}) is obtained from the jump condition and $\lambda_c$ is the characteristic speed of the contact discontinuity. By imposing continuity of the pressure across the contact discontinuity, one finds a quadratic equation for $\lambda_c$~\cite{Mignone:2005ft}:
\begin{align}
 F^\text{HLL}_{\rho_{\rm H}} \lambda_c^2 - \left(\rho^\text{HLL}_{\rm H}+F^\text{HLL}_{J_{(\hat{x})}}\right)\lambda_c + J^\text{HLL}_{(\hat{x})}=0,
\end{align}
where $\rho^\text{HLL}_{\rm H}$, $J^\text{HLL}_{(\hat{x})}$, $F^\text{HLL}_{\rho_{\rm H}}$, and $F^\text{HLL}_{J_{(\hat{x})}}$ denote conserved quantities and fluxes in the HLL state:
\begin{align}
 &J^\text{HLL}_{(\hat{x})} = \frac{\lambda_R J_{(\hat{x})}^R - \lambda_L J_{(\hat{x})}^L+f^{(\hat{x})}_{1,L}-f^{(\hat{x})}_{1,R}}{\lambda_R-\lambda_L},\\
 &\rho^\text{HLL}_{\rm H} = \frac{\lambda_R \rho_{\rm H}^R - \lambda_L \rho_{\rm H}^L+f^{(\hat{x})}_{4,L}-f^{(\hat{x})}_{4,R}}{\lambda_R-\lambda_L},\\
 &F^\text{HLL}_{J_{(\hat{x})}} = \frac{\lambda_R f^{(\hat{x})}_{1,L} - \lambda_L 
 f^{(\hat{x})}_{1,R} + \lambda_R \lambda_L \left( J_{(\hat{x})}^R-J_{(\hat{x})}^L
 \right) }{\lambda_R-\lambda_L},\\
  &F^\text{HLL}_{\rho_{\rm H}} = \frac{\lambda_R f^{(\hat{x})}_{4,L} - \lambda_L 
 f^{(\hat{x})}_{4,R} + \lambda_R \lambda_L \left( \rho_{\rm H}^R-\rho_{\rm H}^L
 \right) }{\lambda_R-\lambda_L}.
\end{align}
Once we obtain the speed of the contact discontinuity $\lambda_c$, the pressure in the intermediate state is determined by
\begin{align}
 P_c \equiv P_{cL} = P_{cR} = - \lambda_c F^\text{HLL}_{\rho_{\rm H}} + F^\text{HLL}_{J_{(\hat{x})}}.
\end{align}
Then the conserved quantities in the $cL$ and $cR$ states are given by
\begin{align}
&D_{cL/cR} = \frac{D_{L/R}\left(\lambda_{L/R}-v^{(\hat{x})}_{L/R}\right)}{\lambda_{L/R}-\lambda_c},\\
&\left(J_{(\hat{\imath})}\right)_{cL/cR}=\frac{1}{\lambda_{L/R}-\lambda_c}\nonumber\\
&\times \left[ \left(J_{(\hat{\imath})}\right)_{L/R}\left(\lambda_{L/R}-v^{(\hat{x})}_{L/R}\right)+(P_c-P_{L/R}){\delta^{(\hat{x})}}_{(\hat{\imath})} \right],\\
&\left(\rho_{\rm H}\right)_{cL/cR}=\frac{\left(\rho_{\rm H}\right)_{L/R}\left(\lambda_{L/R}-v^{(\hat{x})}_{L/R}\right)+P_c\lambda_c-P_{L/R}v^{(\hat{x})}_{L/R}}{\lambda_{L/R}-\lambda_c},
\end{align}
where the subscripts $cL$ and $cR$ on the left-hand side of the equations correspond to $L$ and $R$ on the right-hand side, respectively. These quantities can be used to evaluate the flux in the $cL/cR$ state (\ref{eq:flux_HLLC}) and the flux in the Eulerian frame (see, e.g., Eq.~(\ref{eq:num_flux})). 

For the left and right characteristic speeds $\lambda_{L/R}$, we apply Davis's estimate~\cite{Mignone:2005ft}:
\begin{align}
 &\lambda_L = \min(\lambda^-(q_L),\lambda^-(q_R)),\\
 &\lambda_R = \max(\lambda^+(q_L),\lambda^+(q_R)),
\end{align}
and
\begin{align}
&\lambda^\pm (q_A) = \frac{1}{1-v^2 c_s^2}
\left[v^{(\hat{x})}\left(1-c_s^2\right) \right.\nonumber\\
&~~~~~~~~~~\left.\pm c_s\sqrt{(1-v^2)(1-v^2c_s^2-(1-c_s^2)(v^{(\hat{x})})^2)}\right],
\end{align}
where 
\begin{align}
&v^2 = v^{(\hat{\imath})}v_{(\hat{\imath})},\\
&c_s^2 = \frac{1}{h}\left[\frac{\partial P}{\partial \rho}\Big|_\varepsilon + \frac{P}{\rho^2}\frac{\partial P}{\partial \varepsilon}\Big |_\rho\right].
\end{align}
The equivalent expressions in the $y$- and $z$-directions are given by permutation of the indices $x,y$, and $z$.

\subsection{HLLD solver for relativistic magnetohydrodynamics}\label{sec:HLLD}
In the presence of an electromagnetic field, one choice for the Riemann solver is the HLLD solver proposed in Ref.~\cite{MUB:2009}. For this case, we calculate the HLLD flux, $\left(\tilde{f}^{(\hat{x})}_A\right)_{j+\frac{1}{2},k,l}$ in the tetrad frame by solving the one-dimensional conservation law:
\begin{align}
\partial_{(\hat{t})}q_A+\partial_{(\hat{x})}f^{(\hat{x})}_A = 0,
\end{align}
where
\begin{align}
&q_A=\left(
\begin{array}{c}
   D              \\
   J_{(\hat{\imath})}  \\
   \rho_{\rm H}  \\
   B^{(\hat{\imath})}
\end{array}
\right),\\
&f^{(\hat{x})}_A=\nonumber\\
&\left(
\begin{array}{c}
 D v^{(\hat{x})}\\
 J_{(\hat{\imath})} v^{(\hat{x})} + P_\text{tot}{\delta^{(\hat{x})}}_{(\hat{\imath})} - \frac{B^{(\hat{x})}}{w^2}\left[B_{(\hat{\imath})} + (B^{(\hat{k})} u_{(\hat{k})})u_{(\hat{\imath})}\right]\\
\rho_{\rm H}v^{(\hat{x})} + P_\text{tot}v^{(\hat{x})}-\frac{1}{w}(B^{(\hat{k})}u_{(\hat{k})})B^{(\hat{x})}\\
v^{(\hat{x})} B^{(\hat{\imath})} - v^{(\hat{\imath})} B^{(\hat{x})} \label{eq:MHD-Minkoswki}
\end{array}
\right).
\end{align}
Here, 
$q_A$ has seven components ($A=0, 1, 2, 3, 4, 6, 7$), 
and $P_\text{tot}\equiv P + b^2/2$ is the total pressure (gas plus magnetic). Note that the equation for $B^{(\hat{x})}$ is simply $\partial_{(\hat{t})}B^{(\hat{x})}=0$, and thus, $B^{(\hat{x})}$ is constant for the Riemann problem of the $x$-direction. 
Together with the initial condition given by Eq.~(\ref{eq:initial}) for the relevant components, the full magnetohydrodynamics Riemann fan consists of seven waves separating eight states~\cite{Font:2003}. In the HLLD solver two of these seven waves (the slow magnetosonic waves) are neglected. 
As a result, the Riemann fan with the HLLD solver consists of five waves separating six states (see Fig.~\ref{fig:Riemann_fan}). 
In the HLLD solver, we need to find the total pressure $P_\text{tot}$ which satisfies a jump condition across the five waves. The numerical flux is then given by (see the right-hand panel of Fig.~\ref{fig:Riemann_fan})
\begin{align}
&\left(\tilde{f}^{(\hat{x})}_A\right)_{j+\frac{1}{2}} \nonumber\\
&=\left\{
\begin{array}{lll}
& (f^{(\hat{x})}_A)_L    & \text{if } \lambda_L > v^{(\hat{x})}_\text{interface}\\
& (f^{(\hat{x})}_A)_{aL} & \text{if } \lambda_L < v^{(\hat{x})}_\text{interface} < \lambda_{aL} \\
& (f^{(\hat{x})}_A)_{cL} & \text{if } \lambda_{aL} < v^{(\hat{x})}_\text{interface} <\lambda_c \\
& (f^{(\hat{x})}_A)_{cR} & \text{if } \lambda_c < v^{(\hat{x})}_\text{interface} <\lambda_{aR} \\
& (f^{(\hat{x})}_A)_{aR} & \text{if } \lambda_{aR} < v^{(\hat{x})}_\text{interface} < \lambda_{R} \\
& (f^{(\hat{x})}_A)_R    & \text{if } \lambda_R < v^{(\hat{x})}_\text{interface},\\
\end{array}
\right. \label{eq:num_flux_HLLD}
\end{align}
where 
\begin{align}
\left(f^{(\hat{x})}_A\right)_{L/R} &=f^{(\hat{x})}_A\left(q_{L/R}\right),\\
\left(f^{(\hat{x})}_A\right)_{aL/aR} &= \left(f^{(\hat{x})}_A\right)_{L/R}\nonumber\\
&~+\lambda_{L/R}\left(\left(q_A\right)_{aL/aR}-\left(q_A\right)_{L/R}\right), \label{eq:flux_aL_aR}\\
\left(f^{(\hat{x})}_A\right)_{cL/cR} &= \left(f^{(\hat{x})}_A\right)_{aL/aR}\nonumber\\ 
&~+ \lambda_{aL/aR} \left((q_A)_{cL/cR}-(q_A)_{aL/aR}\right). \label{eq:flux_cL_cR}
\end{align}
The latter two fluxes are obtained from the jump condition. 

In the following subsections, we present specific quantities employed by the HLLD solver: the characteristic speeds of the five waves, and the six states. 


\subsubsection{Characteristic speeds}
For the fast waves, an approximate characteristic speed proposed in Refs.~\cite{Gammie:2003rj,Shibata:2005gp} is given by
\begin{align}
&\lambda_\text{FW}^\pm (q_A) = \frac{1}{1-v^2 \zeta}
\left[v^{(\hat{x})}\left(1-\zeta\right)\right.\nonumber\\
&~~~~~~~~\left.\pm\sqrt{\zeta}\sqrt{(1-v^2)(1-v^2\zeta-(1-\zeta)(v^{(\hat{x})})^2)}\right],
\end{align}
where 
\begin{align}
&v^2 = v^{(\hat{\imath})}v_{(\hat{\imath})},\\
&\zeta = v_A^2 + c_s^2 - v_A^2 c_s^2,\\
&v_A^2 = \frac{b^2}{\rho h + b^2}.
\end{align}
For the Alfv\'{e}n wave, the characteristic speed is given by
\begin{align}
 \lambda_\text{Alf}^\pm(q_A) = \frac{b^{(\hat{x})}\pm u^{(\hat{x})}\sqrt{\rho h + b^2}}{b^{(\hat{t})}\pm u^{(\hat{t})}\sqrt{\rho h + b^2}},
\end{align}
and for the contact wave by
\begin{align}
\lambda_c(q_A)=v^{(\hat{x})}.
\end{align}

\subsubsection{L/R state}\label{subsubsec:L/R}
Given left- and right-state quantities, we first calculate the following quantities which should be preserved when one crosses the fast waves:
\begin{align}
& \left(R_D\right)_{L/R} = \left( \lambda D - f^{(\hat{x})}_0\right)_{L/R},\\
& \left( R_{J_{(\hat{\imath})}} \right)_{L/R} =\left( \lambda J_{(\hat{\imath})} - f^{(\hat{x})}_i\right)_{L/R},\label{eq:RS}\\
& \left(R_{\rho_{\rm H}}\right)_{L/R} = \left(\lambda \rho_{\rm H} - f^{(\hat{x})}_4\right)_{L/R},\\
& \left(R_{B^{(\hat{k})}}\right)_{L/R} =\left( \lambda B^{(\hat{k})} - f^{(\hat{x})}_k\right)_{L/R}, \label{eq:RB}
\end{align}
where $\hat{\imath}=\hat{x},\hat{y},\hat{z}$ for $i=1,2,3$, respectively, in Eq.~(\ref{eq:RS}). Also $\hat{k}=\hat{y},\hat{z}$ for $k=6,7$, respectively, in Eq.~(\ref{eq:RB}). For the above quantities, we employ the characteristic speed defined by
\begin{align}
 &\lambda_L = \min \left(\lambda^-_\text{FW}(q_L),\lambda^-_\text{FW}(q_R)\right),\nonumber\\
 &\lambda_R = \max \left(\lambda^+_\text{FW}(q_L),\lambda^+_\text{FW}(q_R)\right).
\end{align}
\subsubsection{aL/aR state}\label{subsubsec:aL/aR}
Given an initial guess for the unknown total pressure $P_\text{tot}$ (which should be constant inside the Riemann fan), the three velocities in the $aL$ and $aR$ states are given by
\begin{align}
&\left(v^{(\hat{x})}\right)_{aL/aR}\nonumber\\
&=\left(\frac{B^{(\hat{x})}(AB^{(\hat{x})}+\lambda C)-(A+G)(P_\text{tot}+R_{J_{(\hat{x})}})}{X}\right)_{L/R},\label{eq:vx_a}\\
&\left(v^{(\hat{y})}\right)_{aL/aR}\nonumber\\
&=\left(\frac{
Q R_{J_{(\hat{y})}} + R_{B^{(\hat{y})}}[C+B^{(\hat{x})}
(\lambda R_{J_{(\hat{x})}}-R_{\rho_{\rm H}})]}{X}\right)_{L/R},\label{eq:vy_a}\\
&\left(v^{(\hat{z})}\right)_{aL/aR}\nonumber\\
&=\left(\frac{
QR_{J_{(\hat{z})}} + R_{B^{(\hat{z})}}[C+B^{(\hat{x})}
(\lambda R_{J_{(\hat{x})}}-R_{\rho_{\rm H}})]}{X}\right)_{L/R},\label{eq:vz_a}
\end{align}
where
\begin{align}
&A=R_{J_{(\hat{x})}}-\lambda R_{\rho_{\rm H}} + P_\text{tot}(1-\lambda^2),\\
&G=R_{B^{(\hat{y})}}R_{B^{(\hat{y})}}+R_{B^{(\hat{z})}}R_{B^{(\hat{z})}},\\
&C=R_{J_{(\hat{y})}}R_{B^{(\hat{y})}}+R_{J_{(\hat{z})}}R_{B^{(\hat{z})}},\\
&Q=-A-G+(B^{(\hat{x})})^2(1-\lambda^2),\\
&X=B^{(\hat{x})}(A\lambda B^{(\hat{x})}+C)-(A+G)(\lambda P_\text{tot}+R_{\rho_{\rm H}}).
\end{align}
Note that $aL$ and $aR$ on the left-hand side of Eqs.~(\ref{eq:vx_a})--(\ref{eq:vz_a}) correspond\ to $L$ and $R$ for $R_{J_{(\hat{\imath})}}$, $R_{\rho_{\rm H}}$, $R_{B^{(\hat{k})}}$, and $\lambda$ on the right-hand side of the same equations, respectively. With these velocity components, the magnetic field is calculated from the jump condition by 
\begin{align}
\left(B^{(\hat{k})}\right)_{aL/aR}=\frac{\left(R_{B^{(\hat{k})}}\right)_{L/R}-B^{(\hat{x})}\left(v^{(\hat{k})}\right)_{aL/aR}}{\lambda_{L/R}-\left(v^{(\hat{x})}\right)_{aL/aR}}
\end{align}
for $\hat{k}=\hat{y},\hat{z}$. The total enthalpy density is calculated by
\begin{align}
&(\rho h_\text{tot})_{aL/aR} \equiv (\rho h + b^2 )_{aL/aR} \nonumber\\
&= P_\text{tot}+\frac{\left(R_{\rho_{\rm H}}\right)_{L/R}-\left(v^{(\hat{\imath})}\right)_{aL/aR}\left(R_{J_{(\hat{\imath})}}\right)_{L/R}}{\lambda_{L/R}-\left(v^{(\hat{x})}\right)_{aL/aR}}.
\end{align} 
The conserved quantities necessary for the numerical flux in Eq.~(\ref{eq:flux_aL_aR}) and in the Eulerian frame (see Eqs.~(\ref{eq:num_flux})--(\ref{eq:num_flux3})) are calculated by
\begin{align}
&D_{aL/aR} = \frac{\left(R_D\right)_{L/R}}{\lambda_{L/R}-\left(v^{(\hat{x})}\right)_{aL/aR}},\\
&\left(\rho_{\rm H}\right)_{aL/aR}=\nonumber\\
& \frac{\left(R_{\rho_{\rm H}}\right)_{L/R}+P_\text{tot}\left(v^{(\hat{x})}\right)_{aL/aR}-\left(v^{(\hat{k})}B^{(\hat{k})}\right)_{aL/aR}B^{(\hat{x})}}{\lambda_{L/R}-\left(v^{(\hat{x})}\right)_{aL/aR}},\\
&\left(J_{(\hat{\imath})}\right)_{aL/aR}=
 \left((\rho_{\rm H}+P_\text{tot})v^{(\hat{\imath})}-(v^{(\hat{k})}B^{(\hat{k})})B^{(\hat{\imath})}\right)_{aL/aR}.
\end{align}
\subsubsection{cL/cR state}
Following Ref.~\cite{MUB:2009}, we first define 
\begin{align}
&\sigma^{(\hat{\mu})} = \eta u^{(\hat{\mu})} + b^{(\hat{\mu})},\\
&\eta = \pm \text{sgn}(B^{(\hat{x})})\sqrt{\rho h_\text{tot}},
\end{align}
where the plus (minus) sign corresponds to the right (left) state. We then define $K^{(\hat{k})}$ by
\begin{align}
K^{(\hat{k})}\equiv \frac{\sigma^{(\hat{k})}}{\sigma^{(\hat{0})}}
=v^{(\hat{k})}+\frac{B^{(\hat{k})}}{w \sigma^{(\hat{0})}}.
\end{align}
Here $K^{(\hat{x})}$ is nothing other than the Alfv\'{e}n wave speed in the $x$-direction.
From the jump condition one can find that $K^{(\hat{\imath})}$, $\rho h_\text{tot}$, $D/w\sigma^{(\hat{0})}$, and $\eta$ do not change across the Alfv\'{e}n waves. Therefore, $\eta$, $K^{(\hat{\imath})}$, and the total enthalpy density are calculated by
\begin{align}
& \eta_{cL/cR} = \eta_{aL/aR},\\
 &(K^{(\hat{\imath})})_{cL/cR} =     (K^{(\hat{\imath})})_{aL/aR} \nonumber\\
 &=
 \frac{\left(R_{J_{(\hat{\imath})}}\right)_{L/R}+P_\text{tot}\delta_{(\hat{\imath})(\hat{x})}+\left(R_{B^{(\hat{i)}}}\right)_{L/R}\eta_{aL/aR}}{\lambda_{L/R} P_\text{tot} + \left(R_{\rho_{\rm H}}\right)_{L/R} + B^{(\hat{x})}\eta_{aL/aR}},\\
& (\rho h_\text{tot})_{cL/cR} = (\rho h_\text{tot})_{aL/aR},
\end{align}
where $cL$ and $cR$ on the left-hand side of the equations correspond to $aL$ and $aR$ on the right-hand side of the same equations, respectively. 

The magnetic field and the three velocity in the $cL$ and $cR$ states are calculated by
\begin{align}
& (B^{(\hat{k})})_{cL} =  (B^{(\hat{k})})_{cR}  \nonumber\\ 
& = \Big[\left\{B^{(\hat{k})}
(\lambda-v^{(\hat{x})})
+B^{(\hat{x})}v^{(\hat{k})}\right\}_{aR}\nonumber\\
&-\left\{B^{(\hat{k})}
(\lambda-v^{(\hat{x})})+B^{(\hat{x})}v^{(\hat{k})}\right\}_{aL}\Big]
\frac{1}{\lambda_{aR}-\lambda_{aL}},\\
&(v^{(\hat{\imath})})_{cL/cR} = \left(K^{(\hat{\imath})}-\frac{B^{(\hat{\imath})}(1-K^{(\hat{k})}K^{(\hat{k})})}{\eta-K^{(\hat{l})}B^{(\hat{l})}}\right)_{cL/cR},
\end{align}
and the characteristic speed is
\begin{align}
 \lambda_{aL/aR} = K^{(\hat{x})}_{aL/aR}.
\end{align}
We impose the continuity condition on the normal velocity across the contact discontinuity, i.e., $v^{(\hat{x})}_{cL}=v^{(\hat{x})}_{cR}$, by
\begin{align}
 &\Delta K^{(\hat{x})}\left[1-B^{(\hat{x})}\left(Y_R-Y_L\right)\right]=0, \label{eq:non_linear_HLLD}\\
 &Y_{L/R} = \left(\frac{1-K^{(\hat{\imath})}K^{(\hat{\imath})}}{\eta \Delta K^{(\hat{x})} - \Delta K^{(\hat{x})}K^{(\hat{j})}B^{(\hat{j})}}\right)_{cL/cR}, 
\end{align}
where $\Delta K^{(\hat{x})}=K^{(\hat{x})}_{aR}-K^{(\hat{x})}_{aL}$. This equation gives an improved guess of the total pressure in the next iteration step. Then we go back to Eq.~(\ref{eq:vx_a}) and repeat the same procedure until it converges with sufficient accuracy. In practice, we employ the Newton-Raphson method to solve Eq.~(\ref{eq:non_linear_HLLD}).

The conserved quantities necessary for the numerical flux in Eq.~(\ref{eq:flux_cL_cR}) and in the Eulerian frame (see Eqs.~(\ref{eq:num_flux})--(\ref{eq:num_flux3})) are
\begin{align}
&D_{cL/cR} = D_{aL/aR} \frac{\lambda_{aL/aR}-v^{(\hat{x})}_{aL/aR}}{\lambda_{aL/aR}-v^{(\hat{x})}_{cL/cR}},\\
&\left(\rho_{\rm H}\right)_{cL/cR}=\frac{1}{\lambda_{aL/aR}-v^{(\hat{x})}_{cL/cR}}\nonumber\\
&\times\Big[
\lambda_{aL/aR} \left(\rho_{{\rm H}0}\right)_{aL/aR} - \left(J_{(\hat{x})}\right)_{aL/aR} + P_\text{tot}v^{(\hat{x})}_{cL/cR} \nonumber\\ 
&~~~~~-(v^{(\hat{\imath})} B^{(\hat{\imath})})_{cL/cR}B^{(\hat{x})}
\Big],\\
&\left(J_{(\hat{\imath})}\right)_{cL/cR} = \left(\left(\rho_{{\rm H}}\right)_{cL/cR}+P_\text{tot}\right)v^{(\hat{\imath})}_{cL/cR}\nonumber\\
&~~~~~~~~~~~~~~~~~-(v^{(\hat{k})}B^{(\hat{k})})_{cL/cR}B^{(\hat{\imath})}_{cL/cR}.
\end{align}

The equivalent expressions in the $y$- and $z$-directions are given by permutation of the indices $x,y$, and $z$.

\subsection{Electric-field evaluation}
\label{sec:electricfieldevaluation}
The constrained transport method used to enforce the divergence-free condition on the magnetic field requires us to evaluate the electric field defined at the cell edges. Gardiner and Stone~\cite{Gardiner:2007nc} proposed a method for evaluating the electric-field by utilizing the numerical fluxes which are obtained by the Riemann solver. In their method, for example, the $z$-component of the electric field is evaluated by
\begin{align}
&\tilde{E}^{z}_{j+\frac{1}{2},k+\frac{1}{2},l} = \frac{1}{4}\Big(\tilde{E}^z_{j+\frac{1}{2},k,l}+\tilde{E}^z_{j+\frac{1}{2},k+1,l}\nonumber\\
&~~~~~~~~~~~~~~~~~~
+\tilde{E}^z_{j,k+\frac{1}{2},l}+\tilde{E}^z_{j+1,k+\frac{1}{2},l}\Big)\nonumber\\
&+\frac{\Delta y}{8}\left(\left(\frac{\partial E^z}{\partial y}\right)_{j+\frac{1}{2},k+\frac{1}{4},l}-\left(\frac{\partial E^z}{\partial y}\right)_{j+\frac{1}{2},k+\frac{3}{4},l}\right)\nonumber\\
&+\frac{\Delta x}{8}\left(
\left(\frac{\partial E^z}{\partial x}\right)_{j+\frac{1}{4},k+\frac{1}{2},l}-\left(\frac{\partial E^z}{\partial x}\right)_{j+\frac{3}{4},k+\frac{1}{2},l}
\right) \label{eq:CT_GS}
\end{align}
where
\begin{align*}
&\left(\frac{\partial E^z}{\partial y}\right)_{j+\frac{1}{2},k+\frac{1}{4},l}\nonumber\\
&=\left\{
\begin{array}{l}
\frac{2\left(\tilde{E}^z_{j,k+\frac{1}{2},l}-E^z_{j,k,l}\right)}{\Delta y}\text{ for }\tilde{v}^x_{j+\frac{1}{2},k,l} > 0,\\
\frac{2\left(\tilde{E}^z_{j+1,k+\frac{1}{2},l}-E^z_{j+1,k,l}\right)}{\Delta y}\text{ for }\tilde{v}^x_{j+\frac{1}{2},k,l} < 0,\\
\frac{\left(\tilde{E}^z_{j,k+\frac{1}{2},l}-E^z_{j,k,l}
+\tilde{E}^z_{j+1,k+\frac{1}{2},l}-E^z_{j+1,k,l}\right)}{\Delta y}\text{ otherwise,}\\
\end{array}
\right.\nonumber\\
\end{align*}
\begin{align*}
&\left(\frac{\partial E^z}{\partial y}\right)_{j+\frac{1}{2},k+\frac{3}{4},l}\nonumber\\
&=\left\{
\begin{array}{l}
\frac{2\left(E^z_{j,k+1,l}-\tilde{E}^z_{j,k+\frac{1}{2},l}\right)}{\Delta y}\text{ for }\tilde{v}^x_{j+\frac{1}{2},k+1,l} > 0,\\
\frac{2\left(E^z_{j+1,k+1,l}-\tilde{E}^z_{j+1,k+\frac{1}{2},l}\right)}{\Delta y}\text{ for }\tilde{v}^x_{j+\frac{1}{2},k+1,l} < 0,\\
\frac{\left(E^z_{j,k+1,l}-\tilde{E}^z_{j,k+\frac{1}{2},l}+E^z_{j+1,k+1,l}-\tilde{E}^z_{j+1,k+\frac{1}{2},l}\right)}{\Delta y}\text{ otherwise,}\\
\end{array}
\right.\nonumber\\
\end{align*}
\begin{align*}
&\left(\frac{\partial E^z}{\partial x}\right)_{j+\frac{1}{4},k+\frac{1}{2},l}\nonumber\\
&=\left\{
\begin{array}{l}
\frac{2\left(\tilde{E}^z_{j+\frac{1}{2},k,l}-E^z_{j,k,l}\right)}{\Delta x}\text{ for }\tilde{v}^y_{j,k+\frac{1}{2},l} > 0,\\
\frac{2 \left(\tilde{E}^z_{j+\frac{1}{2},k+1,l}-E^z_{j,k+1,l}\right)}{\Delta x}\text{ for }\tilde{v}^y_{j,k+\frac{1}{2},l} < 0,\\
\frac{\left(\tilde{E}^z_{j+\frac{1}{2},k,l}-E^z_{j,k,l}+\tilde{E}^z_{j+\frac{1}{2},k+1,l}-E^z_{j,k+1,l}\right)}{\Delta x}\text{ otherwise,}\\
\end{array}
\right.\nonumber\\
\end{align*}
\begin{align*}
&\left(\frac{\partial E^z}{\partial x}\right)_{j+\frac{3}{4},k+\frac{1}{2},l}\nonumber\\
&=\left\{
\begin{array}{l}
\frac{2\left(E^z_{j+1,k,l}-\tilde{E}^z_{j+\frac{1}{2},k,l}\right)}{\Delta x}\text{ for }\tilde{v}^y_{j+1,k+\frac{1}{2},l} > 0,\\
\frac{2\left(E^z_{j+1,k+1,l}-\tilde{E}^z_{j+\frac{1}{2},k+1,l}\right)}{\Delta x} \text{ for }\tilde{v}^y_{j+1,k+\frac{1}{2},l} < 0,\\
\frac{\left(E^z_{j+1,k,l}-\tilde{E}^z_{j+\frac{1}{2},k,l}+E^z_{j+1,k+1,l}-\tilde{E}^z_{j+\frac{1}{2},k+1,l}\right)}{\Delta x}\text{ otherwise.}\\
\end{array}
\right.
.\nonumber\\
\end{align*}
Here $\tilde{v}^x_{j+\frac{1}{2},k,l}$ and $\tilde{E}^z_{j+\frac{1}{2},k,l}$ are identical to the fluxes
$\left(\tilde{F}^x_0\right)_{j+\frac{1}{2},k,l}$ and $\left(-\tilde{F}^x_6\right)_{j+\frac{1}{2},k,l}$ in Eqs.~(\ref{eq:num_flux}) and (\ref{eq:num_flux2}), which are given by the Riemann solver in the $x$-direction. 
Similarly, $\tilde{v}^y_{j,k+\frac{1}{2},l}$ and $\tilde{E}^z_{j,k+\frac{1}{2},l}$ are given by the Riemann solver in the $y$-direction. $E^z_{j,k,l}$ is calculated from the quantities defined at the cell center, i.e., Eqs.~(\ref{eq:Bcellcenter1})--(\ref{eq:Bcellcenter3}) and the three velocity. Therefore, the accuracy of this constrained transport scheme depends on the accuracy of an employed Riemann solver. Equivalent expressions for the $x$- and $y$-components of the electric field are given by permutation of the indices $x,y,$ and $z$. These electric fields are used to update the magnetic field in Eqs.~(\ref{eq:CT1})--(\ref{eq:CT3}).

\begin{table*}
\caption{Initial conditions used for special relativistic one-dimensional test problems. The third column shows the $\Gamma$ index and the second-to-last column shows the final time of the simulations, $t$.} \label{tab:1D_test_problem}
\begin{tabular}{lcccccccccccc}
\hline\hline
Test problem & State & $\Gamma$ & $\rho$ &  $v^{x}$ & 
$v^{y}$ & $v^{z}$ &
~~~$P$~~~ & ~~~$B^{x}$~~~ & ~~~$B^{y}$~~~ & $B^{z}$ & ~~~~~$t$~~~~~ & ~~~CFL~~~\\
\hline
Problem HD1 & L & 4/3 & 1  & 0.9  & 0  & 0 & 1         & -- & -- & -- & 0.4 & 0.8\\
            & R &     & 1  & 0    & 0  & 0 & 10        & -- & -- & -- &  & \\
Problem HD2 & L & 5/3 & 1  & $-0.6$ & 0  & 0 & 10        & -- & -- & -- & 0.4 & 0.8\\
            & R &     & 10 & 0.5  & 0  & 0 & 20        & -- & -- & -- &  & \\
Problem HD3 & L & 5/3 & 10 & 0    & 0  & 0 & 40        & -- & -- & -- & 0.4 & 0.8\\
            & R &     & 1  & 0    & 0  & 0 & 3         & -- & -- & -- &  & \\
Problem HD4 & L & 5/3 & 1  & 0    & 0  & 0 & $10^3$    & -- & -- & -- & 0.4 & 0.8\\
            & T &     & 1  & 0    & 0  & 0 & $10^{-2}$ & -- & -- & -- & 0.4 & 0.8\\
Problem MHD1 & L & 5/3 & 10 & 0 & 0.7 & 0.2 & 1 & 5 & 1 & 0.5 & 1 & 0.8\\
             & R &    & 1 & 0 & 0.7 & 0.2 & 1 & 5 & 1 & 0.5 &   & \\
Problem MHD2 & L & 5/3 & 1 & 0.4 & $-0.3$ & 0.5 & 1 & 2.4 & 1 & $-1.6$ & 1 & 0.8\\
             & R &     & 1 &0.377347 & $-0.482389$ & 0.424190 &  1 & 2.4 & $-0.1$ & $-2.1728213$ &  & \\
Problem MHD3 & L & 2 & 1 & 0 & 0 & 0 & 1 & 0.5 & 1 & 0 & 0.4 & 0.8\\
             & R &   & 0.125 & 0 & 0 & 0 & 0.1 & 0.5 & $-1$ & 0 &  & \\
Problem MHD4 & L & 5/3 & 1.08 & 0.4  &  0.3 & 0.2 & 0.95 &  2 &  0.3 &  0.3 & 0.55 & 0.8\\
             & R &     & 1    & $-0.45$ & $-0.2$ & 0.2 & 1    & 2 & $-0.7$ & $-0.5$ &      & \\
Problem MHD5 & L & 5/3 & 1 & 0.999 & 0 & 0 & 0.1 & 10 &  7 & 7  & 0.4 & 0.8\\
             & R &     & 1 & $-0.999$ &0 & 0 & 0.1 & 10 & $-7$ & $-7$ &      & \\
Problem MHD6 & L & 5/3 & 1   & 0 & 0.3 & 0.4 & 5   & 1 & 6 & 2  & 0.5 & 0.8\\
             & R &     & 0.9 & 0 & 0   & 0   & 5.3 & 1 & 5 & 2 &      & \\
\hline
\hline
\end{tabular}
\end{table*}

In the rest of this paper, we refer to this particular algorithm for evaluating the electric field as {\tt CT\_GS}. On the other hand, the electric-field evaluation algorithm which was originally implemented in our code, and which is based on HLLE~\cite{DelZanna:2002rv,Shibata:2005gp}, is referred to as {\tt CT\_HLLE}~\cite{Shibata:2005gp}. For the base Riemann solver, we use either {\tt HLLC}, {\tt HLLD}, or {\tt HLLE}. Here the last one is the base Riemann solver which was originally implemented in our code~\cite{Shibata:2005gp}. In the hydrodynamics test simulations shown in the next section, we refer to the particular combination of numerical schemes used in a particular test problem in terms of the base solver, only. In the magnetohydrodynamics test simulations, we describe a simulation both in terms of the base Riemann solver and in terms of the algorithm used for the evaluation of the electric-field. For example, {\tt HLLD-CT\_GS} means that the (base) Riemann solver is {\tt HLLD} and the electric-field evaluation is {\tt CT\_GS}.

\section{Validation of the HLLC and HLLD solvers} \label{sec:validation}
In this section, we introduce various problems designed to test the implementation of the advanced Riemann solvers and constrained transport algorithm discussed in the previous section. We start with a common suite of one-dimensional special relativistic shock-tube problems in both hydrodynamics and magnetohydrodynamics (see Sec.~\ref{SECTION_SR1DTest}). Next, in Sec.~\ref{SECTION_SR2D3DTest}, we turn our attention to multi-dimensional hydrodynamics and magnetohydrodynamics test problems in special relativity (specifically, we consider a two-dimensional hydrodynamical shock, a cylindrical hydrodynamical blast wave, a magnetohydrodynamical current sheet, and the Kelvin-Helmholtz instability in magnetohydrodynamics). In Sec.~\ref{SECTION_GRFixedTest} we consider Bondi flow onto a black hole (in both hydrodynamics and magnetohydrodynamics) as a test problem in general relativity with a static spacetime. 

For all the test problems we assume a $\Gamma$-law equation of state given by
\begin{align}
 P = \left(\Gamma-1\right)\rho\varepsilon.
\end{align}
We also employ a cell-centered grid structure in which the $x$-coordinate~\footnote{
In Ref.~\cite{Kiuchi:2012qv}, we employed a vertex-centered grid structure. We updated the interpolation scheme of the metric and fluid at the refinement boundary for cell-centered grid structure for a simulation in a dynamical spacetime.} is given by
\begin{align}
 x_j = \left( j + \frac{1}{2} \right) \Delta x,
\end{align}
with $j\in[-N_x-1, N_x]$ and grid spacing $\Delta x$ (and likewise for the $y$- and $z$-components). As a time integrator, we employ the fourth-order Runge-Kutta method (RK4) in all our test simulations. For reconstruction of the solution at cell-interfaces, we employ either 1st-order reconstruction or 3rd-order piecewise parabolic method (PPM) \citep{Shibata:2005gp,1984JCoPh..54..174C}. For the PPM reconstruction, we employ the min-mod limiter function with a compression parameter which is generally set to $b=2$~\cite{Shibata:2005gp}, though in some cases we employ different values of $b$.

\begin{figure*}[t]
 	 \includegraphics[width=0.48\linewidth]{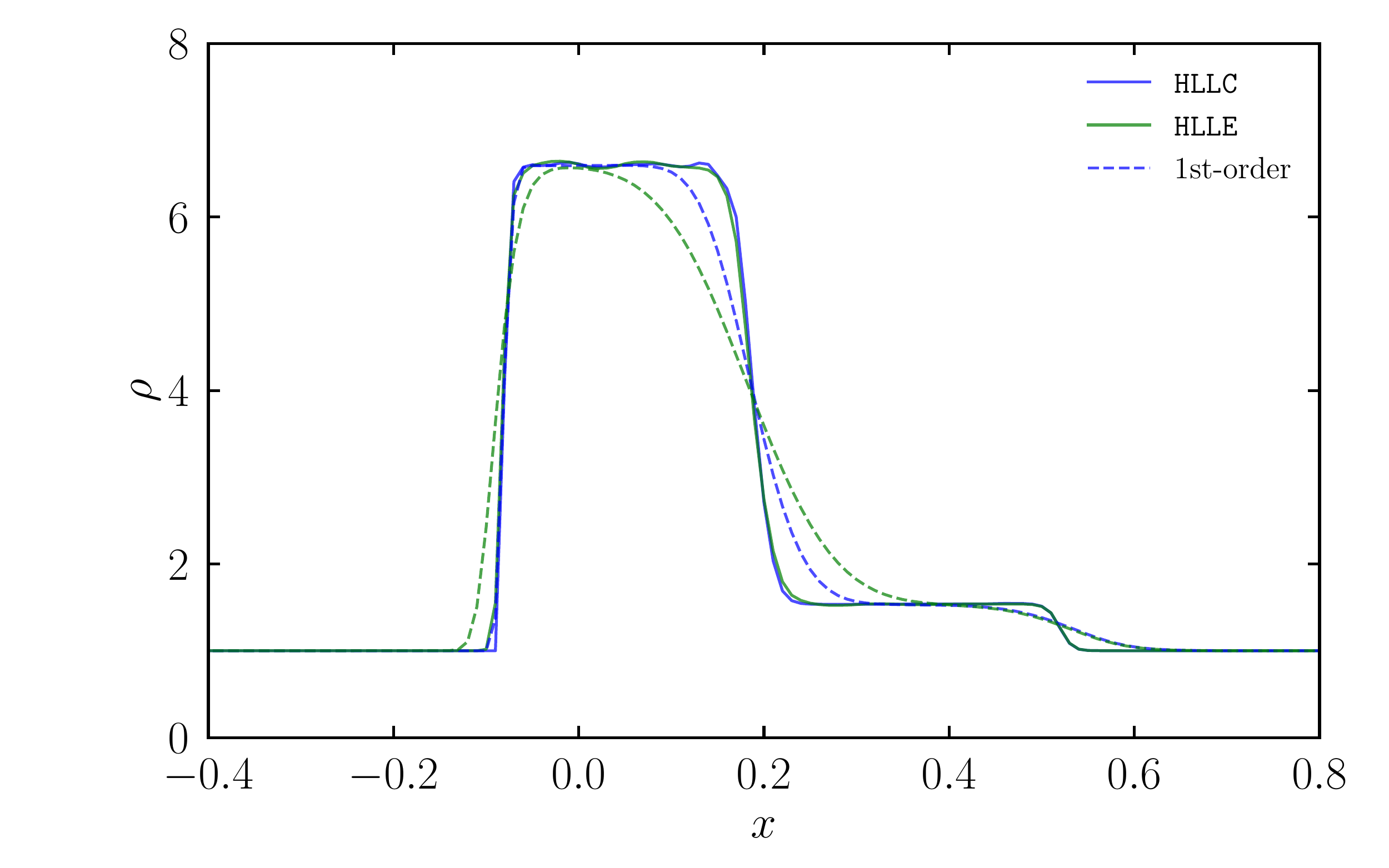}~~
  	 \includegraphics[width=0.48\linewidth]{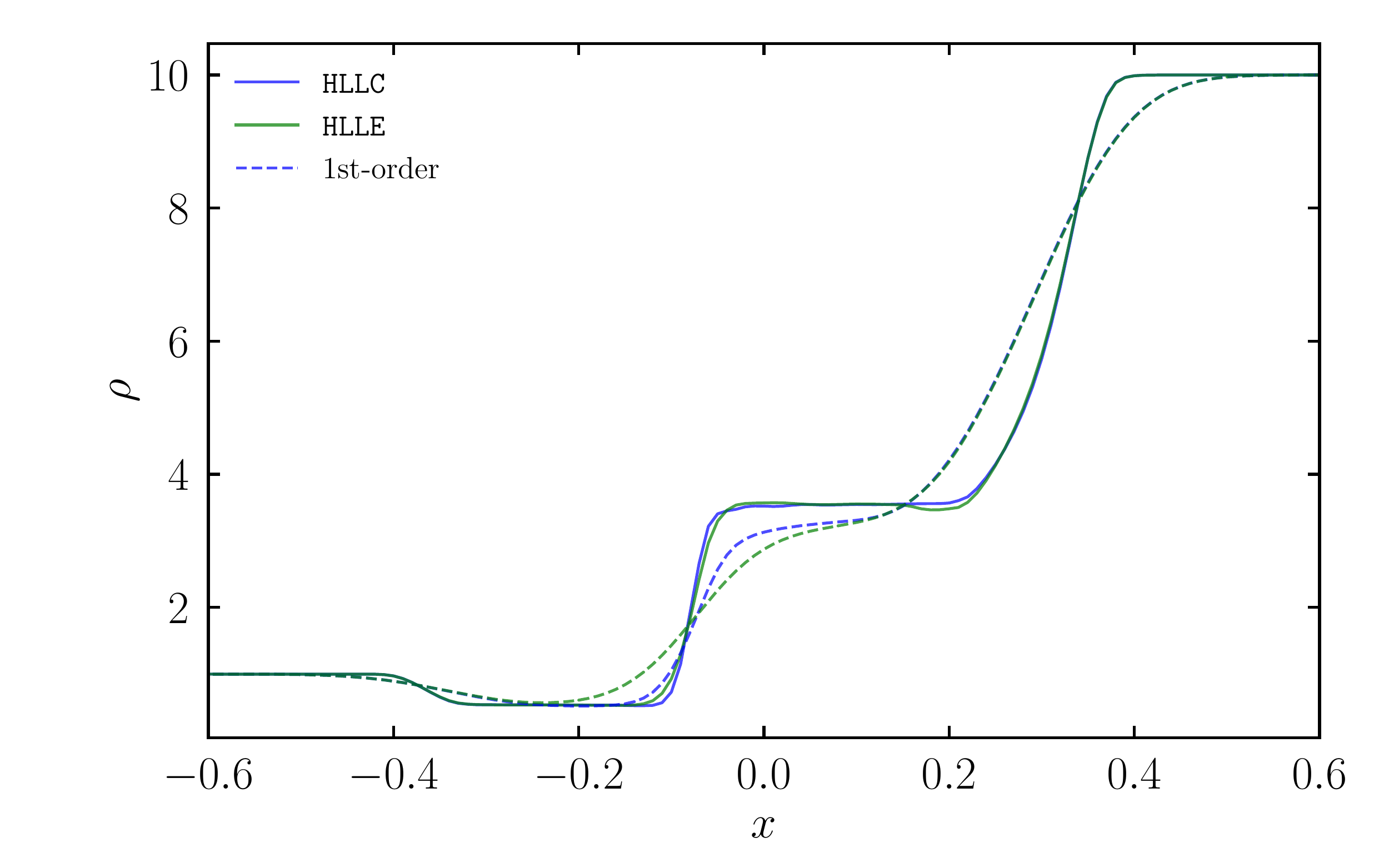}	 
 	 \caption{Left: Rest-mass density profile for Problem HD1 at $t=0.4$ (contact wave located between left- and right-propagating shock waves). The blue and green curves show the results with the {\tt HLLC} and {\tt HLLE} solvers, respectively. The dashed and solid curves show the results with 1st-order reconstruction and 3rd-order (PPM) reconstruction, respectively. Right: Rest-mass density profile for Problem HD2 at $t=0.4$ (contact wave located between left- and right-propagating rarefaction waves). We employ RK4 for the time integration in all the simulations. The blue and green solid curves are indistinguishable on the scale of this plot.
         }\label{fig:PHD1_PHD2}
\end{figure*}

\subsection{Special relativistic one-dimensional problems}
\label{SECTION_SR1DTest}
First, we consider special relativistic problems in one spatial dimension. With this setup, the tetrad basis in Sec.~\ref{subsec:tetrad} is reduced to a coordinate vector in Minkowski spacetime. Thus, the setup is suitable for validating the Riemann solvers described in Sec.~\ref{sec:HLLC} and \ref{sec:HLLD}. 
We assume Minkowski metric, and thus turn off the solver for Einstein's equations in the code. The initial conditions for all the one-dimensional test problems are summarized in Table~\ref{tab:1D_test_problem}. We note that the test suite employed in this paper is the same as that presented in Refs.~\cite{MUB:2009,Mattia:2021bwh}.

\begin{figure*}[t]
 	 \includegraphics[width=0.48\linewidth]{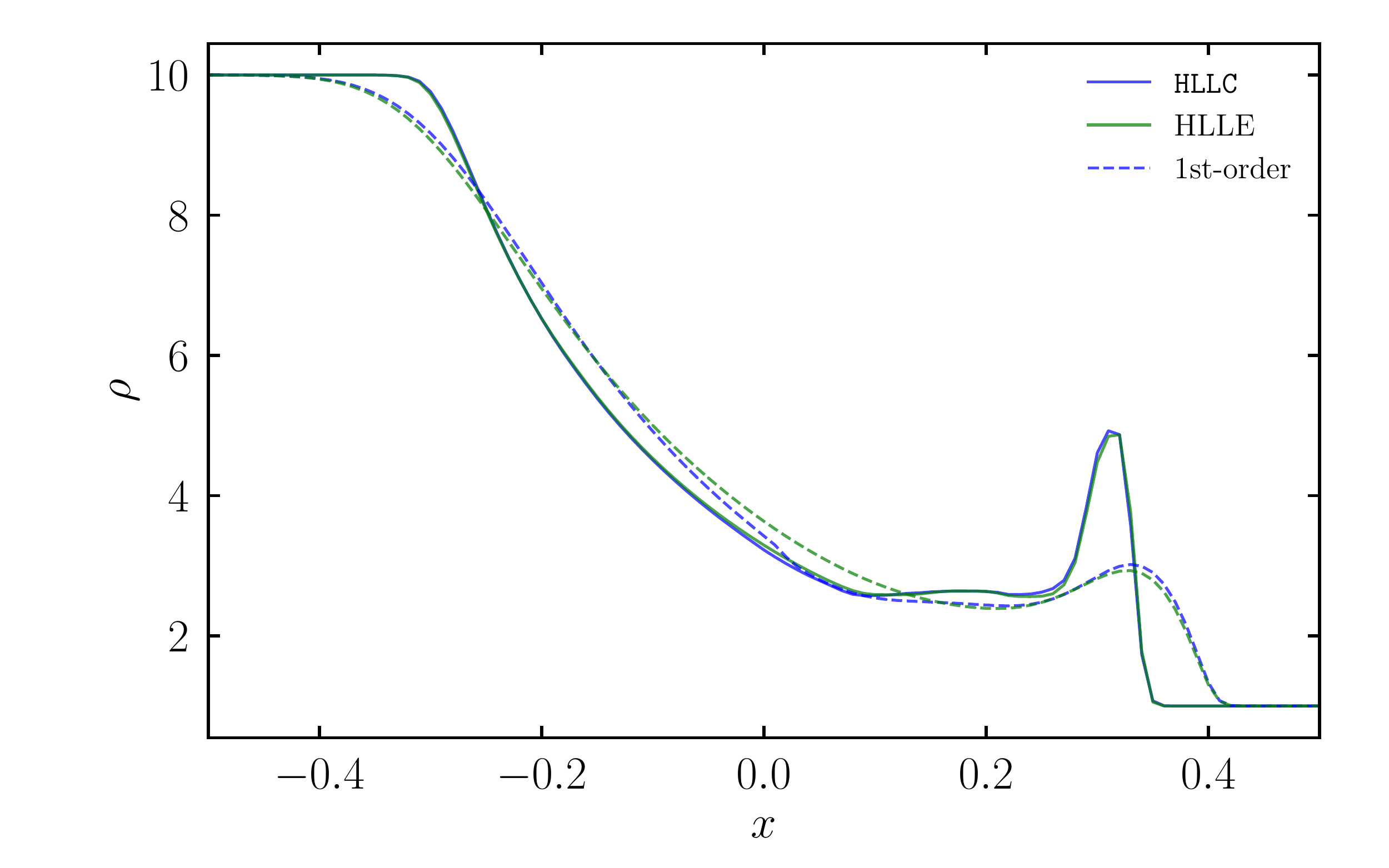}~~
  	 \includegraphics[width=0.48\linewidth]{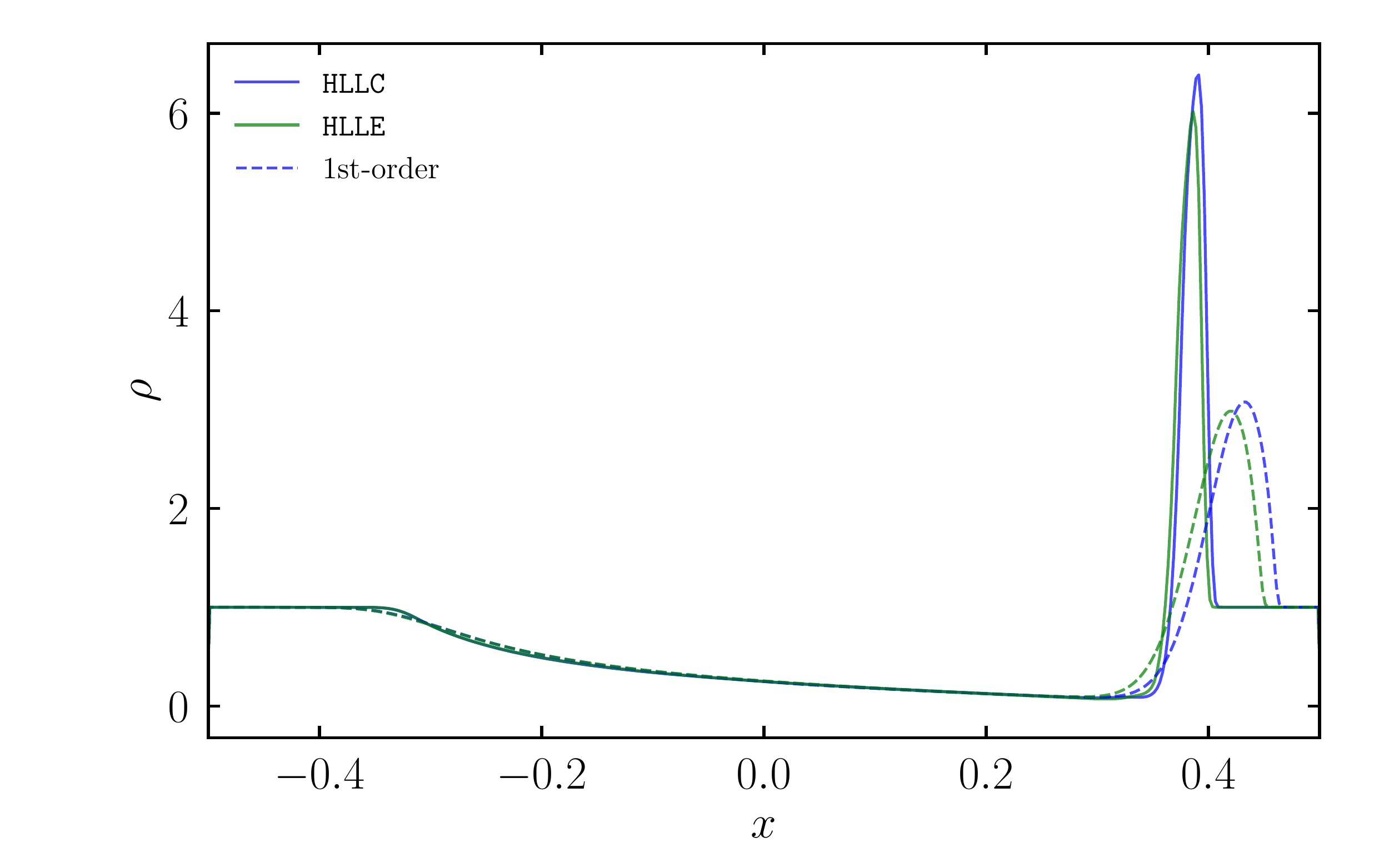}	 
 	 \caption{Same as Fig.~\ref{fig:PHD1_PHD2}, but for Problem HD3 (left) and Problem HD4 (right).
         }\label{fig:PHD3_PHD4}
\end{figure*}
\subsubsection{Hydrodynamics: one-dimensional shock tubes}
The first special relativistic hydrodynamics test (HD1) is the computation of a contact discontinuity. For this we prepare a simulation domain of $x\in[-1,1]$ with $\Delta x =0.01$ and $N_x=100$. We integrate the numerical solution up to $t=0.4$. In the left panel of Fig.~\ref{fig:PHD1_PHD2} we plot the rest-mass density profile at the end of the simulation. In this problem, left- and right-propagating shock waves appear from the initial discontinuity, with a contact discontinuity sandwiched between them. The blue and green curves denote the numerical solution with the {\tt HLLC} and {\tt HLLE} solvers, respectively. The solid and dashed curves denote the simulation results with 3rd-order PPM reconstruction and 1st-order reconstruction, respectively. First, we consider the results obtained with 1st-order reconstruction (dashed curves). With the {\tt HLLC} solver, the contact discontinuity located at $x\approx 0.2$ is more sharply captured than with the {\tt HLLE} solver. This behavior is expected since the {\tt HLLC} solver explicitly restores the contact wave inside the Riemann fan. When we employ 3rd-order reconstruction, however, we find that there is no qualitative difference between the two solvers. This suggests that the weak point of a particular solver may be alleviated by using a high enough resolution. 

\begin{figure*}[t]
 	 \includegraphics[width=0.48\linewidth]{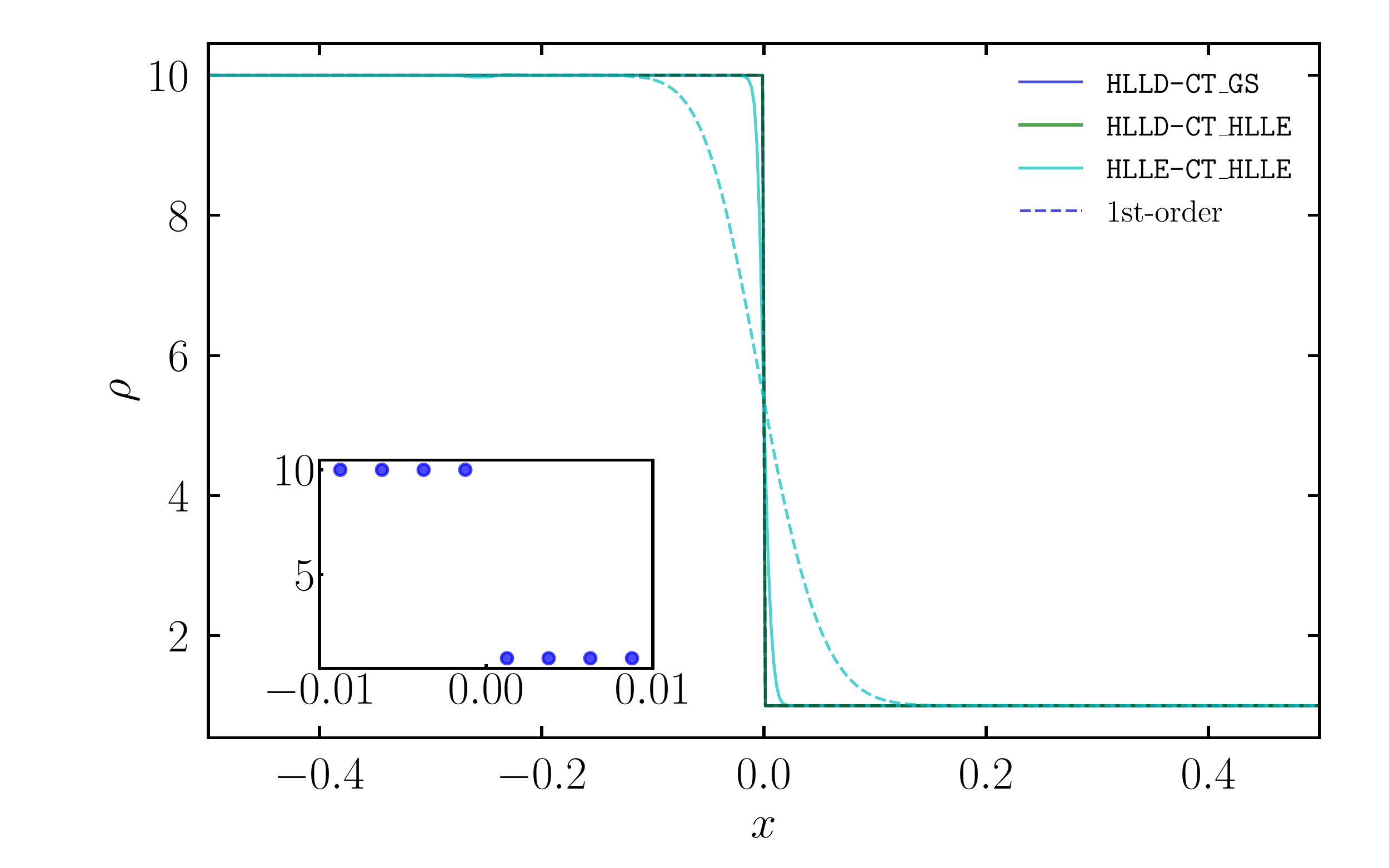}~~
  	 \includegraphics[width=0.48\linewidth]{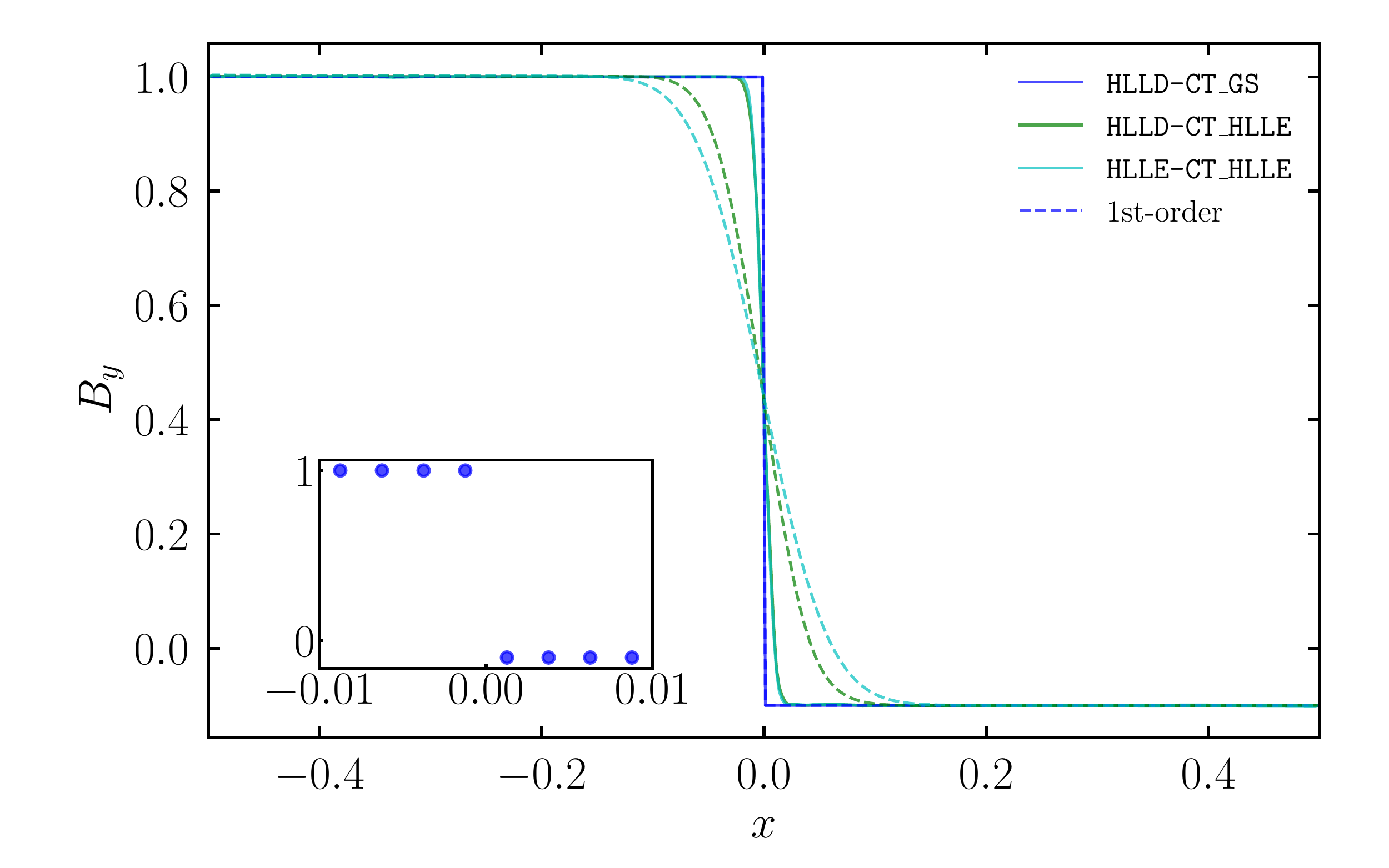}	 
 	 \caption{Left: Rest-mass density profile in Problem MHD1 (a problem with a stationary contact discontinuity) at $t=1$. The blue, green, and cyan curves present the result with the {\tt HLLD-CT\_GS}, {\tt HLLD-CT\_HLLE}, and {\tt HLLE-CT\_HLLE} solvers, respectively. The solid and dashed curves show the results with 3rd-order PPM cell reconstruction and 1st-order cell reconstruction, respectively. Right: Same as the left panel, but for the $B^y$ profile in Problem MHD2 (a problem with a stationary rotational discontinuity). 
 	  The insets are a close-up of the discontinuity with the {\tt HLLD-CT\_GS} solver and 1st-order cell reconstruction.
         }\label{fig:PMHD1_PMHD2}
\end{figure*}


The right panel of Fig.~\ref{fig:PHD1_PHD2} shows the rest-mass density profile for the second test problem (Problem HD2) listed in Table~\ref{tab:1D_test_problem}. The simulation domain and the grid spacing are the same as those in Problem HD1. In this problem, left- and right-going rarefaction waves propagate away from the initial discontinuity, and a contact discontinuity appears between the two and is located at $x\approx -0.1$. As in our first test problem, we find that the contact discontinuity is more sharply captured with the {\tt HLLC} solver than that with the {\tt HLLE} solver when 1st-order reconstruction is used, while we find no qualitative difference between the numerical solutions obtained with the two solvers when we employ 3rd-order-accurate reconstruction. 

The third hydrodynamics test problem (Problem HD3 in Table~\ref{tab:1D_test_problem}) is the often-employed shock-tube problem. Here, the simulation domain spans $x\in[-0.5,0.5]$ with $\Delta x =0.005$ and $N_x=100$. In this problem, the initial discontinuity decays into a left-propagating rarefaction wave and a right-propagating shock wave. The contact discontinuity adjacent to the shock wave also propagates to the right. The left panel of Fig.~\ref{fig:PHD3_PHD4} shows the rest-mass density profile at the end of the simulation for which the contact discontinuity is located at $x\approx 0.25$. In this problem, we find that there is no qualitative difference between the numerical solutions with the two solvers irrespective of the cell reconstruction accuracy. This behaviour is also reported in Ref.~\cite{Mignone:2005ft}. For obtaining an accurate result for this particular shock-tube problem it is necessary to employ an accurate reconstruction method. This suggests that employing an accurate reconstruction method is as important as employing an accurate solver in numerical hydrodynamics at least in the one-dimensional problems.

For the fourth (final) hydrodynamics test problem (Problem HD4), we employ a simulation domain of $x\in[-0.5,0.5]$ with a grid spacing of $\Delta x=0.0025$, i.e., $N_x=200$. The solution consists of a left-propagating rarefaction wave and a right-propagating shock wave. Note that the result differs from that in Problem HD3 as the shock is much stronger compared to the one in Problem HD3 because of the initial large pressure jump (see Table I). A right-propagating contact discontinuity appears adjacent to the shock wave. We plot the rest-mass density profile at $t=0.4$ in the right panel of Fig.~\ref{fig:PHD3_PHD4}.
We find that the contact discontinuity (located at $x\approx 0.35$) is more sharply resolved with the {\tt HLLC} solver than with the {\tt HLLE} solver when we employ 3rd-order reconstruction. We find that the compression parameter $b$ for the min-mod function in the PPM cell reconstruction should be reduced to be 1 in this problem (i.e., a steep limiter function does not work; see, e.g., Ref.~\cite{Shibata:2005gp}). Otherwise, spurious waves appear irrespective of which solver is used (not shown). 

\begin{figure*}[t]
 	 \includegraphics[width=0.46\linewidth]{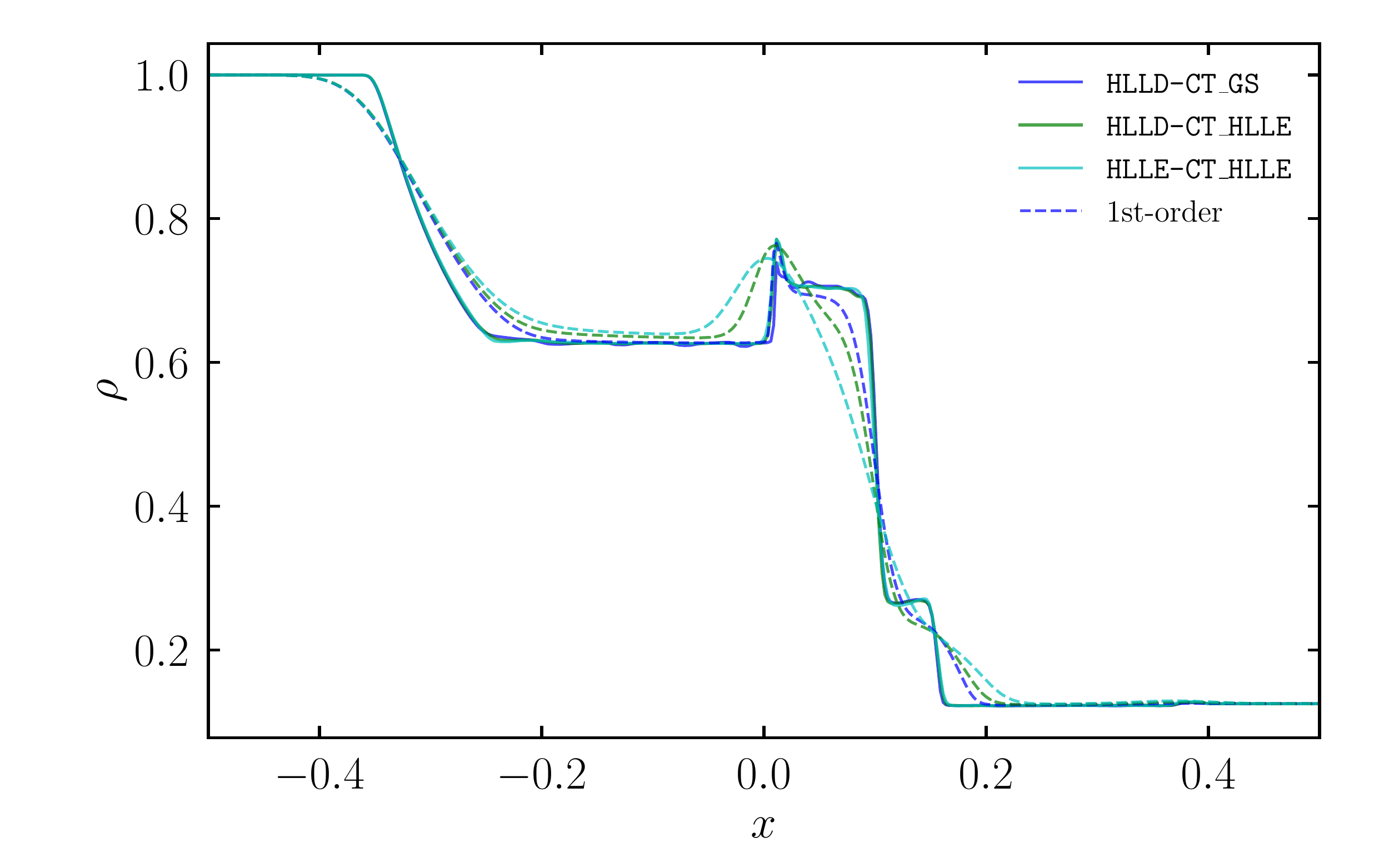}~~
  	 \includegraphics[width=0.46\linewidth]{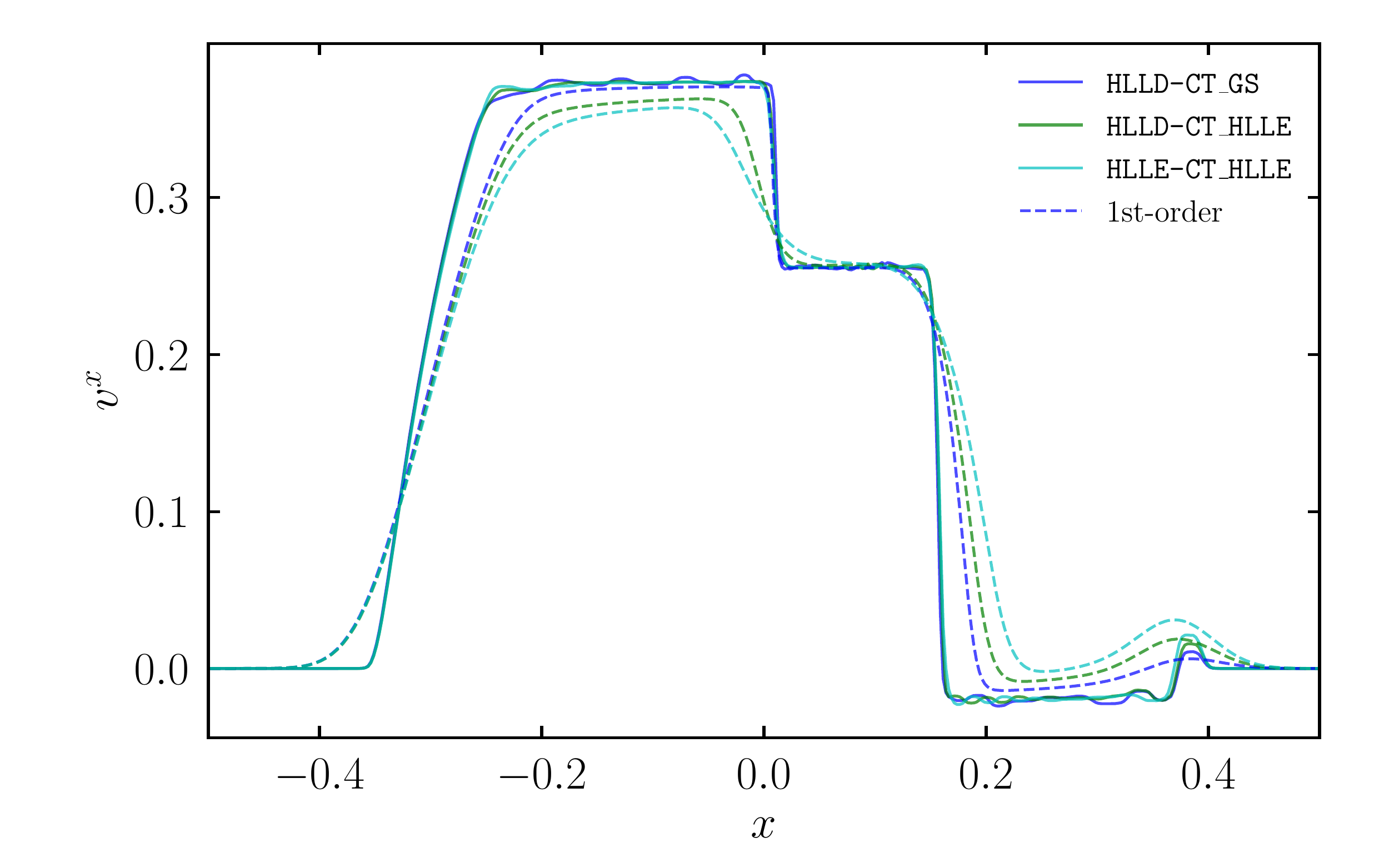}\\
     \includegraphics[width=0.46\linewidth]{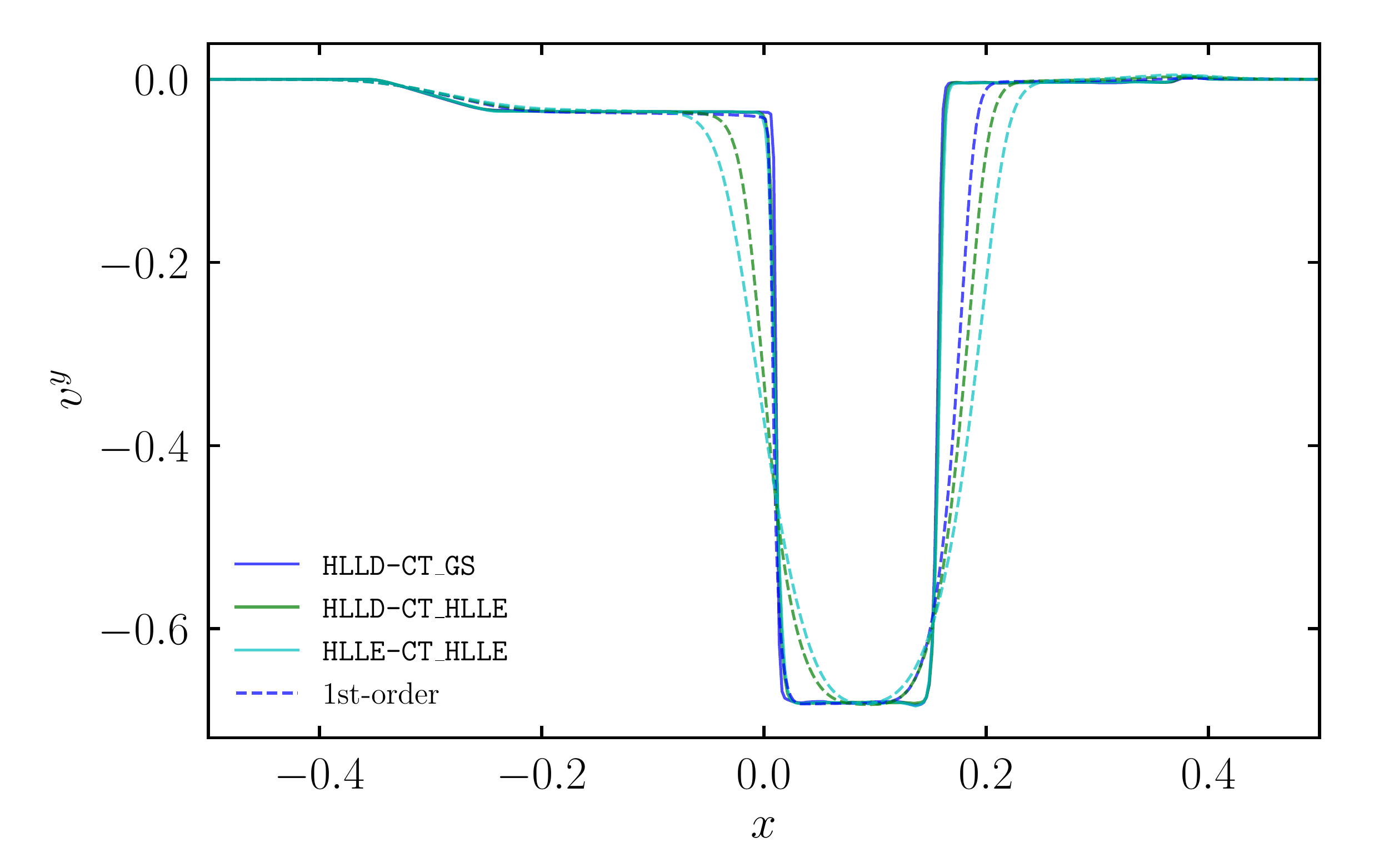}~~
  	 \includegraphics[width=0.46\linewidth]{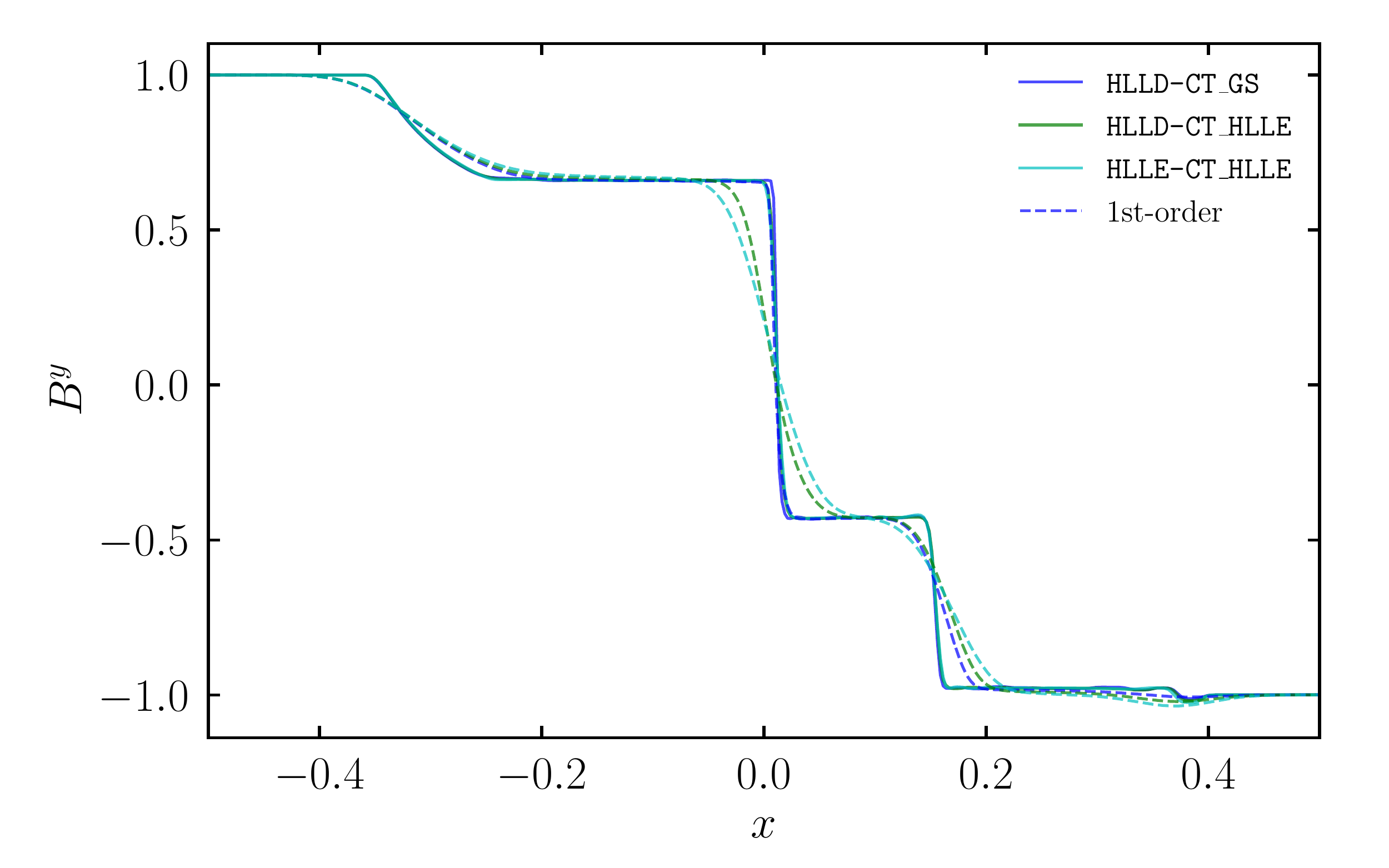}
 	 \caption{Profile of the rest mass density (top-left), the $x$-component of the three velocity (top-right), the $y$-component of the three velocity (bottom-left), and the $y$-component of the magnetic field (bottom-right) at $t = 0.4$ in Problem MHD3. The blue, green, and cyan curves denote the numerical solution with the {\tt HLLD-CT\_GS}, {\tt HLLD-CT\_HLLE}, and {\tt HLLE-CT\_HLLE} solvers, respectively. We employ RK4 with 3rd-order PPM cell reconstruction (solid curves), and also with 1st-order cell reconstruction (dashed curves).}\label{fig:PMHD3_1st}
\end{figure*}

\begin{figure*}[t]
 	 \includegraphics[width=0.46\linewidth]{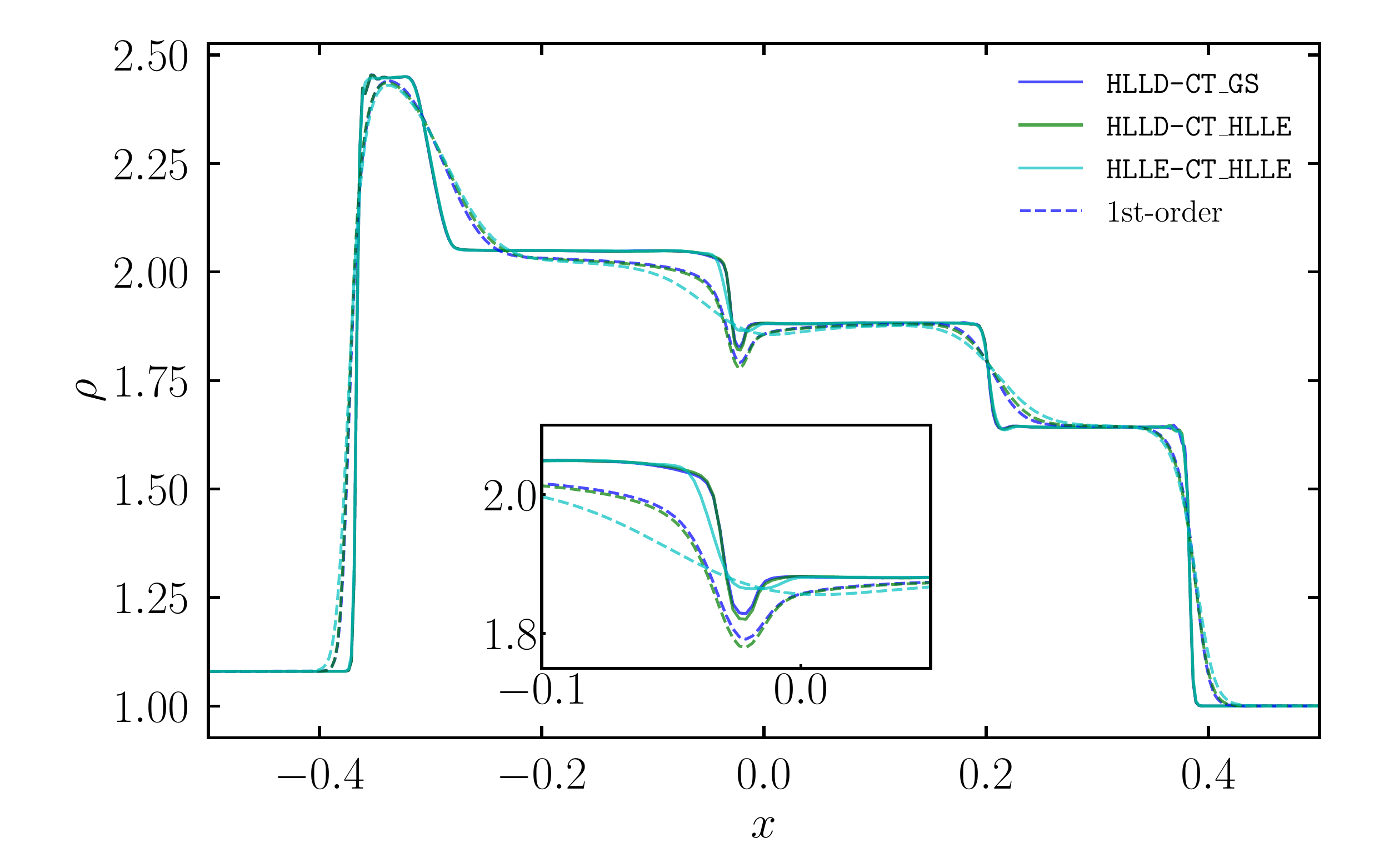}~~
  	 \includegraphics[width=0.46\linewidth]{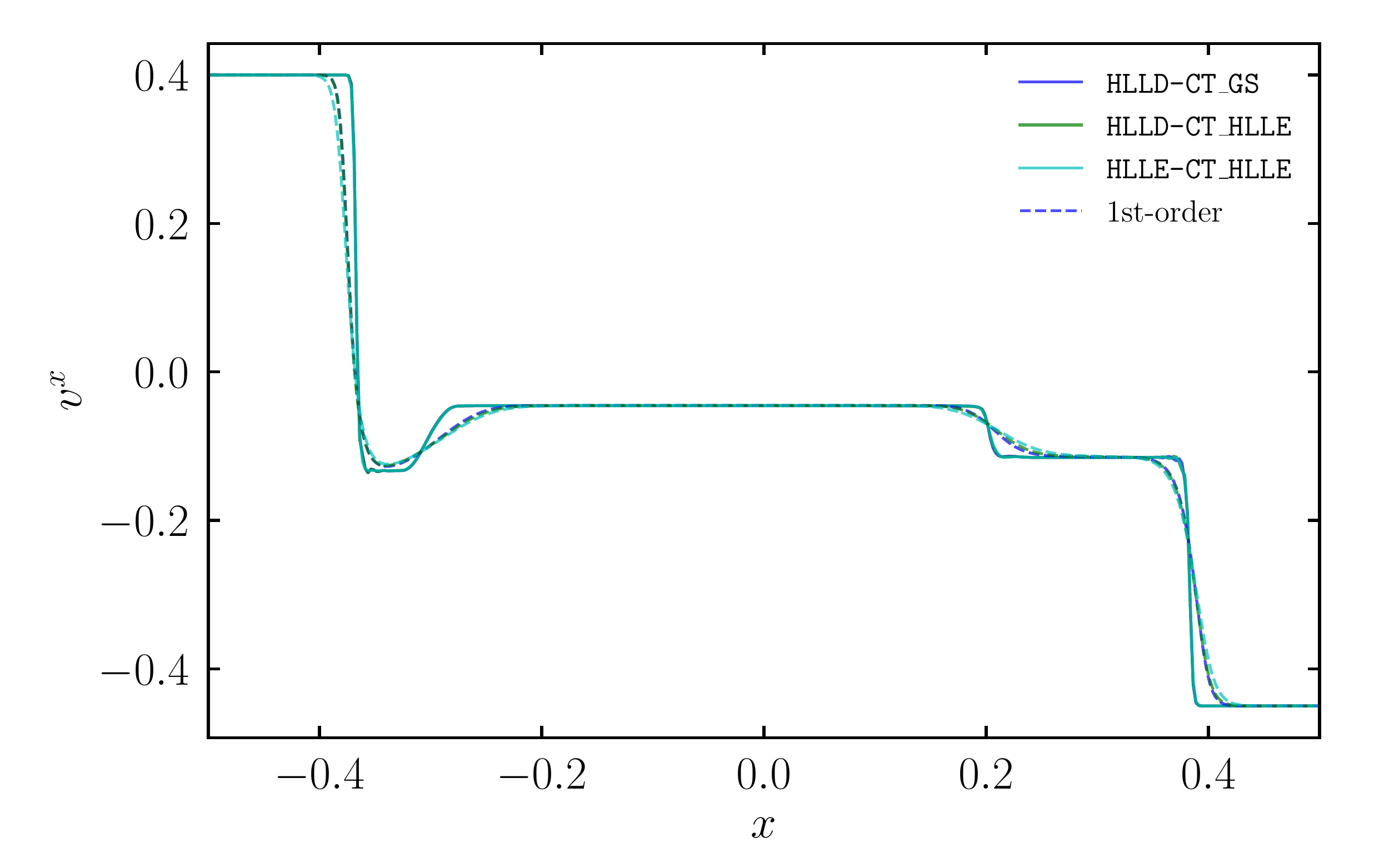}\\
     \includegraphics[width=0.46\linewidth]{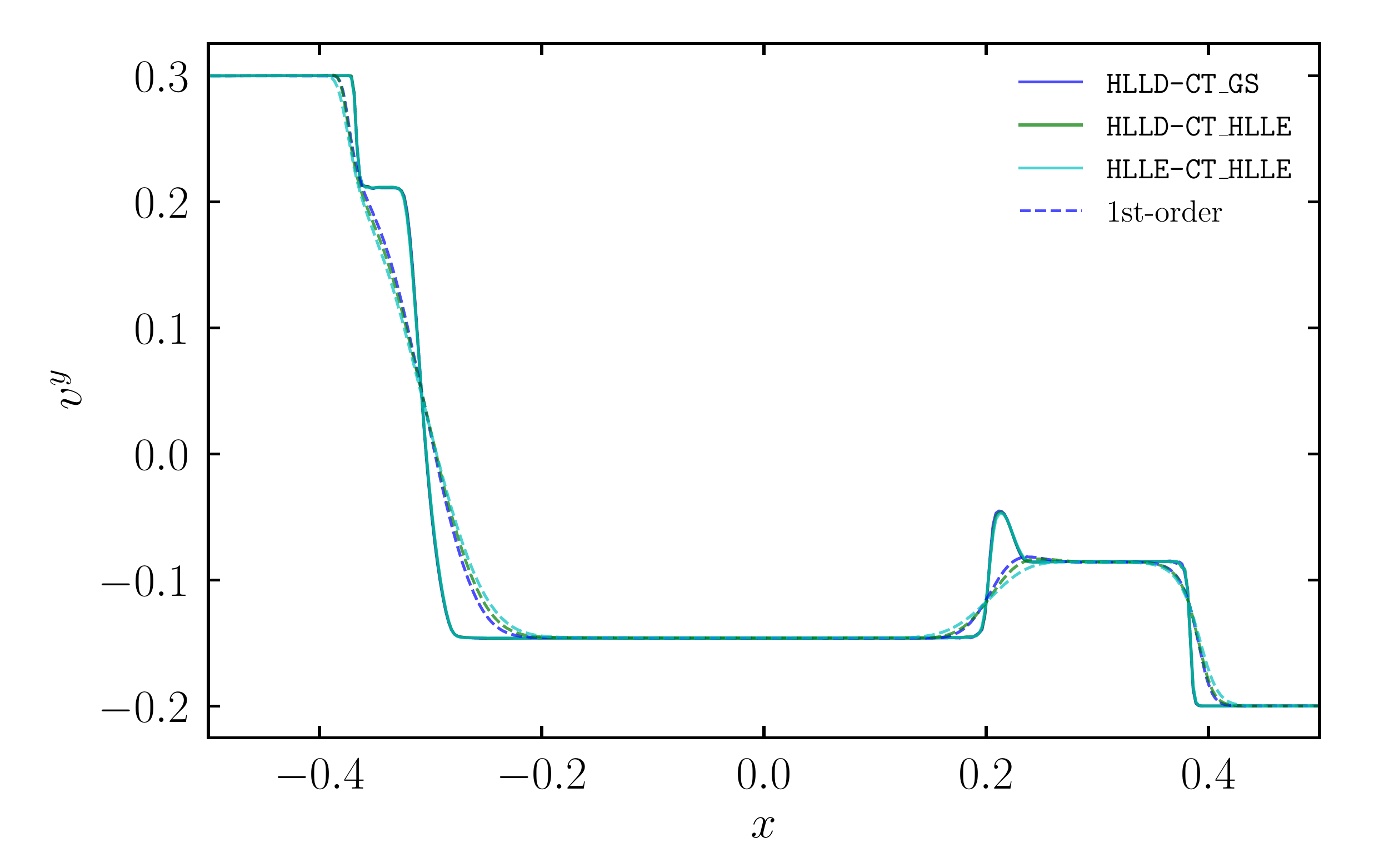}~~
  	 \includegraphics[width=0.46\linewidth]{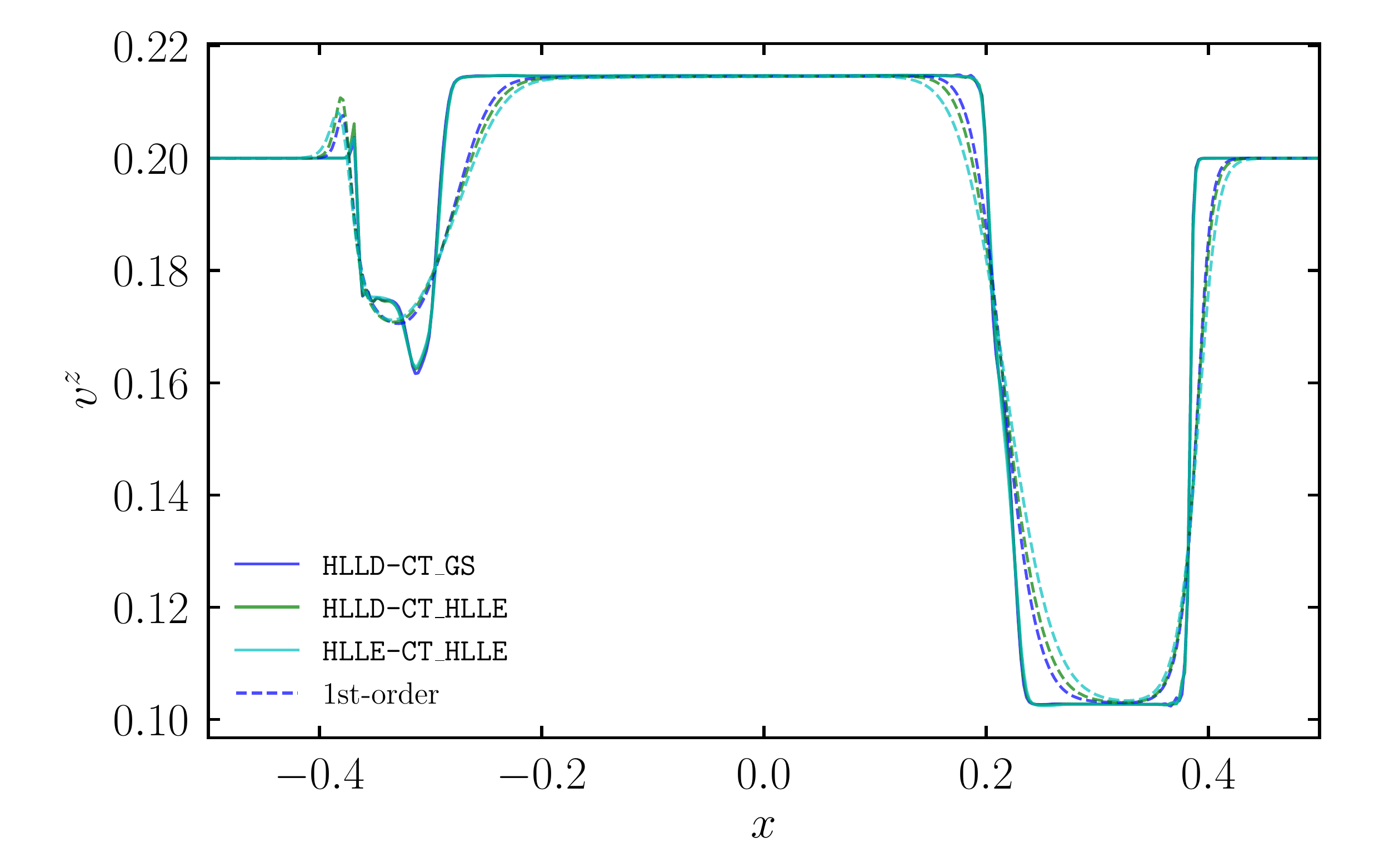}\\
     \includegraphics[width=0.46\linewidth]{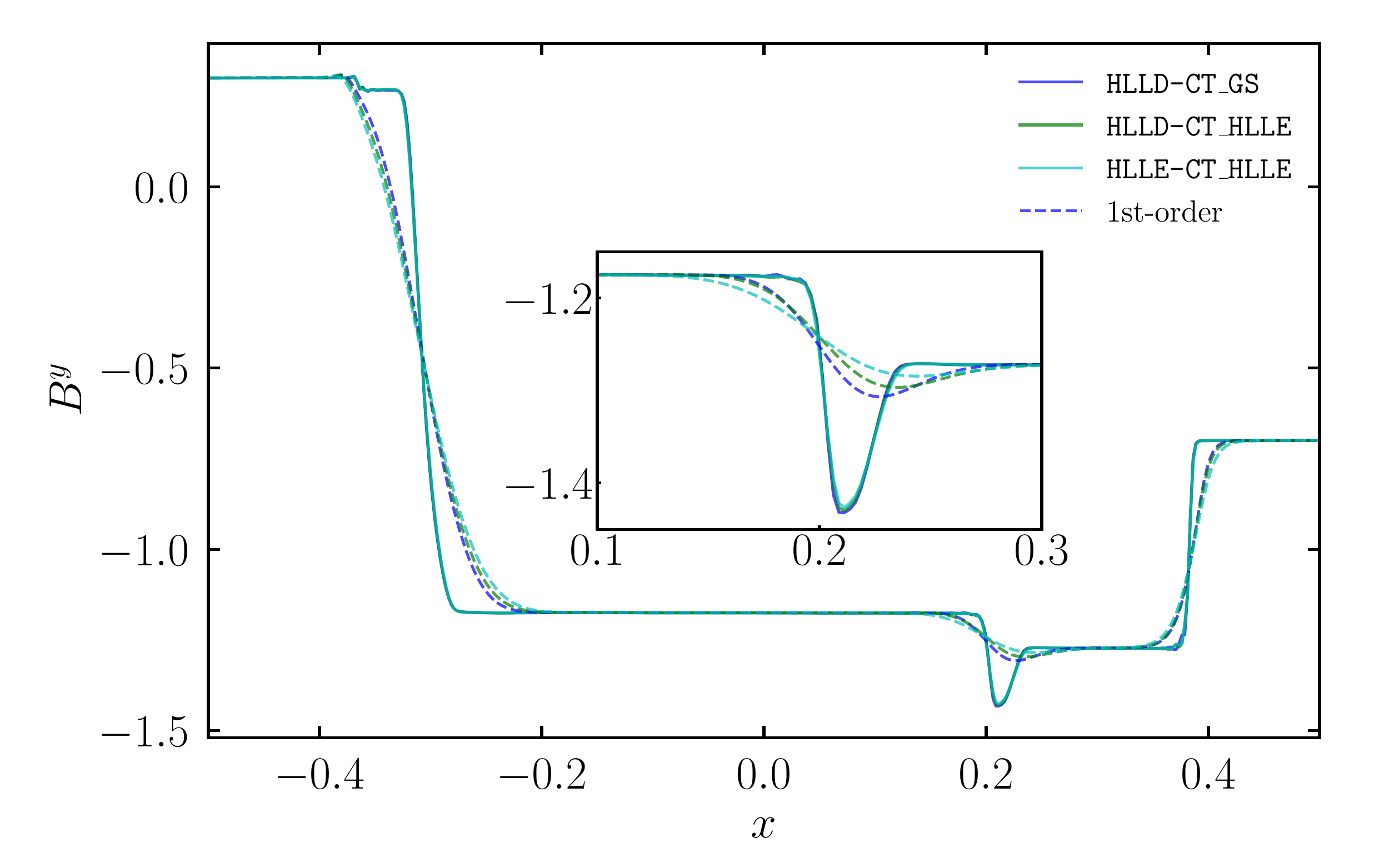}~~
  	 \includegraphics[width=0.46\linewidth]{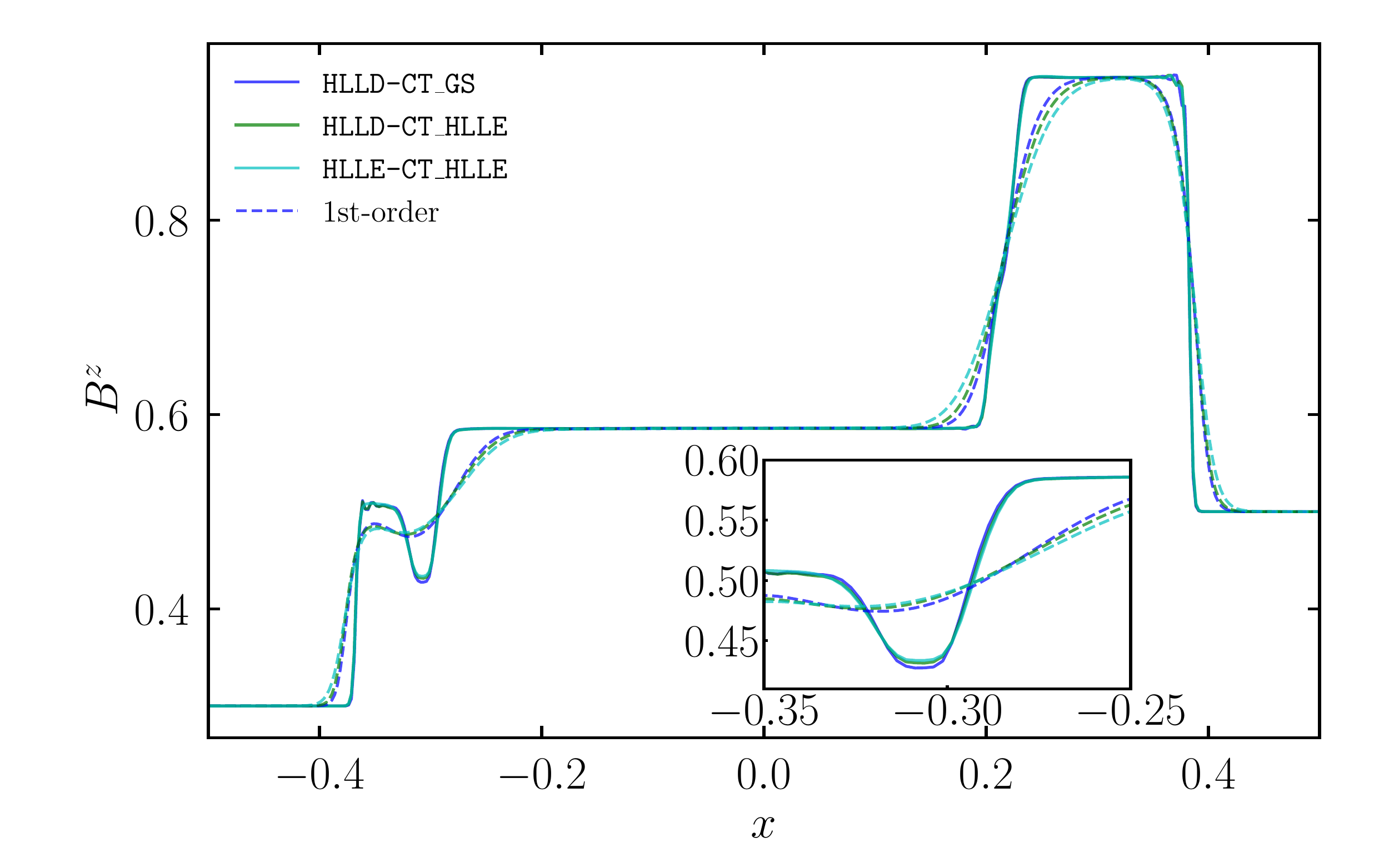}  	
 	 \caption{Profile of the rest mass density (top-left), the $x$-component of the three velocity (top-right), the $y$-component of the three velocity (middle-left), the $z$-component of the three velocity (middle-right), the $y$-component of the magnetic field (bottom-left), and the $z$-component of the magnetic field (bottom-right) at $t = 0.55$ in Problem MHD4. The blue, green, and cyan curves denote the numerical solution with the {\tt HLLD-CT\_GS}, {\tt HLLD-CT\_HLLE}, and {\tt HLLE-CT\_HLLE} solvers, respectively. We employ RK4 with 3rd-order PPM cell reconstruction (solid curves), and also with 1st-order reconstruction (dashed curves).
 	  The insets show a close-up of the discontinuity.
 	 }\label{fig:PMHD4_1st}
\end{figure*}

\begin{figure*}[t]
 	 \includegraphics[width=0.46\linewidth]{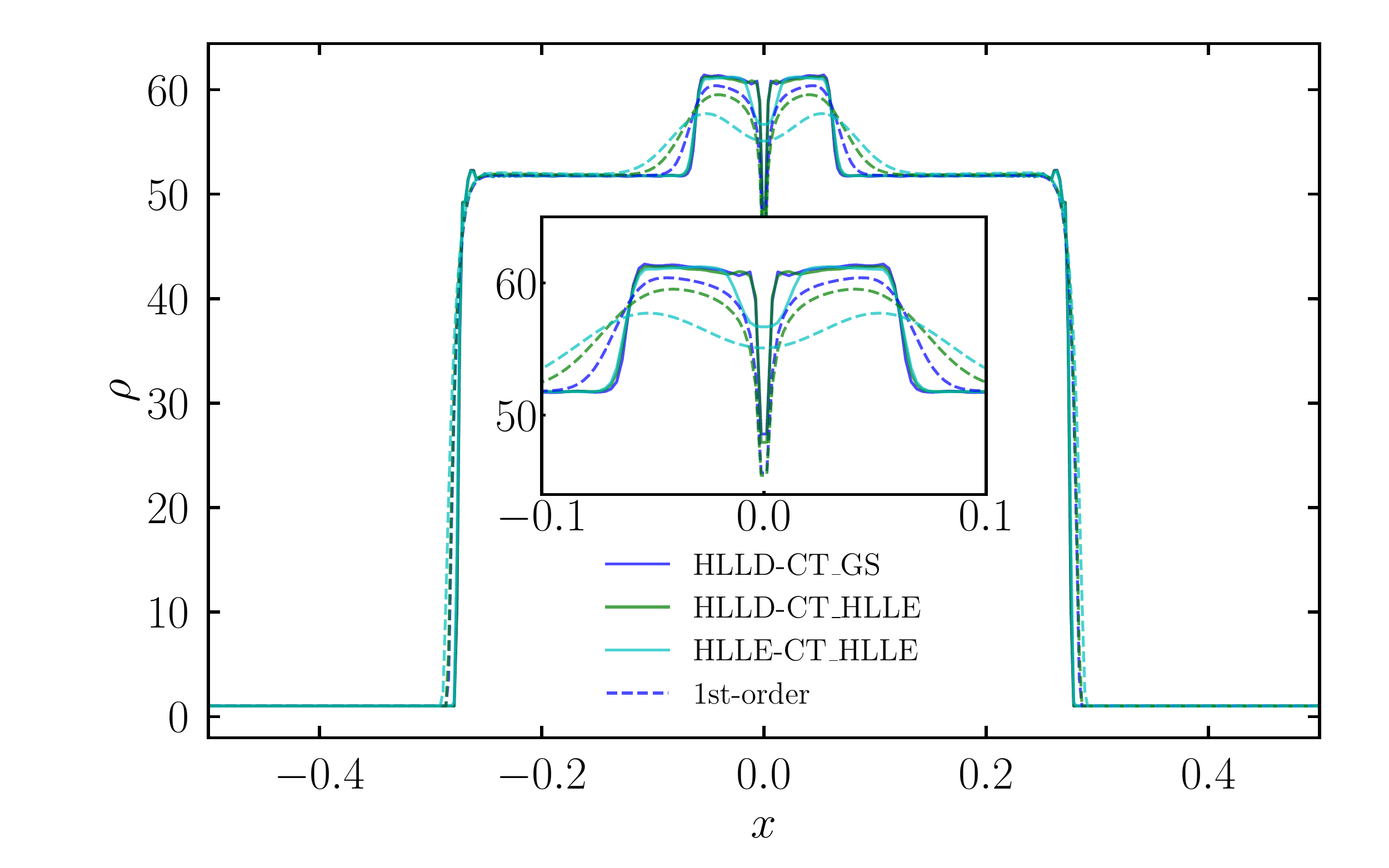}~~
  	 \includegraphics[width=0.46\linewidth]{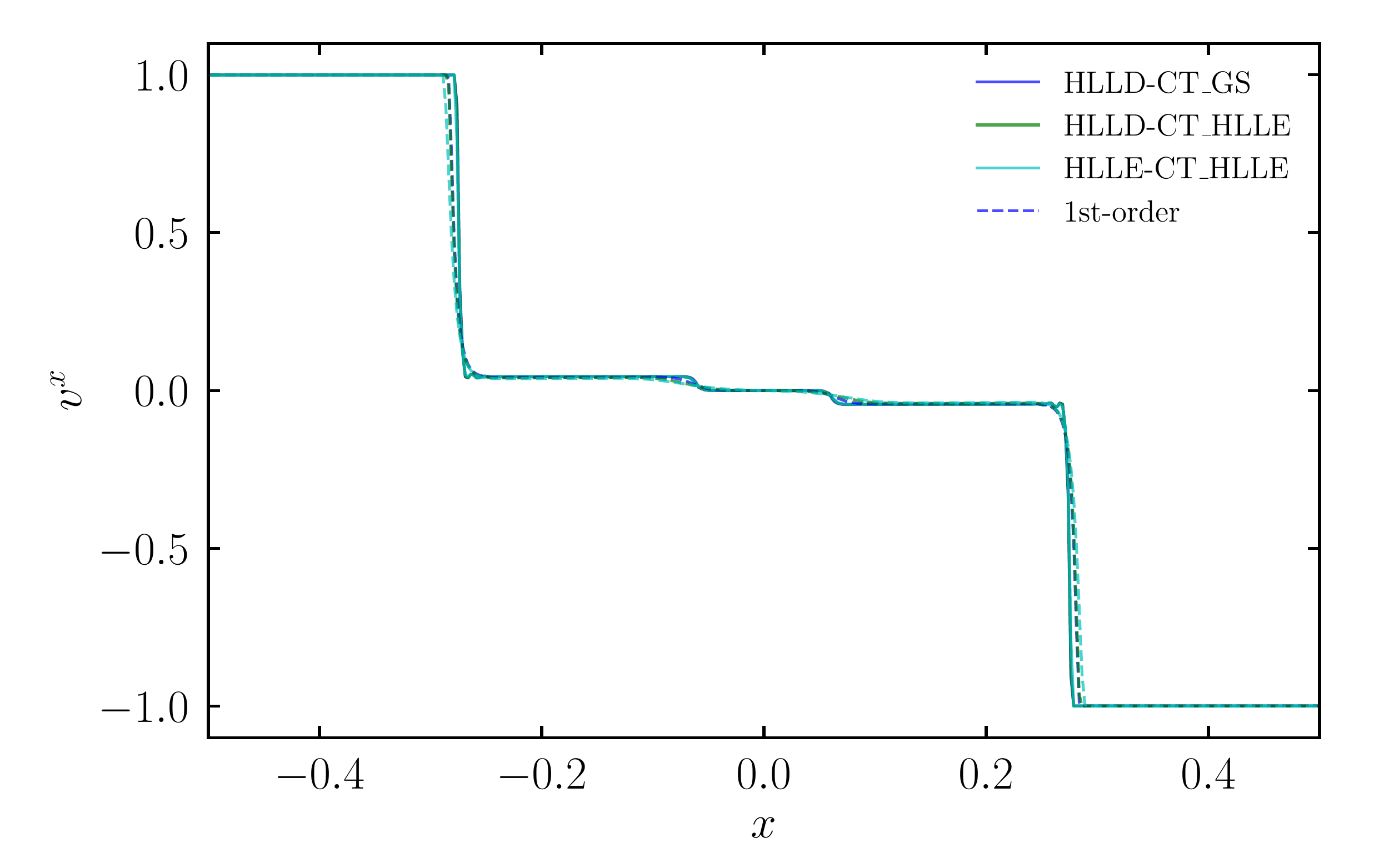}\\
     \includegraphics[width=0.46\linewidth]{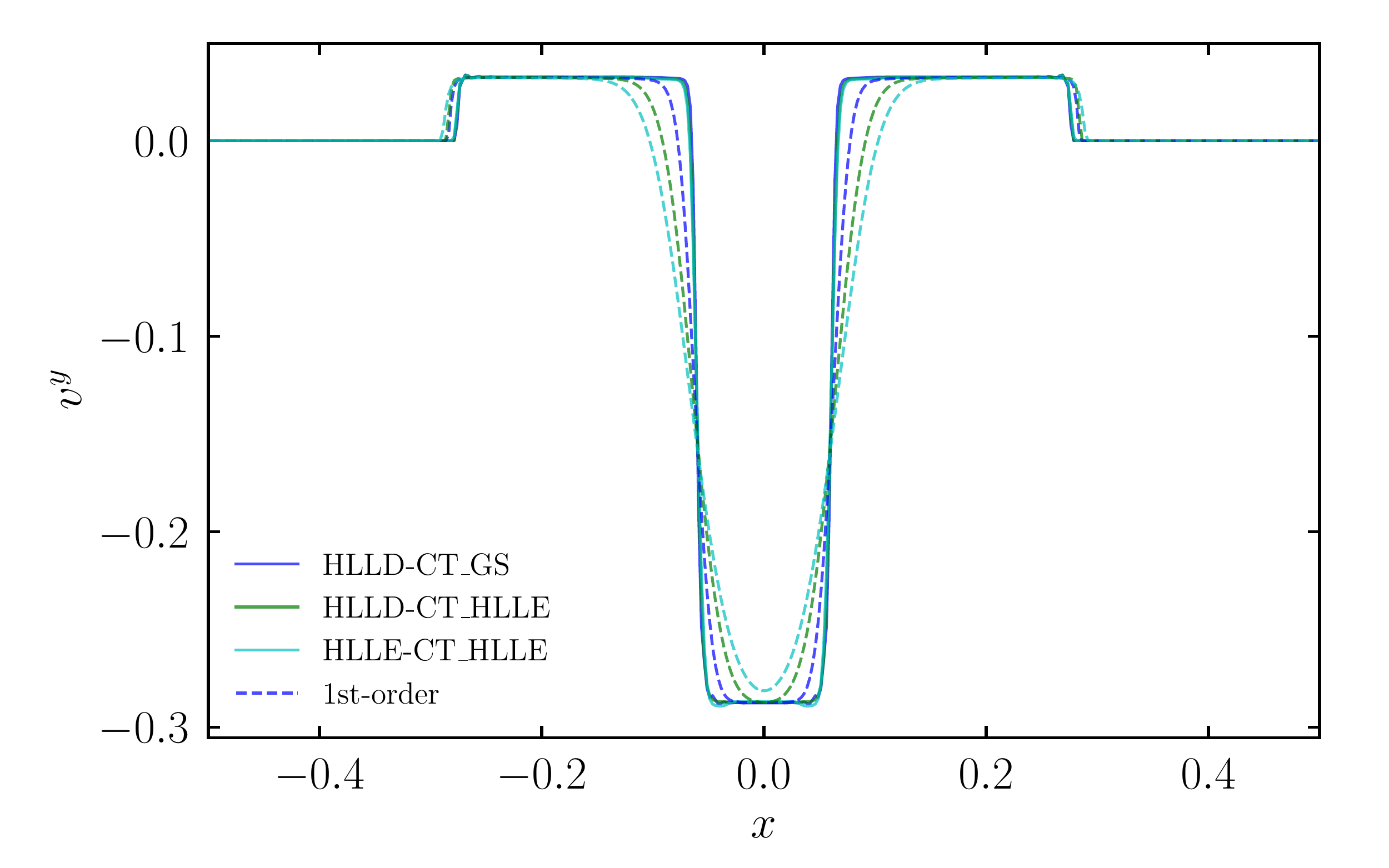}~~
     \includegraphics[width=0.46\linewidth]{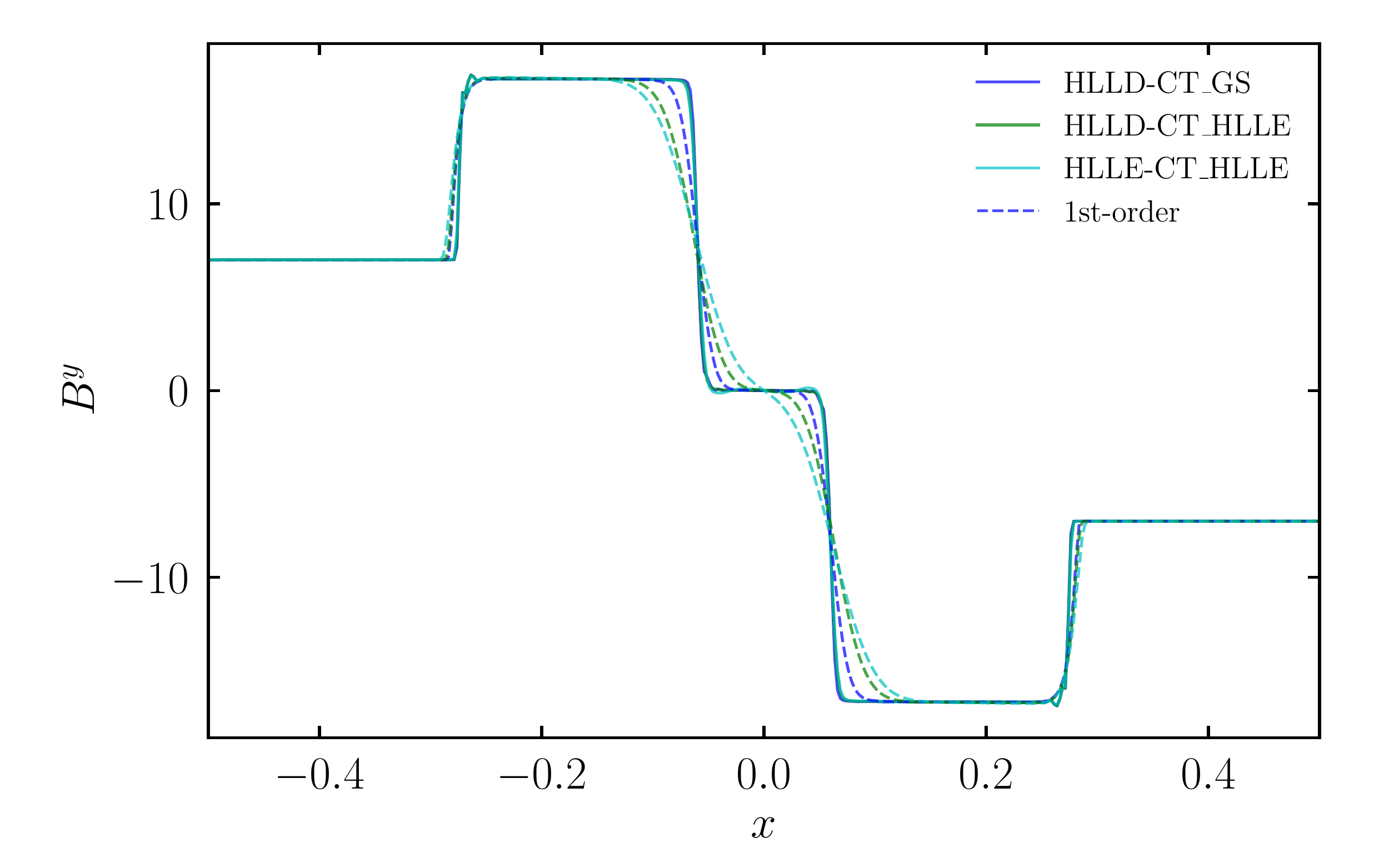}
 	 \caption{Same as Fig.~\ref{fig:PMHD3_1st}, but for Problem MHD5.
         }\label{fig:PMHD5_1st}
\end{figure*}

\begin{figure*}[t]
 	 \includegraphics[width=0.46\linewidth]{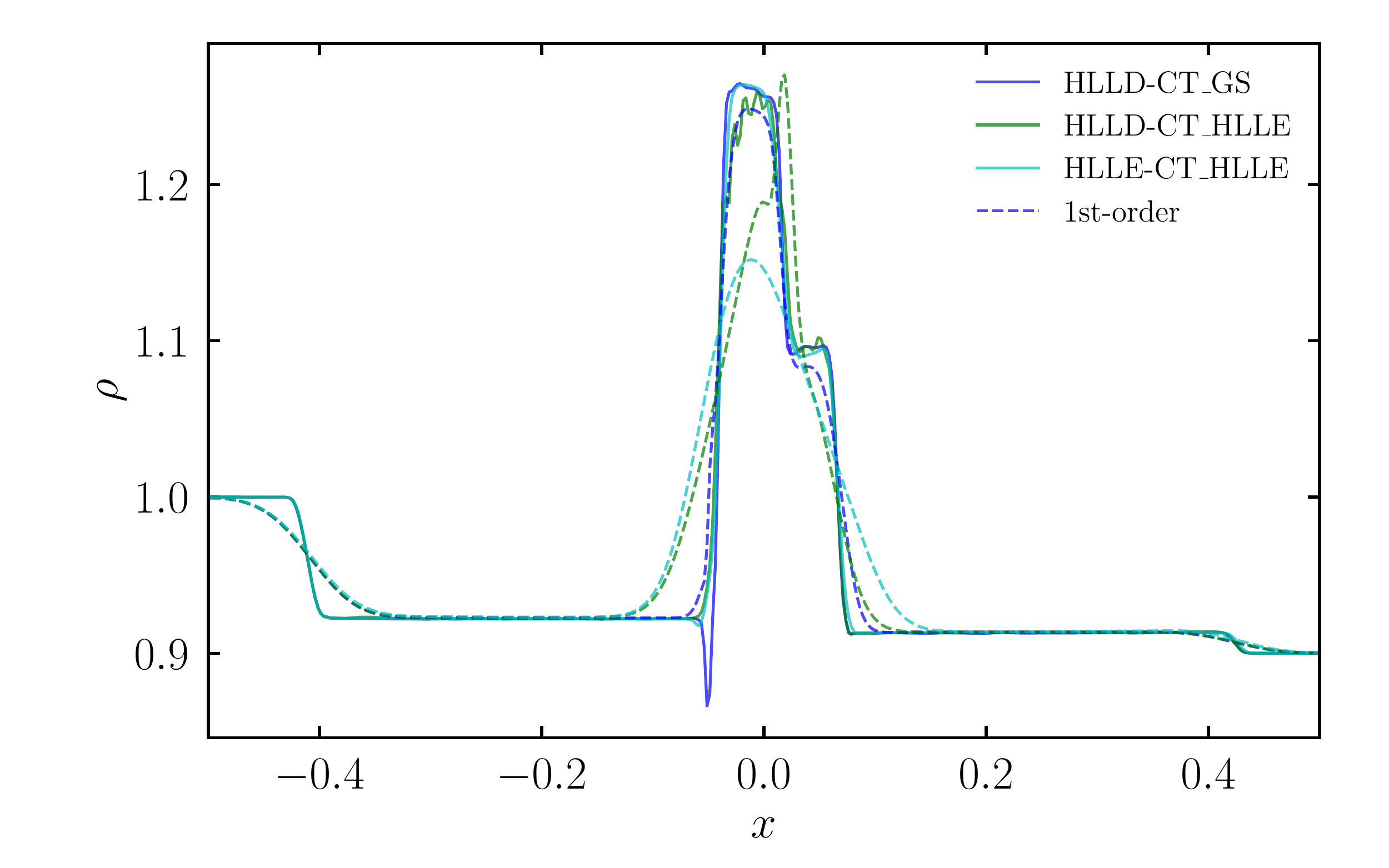}~~
  	 \includegraphics[width=0.46\linewidth]{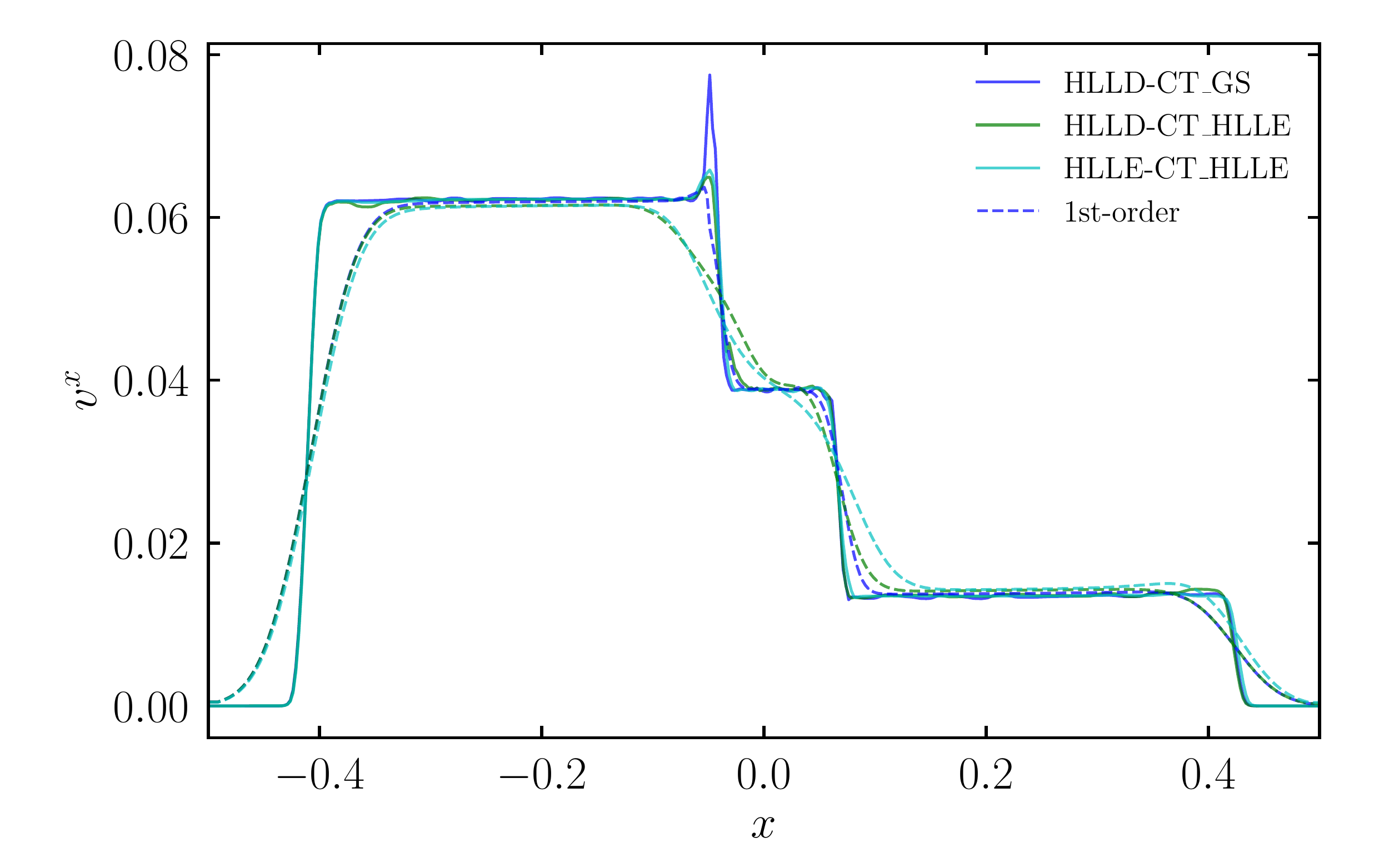}\\
     \includegraphics[width=0.46\linewidth]{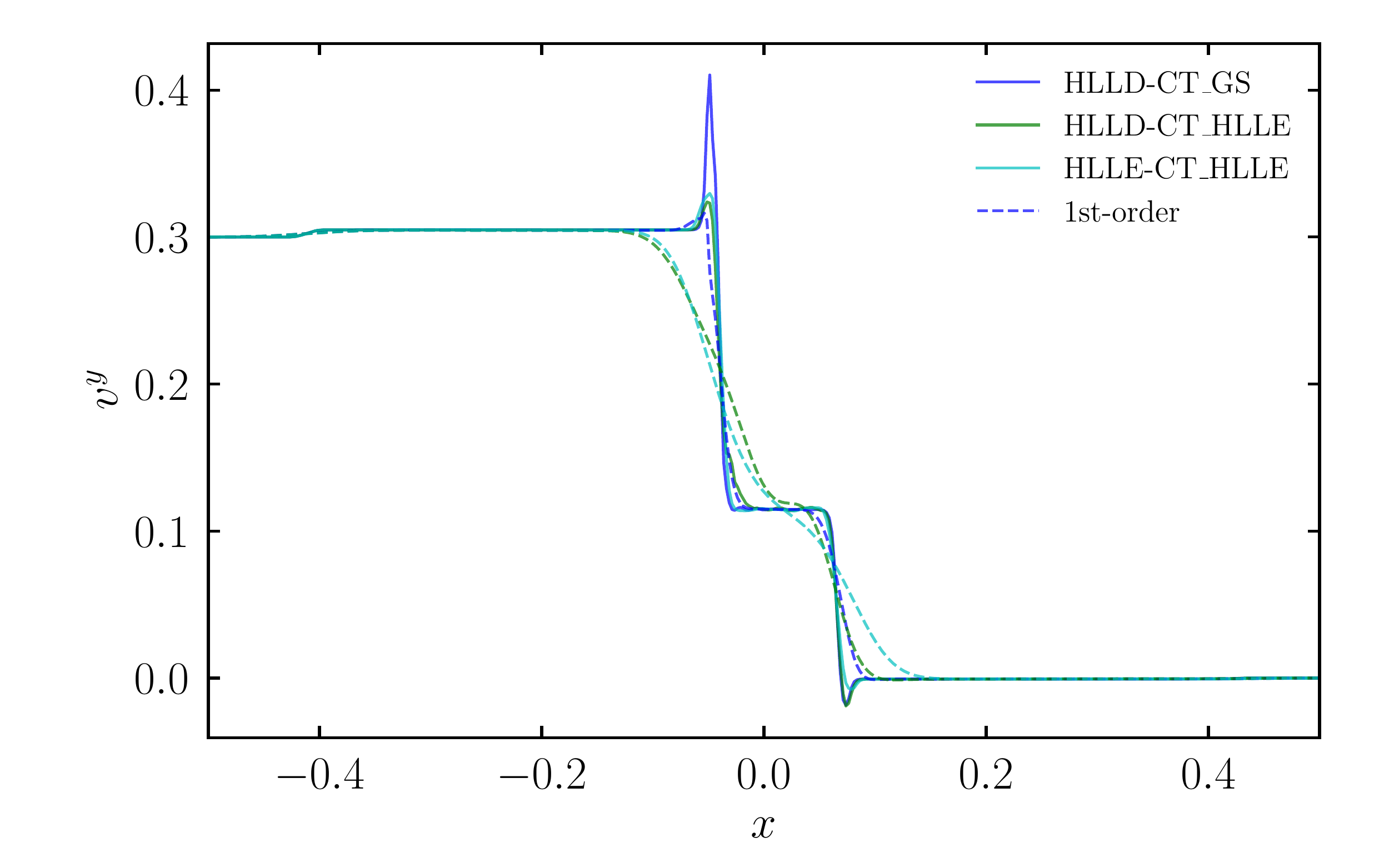}~~
     \includegraphics[width=0.46\linewidth]{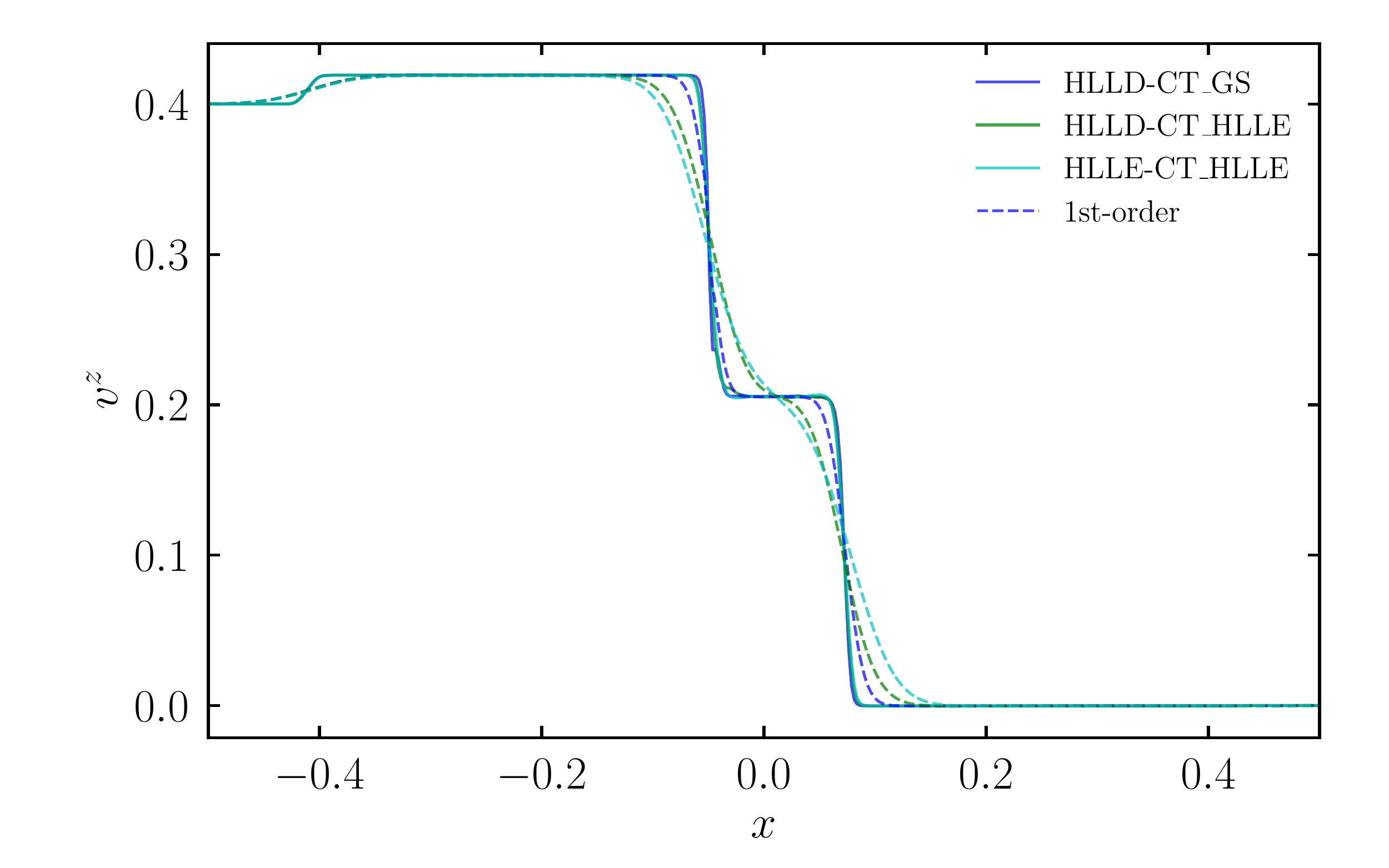}\\
     \includegraphics[width=0.46\linewidth]{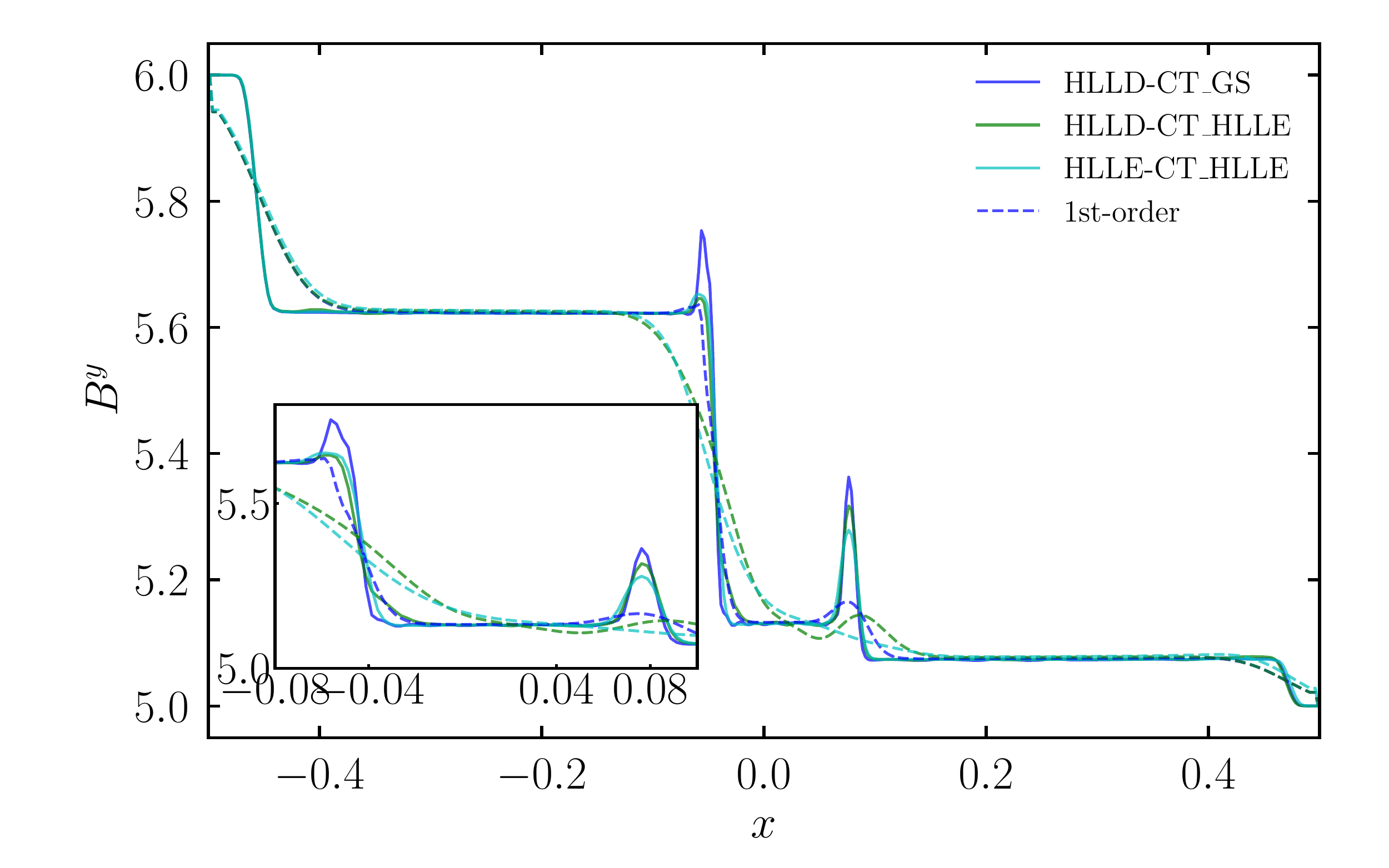}~~
     \includegraphics[width=0.46\linewidth]{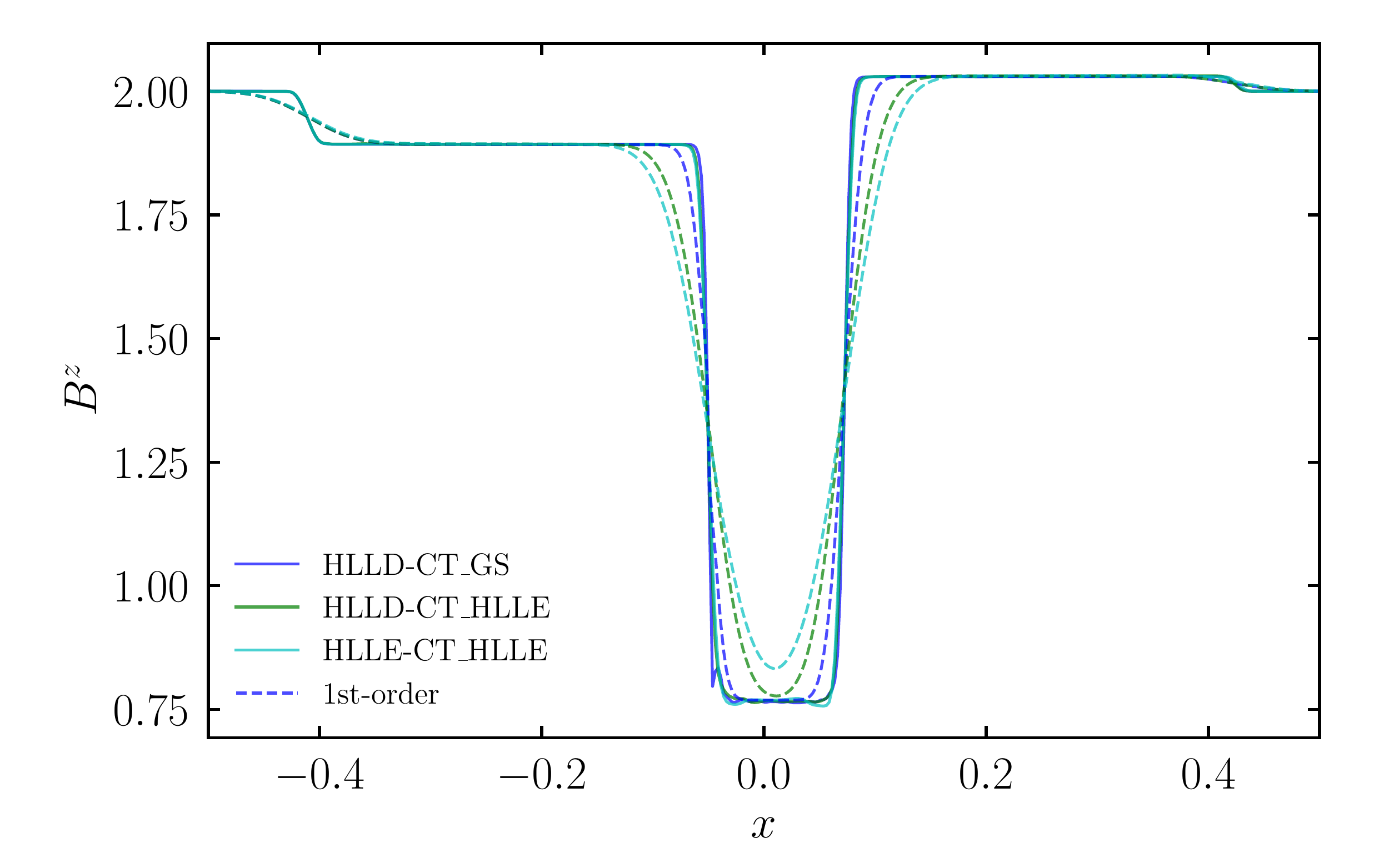}\\
 	 \caption{Same as Fig.~\ref{fig:PMHD4_1st}, but for Problem MHD6.
         }\label{fig:PMHD6_1st}
\end{figure*}
\subsubsection{Magnetohydrodynamics: one-dimensional shock tubes}
In this section we consider six special relativistic magnetohydrodynamics test problems in one spatial dimension. All the test problems except for Problem MHD6 are carried out in a domain of size $x\in[-0.5,0.5]$ with grid spacing $\Delta x=0.005$ (i.e., $N_x=100$). For Problem MHD6, the domain is identical but we employ a higher resolution with $\Delta x=0.0025$ and $N_x=200$. 

In the first problem (Problem MHD1 in Table~\ref{tab:1D_test_problem}) the solution consists of a stationary contact discontinuity. The left panel of Fig.~\ref{fig:PMHD1_PMHD2} plots the rest-mass density profile at the end of the simulation. Because the {\tt HLLD-CT\_GS} and {\tt HLLD-CT\_HLLE} solvers exactly capture the contact discontinuity, the numerical solutions remain stationary even when we employ 1st-order reconstruction (see the inset in the left panel of Fig.~\ref{fig:PMHD1_PMHD2}). On the other hand, with the {\tt HLLE-CT\_HLLE} solver, the initial contact discontinuity is broadened because this solver neglects the contact discontinuity inside the Riemann fan. However when we employ 3rd-order (PPM) reconstruction, this spurious broadening of the contact discontinuity is suppressed, although the contact wave is still not resolved as sharply as it is with the HLLD solver. 

In the second magnetohydrodynamics test problem (Problem MHD2 in Table~\ref{tab:1D_test_problem}) we model a stationary rotational discontinuity (i.e., an Alfv\'en wave). The right panel of Fig.~\ref{fig:PMHD1_PMHD2} presents the profile of the $y$-component of the magnetic field at the end of the simulation. 
When we employ 1st-order cell reconstruction, the {\tt HLLD-CT\_GS} solver reproduces the stationary solution (see the inset in the right panel of Fig.~\ref{fig:PMHD1_PMHD2}). 
This is because the rotational discontinuity is captured exactly by the HLLD solver, and the electric field at the cell edge is evaluated with the numerical flux, i.e., the electric field at the cell interface, given by the {\tt HLLD} solver with the {\tt CT\_GS} scheme (see, e.g., Eq.~(\ref{eq:CT_GS})). With the {\tt HLLD-CT\_HLLE} solver, on the other hand, the rotational discontinuity is broadened because the electric field at the cell interface given by the {\tt HLLD} solver is {\it not} used to evaluate the electric field at the cell edge in the {\tt CT\_HLLE} scheme. With the {\tt HLLE-CT\_HLLE} solver, the rotational discontinuity inside the Riemann fan is not captured. As a result, the initial rotational discontinuity is spuriously broadened. This drawback is improved by employing 3rd-order PPM cell reconstruction in the {\tt HLLD-CT\_HLLE} and {\tt HLLE-CT\_HLLE} solvers. 
Note that for this problem, we employ the compression parameter $b=3$ in the min-mod function for the PPM reconstruction in the {\tt HLLD-CT\_GS} run. Otherwise, we find the over- and under-shoot in the vicinity of the initial rotational discontinuity (not shown) because the default value of $b=2$ is not sufficient to capture the initial steep profile.

The third magnetohydrodynamics test problem (Problem MHD3) is the relativistic extension of the Brio-Wu shock tube~\cite{Brio:1988}. In this problem, the solution consists of a left-propagating rarefaction wave, a right-propagating slow shock wave (located at $x \approx 0.18$ in Fig.~\ref{fig:PMHD3_1st}), and a right-propagating rarefaction wave. In addition there is a right-ward propagating contact discontinuity located at $x\approx 0.15$ in Fig.~\ref{fig:PMHD3_1st} adjacent to the shock wave. Finally at $x\approx 0$, a compound wave appears. When we use 1st-order reconstruction, the contact discontinuity is captured more sharply with the {\tt HLLD-CT\_GS} solver than with the {\tt HLLD-CT\_HLLE} solver, while the {\tt HLLE-CT\_HLLE} solver cannot capture the contact discontinuity at all if the 1st-order reconstruction is used. The slow shock is also captured more sharply with the {\tt HLLD-CT\_GS} solver than with the {\tt HLLD-CT\_HLLE} solver, while with the {\tt HLLE-CT\_HLLE} solver the slow shock wave is significantly broadened. This feature is also found for the compound wave. While the various waves are better captured in 3rd-order PPM reconstruction irrespective of the chosen solvers, we find that the higher order reconstruction method induces artificial oscillatory behavior behind the compound wave (in the regions of $-0.2 \lesssim x \lesssim 0$) in both the rest-mass density and in the $x$-component of the velocity (see also in the regions of $0.2 \lesssim x \lesssim 0.4$ in the $x$-component of the velocity). The amplitude of these oscillations is reduced when we employ the diffusive compression parameter of the PPM reconstruction $b=1$.

The fourth magnetohydrodynamics test problem (Problem MHD4) consists of a left (right)-propagating fast wave located at $x\approx-0.4\,(+0.4)$, a left-propagating rarefaction wave ($x\approx -0.3$), a contact discontinuity ($x\approx -0.04$), a right-propagating slow wave ($x\approx +0.2$), and, finally, a left (right)-propagating Alfv\'{e}n wave (located at $x\approx -0.33\,(+0.22)$). See Fig.~\ref{fig:PMHD4_1st} for the solutions (the inset in the $B^y$ ($B^z$) panel shows a close-up region of the right (left)  Alfv\'{e}n waves). When we employ 1st-order reconstruction, both solvers are able to capture the fast waves, but the contact discontinuity is captured more sharply with the {\tt HLLD-CT\_GS} or {\tt HLLD-CT\_HLLE} solver than with the {\tt HLLE-CT\_HLLE} solver (see the inset in the rest-mass density profile in Fig.~\ref{fig:PMHD4_1st} which shows a close-up of the contact discontinuity). Irrespective of the solvers, it is hard to distinguish the slow wave and the right-propagating Alfv\'{e}n wave, and also between the the rarefaction wave and the left-propagating Alfv\'{e}n wave. When we employ 3rd-order PPM reconstruction, on the other hand, we find no qualitative difference in the numerical solutions between the different solvers. 

For our fifth magnetohydrodynamics test problem (Problem MHD5) we consider the relativistic collision of two streams. Figure~\ref{fig:PMHD5_1st} shows the result at $t=0.4$. In this problem the solution consists of left (right)-propagating fast waves located at $x\approx -0.3\,(+0.3)$, and left (right)-propagating slow waves located at $x\approx -0.06\,(+0.06)$. When we employ 1st-order cell reconstruction, the slow waves are captured more sharply with the HLLD solvers, i.e., the {\tt HLLD-CT\_GS} or {\tt HLLD-CT\_HLLE} solvers than with the {\tt HLLE-CT\_HLLE} solver. On the other hand, the resolution across the outermost fast waves is essentially the same for all solvers. Irrespective of the solver or reconstruction method used, a spurious undershoot in the rest-mass density appears at $x\approx 0$. This is known as the wall-heating problem~\cite{Noh:1987}: it is well-known that Godunov-type schemes cannot avoid this pathological behavior. As reported in Ref.~\cite{MUB:2009}, the undershoot is shallower with the {\tt HLLE-CT\_HLLE} solver due to the solver's larger numerical diffusion. When we employ 3rd-order PPM reconstruction, both the {\tt HLLD} and {\tt HLLE} solvers are equally capable of capturing the slow waves as well as the fast waves. 

In the final problem (Problem MHD6) in our one-dimensional suite, the solution consists of all seven waves~\cite{Giacomazzo:2005jy}. The numerical results are shown in Fig.~\ref{fig:PMHD6_1st}. In this problem, a contact discontinuity appears at $x\approx 0.05$, a rarefaction wave propagates to the left of the contact discontinuity, which can be seen at $x\approx -0.4$, and the rotational discontinuity at $x\approx -0.06$ and the slow shock at $x\approx -0.04$ follow the rarefaction wave (see the inset in the panel for $B^y$ in Fig.~\ref{fig:PMHD6_1st}). To the right of the contact discontinuity, a fast shock propagates up to $x\approx 0.4$. The rotational discontinuity at $x\approx 0.08$ and the slow shock at $x\approx 0.06$ follow the fast shock (again, this is most easily seen in the inset in the panel for $B^y$ in Fig.~\ref{fig:PMHD6_1st}).

When we use 1st-order cell reconstruction (dashed curves), the contact discontinuity is resolved only with the HLLD solvers, i.e. {\tt HLLD-CT\_GS} or {\tt HLLD-CT\_HLLE}, (see the panel for $\rho$ in Fig.~\ref{fig:PMHD6_1st}). With 1st-order reconstruction, however, it is difficult to disentangle the left/right-propagating rotational discontinuities and left/right-propagating slow shocks, even with the HLLD solver (see the dashed curves in the inset in the panel for $B^y$ in Fig.~\ref{fig:PMHD6_1st}). When we employ 3rd-order PPM reconstruction (solid curves), on the other hand, the difference between the various Riemann solvers is striking. With the {\tt HLLD-CT\_GS} solver, the left/right-propagating rotational discontinuities and slow shocks are captured as plotted in the inset in the panel for $B^y$ in Fig.~\ref{fig:PMHD6_1st}. 
With the {\tt HLLD-CT\_HLLE} or {\tt HLLE-CT\_HLLE} solvers, the right-propagating rotational discontinuity and the left/right-propagating slow shock are captured, but the left-propagating rotational discontinuity is not. 
This demonstrates the ability of the {\tt HLLD-CT\_GS} solver to properly capture all seven of the emergent waves.

\begin{figure*}[t]
\hspace{-15mm}\includegraphics[width=0.58\linewidth]{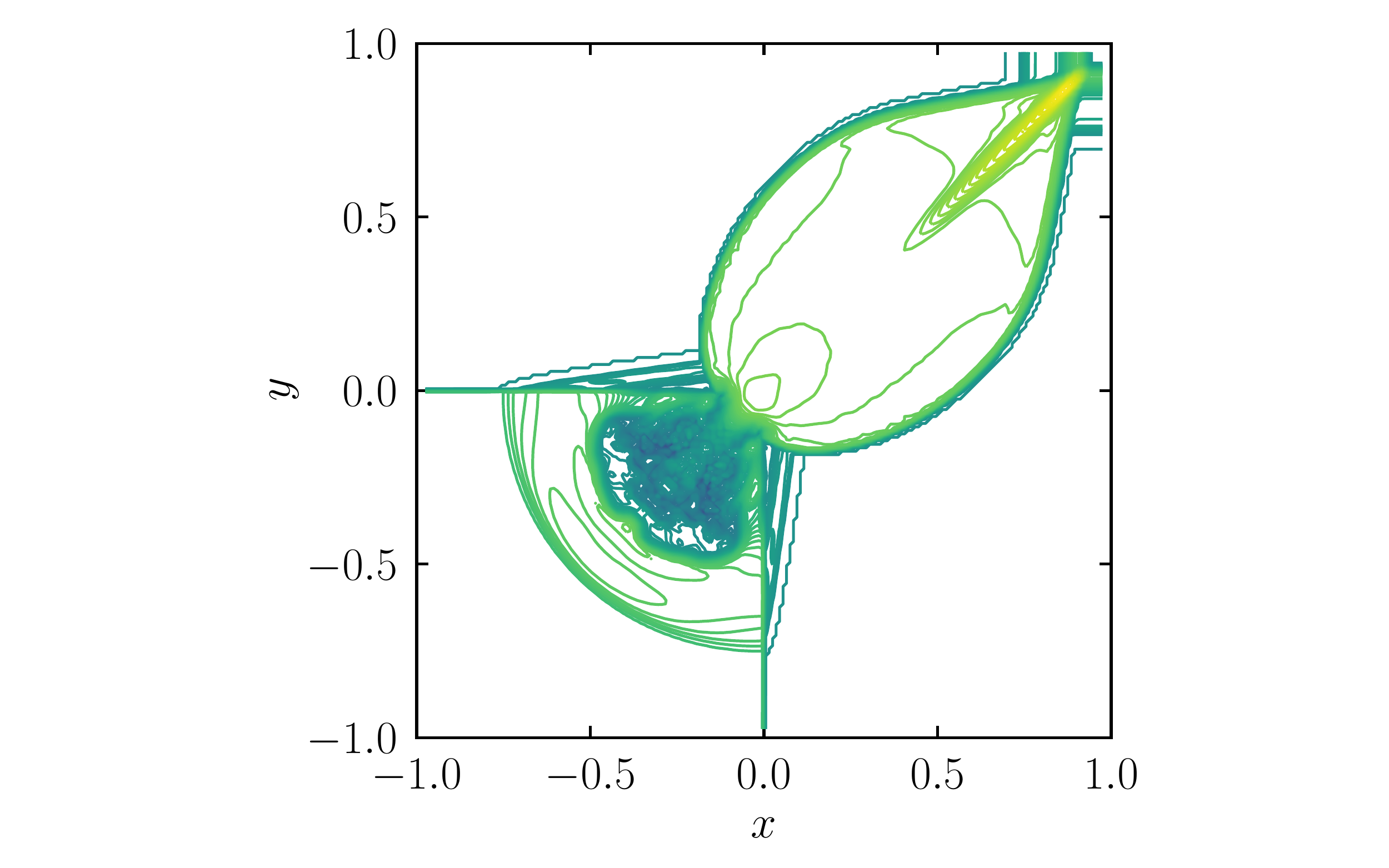}
 	 \hspace{-20mm}
  	 \includegraphics[width=0.58\linewidth]{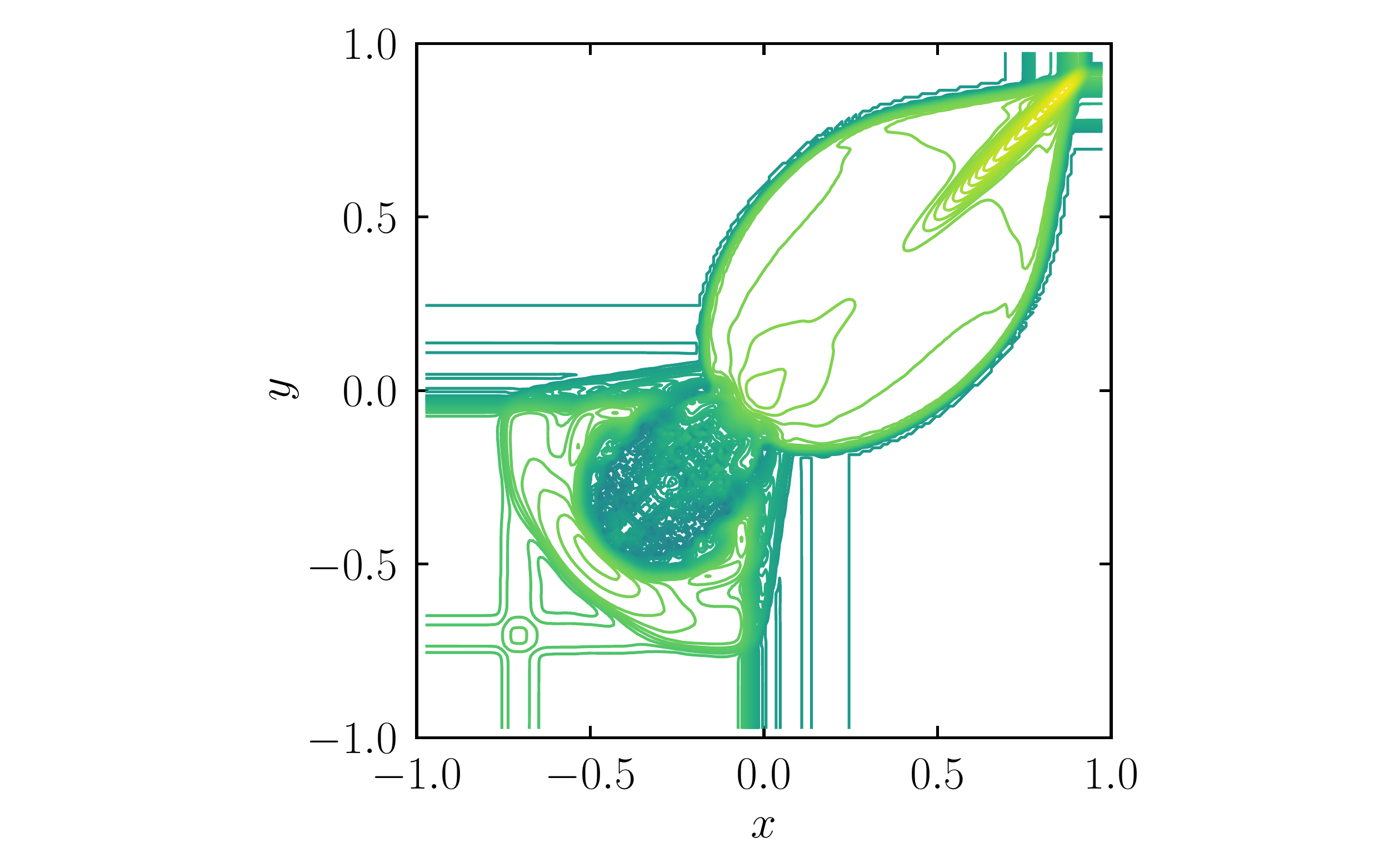}	 
 	 \caption{Logarithmic contour of the rest-mass density in the two-dimensional Riemann problem at $t=0.9$ with the {\tt HLLC} solver (left) and {\tt HLLE} solver (right). We employ RK4 and 3rd-order PPM cell reconstruction. 
         }\label{fig:Shock_Tube_2D}
\end{figure*}

\begin{figure*}[t]
 	 \includegraphics[width=0.55\linewidth]{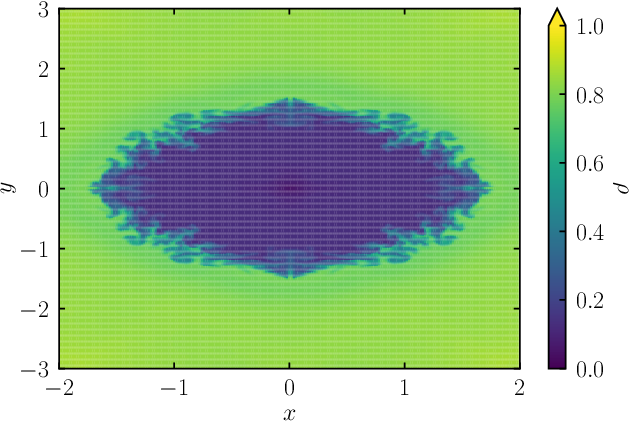}
 	 \hspace{-20mm}
  	 \includegraphics[width=0.55\linewidth]{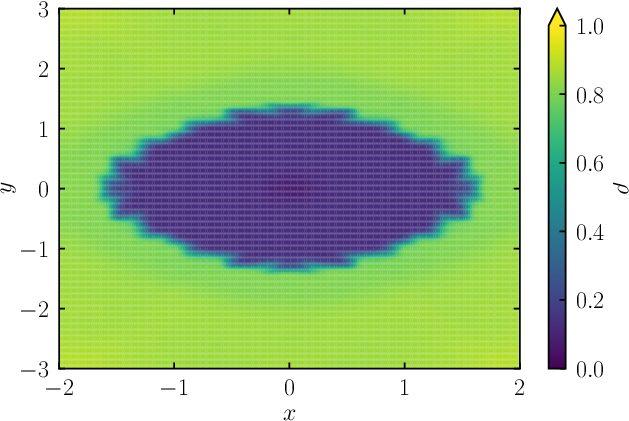}	 
 	 \caption{Rest-mass density profile in the cylindrical blast wave problem at $t=18$ with the {\tt HLLC} solver (left) and {\tt HLLE} solver (right). We employ RK4 and 3rd-order PPM cell reconstruction.
         }\label{fig:2D_Blast_Wave}
\end{figure*}

\begin{figure*}[t]
 	 \includegraphics[width=0.3\linewidth]{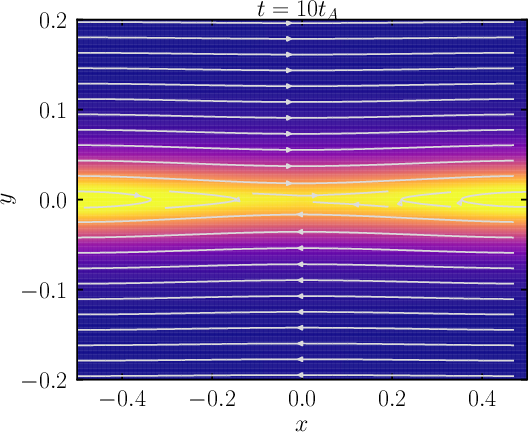}
 	 \hspace{0mm}
  	 \includegraphics[width=0.3\linewidth]{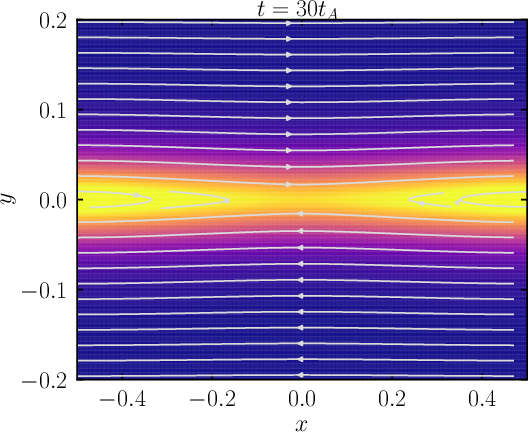}
  	 \hspace{0mm}
  	 \includegraphics[width=0.37\linewidth]{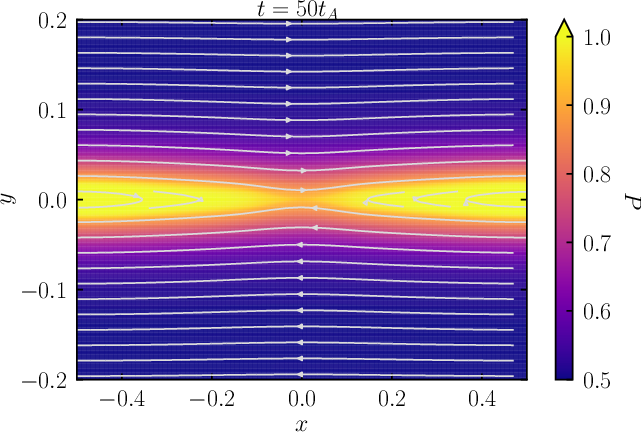}\\
  	 \includegraphics[width=0.3\linewidth]{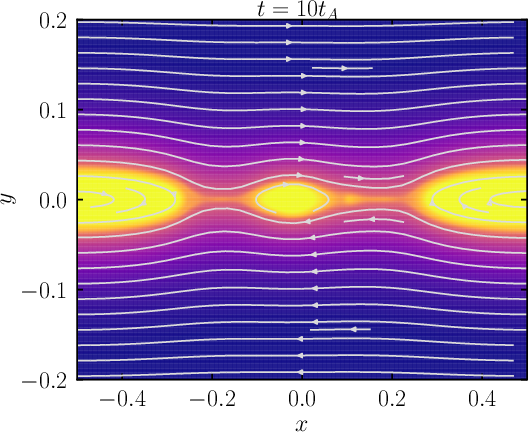}
 	 \hspace{0mm}
  	 \includegraphics[width=0.3\linewidth]{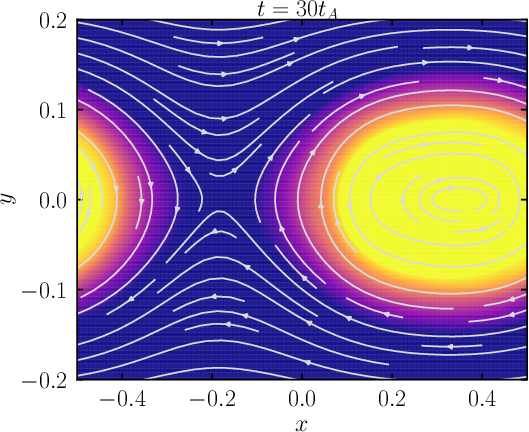}
  	 \hspace{0mm}
  	 \includegraphics[width=0.37\linewidth]{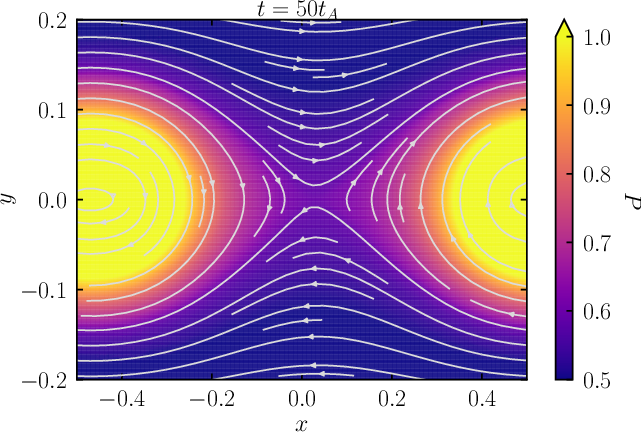}\\
  	 \includegraphics[width=0.3\linewidth]{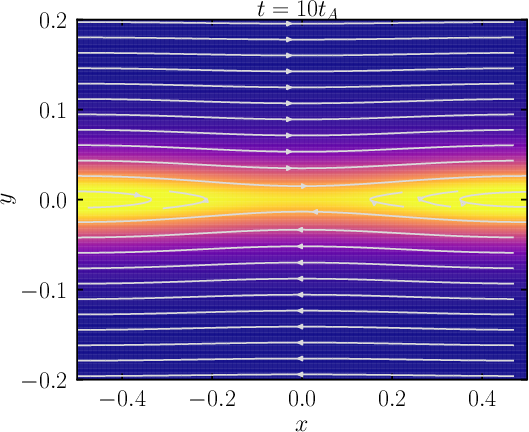}
 	 \hspace{0mm}
  	 \includegraphics[width=0.3\linewidth]{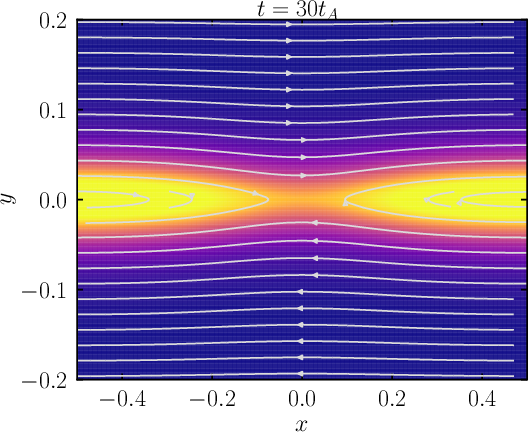}
  	 \hspace{0mm}
  	 \includegraphics[width=0.37\linewidth]{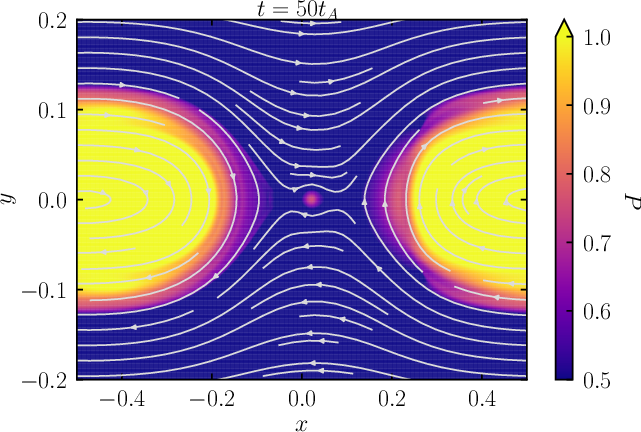}\\
 	 \caption{Thermal pressure profile for the magnetized current sheet problem at three different times: $t=10.03t_A$ (left), $t=30.08t_A$ (middle), and $t=50.06t_A$ (right). Numerical results were obtained with the {\tt HLLD-CT\_GS} (top row), the {\tt HLLD-CT\_HLLE} (middle row), and the {\tt HLLE-CT\_HLLE} solvers (bottom row), respectively. The white curves denote the magnetic-field lines. We employ RK4 with 3rd-order PPM reconstruction and a resolution of $(N_x,N_y)=(512,256)$ in all simulations.}\label{fig:2D_CS}
\end{figure*}
\subsection{Special relativistic multi-dimensional problems}
\label{SECTION_SR2D3DTest}
\subsubsection{Hydrodynamics: two-dimensional shock tube}
For our first multi-dimensional special relativistic test problem, we consider the two-dimensional Riemann problem first proposed in Ref.~\cite{Zanna:2002qr}. The simulation domain spans $\in[-1,1]$ in both the $x$- and $y$-directions. We set $\Delta x=\Delta y=0.01$ and $N_x=N_y=100$. We impose outflow boundary conditions in both directions. We use an adiabatic index of $\Gamma=5/3$. We employ 3rd-order PPM reconstruction and set the CFL number to be 0.45.  Finally, the initial condition is given by

\begin{align}
 \left(\rho,v^x,v^y,P\right)=
 \left\{
 \begin{array}{l}
 (0.1,0,0,0.01)~\text{for}~x,y>0,\\
 (0.1,0.99,0,1)~\text{for}~x<0,~y>0,\\
 (0.5,0,0,1)~\text{for}~x,y<0,\\
 (0.1,0,0.99,1)~\text{for}~x>0,~y<0.
 \end{array}
 \right.
\end{align}

Figure~\ref{fig:Shock_Tube_2D} show the logarithmic contour of the rest-mass density at $t=0.9$ with the {\tt HLLC} solver (left panel) and with the {\tt HLLE} solver (right panel). The most notable difference in the solutions between the two solvers appears around the two tangential discontinuities in the lower-left portion of the simulation domain. With the {\tt HLLC} solver (left panel), the initial tangential discontinuities remain sharp. With the {\tt HLLE} solver, on the other hand, spurious waves propagate along each axis from the initial tangential discontinuities due to numerical diffusion. Unlike the one-dimensional problems, the spurious diffusion out of the initial tangential discontinuities that occurs with the {\tt HLLE} solver cannot be mitigated even when we employ 3rd-order PPM reconstruction. Thus in this multi-dimensional test problem we observe a qualitative difference in the solutions between the {\tt HLLC} and {\tt HLLE} solvers that cannot be removed by resorting to higher-order reconstruction.

\subsubsection{Hydrodynamics: two-dimensional cylindrical explosion}
For the second special relativistic multi-dimensional test problem, we consider a cylindrical blast wave in two dimensions. For this problem, we choose the simulation domain to span $x\in[-2,2]$ and $y\in[-3,3]$, and set $\Delta x=\Delta y=0.02$, i.e.,  $(N_x,N_y)=(100,150)$ in the $x$- and $y$-directions, respectively. Periodic boundary conditions are imposed at the $x$- and $y$-boundaries. We set the adiabatic index to $\Gamma=5/3$, employ 3rd-order PPM reconstruction, and set the CFL number to 0.45. 
The initial condition is given by
\begin{align}\rho = 1,~P=
 \left\{
 \begin{array}{l}
 2.5~\text{for}~\sqrt{x^2+y^2}<0.5,\\
 0.1~\text{for}~\sqrt{x^2+y^2}>0.5.\\
 \end{array}
 \right.
\end{align}

Figure~\ref{fig:2D_Blast_Wave} shows the rest-mass density profile at $t=18$ with the {\tt HLLC} solver (left panel) and the {\tt HLLE} solver (right panel). By this time, the blast wave has intersected itself many times, and consequently a Rayleigh-Taylor-like instability (known in this context as the Richtmyer-Meshkov instability) has developed~\cite{White:2015omx}. With the {\tt HLLC} solver, the Richtmyer-Meshkov instability is well resolved, and as a result the density irregularity around the elliptical figure is sharply captured. By contrast, with the {\tt HLLE} solver, the fine structure around the elliptical figure is not captured well due to the large numerical diffusivity. This demonstrates an \textit{effective} improvement in spatial resolution with the {\tt HLLC} solver compared to that with the {\tt HLLE} solver. 

\subsubsection{Magnetohydrodynamics: two-dimensional magnetized current sheet}

Next we consider a two-dimensional problem in relativistic magnetohydrodynamics: that of a magnetized current sheet, studied recently by Refs.~\cite{Mignone:2021,White:2015omx}. The initial profile for the magnetic field is given by
\begin{align}
B^x = B_0 \tanh\left(\frac{y}{a}\right),
\end{align}
where $B_0=1$ and $a=0.04$. The density is uniform with $\rho=1$ and the fluid is at rest with $v^i=0$. The thermal pressure is determined from the force balance with the magnetic pressure and its profile is given by
\begin{align}
 P = \frac{B_0^2}{2}\left(\beta + 1 \right) - \frac{B_x^2}{2},
\end{align}
where $\beta$ is the initial plasma-beta parameter, which we set to unity. The equilibrium magnetic field is initially perturbed and the perturbation is given by the $z$-component of the vector potential as
\begin{align}
 \delta A_z = \epsilon B_0 \cos\left(\frac{k_yy}{2}\right)\cos\left(k_xx\right),
\end{align}
where $k_x=2\pi/L_x$, $k_y=2\pi/L_y$, $\epsilon=10^{-3}$, and $L_x$ and $L_y$ denote the domain size in the $x$- and $y$-directions, respectively. We employ a simulation domain consisting of $x\in[-0.5,0.5]$ and $y\in [-0.25,0.25]$. To check convergence, we carry out simulations at three different resolutions: $(N_x,N_y) = (512,256)$, $(256,128)$, and $(128,64)$ . We set the CFL number to 0.8 in all simulations. We impose a periodic boundary condition in the $x$-direction, and a reflective boundary condition in the $y$-direction. With this setup, the maximum Alfv\'{e}n wave speed is $\approx 0.557$ and the Alfv\'{e}n timescale is $t_{\rm A}\approx 1.78$.

\begin{figure}[t]
 	 \includegraphics[width=1\linewidth]{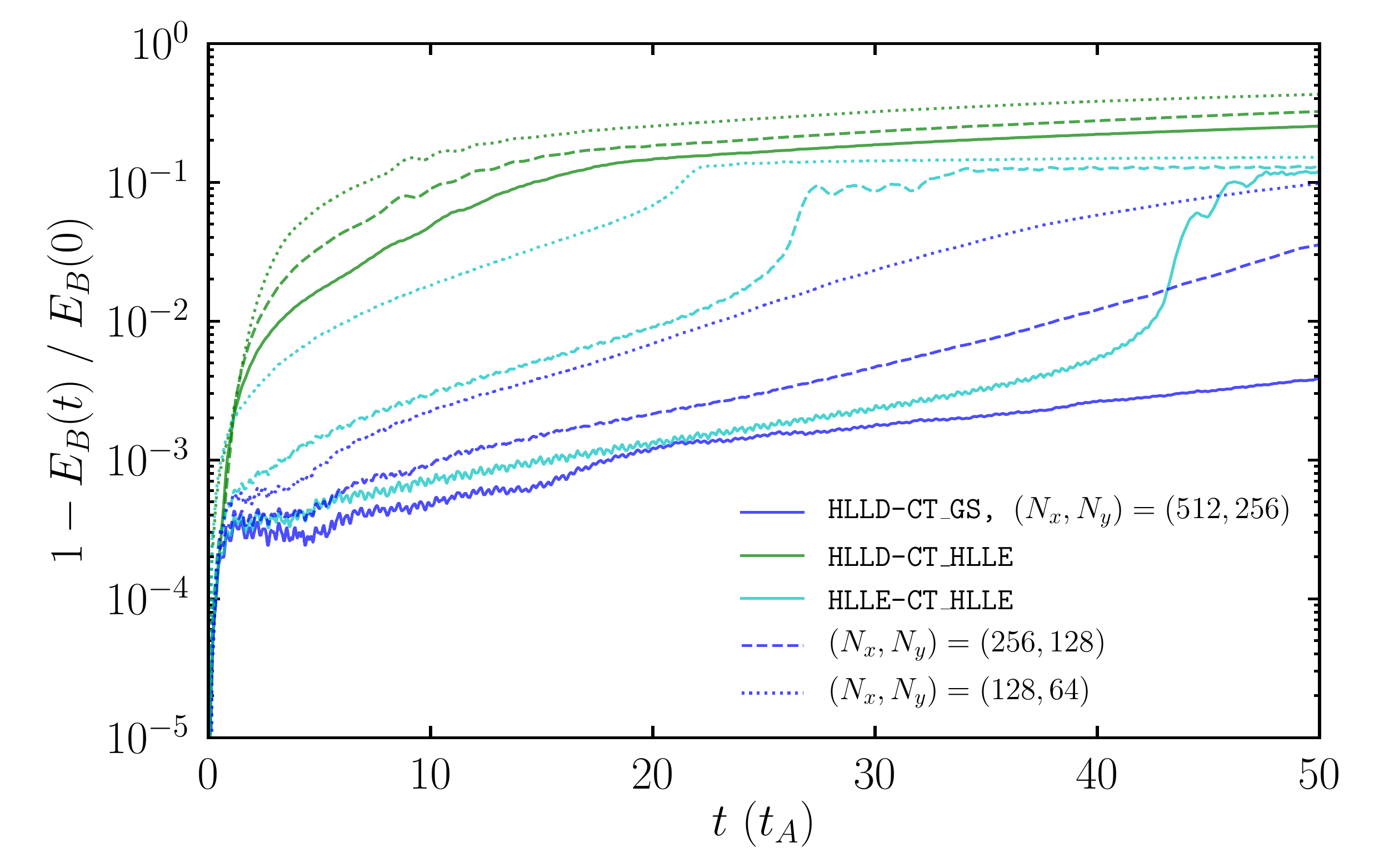}
 	 \caption{Magnetic-field energy dissipation in the magnetized current sheet problem as a function of time. Blue, green, and cyan curves denote  numerical solutions with the {\tt HLLD-CT\_GS}, {\tt HLLD-CT\_HLLE}, and {\tt HLLE-CT\_HLLE} solvers, respectively. The solid, dashed, and dotted curves denote a resolution of $(N_x,N_y)=(512,256)$, $(256,128)$, and $(128,64)$, respectively. We employ RK4 and 3rd-order PPM cell reconstruction in all runs. 
         }\label{fig:CS_MHD}
\end{figure}

Figure~\ref{fig:2D_CS} displays colorplots of the thermal pressure together with the magnetic-field lines at three different times: $t=10.03t_A$ (left panel), $t=30.08t_A$ (center), and $t=50.06t_A$ (right). The top, middle, and bottle panels show the numerical solutions with the {\tt HLLD-CT\_GS}, {\tt HLLD-CT\_HLLE}, and {\tt HLLE-CT\_HLLE} solvers, respectively. The snapshots are all taken from our highest resolution runs with $(N_x,N_y)=(512,256)$.  Magnetic field lines reconnect at $y\approx 0$ due to the numerical resistivity inherent both in the Riemann solvers as well as in the constrained transport scheme. Once reconnection starts, the profile of the magnetic-field lines changes, and as a result, the thermal pressure profile is modified, leading to the formation of island-like structures. 

The timescale of the reconnection depends on how large the numerical resistivity is. Figure~\ref{fig:2D_CS} indicates that the {\tt HLLD-CT\_GS} solver is accompanied with the smallest numerical resistivity because the formation of the islands is delayed. It is found that {\tt HLLD-CT\_HLLE} solver has the largest numerical resistivity, leading to rapid formation of the islands. This does not agree with one's naive expectation, because the {\tt HLLE-CT\_HLLE} solver is actually \textit{less} dissipative than the {\tt HLLD-CT\_HLLE} solver. 
In other words, we observe an {\it unexpected} hierarchy between the {\tt HLLD-CT\_HLLE} and {\tt HLLE-CT\_HLLE} solvers.
This stems from the algorithm of the {\tt CT\_HLLE} solver. In this constrained transport scheme, dissipation terms which are proportional to the maximum absolute value of the characteristic speed appear in the electric-field evaluation (see, e.g., Eq.~(44) in Ref.~\cite{DelZanna:2002rv}). 
This characteristic speed is then obtained from the (global) Riemann solver. We find that the HLLD solver returns a larger characteristic speed than the HLLE solver. As a result, the {\tt HLLD-CT\_HLLE} solver ends up being more diffusive than the {\tt HLLE-CT\_HLLE} solver, as can be seen in this test problem.

Figure~\ref{fig:CS_MHD} shows the fraction of the initial magnetic-field energy that is dissipated as a function of time. With the {\tt HLLD-CT\_GS} solver (blue curves), the magnetic-field energy dissipates only gradually. Also, the dissipation rate is suppressed when we employ higher resolution: the energy increases by an order of magnitude only over $50$ Alfv\'en timescales. This feature is also found for the {\tt HLLE-CT\_HLLE} solver (cyan curves), although the dissipation rate steeply rises at a later time, $t\approx 40t_A$, even in our highest resolution run. With the {\tt HLLD-CT\_HLLE} solver (green curves), magnetic reconnection commences immediately after the simulation starts. We conclude that for problems involving strong magnetic field gradients (current sheets) accurate evolution can be modeled only when the {\tt HLLD} solver is paired with {\tt CT\_GS} for the constrained transport.

\begin{figure}[t]
 \includegraphics[width=1\linewidth]{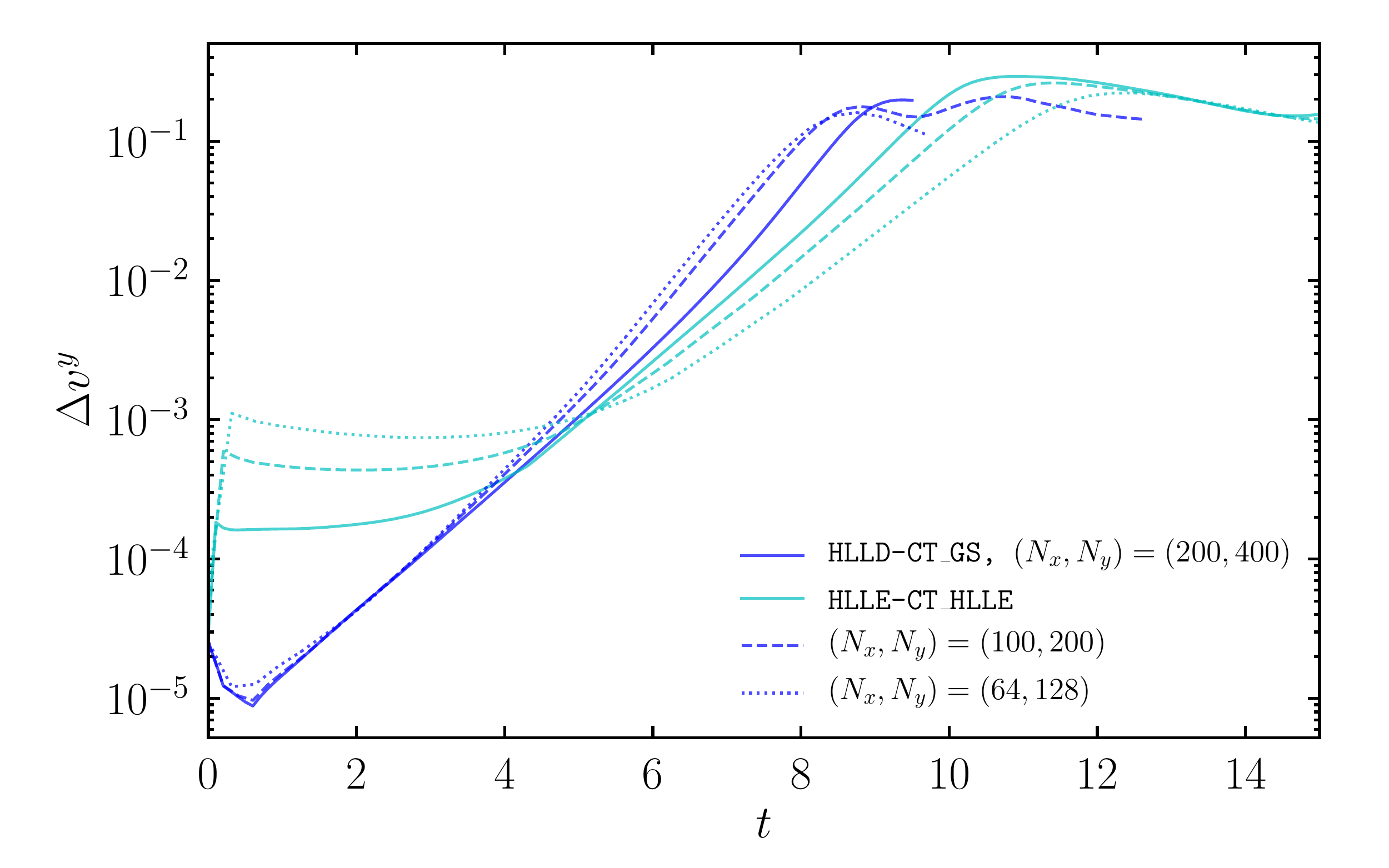}
	 \caption{Perturbed velocity difference $\Delta v^y \equiv (v_\mathrm{max}^y - v_\mathrm{min}^y)/2$ in the special-relativistic magnetohydrodynamical Kelvin-Helmholtz instability as a function of time. The blue curves denote results obtained with the {\tt HLLD-CT\_GS} solver and the cyan curves denote results {\tt HLLE-CT\_HLLE} solver. 
	 The solid, dashed, and dotted curves correspond to resolutions of $(N_x,N_y)=(200,400)$, $(100,200)$, and $(64,128)$, respectively. 
	 }\label{fig:KH_MHD_vy}
\end{figure}

\begin{figure*}[t]
 \includegraphics[width=0.95\linewidth]{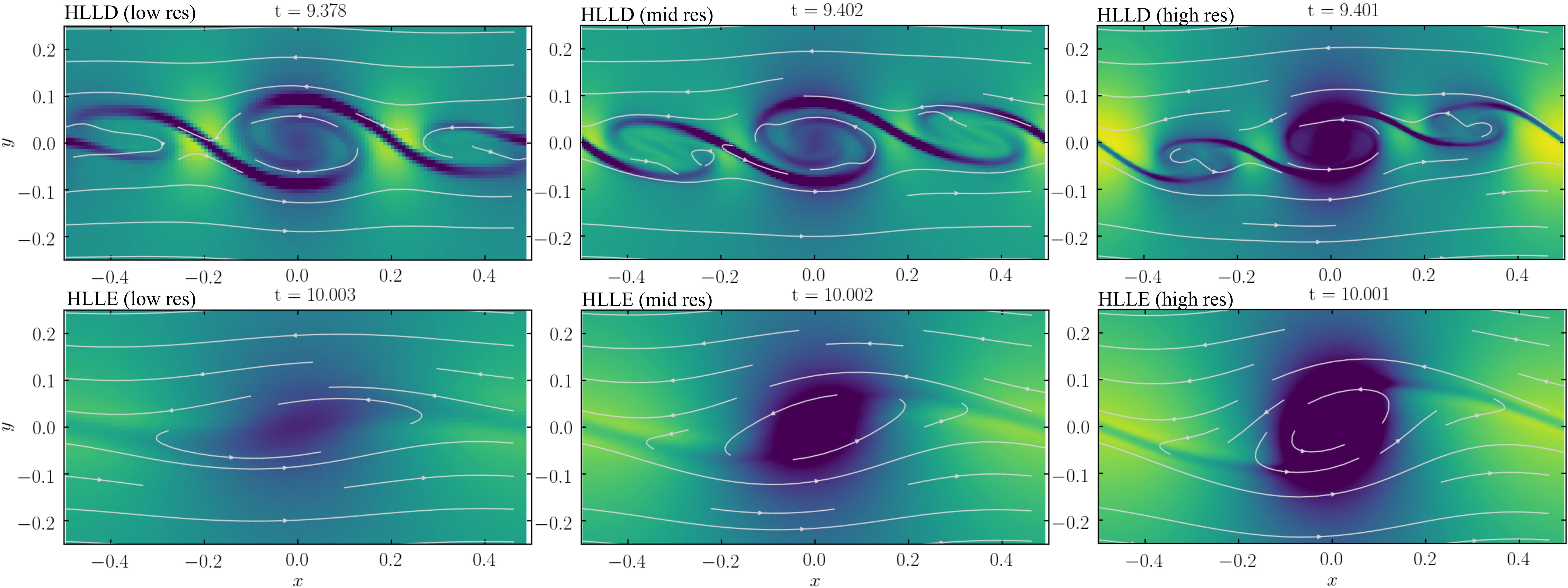}
	 \caption{Density snapshots taken near nonlinear saturation of the special relativistic magnetohydrodynamical Kelvin-Helmholtz instability. Top row: the results with the {\tt HLLD-CT\_GS} solver at three different resolutions. Bottom row: same as the top panel but with the {\tt HLLE-CT\_HLLE} solver. The white lines indicate velocity field streamlines. The left, middle, and right panels show the low, middle, and high resolution runs, respectively.
	 }\label{fig:KH_MHD_flowfield}
\end{figure*}
\subsubsection{Magnetohydrodynamics: two-dimensional Kelvin-Helmoltz instability}
The second two-dimensional problem in special relativistic magnetohydrodynamics is the Kelvin-Helmholtz instability, as proposed in Refs.~\cite{MUB:2009,Bucciantini:2006jg}. For this, we prepare a simulation domain which spans $x\in [-0.5,0.5]$ and $y\in[-1,1]$. To check the convergence, we perform the simulations with three different resolutions: $(N_x,N_y)=(200,400)$ (`high' resolution), $(N_x, N_y)=(100,200)$ (`medium' resolution), and $(N_x,N_y)=(64,128)$ (`low' resolution). 
The simulations are carried out with either the {\tt HLLD-CT\_GS} or {\tt HLLE-CT\_HLLE} solvers, and we employ 3rd-order PPM reconstruction for all the simulations. We impose a periodic boundary condition in the $x$-direction, and an outflow boundary condition in the $y$-direction. The CFL number is set to 0.4 in all the simulations.

As the initial condition, we give a tanh-shaped shear velocity profile for the $x$-component,
\begin{align}
 v^x = - v_{\text{sh}}\tanh(y/a),
\end{align}
where $v_{\text{sh}} = 0.25$ and $a = 0.02$. The thickness $a$ of the shear layer is covered by around $2,4,$ and $8$ grid cells at the low, medium, and high resolutions, respectively. We employ a uniform density of $\rho=1$, and a uniform gas pressure with $P=20$. The adiabatic index is taken to be $4/3$. Note that our setup is different from that employed in the recent test simulation for the Kelvin-Helmholtz instability in special relativistic magnetohydrodynamics of Ref.~\cite{Mattia:2021bwh}, in which the authors employ a non-uniform density field, a smaller shear-layer thickness of $a = 0.01$, and an amplitude of the $x$-component of the velocity ($v_{\text{sh}} = 0.5$) which is twice that used in our runs.

The magnetic field at $t=0$ is given by
\begin{align}
(B^x,B^y,B^z)=\left(\sqrt{2\sigma_\text{pol}P},0,0\right),
\end{align}
i.e. the magnetic field is initially uniform and parallel to the velocity in the lower-half of the $xy$-plane. We set $\sigma_\text{pol}=0.01$. The shear layer is perturbed by the motion in the $y$-direction as
\begin{align}
v^y = \frac{1}{40000}\sin(2\pi x)\exp\left(-100y^2\right),
\end{align}
while $v^z=0$. 

Figure~\ref{fig:KH_MHD_vy} shows the perturbed velocity difference $\Delta v^y \equiv (v_\mathrm{max}^y - v_\mathrm{min}^y)/2$ as a function of time taken from six simulations at three different resolutions and employing either the {\tt HLLD-CT\_GS} or {\tt HLLE-CT\_HLLE} solver. All the simulations start from perturbations of size $\sim 10^{-5}$. We find exponential growth followed by nonlinear saturation at the end of the linear phase at $t \sim 10$. The behavior during the linear phase depends strongly on the solver, particularly at low resolutions. Nonlinear saturation occurs more quickly in the simulations with the {\tt HLLD-CT\_GS} solver than in those with the {\tt HLLE-CT\_HLLE} solver, but the saturation amplitude depends only weakly on the solver and resolution. The growth rate is higher with the less diffusive {\tt HLLD-CT\_GS} solver than with the {\tt HLLE-CT\_HLLE} solver, but the results converge between the two solvers as the resolution is improved. This result is consistent with that in Ref.~\cite{Mattia:2021bwh} (see their Fig.~14). The evolution after the nonlinear saturation is not sensitive to the solver or resolution, although at late times (not shown) the velocity difference decays more quickly in the simulations with the (more diffusive) {\tt HLLE-CT\_HLLE} solver than with the {\tt HLLD-CT\_GS}. 

In Fig.~\ref{fig:KH_MHD_flowfield} we show snapshots of the density at nonlinear saturation $t\sim 10$ from the six simulations. The top row shows results from the low, medium, and high resolution runs using the {\tt HLLD-CT\_GS} solver, while the bottom row shows the corresponding snapshots from runs that employ the {\tt HLLE-CT\_HLLE} solver. Using the {\tt HLLD-CT\_GS} solvers, we observe the formation of a single vortex together with two neighboring, stretched secondary vortices that are well-resolved at all resolutions, whereas with the {\tt HLLE-CT\_HLLE} solver we see the formation of only a single large vortex at the shear interface, mirroring the behaviour of the Kelvin-Helmholtz instability in the simulations of Ref.~\cite{Bucciantini:2006jg} which employed the {\tt HLLE-CT\_HLLE} solver. Our results show that, at least at low resolutions, the {\tt HLLE} solver is not appropriate for studying phenomena in which the Kelvin-Helmholtz instability plays an important role.

\subsection{General relativistic problems in a fixed background spacetime}
\label{SECTION_GRFixedTest}
\begin{figure}[t]
 	 \includegraphics[width=0.88\linewidth]{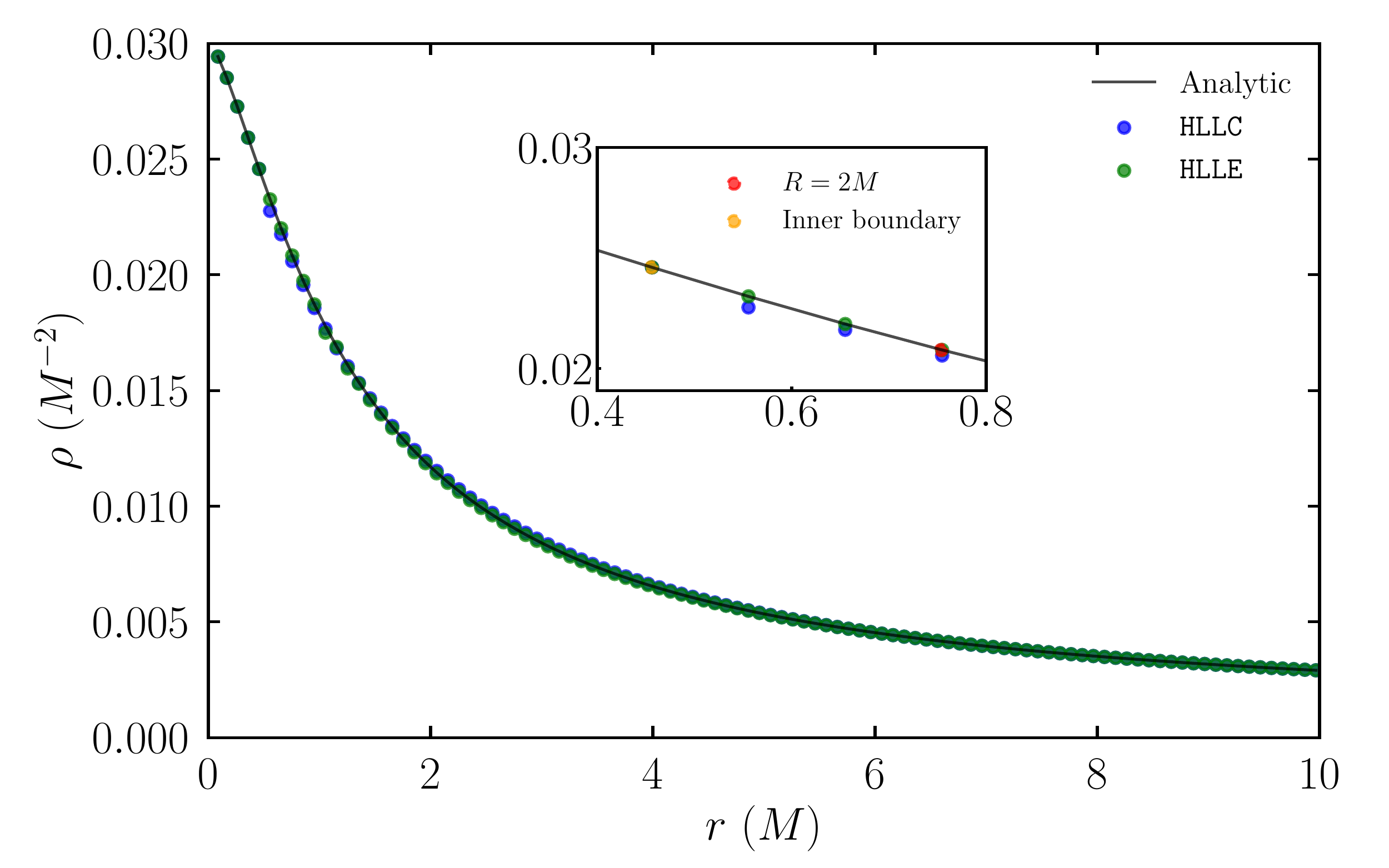}\\
  	 \includegraphics[width=0.88\linewidth]{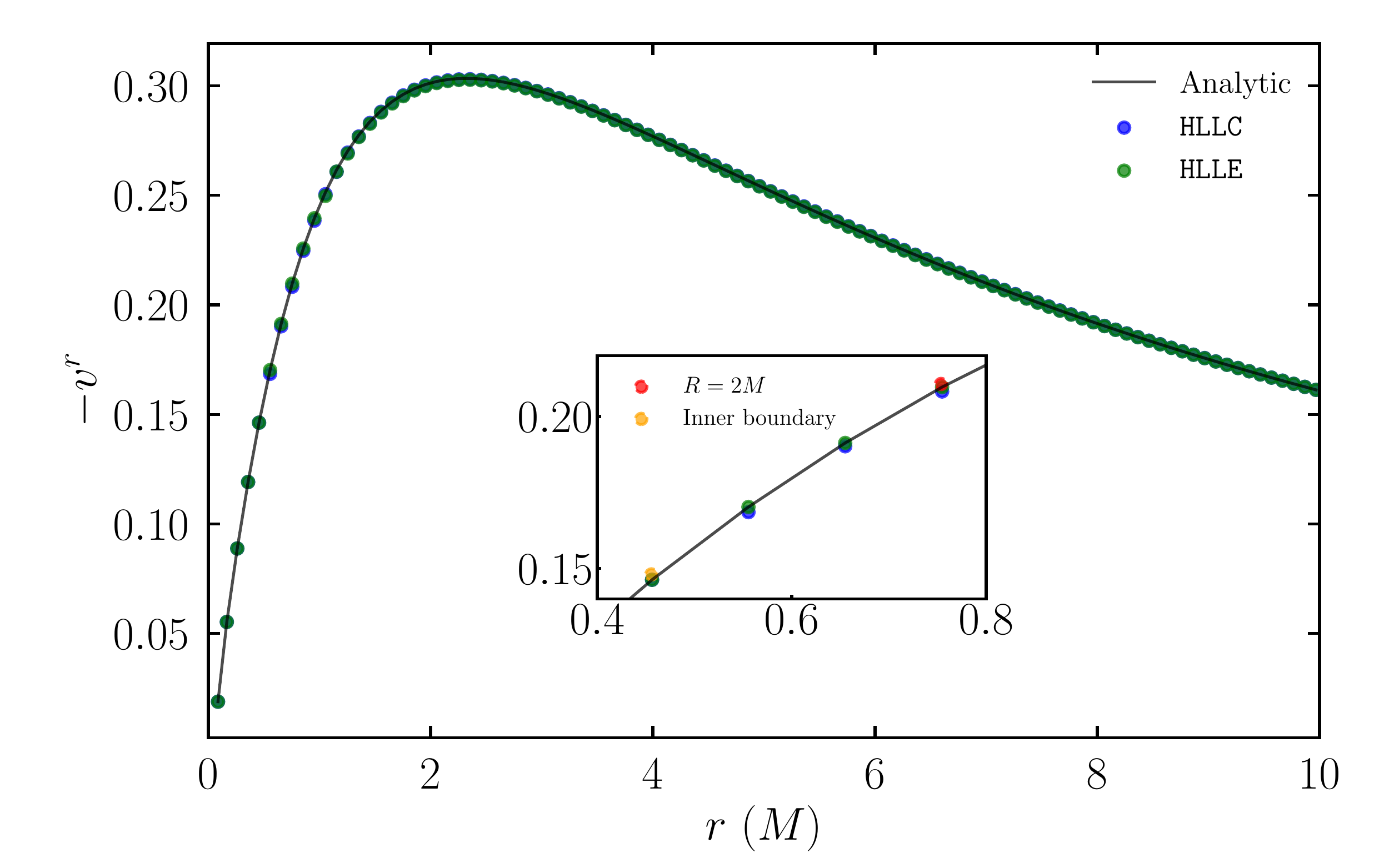}\\
     \includegraphics[width=0.88\linewidth]{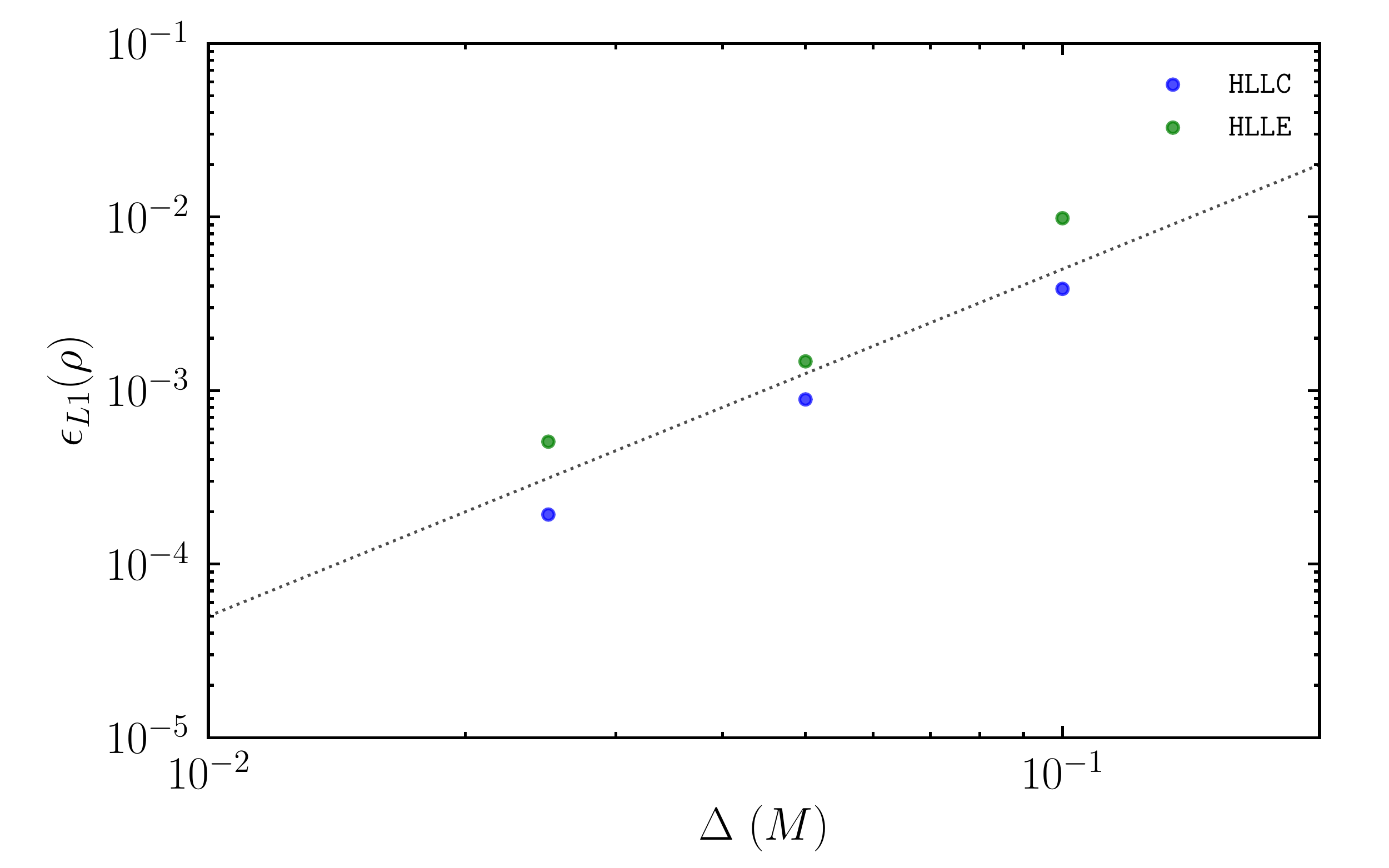}
 	 \caption{Top: Radial rest-mass density profile for (hydrodynamic) Bondi flow in a non-rotating black hole spacetime at $t=22.5M$ (the end of the simulation). The mass accretion rate is fixed at $\dot{M}_\text{acc}=0.797$.
 	  The solid black curve indicates the analytical solution, while the blue and green dots denote the numerical solution obtained with the {\tt HLLC} and {\tt HLLE} solvers, respectively. 
 	  The inset show the solution near the inner boundary. In the inset, the red and yellow dots denote the location of the horizon and of the inner boundary, respectively. 
 	   Middle: Same as the top panel, but showing the radial velocity profile.
 	   Bottom: $L_1$ norm of the error in the rest-mass density as a function of the spatial grid spacing. The blue and green dots denote the error of the numerical solutions with the {\tt HLLC} and {\tt HLLE} solvers, respectively. The dotted line denotes 2nd-order convergence.
         }\label{fig:Bondi}
\end{figure}

\subsubsection{Hydrodynamics: Bondi flow}
As a test problem in a curved (but static) spacetime, we consider spherical accretion (ingoing Bondi flow) onto a non-rotating black hole~\cite{Hawley:1984}. The Bondi flow in Schwarzschild coordinates has been extensively discussed in the literature (see, e.g.,  Ref.~\cite{White:2015omx}). Following previous work~\cite{Gammie:2003rj,Shibata:2005gp}, we adopt the parameters for this problem as follows: an adiabatic index of $\Gamma = 4/3$, an adiabat of $K=1$, and a critical radius of $r_\text{c}=8M$, where $M$ denotes the black hole mass. With this setup, the mass accretion rate $\dot{M}_\text{acc}$ is 0.797. We perform simulations both with the {\tt HLLE} and {\tt HLLC} solvers, and employ 3rd-order PPM reconstruction. 

Our numerical-relativity code employs the so-called puncture formalism, and hence, in the presence of black holes, the black-hole spacetime is foliated in most cases by the so-called limiting hypersurface ~\cite{1973PhRvD...7.2814E}. Thus, for preparing a practical setup in this test problem, a non-rotating black hole should be described in the so-called maximal trumpet geometry rather than in Schwarzschild coordinates or in isotropic coordinates on slices of constant Schwarzschild time~\cite{Baumgarte:2007ht,Miller:2016vkn}.
Note that in both of these latter two coordinate systems, the fluid four-velocity exhibits pathological behavior near the horizon~\cite{Miller:2016vkn}~\footnote{The pathological behavior at the horizon is avoidable if one employs Kerr-Schild coordinates. However, we employ the maximal trumpet geometry in this test problem because of its high compatibility with our numerical relativity code.}. In Appendix~\ref{appendix:coordinate_transformation}, we describe the explicit coordinate transformation from the Schwarzschild coordinates to the maximal trumpet geometry. With this geometry, the radial component of the shift vector is non-zero. Therefore, the tetrad basis (see, e.g., Eq.~(\ref{eq:tetrad_basis})) does not agree any longer with a coordinate basis in the Minkowski spacetime, and the cell interface may be dragged by the shift vector as discussed in Sec.~\ref{subsec:tetrad}.

We employ a simulation domain in Cartesian coordinates spanning $x,y,z\in[0,L]$ with $L=10M$. The grid spacing of the simulation is given by $\Delta=\Delta x=\Delta y = \Delta z =0.1M$ with $N=N_x=N_y=N_z=100$ as the number of grid cells in each direction. We also check convergence by increasing the resolution to $N=200$ and $N=400$, which correspond to grid spacings of $\Delta=0.05M$ and $0.025M$, respectively. We set the CFL number to 0.45 and integrate the numerical solution up to $t=22.5M$. 
We impose a stationary boundary condition at the outer and inner boundaries, with the latter located at $r_\text{in}=0.4M$. Note that the horizon in this geometry is located at $r_\text{BH} \approx 0.78M$. We also impose octant symmetry at the $x,y$, and $z=0$ planes.

Figure~\ref{fig:Bondi} shows radial profiles of the rest-mass density and the radial velocity calculated by the {\tt HLLC} solver with the blue dots and by the {\tt HLLE} solver with the green dots on top of the analytic solution~\cite{Hawley:1984}. The profiles are along the diagonal direction, i.e., $x=y=z$ in the simulation domain. This figure demonstrates that our implementation of the {\tt HLLC} solver in curved spacetime works well. It also shows that, for this particular problem, the {\tt HLLE} solver works as well as the {\tt HLLC} solver because of the smoothness of the accretion flow, as many other previous implementations have shown; e.g., Refs.~\cite{Gammie:2003rj,Shibata:2005gp,Mosta:2013gwu,2010PhRvD..82h4031E}.

In the lower panel of Fig.~\ref{fig:Bondi} we plot the $L_1$ norm of the error in the rest-mass density as a function of the spatial grid spacing. The convergence order of the $L_1$ norm of the error is $\approx 2$ both for the {\tt HLLC} and {\tt HLLE} solvers, because our Riemann solver is 2nd-order accurate. 
One likely reason for the slight deviation from the expected accuracy is that spherical symmetry of the accretion flow is not perfectly preserved during the evolution because we simulate it in the Cartesian geometry. This plot also shows that the numerical solution with the {\tt HLLC} solver is more accurate than that with the {\tt HLLE} solver. Our interpretation of this is that with the tetrad transformation (see  Sec.~\ref{subsec:tetrad}) the frame-dragging effect of the cell interface is taken into account with a better accuracy (see also Fig.~\ref{fig:Riemann_fan})~\footnote{Note that our HLLE solver (which was the only Riemann solver present in our original formulation), does not employ the tetrad transformation, but is instead formulated directly in a curved spacetime. See Ref.~\cite{Shibata:2005gp} for details.}.

\begin{figure*}[t]
 	 \includegraphics[width=0.46\linewidth]{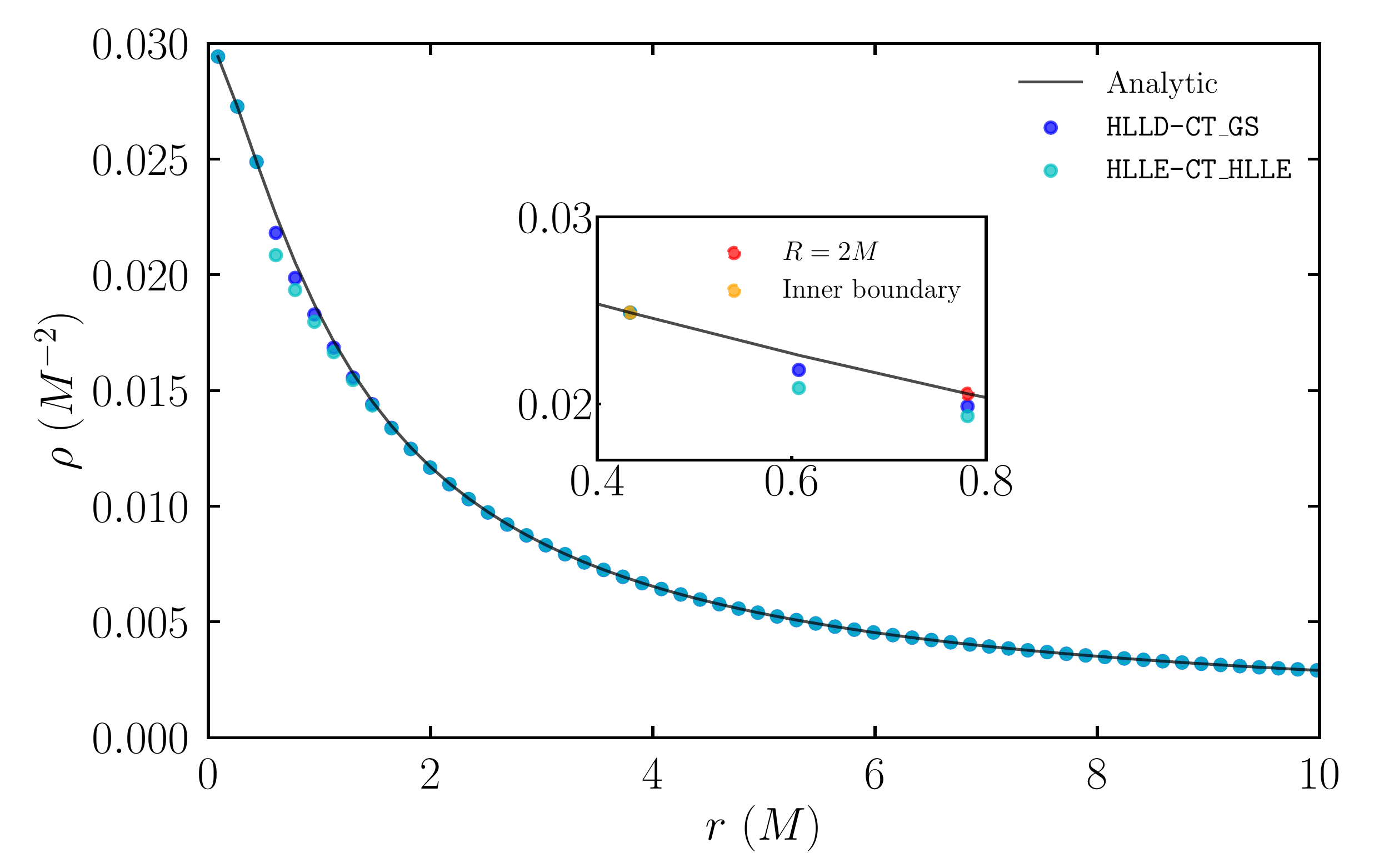}~~
  	 \includegraphics[width=0.46\linewidth]{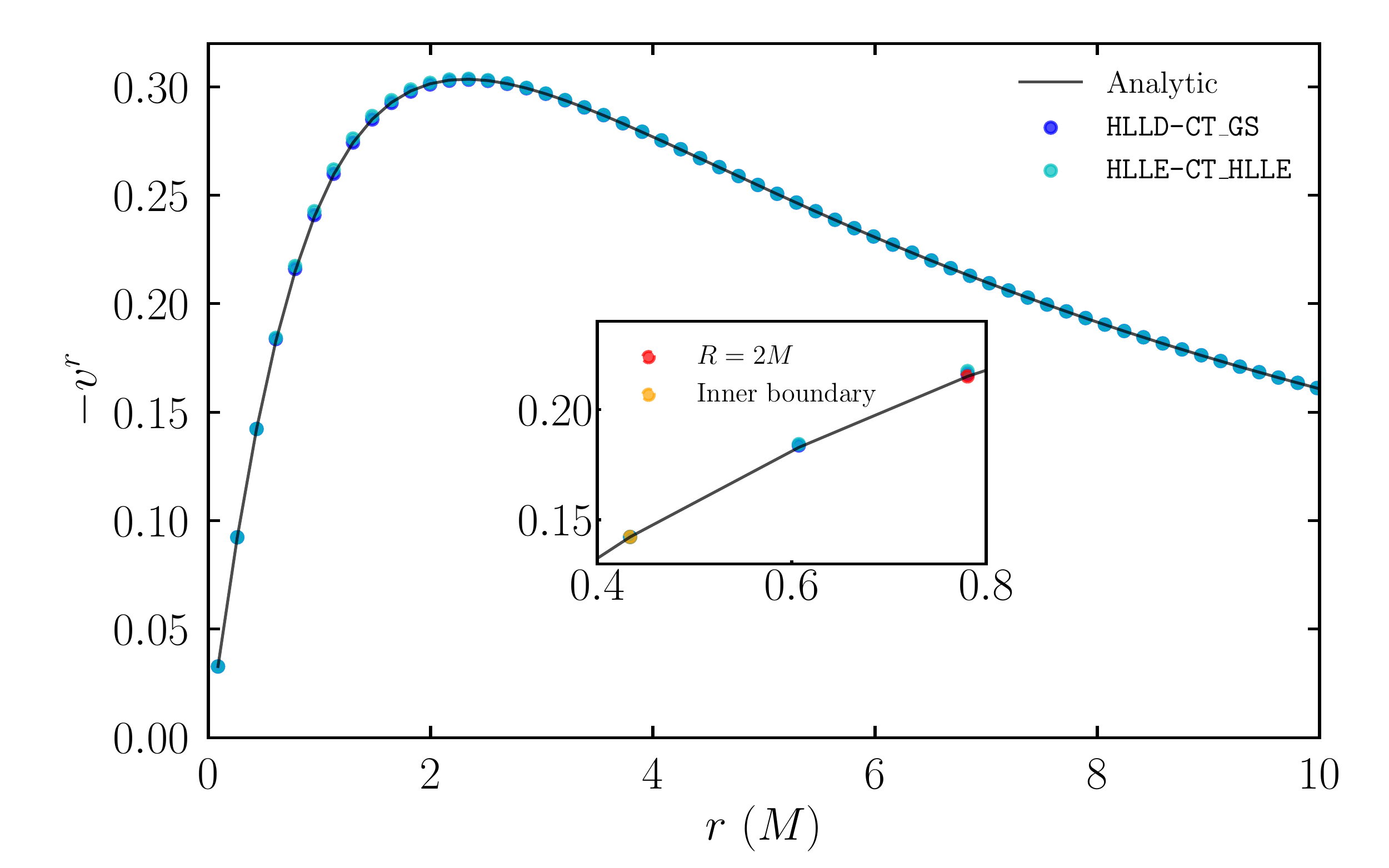}\\
     \includegraphics[width=0.46\linewidth]{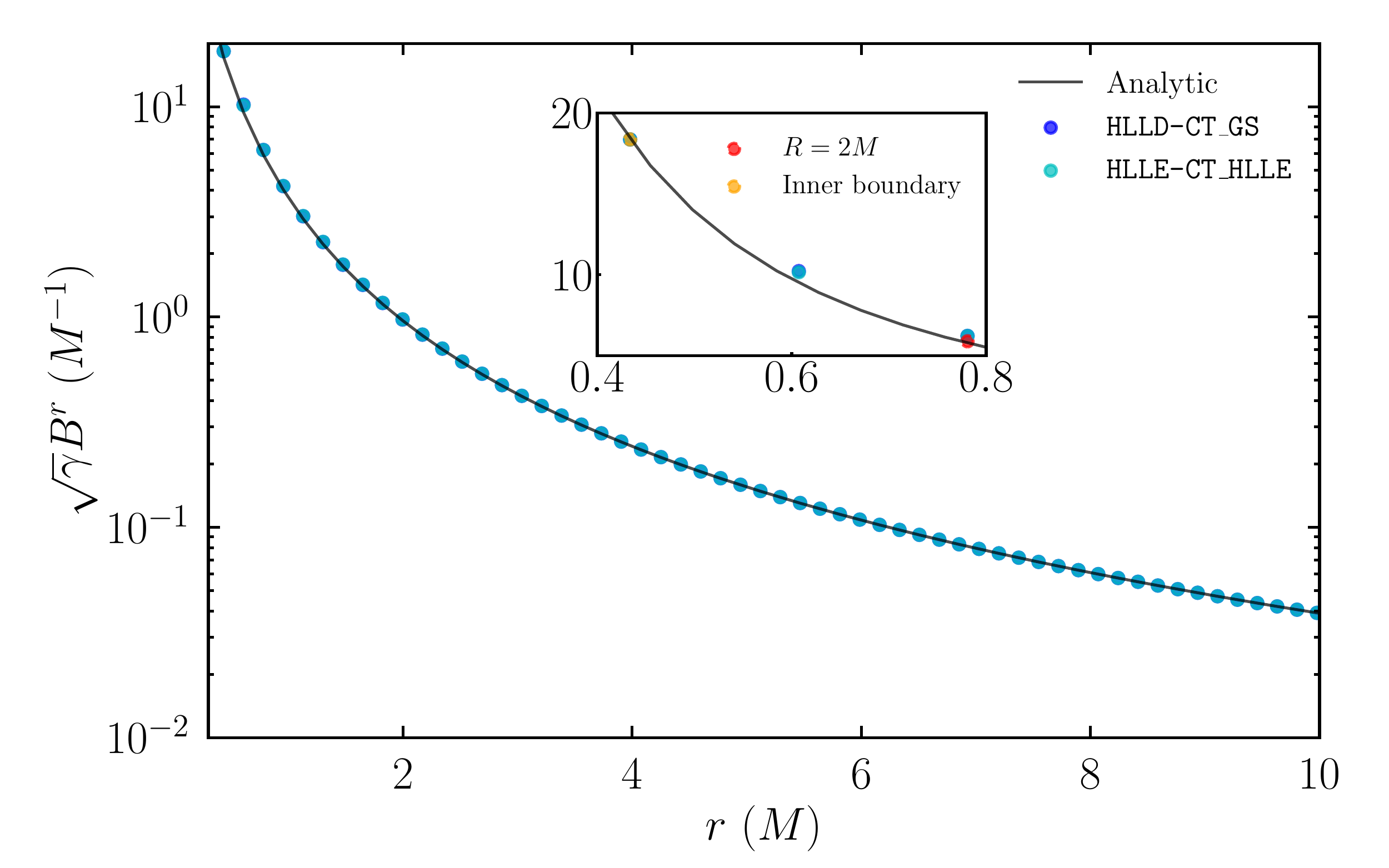}~~
     \includegraphics[width=0.46\linewidth]{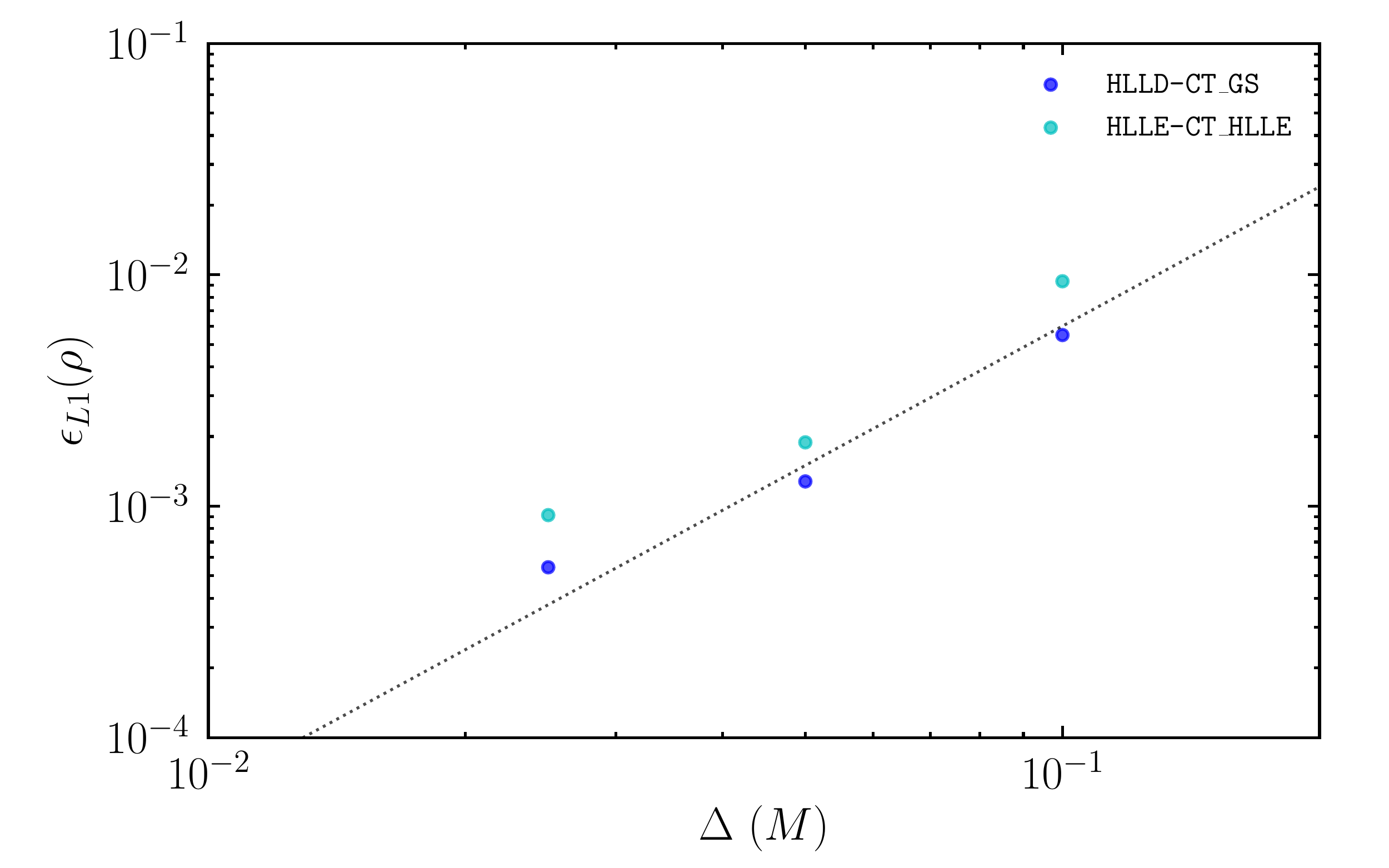}
 	 \caption{Same as Fig.~\ref{fig:Bondi}, but for a magnetized Bondi flow in a non-rotating black hole spacetime at $t=22.5M$. The blue and cyan dots denote numerical solutions with the {\tt HLLD-CT\_GS} and {\tt HLLE-CT\_HLLE} solvers, respectively. The bottom-left panel shows the radial magnetic field profile.
         }\label{fig:Mag_Bondi}
\end{figure*}

\subsubsection{Magnetohydrodynamics: Magnetized Bondi flow}
The next test problem in a curved spacetime is magnetized Bondi flow onto a non-rotating black hole. It is known that a purely radial magnetic field does not alter the flow profile of non-magnetized Bondi flow~\cite{Shibata-textbook}. 
Therefore, we employ the same flow profile used in the previous section. From the divergence-free condition~(\ref{eq:divB}), the radial magnetic field should be $B^R \propto f/R^2$ in Schwarzschild coordinates (see Appendix~\ref{appendix:coordinate_transformation} for the definition of $f$ and the transformation to the maximal trumpet geometry). The amplitude of the magnetic field is chosen to be such that $b^2/\rho=1$ at $R=3M$. 
We perform two simulations, one with the {\tt HLLD-CT\_GS} solver and the other with the {\tt HLLE-CT\_HLLE} solver. We employ RK4 and 3rd-order PPM reconstruction in both cases.

We employ a simulation domain in Cartesian coordinates spanning $x,y,z\in [0,L]$ with $L=12.8M$. The grid spacing of the simulation is $\Delta = \Delta x=\Delta y=\Delta z=0.1M$ with $N=N_x=N_y=N_z=128$ being the number of the grid cells in each direction. 
To check convergence, we perform better-resolved simulations with $N=256$ and $N=512$, i.e., $\Delta =0.05M$ and $0.025M$, respectively. We impose octant symmetry at the $x$, $y$, and $z=0$ planes, and a stationary condition at the outer and inner boundaries, with the latter located at $r_\text{in}=0.4M$. Numerical simulations are performed up to $t=22.5M$.

Figure~\ref{fig:Mag_Bondi} shows the radial profiles of the rest-mass density (top-left), the radial velocity (top-right), and the radial magnetic field (bottom-left). Numerical solutions with the {\tt HLLD-CT\_GS} solver are indicated by the blue dots, while those with the {\tt HLLE-CT\_HLLE} solver are indicated by the cyan dots. As in the non-magnetized cases, the flow profiles agree with the analytic solution~\cite{Hawley:1984} (see also the insets in Fig.~\ref{fig:Mag_Bondi} which show the solution close to the inner boundary). The rest mass density inside the horizon slightly deviates from the analytic solution. However, the deviation decreases as the spatial resolution is increased. This demonstrates that our HLLD solver works just as well as our HLLC solver in a curved spacetime.  
As in the hydrodynamic case, we find no qualitative difference in the numerical solutions between the {\tt HLLD-CT\_GS} and {\tt HLLE-CT\_HLLE} solvers because of the smoothness of the accretion flow. The bottom-right panel in Fig.~\ref{fig:Mag_Bondi} plots the $L_1$ norm of the error in the rest-mass density as a function of the spatial grid spacing. It shows that (i) the numerical solution with the {\tt HLLD-CT\_GS} solver is more accurate than that with the {\tt HLLE-CT\_HLLE} solver, 
and (ii) the order of the convergence is $\approx 2$. 
These results are essentially the same as those in the previous subsection. Again, the deviation from the formal accuracy of the Riemann solver is likely to be an artifact of the Cartesian geometry which we employ.


\section{Application to a dynamical spacetime}
\label{sec:dynamical_spacetime}

Finally, we apply our new Riemann solvers in general relativity to a dynamical spacetime. We simulate a binary neutron star merger, both with and without magnetic fields. We turn on the solver for Einstein's equations and the neutrino-radiation hydrodynamics solver in the simulations shown in this section (see Eqs.~(\ref{eq:MHD})--(\ref{eq:MHD3})).

\subsection{Hydrodynamics: binary neutron star merger}

\subsubsection{Setup}

First, we consider non-magnetized asymmetric binary neutron stars with masses of  $1.2$ and $1.5M_\odot$. We utilize the spectral method library LORENE~\cite{LORENE,Gourgoulhon:2000nn,Taniguchi:2010kj,Taniguchi:2002ns,Taniguchi:2003hx} to generate a quasi-equilibrium configuration of the irrotational binary neutron star. We also employ an eccentricity reduction prescription to generate an initial condition that has low orbital eccentricity~\cite{Kyutoku:2014yba}. The initial orbital angular velocity is set to be $m_0\Omega_0=0.028$ where $m_0=2.7M_\odot$ is the total mass of the binary. 

Our solver for Einstein's equations implements the BSSN-puncture formulation~\cite{Shibata:1995,Baumgarte:1998te,Campanelli:2005dd,Baker:2005vv}, locally incorporating the Z4c prescription for constraint propagation~\cite{Hilditch:2012fp}. We employ 4th-order centered finite differencing for the spatial derivative of the metric, a lop-sided finite difference for the advection term associated with the shift vector, and 4th-order Runge-Kutta for the time integrator. For the relativistic hydrodynamics solver, we employ either the {\tt HLLC} or {\tt HLLE} solver, together with 3rd-order PPM cell reconstruction. 

We employ the SFHo equation of state for relatively high-density nuclear matter~\cite{Steiner:2012rk}, and the Timmes (Helmholtz) equation of state for the low-density part \cite{Timmes}. Because high-resolution shock-capturing schemes cannot treat the vacuum state, we need to implement an artificial atmosphere outside the neutron stars. In this simulation, we set a constant atmospheric density of $\rho_\text{atm} = 10^3~{\rm g/cm^3}$ for the inner part of the finest fixed mesh refinement (FMR) domain, for which the refinement boundary along each axis is typically located at $L_\text{fin} = 38.7\,{\rm km}$ (see below for the FMR setup in detail). We also set a power-law profile of the atmospheric density of $\rho_\text{atm}=10^3 (L_\text{fin}/r)^3\,{\rm g/cm^3}$ for $r>L_\text{fin}$ and as far as the atmospheric density is larger than the floor value which is determined by the employed equation of state. In our present table for the equation of state, this floor is $\approx 0.17~{\rm g/cm^3}$ and if $\rho_\text{atm}$ becomes smaller than this value, we set the the atmospheric density to the floor value. The atmospheric temperature is set to be $10^{-3}~{\rm MeV}$. 

We also explicitly solve the radiation-hydrodynamics equations for neutrinos in time using an approximate neutrino-transfer scheme based on a leakage scheme~\cite{Sekiguchi:2010ep} and the truncated moment formalism~\cite{Thorne:1981,Shibata:2011kx}. The cooling source terms are computed using a general-relativistic leakage scheme~\cite{Sekiguchi:2012uc}, and heating source terms due to neutrino capture processes are computed by the method presented in Ref.~\cite{Fujibayashi:2017xsz}.

The computational region consists of $13$ levels of FMR half-cubic domains. The size of each FMR domain is $\in [-L/2^{l-1},L/2^{l-1}]$ for $x$ and $y$, and $z\in [0,L/2^{l-1}]$ with $l=1,2,\cdots,13$. Note that in the $z$-direction we impose reflection symmetry with respect to the equatorial plane, $z=0$. We set the overall domain size to $L\approx 158,000~{\rm km}$ and $N=N_x=N_y=N_z=258$. Thus the grid spacing of the finest FMR domain is $\Delta x_{13}=\Delta y_{13}=\Delta z_{13}=150$\,m. To check convergence, we also perform simulations with lower resolutions of $N=196$ and $N=158$, for which the grid spacing of the finest FMR domain is $\Delta x_{13}=200$\,m and $\Delta x_{13}=250$\,m, respectively. For the {\tt HLLC} solver, we perform an additional simulation with $N=377$ and $\Delta x_{13}=100$\,m. By virtue of the cell-centered grid structure, the cell interface of the parent FMR domain coincides with that of the child FMR domain. We employ the {\it reflux} prescription during time marching of the Berger-Oliger type mesh refinement algorithm to ensure the conservation of baryonic mass.


\subsubsection{Inspiral phase}

\begin{figure*}[t]
 	 \includegraphics[width=0.46\linewidth]{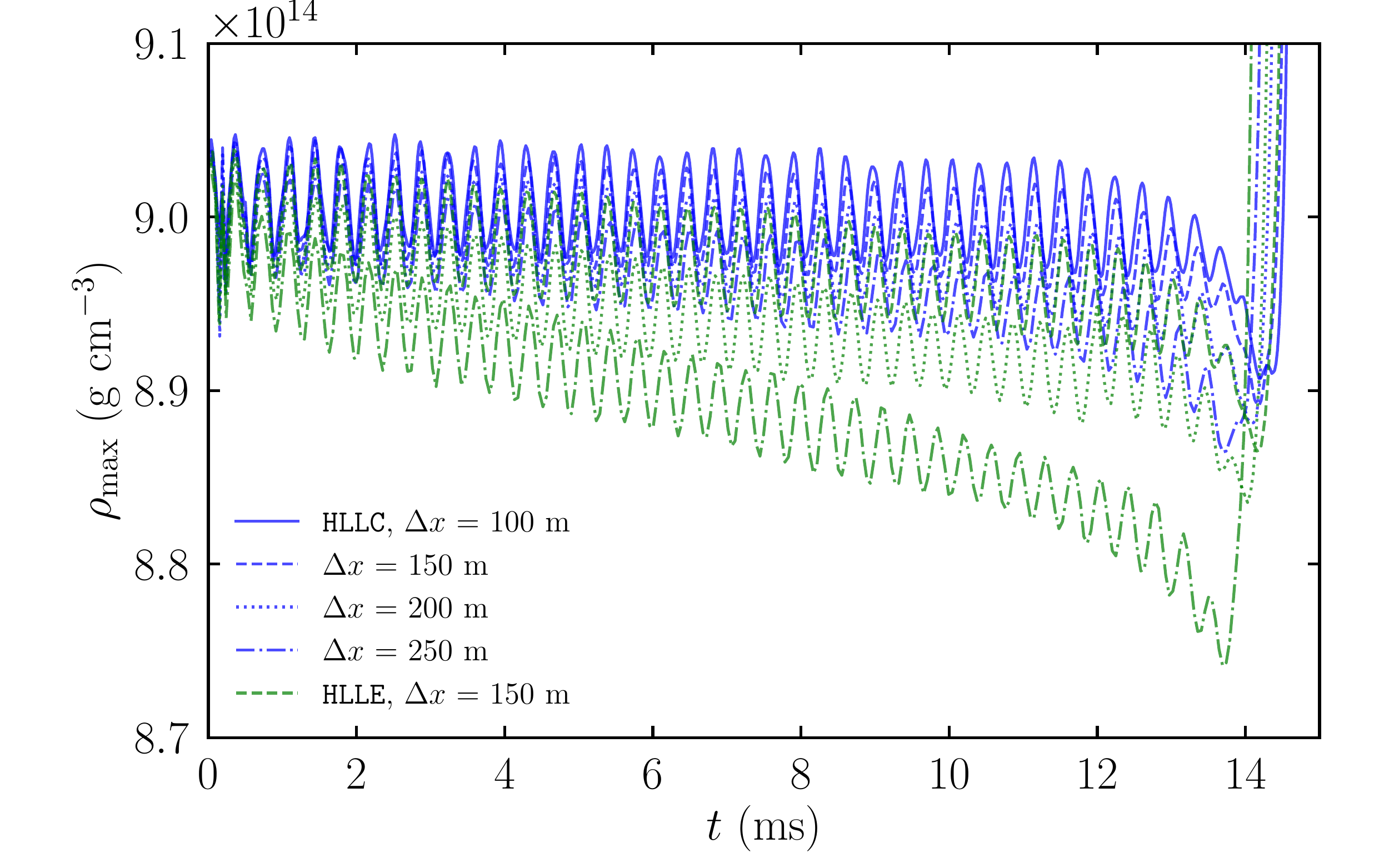}~~
  	 \includegraphics[width=0.46\linewidth]{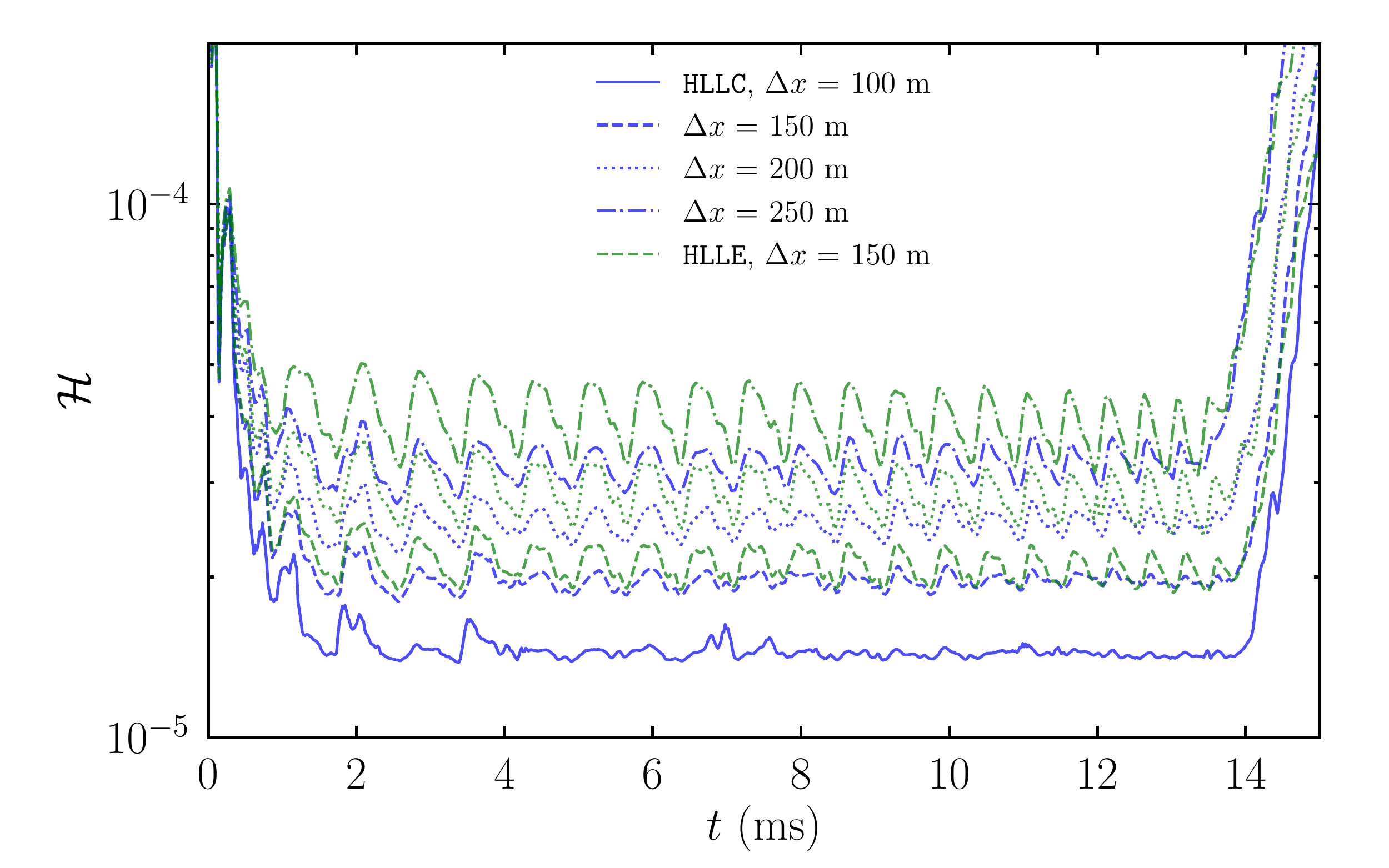}	 
 	 \caption{Maximum rest-mass density (left) and density-weighted Hamiltonian constraint violation (right) as functions of time in non-magnetized simulations of inspiraling binary neutron stars. The blue and green curves denote the results with the {\tt HLLC} and {\tt HLLE} solvers, respectively. The solid, dashed, dotted, and dot-dashed curves denote the results with (finest-level) grid spacings of $\Delta x_{13}=100$\,m, $150$\,m, $200$\,m, and $250$\,m, respectively.
 	 }\label{fig:non_magnetized_BNS_inspiral}
\end{figure*}

\begin{figure*}[t]
 	 \includegraphics[width=0.46\linewidth]{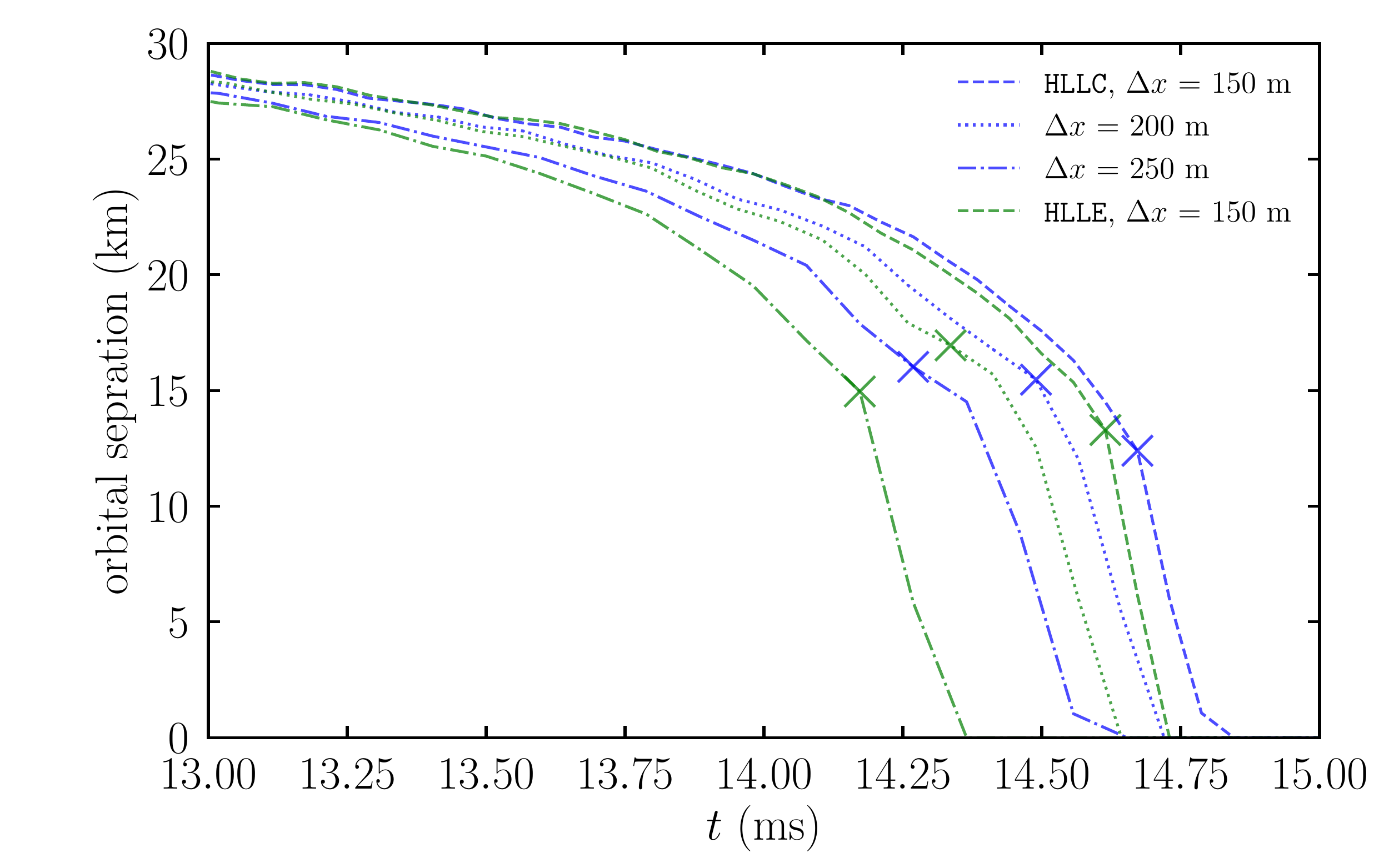}\\
  	 \includegraphics[width=0.46\linewidth]{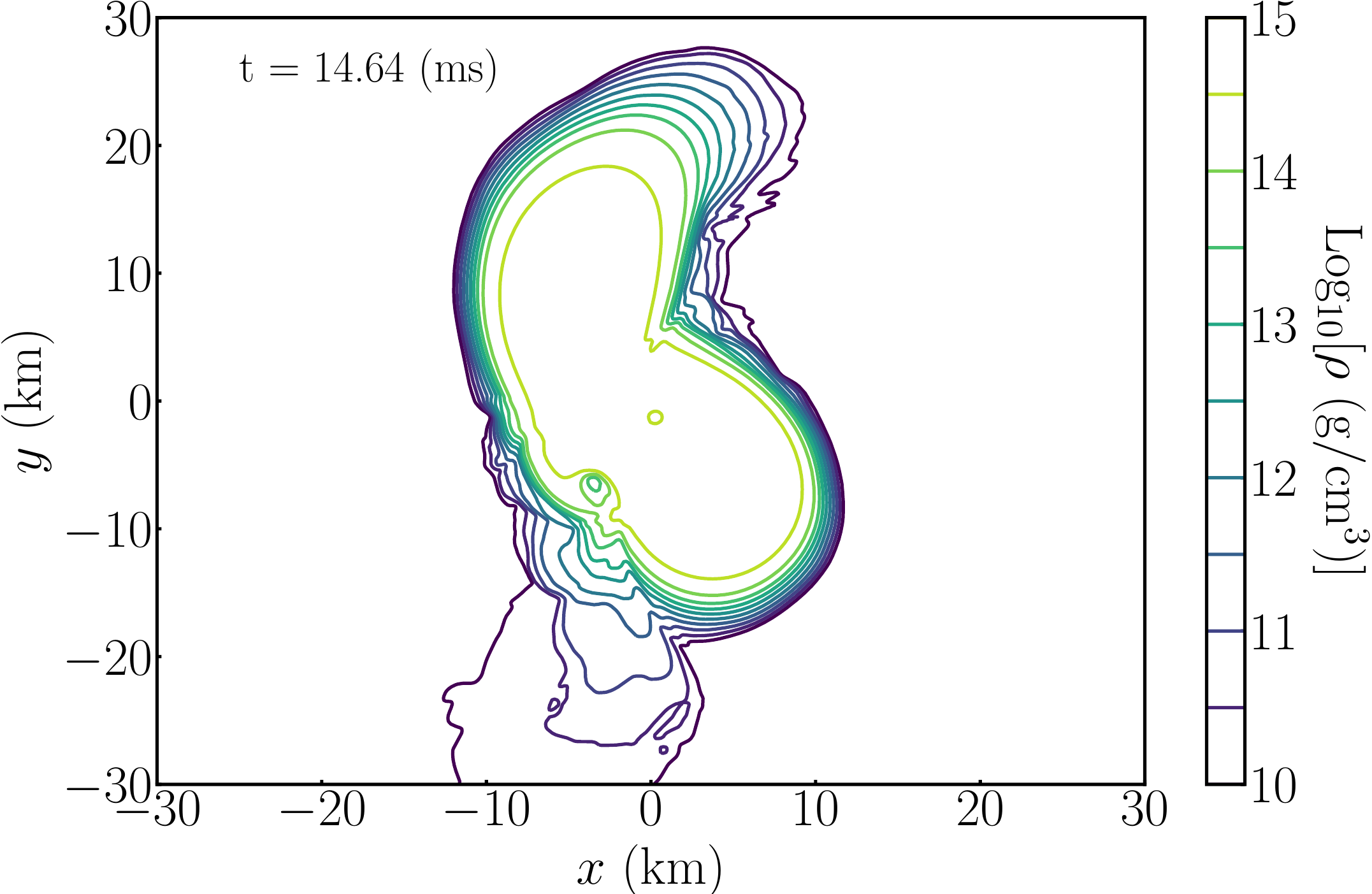}~~
     \includegraphics[width=0.46\linewidth]{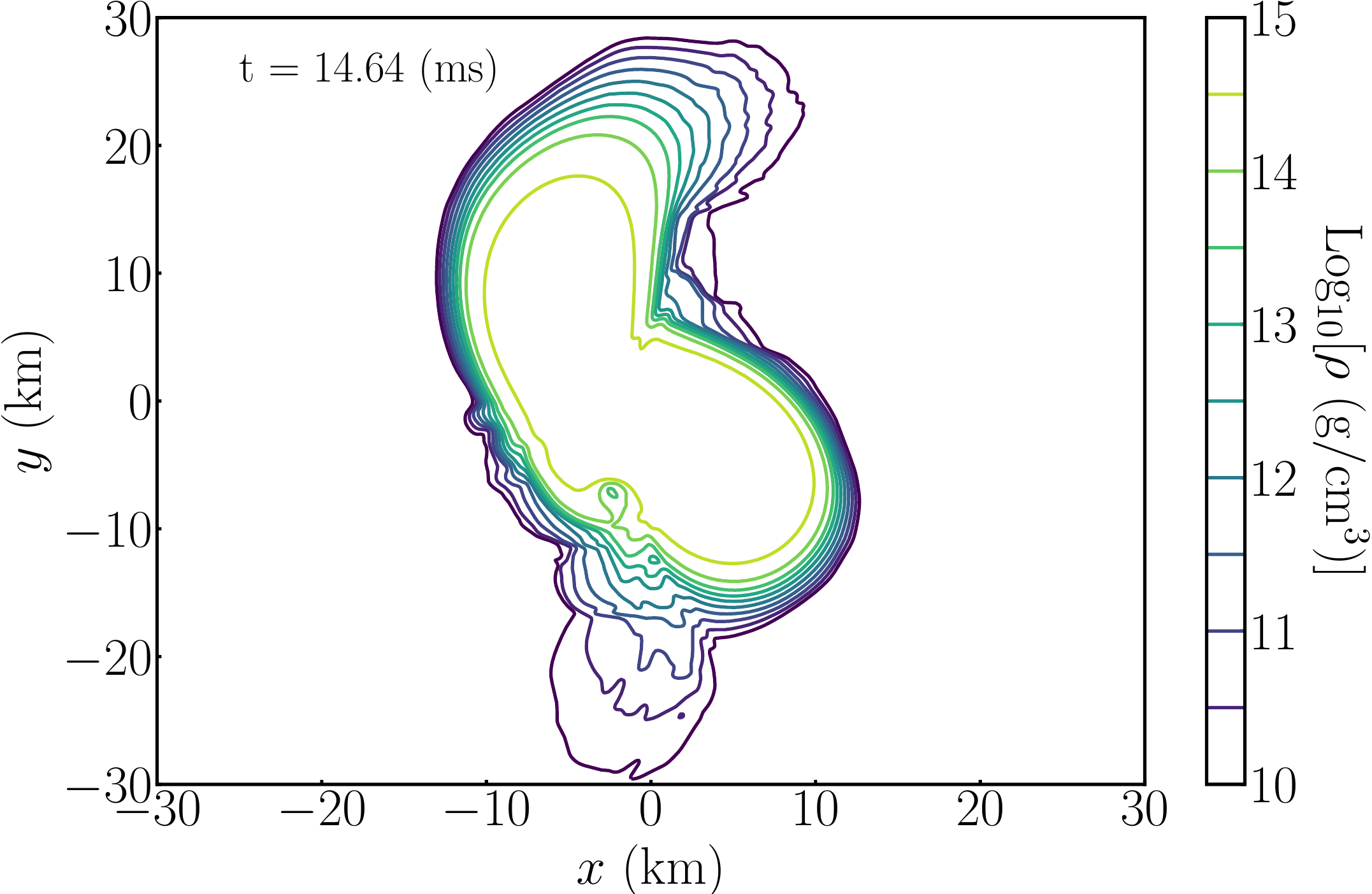} 
 	 \caption{(Top) Orbital separation as a function of time in hydrodynamic simulations of a binary neutron star merger. 
 	  The color code and line styles have the same meaning as in Fig.~\ref{fig:non_magnetized_BNS_inspiral}. The cross symbols denote the final moment at which two density maxima can be identified. 
 	 (Bottom) Colorplots of the rest-mass density in the orbital plane at $t=14.64$\,ms with the {\tt HLLC} solver (left panel) and with the {\tt HLLE} solver (right panel). Both simulations have been run with a (finest-level) grid-spacing of $\Delta x_{13}=150$\,m.
 	 }\label{fig:non_magnetized_BNS_inspiral2}
\end{figure*}

\begin{figure*}[t]
 	 \includegraphics[width=0.46\linewidth]{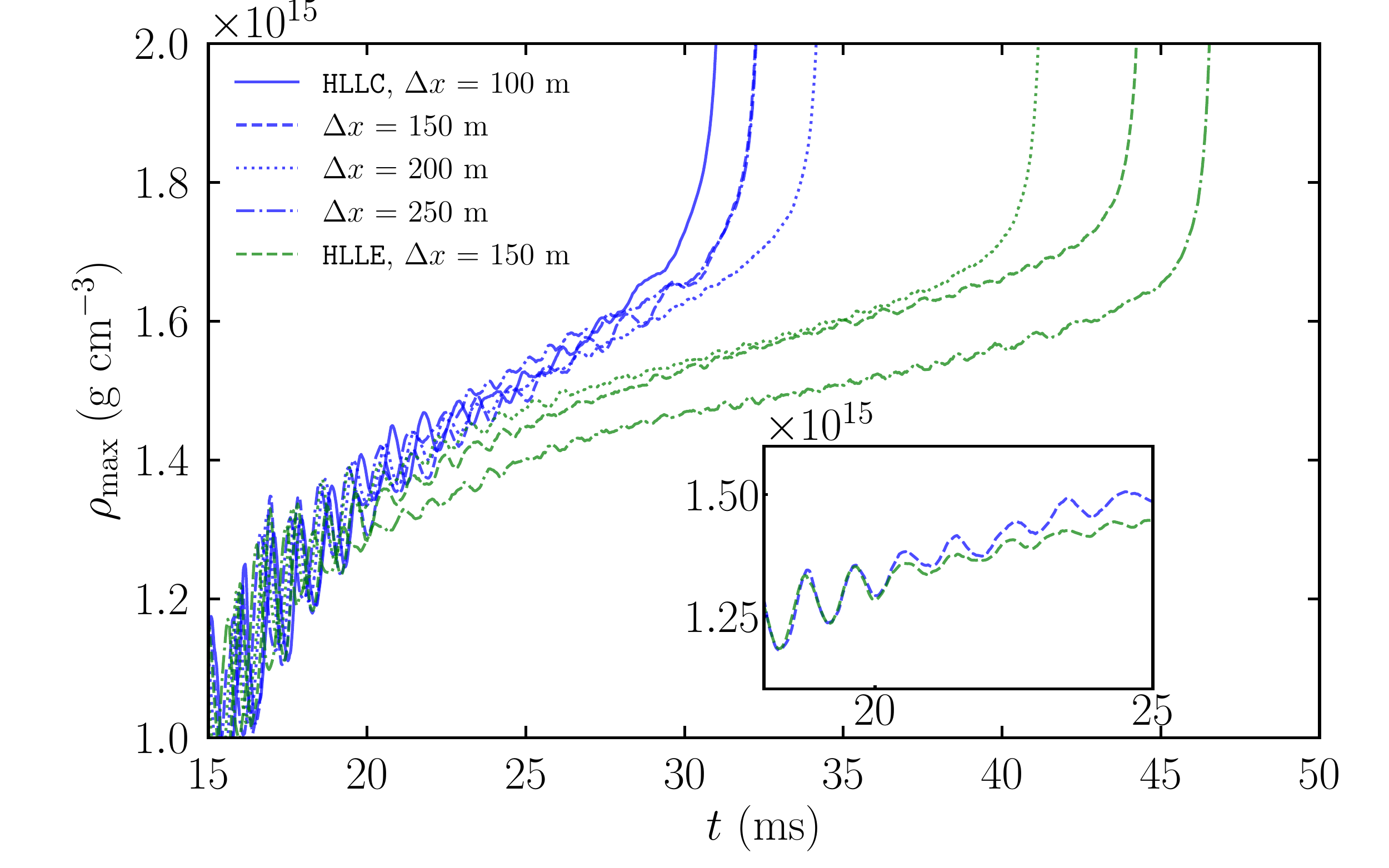}~~
  	 \includegraphics[width=0.46\linewidth]{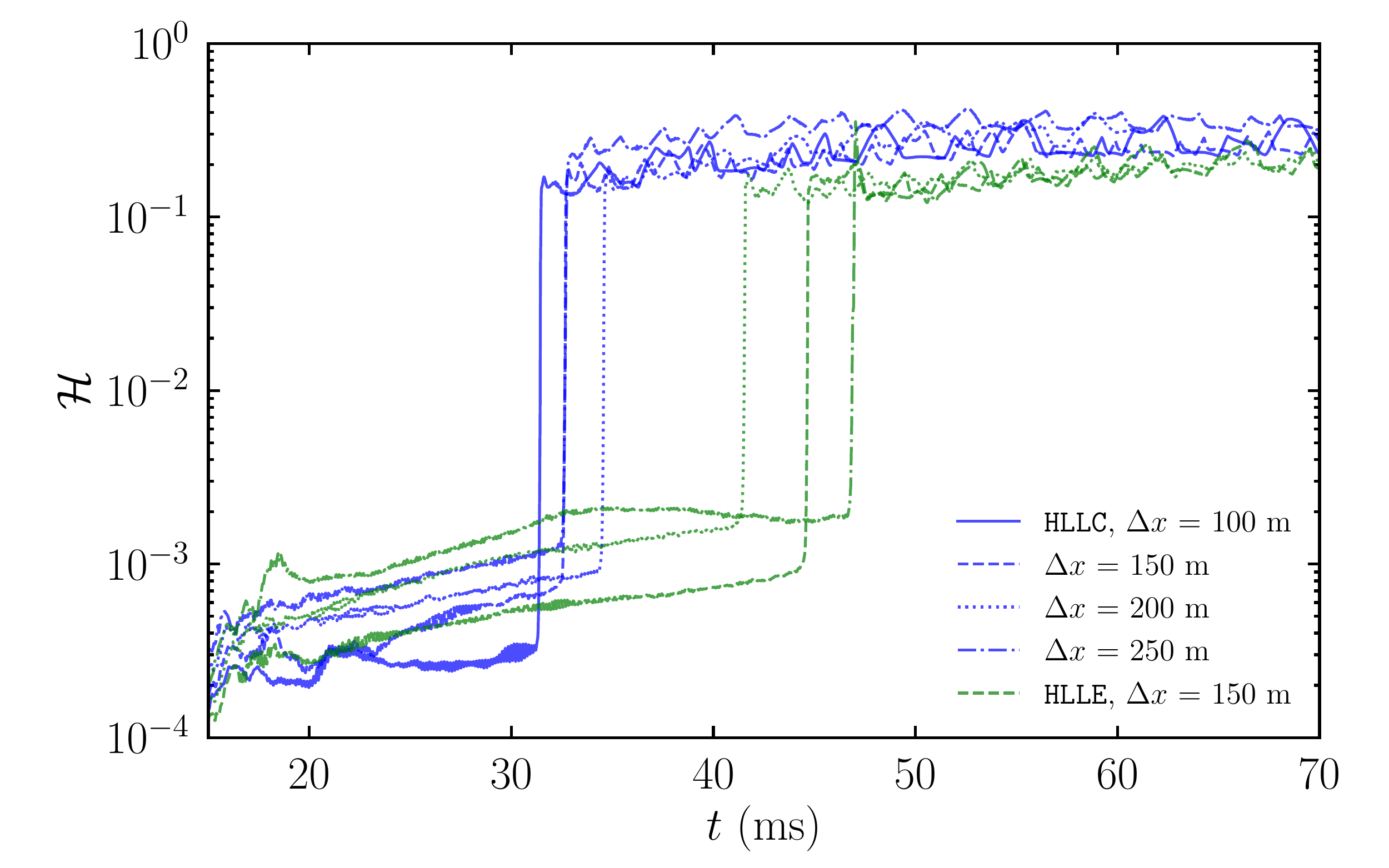}	 
 	 \caption{Same as Fig.~\ref{fig:non_magnetized_BNS_inspiral}, but for $t\ge15$\,ms. 
 	  The inset in the left panel shows a close-up over a short duration of the solution obtained with the {\tt HLLC} and {\tt HLLE} solvers with $\Delta x_{13}=150$\,m. 
 	 }\label{fig:non_magnetized_BNS_post_merger}
\end{figure*}

The left and right panels of Fig.~\ref{fig:non_magnetized_BNS_inspiral}, respectively, show the time-evolution (during the inspiral phase) of the maximum rest-mass density and the density-weighted Hamiltonian constraint violation (see Eqs.~(29) and (30) in Ref.~\cite{Shibata:2005gp} for definitions). The blue and green curves denote the results with the {\tt HLLC} and {\tt HLLE} solvers, respectively, and the solid, dashed, dotted, and dot-dashed curves indicate the resolution (i.e. $\Delta x_{13}=100$\,m, $150$\,m, $200$\,m, and $250$\,m, respectively). During inspiral, the maximum rest-mass density oscillates due to numerical error regardless of which solver is used. It also decreases partly due to the numerical error. However, the degree of the decrease is much more prominent with the {\tt HLLE} solver than it is with the {\tt HLLC} solver, especially at the coarsest resolution. This is due to the large numerical diffusion inherent in the {\tt HLLE} solver. Specifically, this solver is more subject to spurious broadening of the density profile near the stellar surface (not shown), leading to a higher degree of spurious neutron-star expansion and to a resultant decrease in the maximum rest-mass density. However, this artifact is mitigated with the {\tt HLLC} solver, because of its stronger capability of capturing irregular surfaces, i.e., the stellar surface.

The right panel of Fig.~\ref{fig:non_magnetized_BNS_inspiral} shows that, for a given grid resolution, 
the time-averaged value of the constraint violation during the inspiral phase is smaller with the {\tt HLLC} solver than with the {\tt HLLE} solver. This demonstrates that the numerical result with the {\tt HLLC} solver is more accurate than that with the {\tt HLLE} solver. We find that the order of convergence of the density-weighted Hamiltonian constraint violation is $1.7$--$1.8$, irrespective of which Riemann solver is used. Note that this convergence is slow compared to that achieved using a higher-order {\it finite difference} scheme~\cite{Bernuzzi:2016pie,Radice:2013hxh,Most:2019kfe}, but could be improved if we were to employ a more accurate reconstruction scheme such as MP5~\cite{Suresh:1997}. However, the implementation of such a scheme is beyond the scope of this paper. 

The top panel of Fig.~\ref{fig:non_magnetized_BNS_inspiral2} shows the orbital separation of the binary as a function of time.
Here we define `orbital separation' as the coordinate distance in the orbital plane between the two rest-mass density maxima. The cross symbols denote the final moment at which we can unambiguously identify the two rest-mass density maxima. At this point the less massive neutron star has been significantly tidally elongated, and we define this time as being the time of onset of the merger. This plot shows that the merger time found in the simulation with the {\tt HLLC} solver is later than that with the {\tt HLLE} solver (the reason for this will be described shortly). 
The bottom panels display contour plots of the rest-mass density in the orbital plane at the moment of merger for the runs with the {\tt HLLC} solver (left panel) and the {\tt HLLE} solver (right panel) (for both cases, $\Delta x_{13}=150$\,m). The orbital phase with the {\tt HLLE} solver slightly larger compared to that with the {\tt HLLC} solver. This implies that the neutron star simulated with the {\tt HLLE} solver is more subject to artificial tidal deformation than the neutron star with the {\tt HLLC} solver, because the {\tt HLLE} solver (since it cannot accurately resolve the irregularities at the stellar surface) results in a larger  spurious expansion of the neutron star. Note that the tidal elongation of the low-density part of the less massive neutron star is more enhanced with the {\tt HLLE} solver than with the {\tt HLLC} solver, as found from the comparison of the two contour plots. It is this enhanced (but artificial) tidal elongation with the {\tt HLLE} solver that ultimately results in the earlier merger time observed when when we employ that solver.

\begin{figure*}[t]
  	 \includegraphics[width=0.46\linewidth]{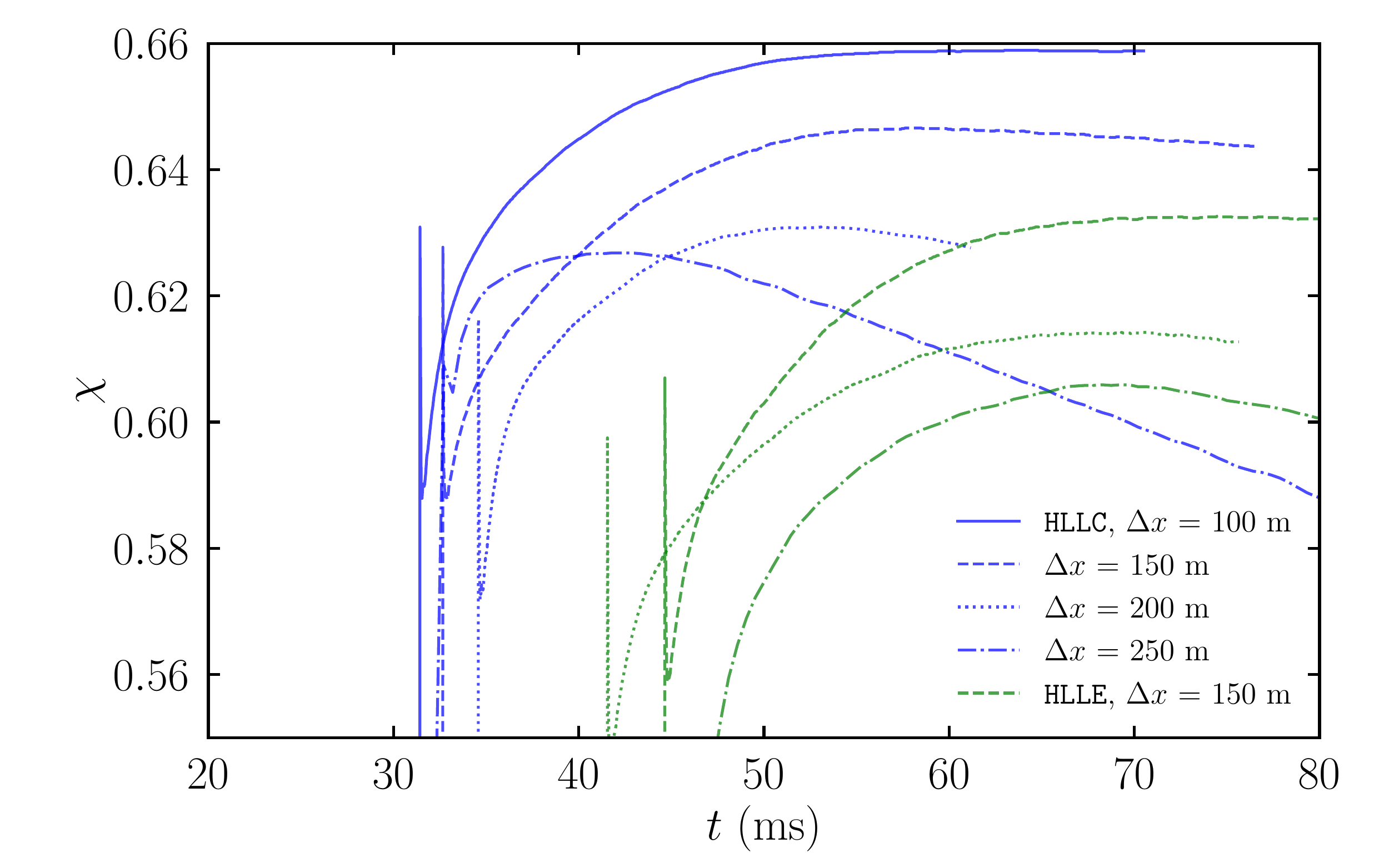} ~~
 	 \includegraphics[width=0.46\linewidth]{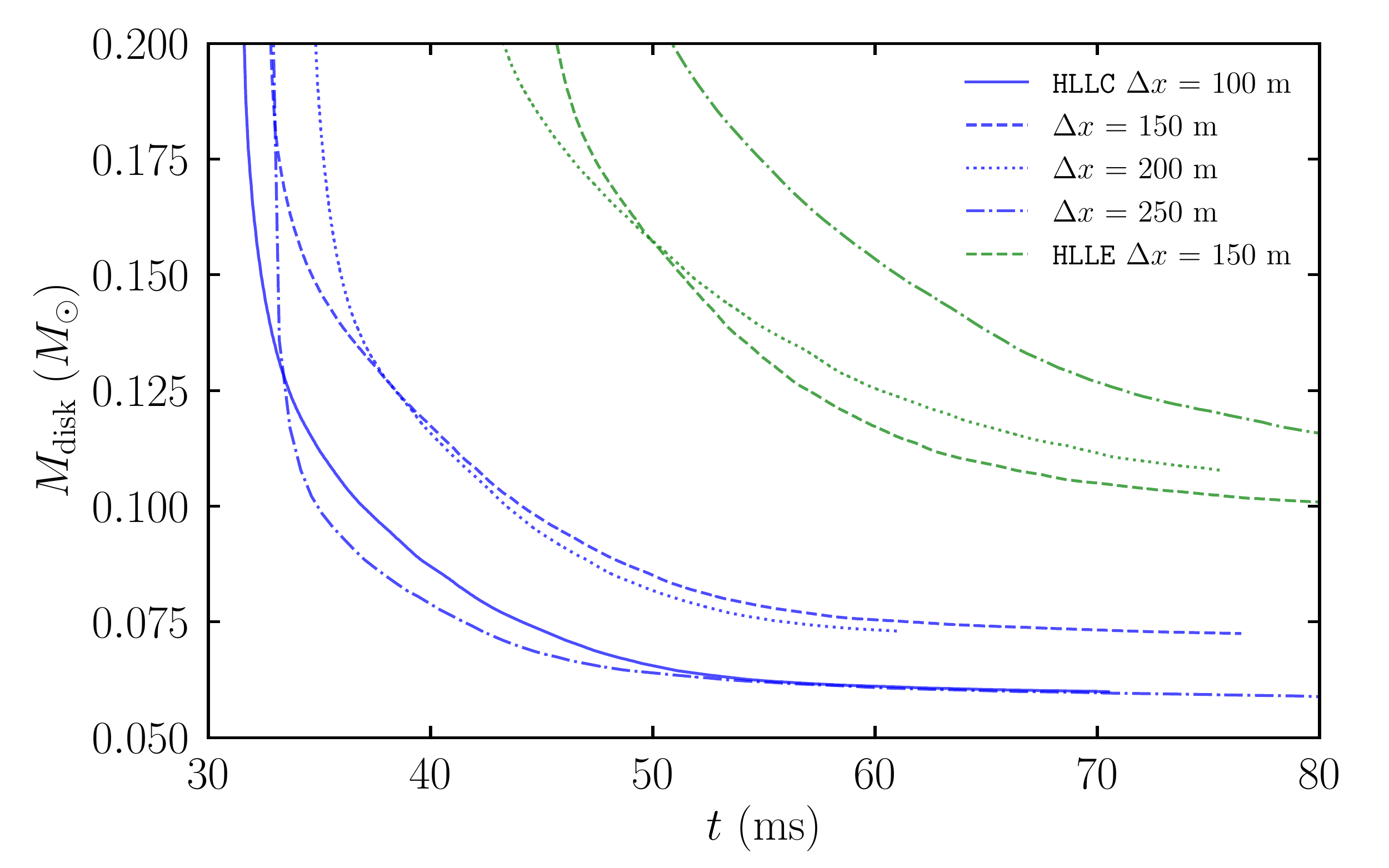}
 	 \caption{(Left) Dimensionless spin of the black hole as a function of time in non-magnetized binary neutron star merger simulations. 
 	 (Right) Gravitationally bound baryonic mass outside the apparent horizon as a function of time in non-magnetized binary neutron star merger simulations. 
 	  The color code and the line style of the legend are the same as in Fig.~\ref{fig:non_magnetized_BNS_inspiral}. 
 	 }\label{fig:non_magnetized_BNS_Mdisk_spin}
\end{figure*}

\begin{figure*}[t]
 	 \includegraphics[width=0.46\linewidth]{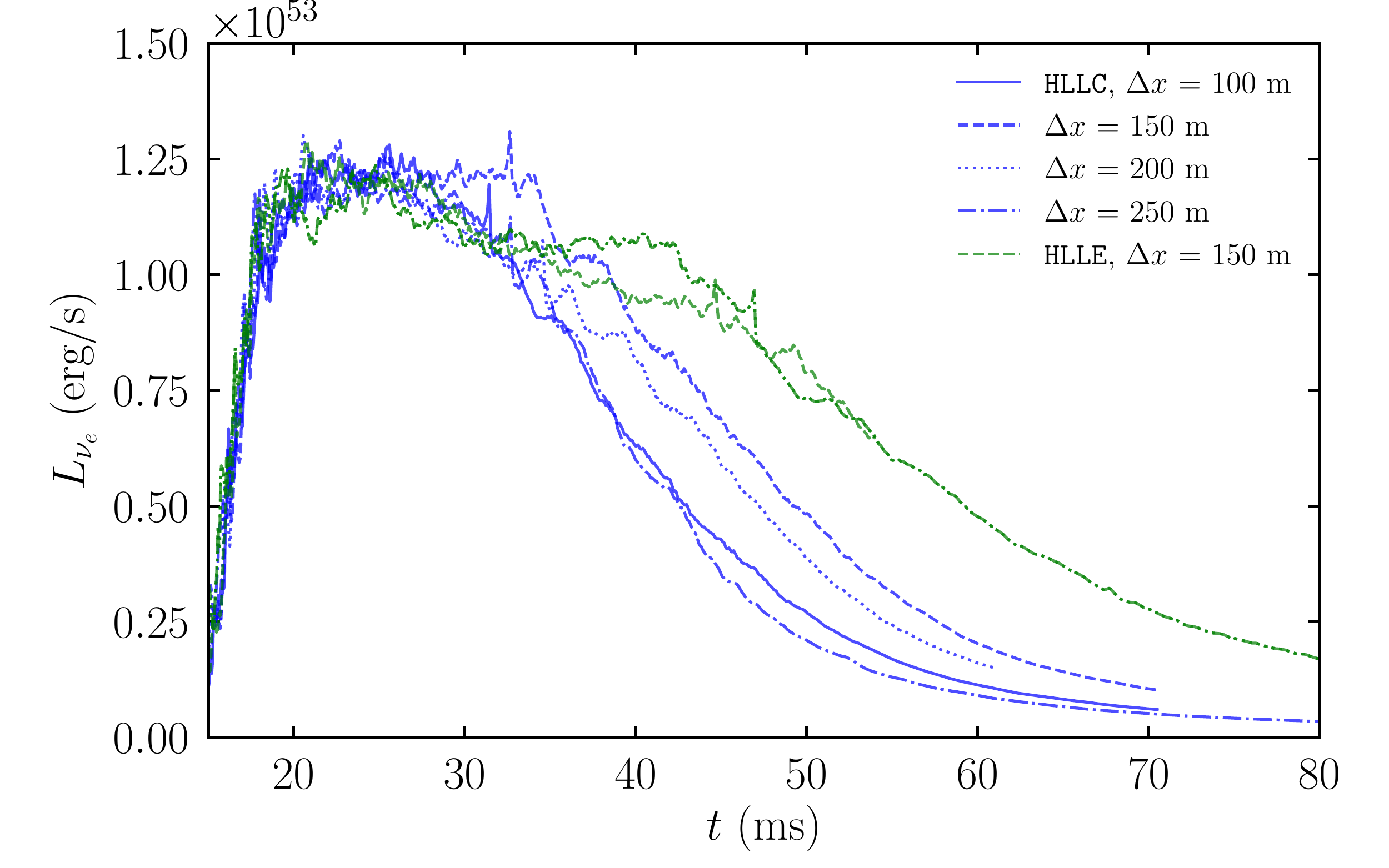}~~
  	 \includegraphics[width=0.46\linewidth]{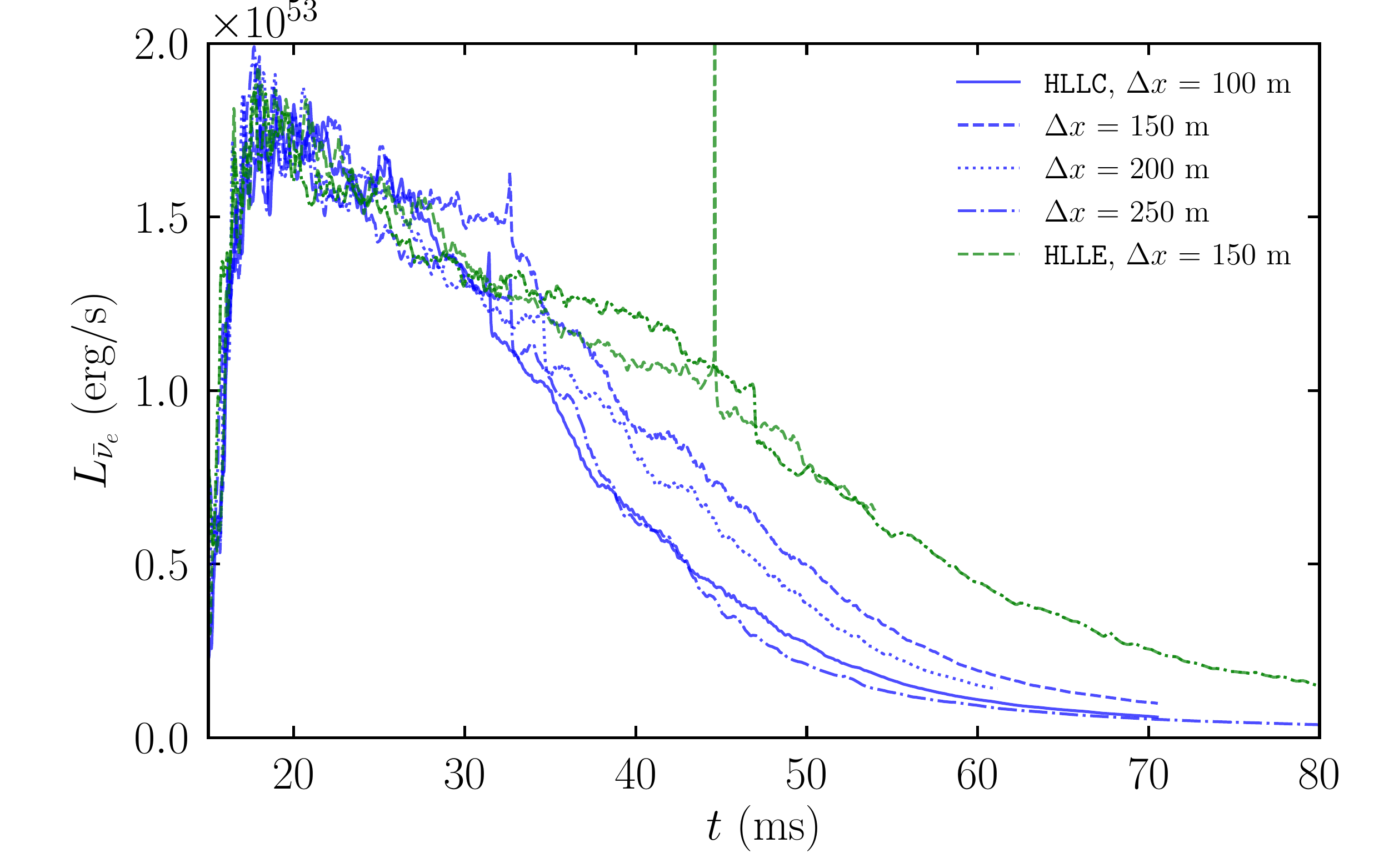}
 	 \caption{Luminosity of electron neutrinos (left) and electron antineutrinos (right) as functions of time in non-magnetized binary neutron star mergers. The color code and the line style of the legend are the same as in  Fig.~\ref{fig:non_magnetized_BNS_inspiral}.
 	 }\label{fig:non_magnetized_BNS_post_merger2}
\end{figure*}

\begin{figure*}[t]
 	 \includegraphics[width=0.6\linewidth]{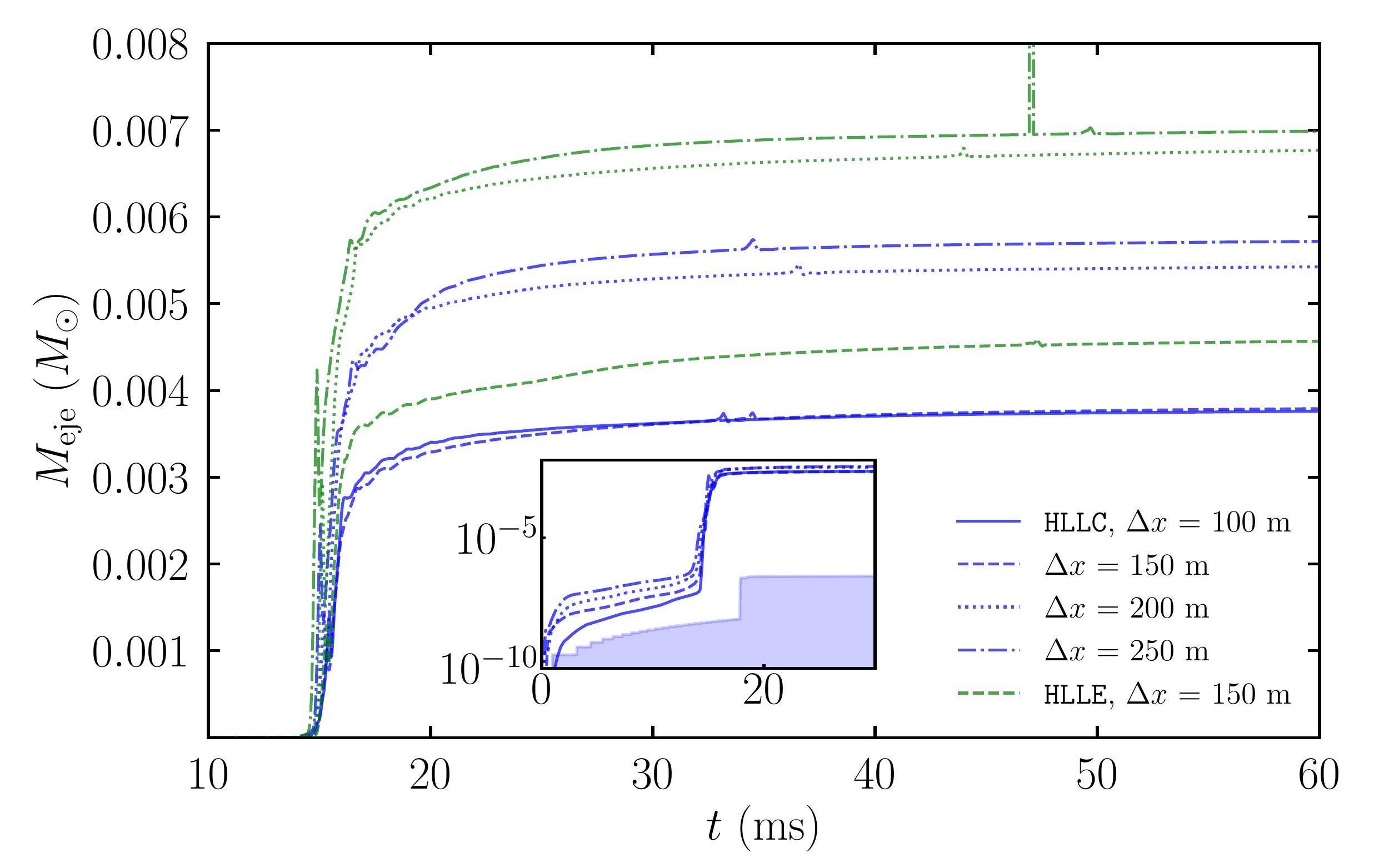}
 	 \caption{Dynamical ejecta mass as a function of time. In the inset, the vertical axis is shown using a logarithmic scale and the shaded region represents the violation of the baryonic mass conservation in a run using the {\tt HLLC} solver and at a resolution of $\Delta x_{13}=100$\,m. The color code and the line style of the legend are the same as in Fig.~\ref{fig:non_magnetized_BNS_inspiral}.
 	 }\label{fig:non_magnetized_BNS_Meje}
\end{figure*}
\subsubsection{Post-merger phase}
Having presented various diagnostics from the inspiral phase, we now turn our attention to the post-merger phase. Figure~\ref{fig:non_magnetized_BNS_post_merger} shows the maximum rest-mass density and the density-weighted Hamiltonian constraint violation as functions of time during the post-merger phase. The existence of oscillations in the density after the merger indicates the formation of a massive neutron star remnant rather than a direct collapse to a black hole. The remnant massive neutron star gradually contracts due to angular momentum transport by the gravitational torque, neutrino cooling, and gravitational-wave emission. Eventually the remnant collapses to a black hole, indicated by the blow-up of the maximum rest-mass density (see the figure at $30$--$40$\,ms). Because the remnant massive neutron star is a meta-stable object, its stability is influenced significantly by the numerical truncation error, by the randomness with which the collapse time does not become a monotonic function of the spatial grid spacing. (The non-monotonic behavior of the black-hole formation time with respect to the grid spacing is also reported in the literature (see, e.g, Ref.~\cite{2017RPPh...80i6901B}).)
Specifically, with the {\tt HLLC} solver, the collapse time of the remnant coincides for both the $\Delta x_{13}=150$\,m and $\Delta x_{13}=250$\,m runs, is earliest for the $\Delta x_{13}=100$\,m run, and is latest for the $\Delta x_{13}=200$\,m run. With the {\tt HLLE} solver, the collapse time is earliest for the $\Delta x_{13}=200$\,m run and the latest for $\Delta x_{13}=250$\,m run. The inset in the left panel of Fig.~\ref{fig:non_magnetized_BNS_post_merger} shows a close-up of the results with the {\tt HLLC} and {\tt HLLE} solvers for $\Delta x_{13}=150$\,m. 

However, the collapse time of the remnant is systematically earlier for runs with the {\tt HLLC} solver. This is related to the evolution of the oscillation amplitude of the remnant neutron star. For $t \lesssim 20$\,ms, the oscillation amplitude of the maximum rest-mass density is approximately identical for the two solvers (see the left-hand panel of Fig.~\ref{fig:non_magnetized_BNS_post_merger}). After that, however, the oscillations are noticeably damped when we use the {\tt HLLE} solver. This implies that the oscillation energy is dissipated by the numerical diffusion inherent in the {\tt HLLE} solver. Thus, the lifetime of the remnant massive neutron star is significantly overestimated when the more diffusive {\tt HLLE} solver is used. 

The right panel of Fig.~\ref{fig:non_magnetized_BNS_post_merger} shows that the density-weighted Hamiltonian constraint violation is of order $10^{-4}$ during the remnant massive neutron star phase and of order $10^{-1}$ after the black hole formation. The constraint violation only slowly decreases with increased resolution in the post-merger phase, and does so regardless of which solver is used. The reason for this is that during the merger phase, shocks are formed inside a large portion of the neutron stars. Because shocks are always computed with first-order accuracy in numerical hydrodynamics, the overall accuracy of the solution deteriorates and the convergence becomes slow. 

The left panel of Fig.~\ref{fig:non_magnetized_BNS_Mdisk_spin} shows the evolution of the dimensionless spin of the remnant black hole~\footnote{The methods we use to estimate the non-dimensional spin and the mass of the black hole are the same as in Refs.~\cite{Kiuchi:2009jt,Shibata:2007zm}.}.
We find spurious spin-down of the black hole due to numerical diffusion, in particular when the simulations are performed at low resolutions with $\Delta x_{13}=200$\,m or $250$\,m. 
We measure the spin-down rate in the {\tt HLLC} run and estimate that the dimensionless spin decreases by $\agt 0.1$ in 1\,s if $r_{\rm AH}/\Delta x_{13} \alt 15$ where $r_{\rm AH}$ denotes the minimum radius of the apparent horizon.
However, the spurious spin-down rate decreases approximately at the 4th order, reflecting the order of the accuracy in the solver for Einstein's equations. This implies that the spurious decrease of the dimensionless spin will be suppressed to the required level if we perform a simulation with a sufficiently high resolution. 
In low-resolution runs, however, the spurious spin down will influence the evolution of the disk because the specific angular momentum at the inner stable circular orbit will increase as a result of the spin-down, which in turn will result in spurious mass accretion. Thus the grid resolution must be chosen carefully when the main aim is to quantitatively explore the evolution of the disk and subsequent mass ejection.


The right panel of Fig.~\ref{fig:non_magnetized_BNS_Mdisk_spin} shows the gravitationally bound baryonic mass outside the apparent horizon~\footnote{We employ the geodesic criterion $u_t>-1$ to identify gravitationally bound fluid (conversely, $u_t<-1$ corresponds to unbound fluid elements)~\cite{Hotokezaka:2012ze}.}.
Irrespective of which Riemann solver we employ, the bounded baryonic mass is not a monotonic function of the grid spacing. Before the formation of the black hole, the non-axisymmetric density structure of the remnant massive neutron star exerts a gravitational torque on the fluid elements. As a result, angular momentum is transported outwards. Thus the longer lifetime of the remnant massive neutron star results in the formation of a more massive torus after the neutron star remnant collapses to the black hole. Because the lifetime of the remnant massive neutron star is not a monotonic function of the grid spacing it is a natural consequence that we find that the baryonic mass of bound material does not converge as the resolution is increased.
Nevertheless, in the simulations with the {\tt HLLC} solver, the gravitationally bound baryonic mass is found to lie in a narrow range, between $0.055M_\odot$ and $0.075M_\odot$, at the time of formation of the black hole (at $t \sim 30$\,ms). When we employ the {\tt HLLE} solver, the bound baryonic mass is systematically larger than that with the {\tt HLLC} solver (between $0.100M_\odot$ and $0.125M_\odot$). This is because the lifetime of the remnant massive neutron star is systematically longer in the simulations with the {\tt HLLE} solver than with the {\tt HLLC} solver, as already mentioned.
Therefore, when one employs the {\tt HLLE} solver, one should keep in mind that the bound baryonic mass could be overestimated with a systematic error of $O(10^{-2}M_\odot)$.


Figure~\ref{fig:non_magnetized_BNS_post_merger2} shows the time-evolution of the luminosity of electron neutrinos (left panel) and of electron antineutrinos (right panel). These plots show that the luminosity increases quickly after merger, reaching a peak value of $\approx 1.2\times10^{53}$\,erg/s for the electron neutrinos and $\approx 1.9\times10^{53}$\,erg/s for the electron antineutrinos at $t\approx 20$\,ms. These values agree broadly with our previous results~\cite{Sekiguchi:2016bjd}. After the formation of the black hole, the luminosity quickly decreases because the high density and temperature regions of the remnant massive neutron star are swallowed into the black hole~\cite{Sekiguchi:2011zd,Sekiguchi:2016bjd}. Note that the overall evolution of the neutrino luminosity in the remnant massive neutron star phase does not significantly depend either on the Riemann solver, nor on the spatial grid spacing. 

Finally, Fig.~\ref{fig:non_magnetized_BNS_Meje} shows the time-evolution of the gravitationally \textit{unbound} baryonic mass, i.e. the ejecta mass. In this model (i.e., the model with appreciable mass asymmetry in the binary), mass ejection is driven primarily by the tidal force from the heavier component to the lighter one. The blue and green curves denote results with the {\tt HLLC} and {\tt HLLE} solvers, respectively. The solid, dashed, dotted, and dot-dashed curves denote the results with grid spacings of $\Delta x_{13}=100$\,m, $150$\,m, $200$\,m, and $250$\,m, respectively. The inset depicts the ejecta-mass evolution on a logarithmic scale along the vertical axis, and the shaded region denotes the violation of baryonic mass conservation. We find that the spurious mass ejection during the inspiral phase is $O(10^{-7}M_\odot)$, and it decreases as the resolution is enhanced. We also find that the error in baryonic mass conservation is below $10^{-7}M_\odot$ even after the merger. 
This figure shows that the ejecta mass decreases as the grid spacing is improved from $250$\,m to $150$\,m. 
This is likely to be related to the spurious expansion of the less massive neutron star during the inspiral phase, which we discussed above. This spurious expansion is enhanced in the lower resolution runs. When we employ $\Delta x_{13}=100$\,m for the {\tt HLLC} solver, the ejecta mass is approximately identical to that with $\Delta x_{13}=150$\,m. Therefore, the convergence for the ejecta mass is approximately achieved in this model. 

Figure~\ref{fig:non_magnetized_BNS_Meje} also shows that the amount of ejecta mass in the simulation with the {\tt HLLC} solver is smaller than that with the {\tt HLLE} solver for a given grid spacing. Quantitatively, the ejecta mass difference due to the Riemann solver is $\approx 10^{-3}M_\odot$ for $\Delta x_{13}=150$\,m in this model. 
This difference arises from how accurately the employed Riemann solver can capture the neutron-star shape during the late inspiral phase. As we have already emphasized, with the {\tt HLLE} solver the neutron star spuriously expands during the inspiral phase. As a result, the less massive neutron star is more subject to (partly artificial) tidal deformation, thereby ultimately increasing the tidally-driven ejecta mass. When we employ the {\tt HLLC} solver together with a high grid resolution this artifact is mitigated. This is one of the advantages of using a more sophisticated Riemann solver in this problem.

We conclude that the {\tt HLLC} solver is superior to the {\tt HLLE} solver both during the inspiral and post-merger phases of the binary neutron star merger. In particular, we note that for the purpose of obtaining accurate and high-precision gravitational waveforms during the late inspiral phase over more than 10 orbits, the {\tt HLLE} solver is likely not an appropriate choice.

\subsection{Magnetohydrodynamics: binary neutron star merger (evolution of remnant)}

\begin{figure*}[t]
 	 \includegraphics[width=0.46\linewidth]{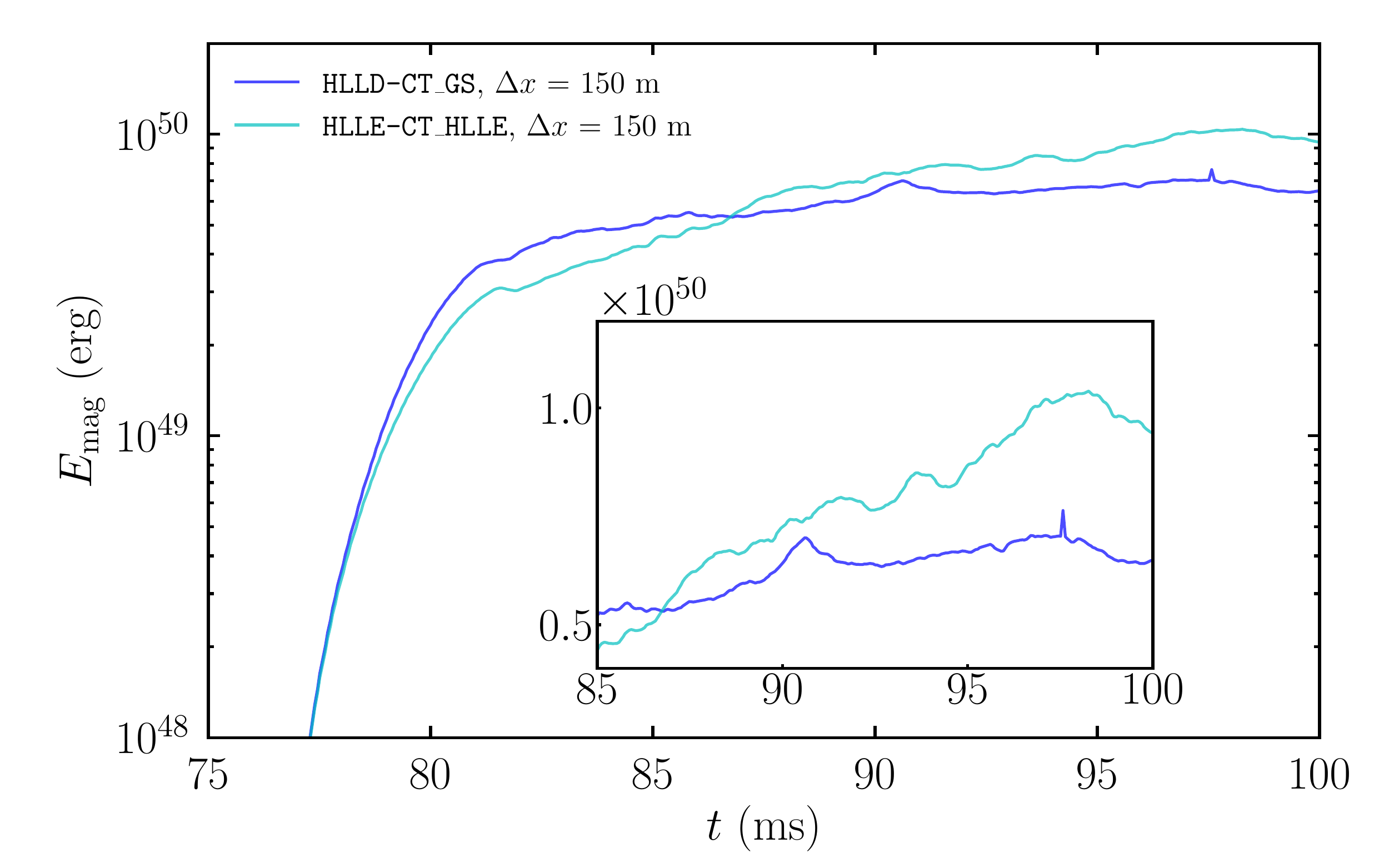}~~
     \includegraphics[width=0.46\linewidth]{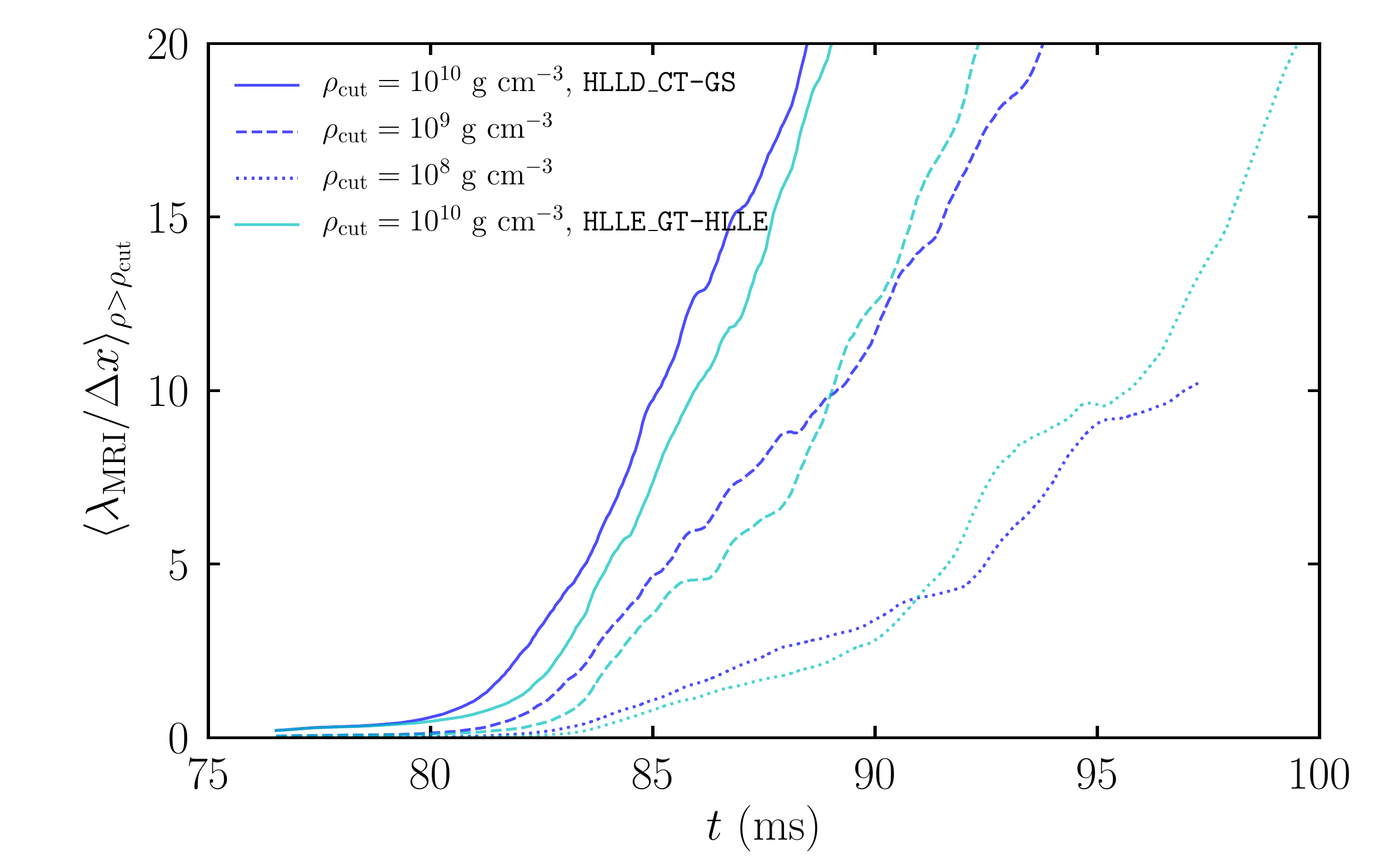}\\
 	 \caption{Time-evolution of electromagnetic energy (left) and magnetorotational-instability quality factor (right) from the magnetohydrodynamic simulations of the remnant formed after the merger of two binary neutron stars. In calculating the magnetorotational-instability quality factor, low-density regions have been excluded by employing a cut-off density $\rho_\mathrm{cut}$. The blue and cyan curves denote results for  runs with the {\tt HLLD-CT\_GS} and {\tt HLLE-CT\_HLLE} solvers, respectively. The grid spacing is $\Delta x_{13}=150$m in both runs. 
 	  The inset in the left panel shows a zoom-in of the electromagnetic energy evolution (using a linear scale along the vertical axis) over the interval $t=85$--$100$ ms.
 	 }\label{fig:magnetized_BNS_Emag}
\end{figure*}

\begin{figure*}[t]
 	 \includegraphics[width=0.46\linewidth]{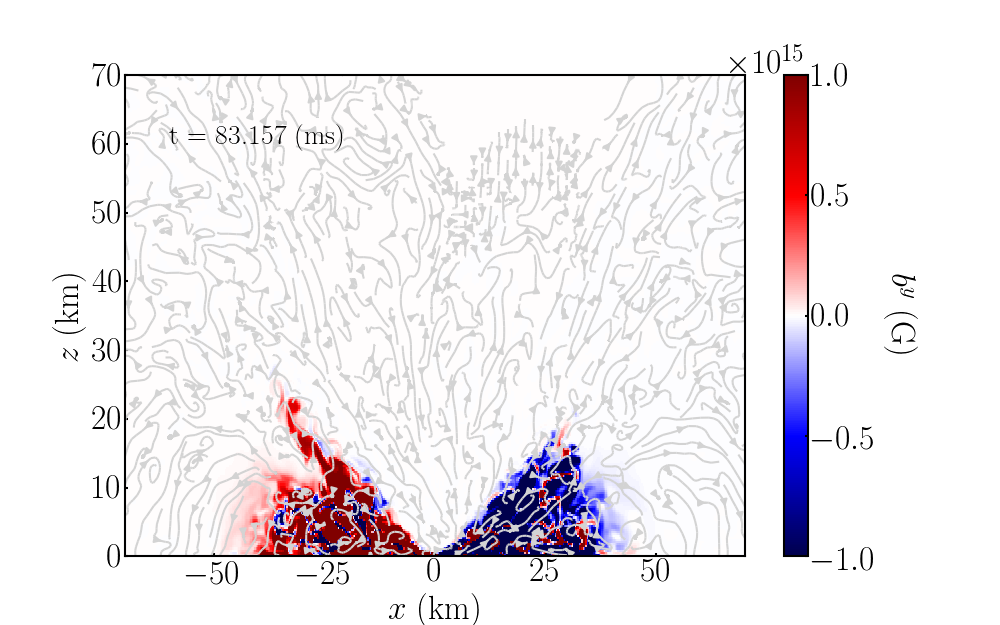}
     \includegraphics[width=0.46\linewidth]{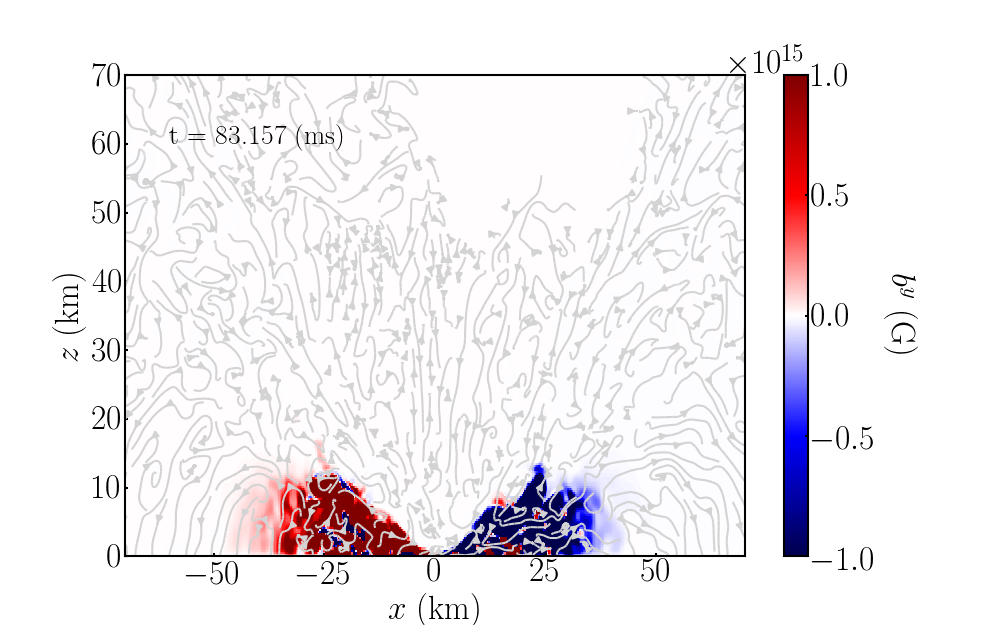}
     \includegraphics[width=0.46\linewidth]{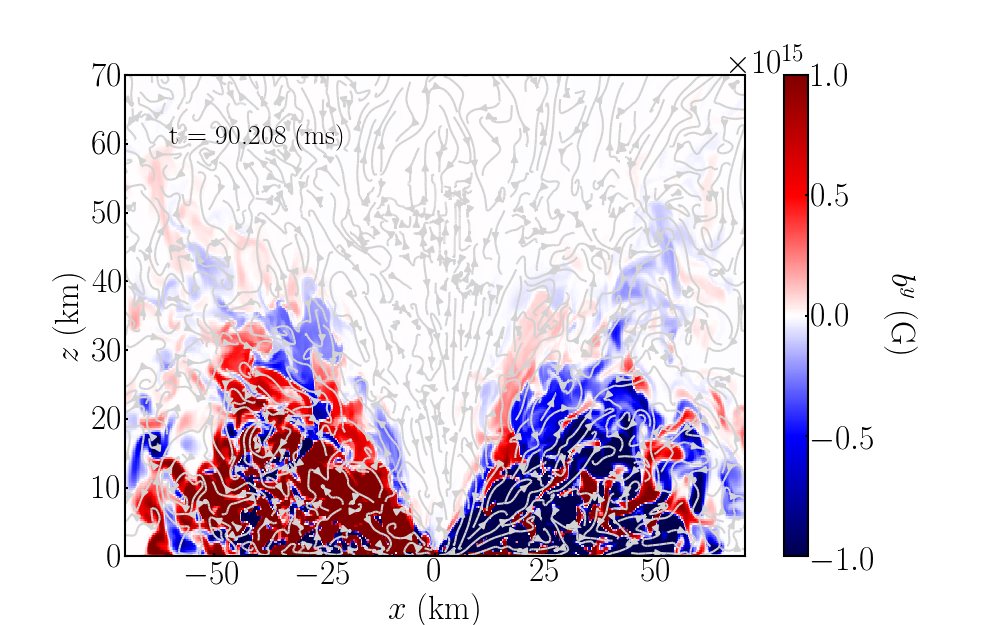}
     \includegraphics[width=0.46\linewidth]{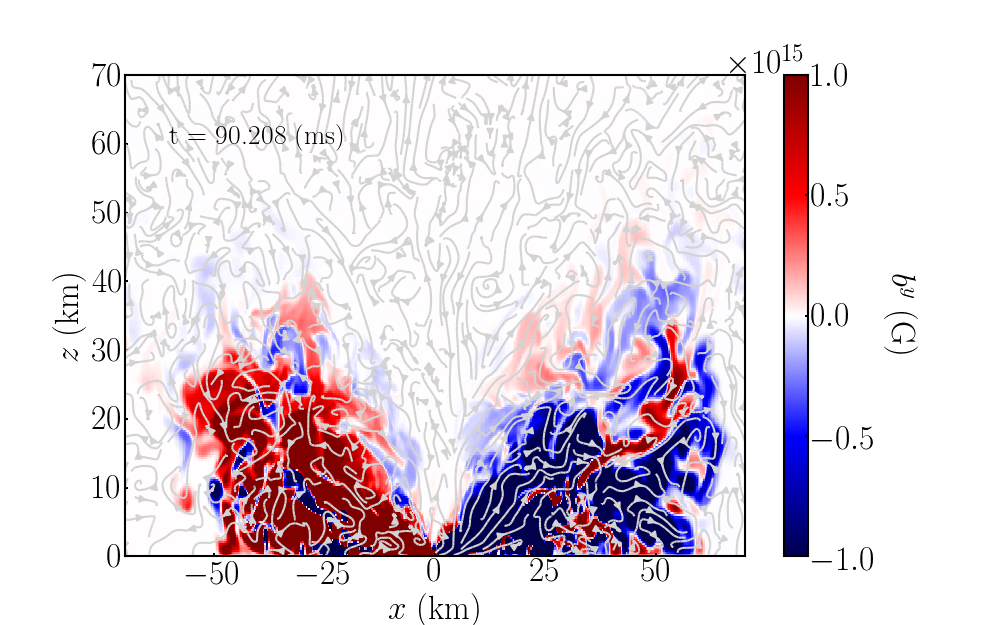}
     \includegraphics[width=0.46\linewidth]{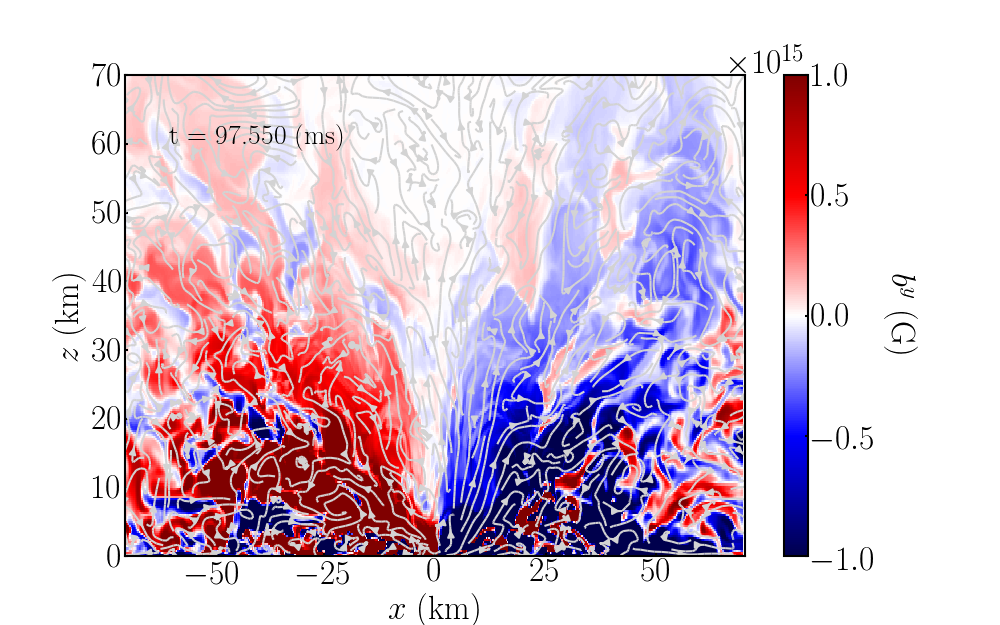}
     \includegraphics[width=0.46\linewidth]{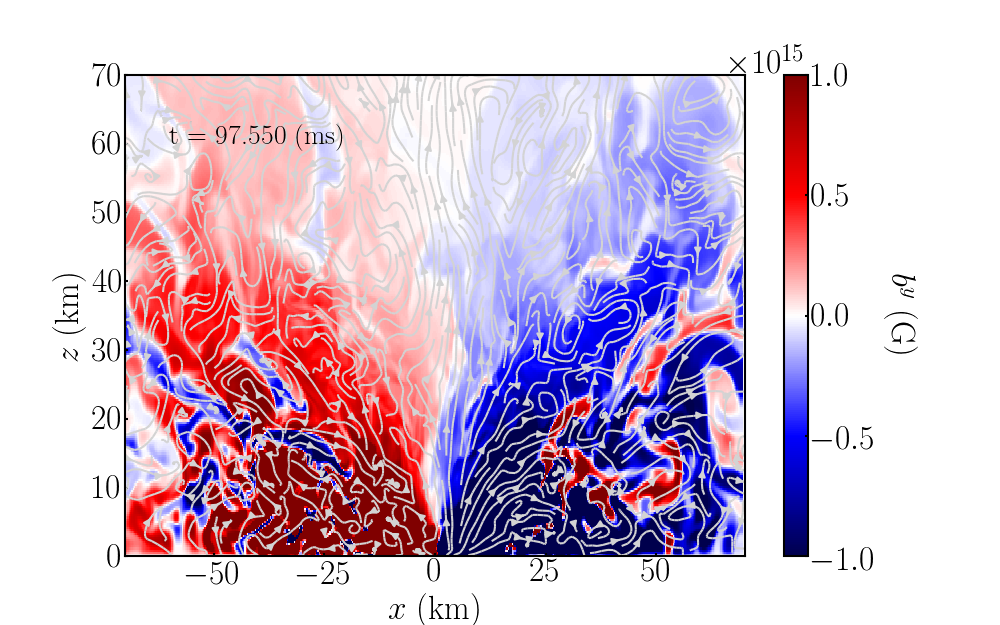}
 	 \caption{Magnetic-field structure for the magnetized binary neutron star merger remnant simulations with the {\tt HLLD-CT\_GS} solver (left) and with the {\tt HLLE-CT\_HLLE} solver (right). The gray curves show the poloidal magnetic-field lines and the color contours indicate the toroidal magnetic-field strength. 
 	 The top, middle, and bottle panels show the numerical solutions at times $t\approx 83$\,ms, $90$\,ms, and $98$ ms, respectively. 
 	 }\label{fig:magnetized_BNS_Bxz}
\end{figure*}

\begin{figure}[t]
 	 \includegraphics[width=1\linewidth]{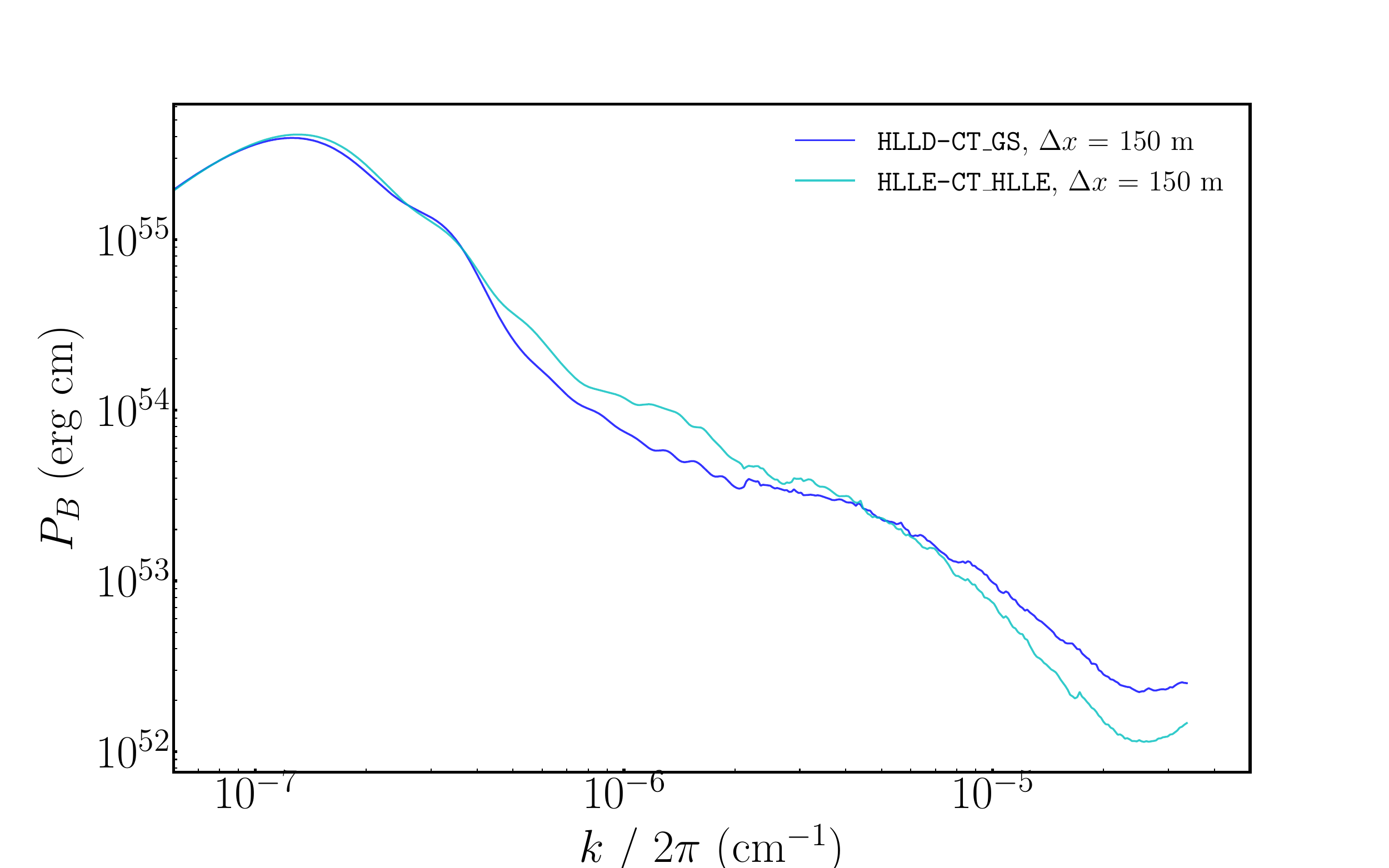}
 	 \caption{Power spectrum density of the electromagnetic energy from simulations of a magnetized binary neutron star merger remnant at $t\approx 90.2$\,ms with the {\tt HLLD-CT\_GS} solver (blue) and the {\tt HLLE-CT\_HLLE} solver (cyan). 
 	 }\label{fig:PSD}
\end{figure}

\begin{figure*}[t]
 	 \includegraphics[width=0.45\linewidth]{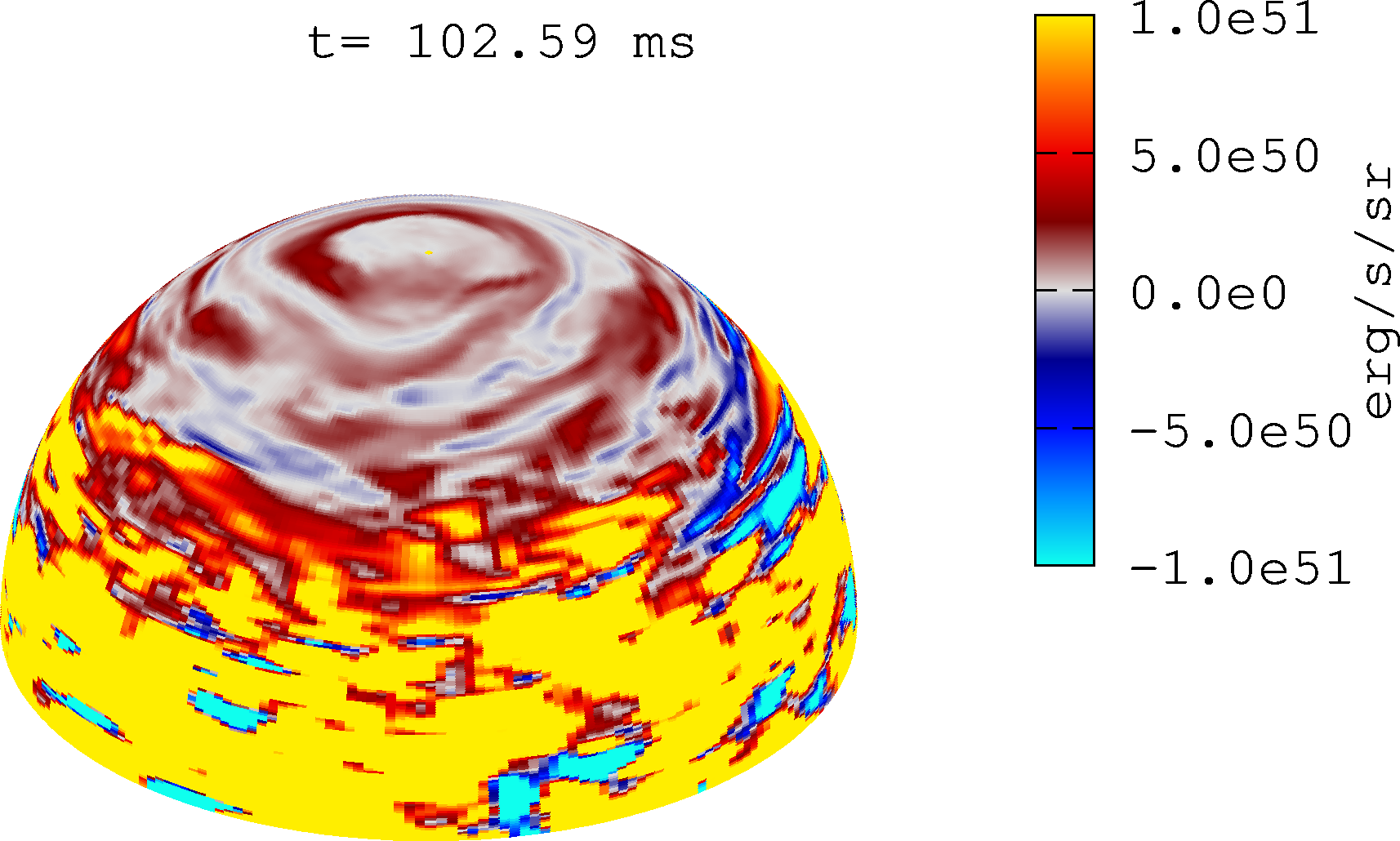}
 	 \includegraphics[width=0.45\linewidth]{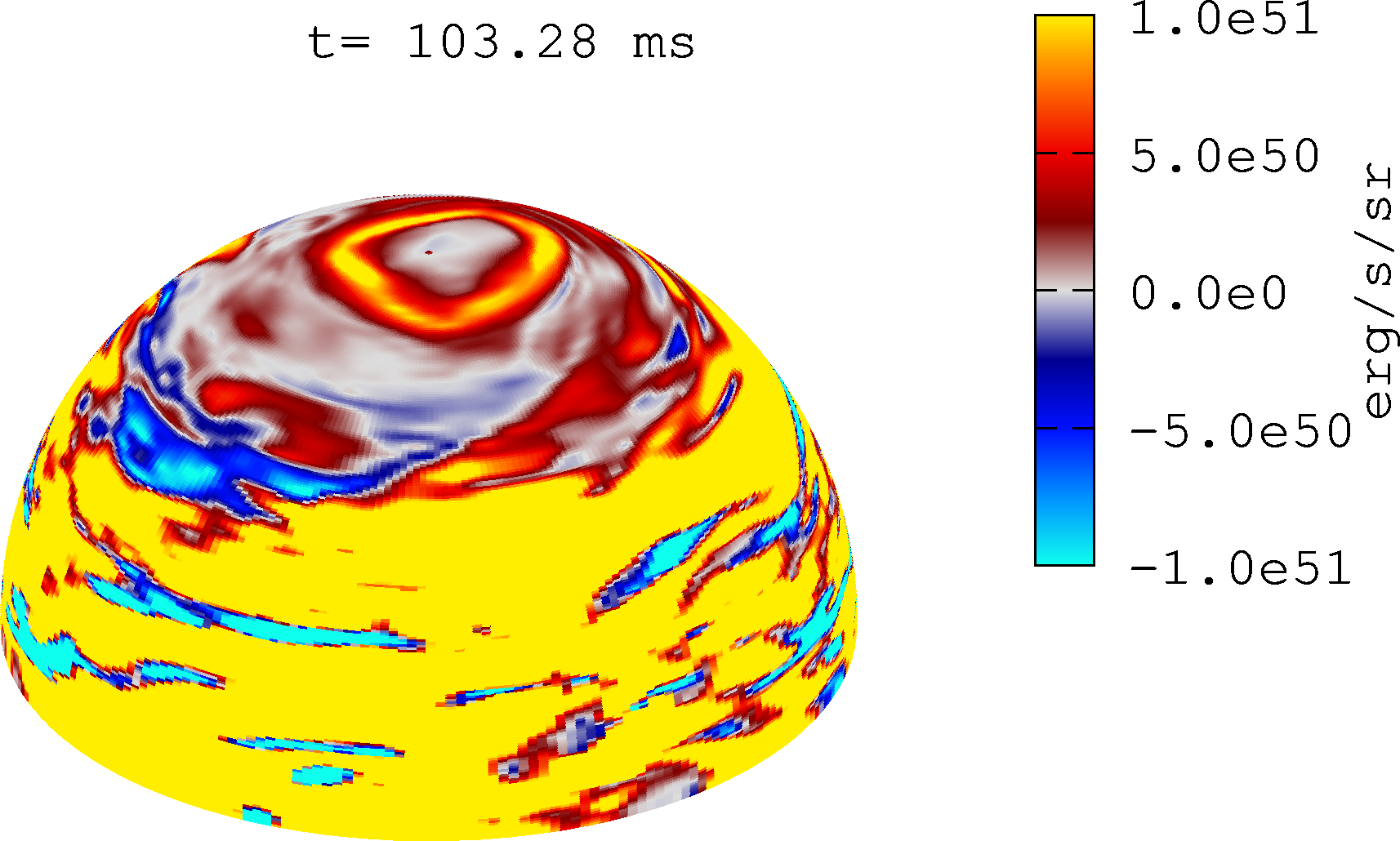}
 	 \caption{Angular distribution of the Poynting flux on a sphere of $r\approx 50$ km at $t\approx 103$ ms with the {\tt HLLD-CT\_GS} solver (left) and the {\tt HLLE-CT\_HLLE} solver (right).
 	 }\label{fig:BZ_Lum}
\end{figure*}

\subsubsection{Setup}
As an application of the new Riemann solvers (paired with our new implementation of the constrained transport scheme) to relativistic magnetohydrodynamics in a dynamical spacetime, we consider the evolution of the magnetized torus surrounding a black hole formed after a binary neutron star merger. The initial condition is taken from the final moment of the hydrodynamics simulation for a binary neutron star merger presented in the previous section. Specifically, our initial condition is taken from the result of the simulation run with the {\tt HLLC} solver at a resolution of $\Delta x_{13}=150$\,m and at $t\approx 76$\,ms. The grid setup is exactly the same as in the hydrodynamics simulation for the binary neutron star merger.

We initialize the magnetic field inside the torus with a vector potential of the form  
\begin{align}
&A_i = \left[ - ( y - y_{\rm BH} )\delta^x_{~i} + ( x - x_{\rm BH} ) \delta^y_{~i}
\right]\nonumber\\
&~~~~~~~~\times A_b\max(P-10^{-2}P_\mathrm{max},0)^2,
\end{align}
where $i=x$ or $y$, $x_{\rm BH}$ and $y_{\rm BH}$ denote the $x$- and $y$-coordinates of the central black hole, $P$ is the gas pressure, and $P_\mathrm{max}$ is its maximum. We choose the amplitude $A_b$ such that the initial maximum magnetic field strength is $10^{15}$G. We employ the {\tt HLLD-CT\_GS} and {\tt HLLE-CT\_HLLE} solvers and compare the results. We also employ Balsara's method to ensure the divergence-free condition and magnetic flux conservation in the refinement boundary~\cite{Balsara:2001,Balsara:2009,Kiuchi:2012qv}. In our implementation, not only is the divergence-free condition preserved to machine-precision, but the magnetic flux is also preserved across the refinement boundary. We note that the vector potential method~\cite{Etienne:2010ui}, which has been widely implemented in  numerical relativity codes, does not ensure the latter property~\cite{Etienne:2015cea,Cipolletta:2020kgq,Most:2019kfe,Mosta:2013gwu}. 

\subsubsection{Post-merger evolution}
The left and right panels of Fig.~\ref{fig:magnetized_BNS_Emag} show the time evolution of the electromagnetic energy and the time evolution of the magnetorotational-instability quality factor, respectively. 
The electromagnetic energy is defined by~\cite{Kiuchi:2008ss}
\begin{align}
E_\text{mag}\equiv \frac{1}{2}\int b^2w\sqrt{\gamma}d^3x.
\end{align}
The origin of the time-axis is the same as in Fig.~\ref{fig:non_magnetized_BNS_inspiral}. The blue and cyan curves denote results with the {\tt HLLD\_CT-GS} and {\tt HLLE\_CT-HLLE} solvers, respectively. The magnetorotational-instability quality factor is defined by
\begin{align}
 \langle\lambda_\mathrm{MRI}\rangle_{\rho_\mathrm{cut}} \equiv \frac{\displaystyle \int_{\rho \ge \rho_\mathrm{cut}} \lambda_\mathrm{MRI} d^3x}{\displaystyle \int_{\rho \ge \rho_\mathrm{cut}} d^3x},
\end{align}
where
\begin{align}
\lambda_\mathrm{MRI}=\frac{b^z}{\sqrt{\rho h + b^2}}\frac{2\pi}{\Omega}
\end{align}
is the wavelength of the fastest growing mode of the axisymmetric magnetorotational instability~\cite{Balbus-Hawley:1991,Hawley:1995}.
Note that we introduce a cut-off density in the quality factor 
to determine in which part of the torus the magnetorotational instability is resolved.

These panels show that the electromagnetic 
energy is amplified during the initial stage of $t\lesssim 84$--85 ms primarily due to magnetic winding rather than the magnetorotational instability, because the fastest growing mode of the magnetorotational instability in the high-density regions of the torus is not well resolved at these early times (see the solid curves in the right panel with a cut-off density of $\rho_\mathrm{cut}=10^{10}~\rm{g~cm^{-3}}$). During this stage, the electromagnetic energy with the {\tt HLLD-CT\_GS} solver is larger than that with the {\tt HLLE-CT\_HLLE} solver because the large numerical diffusion inherent in the {\tt HLLE-CT\_HLLE} solver results in the diffusion of magnetic field lines. In addition, in the orbital plane magnetic fields are forced to reconnect because we impose plane symmetry with respect to the equatorial plane. For the {\tt HLLE-CT\_HLLE} solver, reconnection in this plane is also enhanced due to numerical diffusion, and thus reduces the electromagnetic energy even further (see also the magnetized current sheet problem in Fig.~\ref{fig:2D_CS}). 


After $t\approx 84$--85 ms by which the poloidal magnetic-field strength has been enhanced nearly to saturation level due to winding and subsequent outgoing motion resulting from the enhanced magnetic-field pressure, magnetorotational instability-driven turbulence begins to develop in the high-density region of the torus, because the fastest growing mode is now resolved by more than ten grid points (see the right panel of Fig.~\ref{fig:magnetized_BNS_Emag}). This then establishes a turbulent state. 

Figure~\ref{fig:magnetized_BNS_Bxz} displays the magnetic-field structure in the $x$-$z$ plane. This figure shows that by the time the magnetic-field strength has saturated, turbulence has developed and an outflow associated with the turbulent activity is driven from the disk. The middle panels of Fig.~\ref{fig:magnetized_BNS_Bxz} show the magnetic-field structure at $t \approx 90$\,ms. With the {\tt HLLD-CT\_GS} solver, the inside of the torus exhibits smaller-scale turbulence than that with the {\tt HLLE-CT\_HLLE} solver (see, e.g., the region of $x\in [20,50]$\,km and $z\in[0,20]$\,km).
The larger structures seen in the colormap also suggest that magnetic-field lines are more {\it coherent} with the {\tt HLLE-CT\_HLLE} solver than they are with the {\tt HLLD-CT\_GS} solver. Our explanation for this is that with the {\tt HLLE-CT\_HLLE} solver, the magnetorotational instability is less resolved, and thus, the small-scale turbulent structure is less developed. As a result, large-scale magnetic fields appear to be spuriously enhanced with {\tt HLLE-CT\_HLLE} compared to {\tt HLLD-CT\_GS}. 

As evidence for this explanation, we calculate the power spectrum density of the electromagnetic energy defined by
\begin{align}
P_B(k)= \frac{1}{2} \int \tilde{b}(k_i) \tilde{b}^*(k_i) k^2 d\Omega_k,
\end{align}
where $\tilde{b}(k_i)$ is the Fourier component of the magnetic-field strength (in the frame comoving with the fluid), $b=|b^2|^{1/2}$, calculated by
\begin{align}
\tilde{b}(k_i) = \int b(x^i) {\rm e}^{ik_i x^i}d^3 x,
\end{align}
and $\tilde{b}^*(k_i)$ is its complex conjugate. Here, $k_i$ is the wave vector with $i=x,y,z$ and $k^2=\sum_i k_i^2$. $d\Omega_k$ is a solid angle in $k$-space. 
We employ the Python package {\tt fiNUFFT}~\cite{2018arXiv180806736B,2020arXiv200109405B} to perform a non-uniform Fast Fourier Transformation in our FMR domain. Practically, we employ the first five finest domains, which span from $L_{13} \in [-38.7\,{\rm km},38.7\,{\rm km}]^2\times [0\,{\rm km},38.7\,{\rm km}]$ to $L_{9}\in [-619.2\,{\rm km},619.2\,{\rm km}]^2 \times [0\,{\rm km},619.2\,{\rm km}]$, in this analysis. 

Figure~\ref{fig:PSD} plots the power spectrum density of the magnetic-field energy at $t\approx 90.2$\,ms. The blue and cyan curves denote solutions with the {\tt HLLD-CT\_GS} and {\tt HLLE-CT\_HLLE} solvers, respectively. With the help of the non-uniform Fast Fourier Transformation, we obtain a power spectrum density that spans three orders of magnitude. It clearly shows that the power spectrum amplitude around $k/(2\pi) = 10^{-6}{\rm cm}^{-1}$ is larger in the {\tt HLLE-CT\_HLLE} than in the {\tt HLLD-CT\_GS} run. This implies that a relatively large-scale magnetic field with a scale of $\approx 10^6{\rm cm}$ is generated in the {\tt HLLE-CT\_HLLE} run compared to the {\tt HLLD-CT\_GS} run. On the other hand, at small scales (i.e. with $k/(2\pi) \gtrsim 10^{-5}{\rm cm}^{-1}$), the power spectrum density is higher in the {\tt HLLD-CT\_GS} run than in the {\tt HLLE-CT\_HLLE} run. This shows the {\tt HLLD-CT\_GS} solver is able to sustain smaller-scale magnetorotational instability-driven turbulence than the {\tt HLLE-CT\_HLLE} solver. 

Figure~\ref{fig:magnetized_BNS_Emag} indicates that the electromagnetic energy is still increasing for $t \gtrsim 90$ ms. We find that (i) the growth is not exponential, and (ii) the growth rate with the {\tt HLLE-CT\_HLLE} solver is higher than with the {\tt HLLD-CT\_GS} solver. This indicates that magnetic winding of a {\it coherent} poloidal magnetic field proceeds more efficiently (though spuriously) in the simulation with the {\tt HLLE-CT\_HLLE} solver than with the {\tt HLLD-CT\_GS} solver. This in turn enhances the launch of a magnetic tower outflow in the polar direction, as shown in the bottom panels of Fig.~\ref{fig:magnetized_BNS_Bxz}. While this outflow is observed regardless of which solver is used, we observe a more powerful magnetic tower outflow with the {\tt HLLE-CT\_HLLE} solver, which reflects the greater (but spurious) coherency of the magnetic-field lines when we use of this solver. 

To quantify how powerful the magnetic tower outflow is, we plot the angular distribution of the Poynting flux  $-\sqrt{-g}({T^r}_t)^\text{(EM)}=-\sqrt{-g}(b^2u^r u_t-b^r b_t)$ on a sphere of $r\approx 50$\,km in Fig.~\ref{fig:BZ_Lum}. The snapshot is taken at $t\approx 103$\,ms. In the polar region, the Poynting flux with the {\tt HLLE-CT\_HLLE} solver is much stronger than with the {\tt HLLD-CT\_GS} solver. This plot suggests that the power of the magnetic tower outflow is overestimated when we employ the {\tt HLLE-CT\_HLLE} solver. 

\section{Summary and conclusion}\label{sec:conclusion}
We implemented the advanced Riemann solvers HLLC~\cite{Mignone:2005ft} and HLLD~\cite{MUB:2009} in our numerical relativity neutrino-radiation magnetohydrodynamics code. We validated our implementation by performing one- and multi-dimensional test problems in both Minkowski spacetime and in a fixed background spacetime, both in relativistic hydrodynamics and relativistic (ideal) magnetohydrodynamics. In the relativistic hydrodynamics test problems, we found that the {\tt HLLC} solver is always superior to the {\tt HLLE} solver, in particular, for the multi-dimensional case: the spurious waves associated with the {\tt HLLE} solver disappear, and the grid resolution is {\it effectively} improved, when we employ the {\tt HLLC} solver. For relativistic magnetohydrodynamics test problems, we also found that the performance of the {\tt HLLD} solver together with the constrained transport method proposed by Gardiner and Stone~\cite{Gardiner:2007nc}, which relies on the accuracy of a Riemann solver, is the best for both one-dimensional as well as multi-dimensional test problems. 

We also performed simulations of a non-magnetized asymmetric binary neutron star merger in a dynamical spacetime with the {\tt HLLC} and {\tt HLLE} solvers. We found that spurious broadening of the neutron star surface during the inspiral phase can be mitigated by employing the {\tt HLLC} solver. As a result, the less massive companion of the binary is less subject to tidal elongation during the late inspiral phase than when the {\tt HLLE} solver is used. This point is particularly important for deriving a high-precision gravitational waveform during the late inspiral phase, because one has to compute the orbital evolution precisely, i.e. excluding spurious numerical effects for this problem. The solution with the {\tt HLLC} solver also differs from that with the {\tt HLLE} solver in the subsequent post-merger evolution. For example, the amount of dynamical ejecta driven by the tidal interaction of the two stars and the lifetime of the remnant massive neutron star are overestimated when we employ the {\tt HLLE} solver. 

The neutron-rich dynamical ejecta and post-merger ejecta, the latter of which is launched from the merger remnant by an effective turbulent viscosity due to the magnetorotational instability~\cite{Fujibayashi:2020qda,Fujibayashi:2017xsz,Fujibayashi:2020dvr,Fujibayashi:2020jfr,Christie:2019lim,Fernandez:2018kax}, will shine by means of radioactive decay of $r$-process elements which have been freshly synthesized in the ejecta (see, e.g., \cite{Metzger:2010,Wanajo:2014wha,Metzger:2014ila}). One of the most important aims in the observation of binary neutron star mergers is to observe this signal and to infer the binary parameters by comparing the observational results with the theoretical prediction from numerical relativity simulations. 
Therefore, we conclude that employing a better solver (i.e., the {\tt HLLC} solver rather than the {\tt HLLE} solver) is crucial for reliable modeling of electromagnetic counterparts from binary neutron star mergers.

We also performed simulations of the binary neutron star merger remnant, i.e. a black hole surrounded by a massive torus, in the framework of neutrino-radiation magnetohydrodynamics. We embedded a purely poloidal magnetic-field loop inside the torus and performed simulations with the {\tt HLLD-CT\_GS} and {\tt HLLE-CT\_HLLE} solvers. We found that (i) artificial magnetic-field dissipation is suppressed, and (ii) a well-resolved magneto-turbulent state is reproduced, when we employed the {\tt HLLD-CT\_GS} solver. On the other hand, when we employed the dissipative {\tt HLLE-CT\_HLLE} solver, the coherency of the magnetic-field lines is artificially enhanced, resulting in the launch of a powerful magnetic tower outflow due to magnetic winding of this coherent poloidal field. The emergence of a Poynting flux-dominated outflow from the black hole-torus system could be a key ingredient for driving a short gamma-ray burst from the compact binary merger remnant~\cite{Hayashi:2021oxy}. Therefore, we conclude that employing the {\tt HLLD} solver paired with the constrained transport method proposed by Gardiner and Stone~\cite{Gardiner:2007nc} is crucial for reliable modeling of the central engine of short gamma-ray bursts. 

As a future project, we plan to perform long-term simulations of binary neutron star mergers and black hole-neutron star binary mergers, employing the advanced Riemann solvers which we have implemented in our code. 

\acknowledgments 
We thank Tsz Lok Lam, Sho Fujibayashi, Shinya Wanajo, and the members of the Computational Relativistic Astrophysics division at the AEI for useful discussions. Kenta Kiuchi thanks Koutarou Kyutoku for providing the initial data for the binary neutron star simulations, and for checking the manuscript. Kenta Kiuchi also thanks Kota Hayashi for providing the script used to generate Figure \ref{fig:BZ_Lum}. Numerical simulations were performed on the Sakura, Cobra, and Raven clusters at the Max Planck Computing and Data Facility and on the Cray XC50 at CfCA of the National Astronomical Observatory of Japan. This work was in part supported by the Grant-in-Aid for Scientific Research (grant Nos. 18H01213, 19K14720, and 20H00158) of Japan MEXT/JSPS. 


\appendix

\section{Tetrad basis in the $y$- and $z$-directions}
For convenience, we explicitly show the tetrad basis for the Riemann problem in the $y$- and $z$-directions.
\subsubsection{y-direction}
In the $y$-direction, the contravariant components of the tetrad basis are:
\begin{align}
    &{e_{(\hat{t})}}^\mu = n^\mu,\\
    &{e_{(\hat{x})}}^\mu = \hat{C}\left(0,1,0,0\right),\\
    &{e_{(\hat{y})}}^\mu = \hat{B}\left(0,\gamma^{yi}\right),\\
    &{e_{(\hat{z})}}^\mu = \hat{D}\left(0,-\gamma_{xz},0,\gamma_{xx}\right),
\end{align}
where
\begin{align}
&\hat{B} = \frac{1}{\sqrt{\gamma^{yy}}},\\
&\hat{C} = \frac{1}{\sqrt{\gamma_{xx}}},\\
&\hat{D} = \frac{1}{\sqrt{\gamma_{xx}\left(\gamma_{xx}\gamma_{zz}-\gamma_{xz}^2\right)}}.
\end{align}
The covariant components of the tetrad basis are given by
\begin{align}
&e_{(\hat{t})\mu} = n_\mu,\\
&e_{(\hat{x})\mu} = \hat{C}\left(\beta_x,\gamma_{xi}\right).\\
&e_{(\hat{y})\mu} = \hat{B}\left(\beta^y,{\delta_i}^y\right),\\
&e_{(\hat{z})\mu} = \hat{D}\left(\beta_z\gamma_{xx}-\beta_x\gamma_{xz},0,\gamma_{xx}\gamma_{yz}-\gamma_{xy}\gamma_{xz},\gamma_{xx}\gamma_{zz}-\gamma_{xz}^2 \right).
\end{align}
The components of the numerical flux at the $y$-interface in the Eulerian frame are 
\begin{align}
\label{eq:num_flux_y}
&({\tilde{F}^y}_0)_{j,k+\frac{1}{2},l} = \left( D v^y \right)_{j,k+\frac{1}{2},l}
\nonumber\\
&=\left(\alpha
\left({e_{(\hat{t})}}^yD+{e_{(\hat{y})}}^y\tilde{f}^{(\hat{y})}_0
\right)
\right)_{j,k+\frac{1}{2},l},\\
&({\tilde{F}^y}_1)_{j,k+\frac{1}{2},l}
=\left(\alpha{T^y}_x\right)_{j,k+\frac{1}{2},l}
\nonumber\\
&=\Big(\alpha
\Big(
{e_{(\hat{t})}}^y e_{(\hat{x})x}J_{(\hat{x})}+{e_{(\hat{y})}}^y e_{(\hat{x})x}\tilde{f}^{(\hat{y})}_1
\Big)
\Big)_{j,k+\frac{1}{2},l},\\
&({\tilde{F}^y}_2)_{j,k+\frac{1}{2},l}=\left(\alpha {T^y}_y\right)_{j,k+\frac{1}{2},l}
\nonumber\\
&=\Big(\alpha
\Big(
{e_{(\hat{t})}}^y e_{(\hat{\imath})y}J_{(\hat{\imath})}+{e_{(\hat{y})}}^y e_{(\hat{\imath})y}\tilde{f}^{(\hat{y})}_i
\Big)
\Big)_{j,k+\frac{1}{2},l},\\
&({\tilde{F}^y}_3)_{j,k+\frac{1}{2},l}
=\left(\alpha{T^y}_z\right)_{j,k+\frac{1}{2},l}
\nonumber\\
&=\Big(\alpha
\Big(
{e_{(\hat{t})}}^y e_{(\hat{\imath})z}J_{(\hat{\imath})}+{e_{(\hat{y})}}^y e_{(\hat{\imath})z}\tilde{f}^{(\hat{y})}_i
\Big)
\Big)_{j,k+\frac{1}{2},l},
\end{align}
\begin{align}
&({\tilde{F}^y}_4)_{j,k+\frac{1}{2},l}
=\left(-\alpha{T^y}_\mu n^\mu\right)_{j,k+\frac{1}{2},l}
\nonumber\\
&=\left(\alpha
\left(
{e_{(\hat{t})}}^y \rho_{\rm H}+{e_{(\hat{y})}}^y \tilde{f}^{(\hat{y})}_4
\right)
\right)_{j,k+\frac{1}{2},l},\\
&(\tilde{F}^y_5)_{j,k+\frac{1}{2},l}=
\left(\tilde{E}_z\right)_{j,k+\frac{1}{2},l}
=\left(\alpha{^*F}^{xy}\right)_{j,k+\frac{1}{2},l}\nonumber\\
&=\Big(\alpha\Big(
{e_{(\hat{t})}}^y{e_{(\hat{\imath})}}^x\bar{B}^{(\hat{\imath})}
-{e_{(\hat{t})}}^x{e_{(\hat{y})}}^y\bar{B}^{(\hat{y})}\nonumber\\
&+{e_{(\hat{x})}}^x{e_{(\hat{y})}}^y\tilde{f}^{(\hat{y})}_5
+{e_{(\hat{z})}}^x{e_{(\hat{y})}}^y\tilde{f}^{(\hat{y})}_7
\Big)\Big)_{j,k+\frac{1}{2},l},\\
&(\tilde{F}^y_6)_{j,k+\frac{1}{2},l}=0,\\
&(\tilde{F}^y_7)_{j,k+\frac{1}{2},l}=
\left(-\tilde{E}_x\right)_{j,k+\frac{1}{2},l}
=
\left(\alpha{^*F}^{zy}\right)_{j,k+\frac{1}{2},l}\nonumber\\
&=\Big(\alpha\Big(
{e_{(\hat{t})}}^y{e_{(\hat{\imath})}}^z\bar{B}^{(\hat{\imath})}-{e_{(\hat{t})}}^z{e_{(\hat{y})}}^y\bar{B}^{(\hat{y})}\nonumber\\
&+{e_{(\hat{z})}}^z{e_{(\hat{y})}}^y\tilde{f}^{(\hat{y})}_7
\Big)\Big)_{j,k+\frac{1}{2},l}.
\end{align}
The interface velocity is
\begin{align}
v^{(\hat{y})}_\text{interface}=\frac{d\hat{y}}{d\hat{t}}=\frac{\beta^y}{\alpha\sqrt{\gamma^{yy}}}.
\end{align}
\subsubsection{z-direction}
In the $z$-direction, the contravariant components of the tetrad basis are:
\begin{align}
    &{e_{(\hat{t})}}^\mu = n^\mu,\\
    &{e_{(\hat{x})}}^\mu = \hat{D}\left(0,\gamma_{yy},-\gamma_{xy},0\right),\\
    &{e_{(\hat{y})}}^\mu = \hat{C}\left(0,0,1,0\right),\\
    &{e_{(\hat{z})}}^\mu = \hat{B}\left(0,\gamma^{zi}\right),
\end{align}
where
\begin{align}
&\hat{B} = \frac{1}{\sqrt{\gamma^{zz}}},\\
&\hat{C} = \frac{1}{\sqrt{\gamma_{yy}}},\\
&\hat{D} = \frac{1}{\sqrt{\gamma_{yy}\left(\gamma_{xx}\gamma_{yy}-\gamma_{xy}^2\right)}}.
\end{align}
The covariant components of the tetrad basis are
\begin{align}
&e_{(\hat{t})\mu} = n_\mu,\\
&e_{(\hat{x})\mu} = \hat{D}\left(\beta_x\gamma_{yy}-\beta_y\gamma_{xy},\gamma_{xx}\gamma_{yy}-\gamma_{xy}^2,0,\gamma_{yy}\gamma_{xz}-\gamma_{yz}\gamma_{xy} \right)\\
&e_{(\hat{y})\mu} = \hat{C}\left(\beta_y,\gamma_{yi}\right).\\
&e_{(\hat{z})\mu} = \hat{B}\left(\beta^z,{\delta_i}^z\right).
\end{align}
The components of the numerical flux at the $z$-interface in the Eulerian frame are given by
\begin{align}
\label{eq:num_flux_z}
&({\tilde{F}^z}_0)_{j,k,l+\frac{1}{2}} = \left( D v^z \right)_{j,k,l+\frac{1}{2}}
\nonumber\\
&=\left(\alpha
\left({e_{(\hat{t})}}^zD+{e_{(\hat{z})}}^z\tilde{f}^{(\hat{z})}_0
\right)
\right)_{j,k,l+\frac{1}{2}},\\
&({\tilde{F}^z}_1)_{j,k,l+\frac{1}{2}}
=\left(\alpha{T^z}_x\right)_{j,k,l+\frac{1}{2}}
\nonumber\\
&=\Big(\alpha
\Big(
{e_{(\hat{t})}}^z e_{(\hat{\imath})x}J_{(\hat{\imath})}+{e_{(\hat{z})}}^z e_{(\hat{\imath})x}\tilde{f}^{(\hat{z})}_i
\Big)
\Big)_{j,k,l+\frac{1}{2}},\\
&({\tilde{F}^z}_2)_{j,k,l+\frac{1}{2}}
=\left(\alpha{T^z}_y\right)_{j,k,l+\frac{1}{2}}
\nonumber\\
&=\Big(\alpha
\Big(
{e_{(\hat{t})}}^z e_{(\hat{y})y}J_{(\hat{y})}+{e_{(\hat{z})}}^z e_{(\hat{y})y}\tilde{f}^{(\hat{z})}_2
\Big)
\Big)_{j,k,l+\frac{1}{2}},\\
&({\tilde{F}^z}_3)_{j,k,l+\frac{1}{2}}=\left(\alpha {T^z}_z\right)_{j,k,l+\frac{1}{2}}
\nonumber\\
&=\Big(\alpha
\Big(
{e_{(\hat{t})}}^z e_{(\hat{\imath})z}J_{(\hat{\imath})}+{e_{(\hat{z})}}^z e_{(\hat{\imath})z}\tilde{f}^{(\hat{z})}_i
\Big)
\Big)_{j,k,l+\frac{1}{2}},
\end{align}
\begin{align}
&({\tilde{F}^z}_4)_{j,k,l+\frac{1}{2}}
=\left(-\alpha{T^z}_\mu n^\mu\right)_{j,k,l+\frac{1}{2}}
\nonumber\\
&=\left(\alpha
\left(
{e_{(\hat{t})}}^z \rho_{\rm H}+{e_{(\hat{z})}}^z \tilde{f}^{(\hat{z})}_4
\right)
\right)_{j,k,l+\frac{1}{2}},\\
&(\tilde{F}^z_5)_{j,k,l+\frac{1}{2}}=
\left(-\tilde{E}_y\right)_{j,k,l+\frac{1}{2}}
=
\left(\alpha{^*F}^{xz}\right)_{j,k,l+\frac{1}{2}}\nonumber\\
&=\Big(\alpha\Big(
{e_{(\hat{t})}}^z{e_{(\hat{\imath})}}^x\bar{B}^{(\hat{\imath})}-{e_{(\hat{t})}}^x{e_{(\hat{z})}}^z\bar{B}^{(\hat{z})}\nonumber\\
&+{e_{(\hat{x})}}^x{e_{(\hat{z})}}^z\tilde{f}^{(\hat{z})}_5
\Big)\Big)_{j,k,l+\frac{1}{2}},\\
&(\tilde{F}^z_6)_{j,k,l+\frac{1}{2}}=
\left(\tilde{E}_x\right)_{j,k,l+\frac{1}{2}}
=\left(\alpha{^*F}^{yz}\right)_{j,k,l+\frac{1}{2}}\nonumber\\
&=\Big(\alpha\Big(
{e_{(\hat{t})}}^z{e_{(\hat{\imath})}}^y\bar{B}^{(\hat{\imath})}
-{e_{(\hat{t})}}^y{e_{(\hat{z})}}^z\bar{B}^{(\hat{z})}\nonumber\\
&+{e_{(\hat{x})}}^y{e_{(\hat{z})}}^z\tilde{f}^{(\hat{z})}_5
+{e_{(\hat{y})}}^y{e_{(\hat{z})}}^z\tilde{f}^{(\hat{z})}_6
\Big)\Big)_{j,k,l+\frac{1}{2}},\\
&(\tilde{F}^z_7)_{j,k,l+\frac{1}{2}}=0.
\end{align}
The interface velocity is
\begin{align}
v^{(\hat{z})}_\text{interface}=\frac{d\hat{z}}{d\hat{t}}=\frac{\beta^z}{\alpha\sqrt{\gamma^{zz}}}.
\end{align}

\section{Coordinate transformation to the maximal trumpet black hole puncture solution}\label{appendix:coordinate_transformation}
Bondi flow is usually described in Schwarzschild coordinates. However, 
our numerical relativity code has a high affinity with puncture coordinates because the solver for Einstein's equations handles a black hole with the moving puncture gauge. In numerical relativity with this gauge condition, black holes relax to a stationary solution in the so-called limit hypersurface. This implies that the code test should be done employing this special stationary hypersurface. 
To do this, one needs to seek a coordinate transformation from the Schwarzschild coordinates to the puncture coordinates (i.e., the coordinates of the limit hypersurface). One simple way of doing this is to describe a black hole as the maximal trumpet black hole puncture solution described in Ref.~\cite{Baumgarte:2007ht}. In these coordinates, the fluid quantities are well-behaved on the horizon. 

\subsection{Maximal trumpet black hole puncture}
The stationary solution of the Schwarzschild spacetime in the limiting hypersurface can be written by
\begin{align}
ds^2 = - (\alpha^2 - \beta_R \beta^R )dt^2 + 2 \beta_R dt dR + f^{-2} dR^2 + R^2 d\Omega^2,
\end{align}
where
\begin{align}
&f = \left( 1  - \frac{2M}{R} + \frac{C^2}{R^4} \right)^{1/2},\\
&\alpha = f,\\
&\beta^R = \frac{Cf}{R^2}.
\end{align}
Here, $C$ is the integration constant and $R$ is the circumferential radius. A number of numerical relativity simulations of a single black hole spacetime using the moving puncture gauge showed that the numerical solution settles down to a member of the family with $C=\frac{3\sqrt{3}M^2}{4}$, which has a limiting surface at $R=3M/2$~\cite{1973PhRvD...7.2814E}. If we consider a transformation of this solution into the isotropic coordinates by identifying the spatial metric in both coordinates as
\begin{align}
f^{-2}dR^2 + R^2 d\Omega^2 = \psi^4 (dr^2 + r^2d\Omega^2), \label{eq:metric}
\end{align}
one may find a solution for $r$ and $\psi$ as~\cite{Baumgarte:2007ht}
\begin{align}
&r = \left[\frac{2R+M+(4R^2+4MR+3M^2)^{1/2}}{4}\right]\nonumber\\
&\times\left[
\frac{(4+3\sqrt{2})(2R-3M)}{8R+6M+3(8R^2+8MR+6M^2)^{1/2}}
\right]^{1/\sqrt{2}},\\
&\psi^2=\frac{R}{r},
\end{align}
where we assumed $C=3\sqrt{3}M^2/4$. 
The lapse function, shift vector, and non-zero components of the extrinsic curvature are given by
\begin{align}
&\alpha=\sqrt{1-\frac{2M}{R}+\frac{27M^4}{16R^4}},\\
&\beta^r=\frac{3\sqrt{3}M^2r}{4R^3},\\
&K_{rr}=-\frac{6\sqrt{3}M^2\psi^4}{4R^3},\\
&K_{\theta\theta}=\frac{K_{\phi\phi}}{\sin^2\theta}=\frac{3\sqrt{3}M^2}{4R}. 
\end{align}

\subsection{Velocity field and magnetic field of Bondi flow}
The velocity field of Bondi flow in Schwarzschild coordinates should be transformed into the isotropic coordinates described in the previous section. The radial component is obtained by
\begin{align}
 u^r = \frac{u^R}{\psi^2 f},    
\end{align}
where $u^R$ is the radial velocity of Bondi flow in Schwarzschild coordinates (see, e.g.,  Ref.~\cite{White:2015omx}). The time component is obtained by the normalization of the four velocity:
\begin{align}
 &u^t = -\frac{C\psi^2u^R}{R(R-2M)}\nonumber\\
&\times \left[
-1+\left(1+\frac{R^3(R-2M)}{C^2\psi^4 \left(u^R\right)^2}\left(\psi^4\left(u^R\right)^2+1\right)\right)^{1/2}
\right].\label{eq:ut}
\end{align}
Note that the four velocity in these coordinates does not exhibit pathological behavior on the horizon, which can be confirmed by a Taylor expansion of Eq.~(\ref{eq:ut}) near the horizon~\cite{Miller:2016vkn}. Note also that the lower components of the four velocity, $u_t$ and $u_r$, are well-behaved at the horizon because the metric has a regular form in the maximal trumpet geometry~(\ref{eq:metric}).

For magnetized Bondi flow, the radial component of the magnetic field in the maximal trumpet geometry is given by
\begin{align}
B^r = \frac{B^R}{\psi^2 f},
\end{align}
where $B^R$ is the radial component of the magnetic field in Schwarzschild coordinates. In the case of a purely radial magnetic field, the divergence-free condition~(\ref{eq:divB}) requires the radial component of the magnetic field in Schwarzschild coordinates be
\begin{align}
B^R \propto 1/\sqrt{\gamma} \propto \frac{f}{R^2}.
\end{align}

\bibliography{reference}

\end{document}